\documentclass[acmlarge, table]{acmart}

\AtBeginDocument{%
  \providecommand\BibTeX{{%
    \normalfont B\kern-0.5em{\scshape i\kern-0.25em b}\kern-0.8em\TeX}}}

\acmVolume{0}
\acmNumber{0}
\acmArticle{0}
\acmMonth{0}

\usepackage{natbib}
\usepackage{color}
\usepackage{hyperref}
\usepackage{comment}
\usepackage{supertabular}
\usepackage{longtable}
\usepackage{footnote}
\makesavenoteenv{tabular}
\usepackage{threeparttable}\usepackage{footnote}
\usepackage{graphicx}
\usepackage{xcolor}
\usepackage{titlesec}
\usepackage{multirow}

\usepackage{booktabs} % For formal tables
\usepackage{listings}
\usepackage{xspace}
\usepackage{paralist}
\usepackage{tabularx,booktabs}
\usepackage{comment}
\usepackage{multirow}
\usepackage{url}
\usepackage{lipsum}
\usepackage{array}
\usepackage[]{natbib}
\usepackage{amsmath,amsfonts}
\usepackage{algorithmic}
\usepackage{graphicx}
\usepackage{textcomp}
\usepackage{siunitx}
\usepackage{tcolorbox}
\usepackage{chemformula}
\usepackage[inline]{enumitem}
\usepackage{xcolor,colortbl}
\usepackage{pifont}

\titleformat{\paragraph}[runin]
{\normalfont\normalsize\itshape}{\theparagraph}{0.5em}{}
\titlespacing{\paragraph}
{0pt}{1.25ex plus 1ex minus .2ex}{0.5ex plus .2ex}

\newcommand{\ie}{\textit{i.e.,}}

\newcommand{\ml}{{\sc ml}}
\newcommand{\dl}{{\sc dl}}

\newcommand{\RNN}{{\sc rnn}}
\newcommand{\CNN}{{\sc cnn}}
\newcommand{\LSTM}{{\sc lstm}}
\newcommand{\GRU}{{\sc gru}}
\newcommand{\SVM}{{\sc svm}}
\newcommand{\TFIDF}{{\sc tf-idf}}
\newcommand{\rtnn}{\textit{Reverse neural network}}
\newcommand{\DNN}{{\sc dnn}}
\newcommand{\ANN}{{\sc ann}}
\newcommand{\NMT}{{\sc nmt}}
\newcommand{\MLP}{{\sc mlp}} %multi-layer perceptron
\newcommand{\NB}{{\sc nb}} 
\newcommand{\RF}{{\sc rf}} 
 
\newcommand{\NLP}{{\sc nlp}}

\newcommand{\GNN}{{\sc gnn}}

\newcommand{\bn}{\textit{Bayes Network}}
\newcommand{\dt}{\textit{Decision Tree}}
\newcommand{\nb}{\textit{Naive Bayes}}
\newcommand{\svm}{\textit{Support Vector Machine}}
\newcommand{\lda}{\textit{Linear Discriminant Analysis}}
\newcommand{\ibk}{\textit{Instance-Based Learner}}
\newcommand{\lkm}{\textit{Lazy K-means}}
\newcommand{\mlp}{\textit{Multilayer Perceptron}}
\newcommand{\lr}{\textit{Linear Regression}}
\newcommand{\logr}{\textit{Logistic Regression}}
\newcommand{\poly}{\textit{Polynomial Regression}}
\newcommand{\rf}{\textit{Random Forest}}
\newcommand{\lb}{\textit{Logit Boost}}
\newcommand{\ada}{\textit{AdaBoost}}
\newcommand{\bg}{\textit{Bagging}}
\newcommand{\knn}{\textit{K Nearest Neighbors}}
\newcommand{\ENDE}{\textit{Encoder-Decoder}}

\newcommand{\abst}{{\sc ast}}
\newcommand{\CFG}{{\sc cfg}}
\newcommand{\task}[1]{ \vspace{2mm} \noindent{\bf #1:}}
\newcommand{\subtask}[1]{ \vspace{1mm} \noindent{\em #1:}}

\newboolean{showcomments}
\setboolean{showcomments}{true} % comment this line to deactivate comments
\ifthenelse{\boolean{showcomments}}{
	\newcommand{\nbc}[3]{
		{\colorbox{#3}{\bfseries\sffamily\scriptsize\textcolor{white}{#1}}}%
		{\textcolor{#3}{\sf\small$\blacktriangleright$\textit{#2}$\blacktriangleleft$}}}
	\newcommand{\todo}[1]{\nbc{TODO}{#1}{red}\xspace}
	\newcommand{\stef}[1]{\nbc{Stefanos}{#1}{blue}\xspace}
	\newcommand{\maria}[1]{\nbc{Maria}{#1}{cyan}\xspace}
	\newcommand{\tushar}[1]{\nbc{Tushar}{#1}{brown}\xspace}
	\newcommand{\fe}[1]{\nbc{Fe}{#1}{blue}\xspace}
}{
	\newcommand{\nbc}[3]{}
	\newcommand{\todo}[1]{}
	\newcommand{\stef}[1]{}
	\newcommand{\maria}[1]{}
	\newcommand{\tushar}[1]{}
	\newcommand{\fe}[1]{}
}

\hypersetup{%
	colorlinks=true,% hyperlinks will be coloured
	linkcolor=red,% hyperlink text will be green
	linkbordercolor=blue,% hyperlink border will be red
}
\makeatletter
\Hy@AtBeginDocument{%
	\def\@pdfborder{0 0 1}% Overrides border definition set with colorlinks=true
	\def\@pdfborderstyle{/S/U/W 1}% Overrides border style set with colorlinks=true
}
\makeatother

\begin{document}

\title{A Survey on Machine Learning Techniques for Source Code Analysis}

\author{Tushar Sharma}
\email{tushar@dal.ca}
\affiliation{%
  \institution{Dalhousie University}
  \city{Halifax}
  \country{Canada}
}

\author{Maria Kechagia}
\email{m.kechagia@ucl.ac.uk}
\affiliation{%
	\institution{University College London}
	\city{London}
	\country{UK}
}

\author{Stefanos Georgiou}
\email{stefanos.georgiou@queensu.ca}
\affiliation{%
	\institution{Queen's University}
	\city{Kingston}
	\country{Canada}
}

\author{Rohit Tiwari}
\email{rohit.beawar@gmail.com}
\affiliation{%
	\institution{DevOn}
	\city{Bangalore}
	\country{India}
}

\author{Indira Vats}
\email{indiravats@gmail.com}
\affiliation{%
	\institution{J.S.S. Academy of Technical Education}
	\city{Noida}
	\country{India}
}

\author{Hadi Moazen}
\email{hadimoazen@ymail.com}
\affiliation{%
	\institution{Dalhousie University}
	\city{Halifax}
	\country{Canada}
}
\author{Federica Sarro}
\email{f.sarro@ucl.ac.uk}
\affiliation{%
	\institution{University College London}
	\city{London}
	\country{UK}
}

\setcopyright{none}


\renewcommand{\shortauthors}{Sharma et al.}

\begin{abstract}

The advancements in machine learning techniques have encouraged researchers to apply these techniques to a myriad of software engineering tasks that use source code analysis, such as testing and vulnerability detection. Such a large number of studies hinders the community from understanding the current research landscape.
This paper aims to summarize the current knowledge in applied machine learning for source code analysis.
We review studies belonging to twelve categories of software engineering tasks and corresponding machine learning techniques, tools, and datasets that have been applied to solve them. To do so, we conducted an extensive literature search and identified 479 primary studies published between 2011 and 2021. We summarize our observations and findings with the help of the identified studies. 
Our findings suggest that the use of machine learning techniques for source code analysis tasks is consistently increasing. We synthesize commonly used steps and the overall workflow for each task and summarize machine learning techniques employed. We identify a comprehensive list of available datasets and tools useable in this context. Finally, the paper discusses perceived challenges in this area, including the availability of standard datasets, reproducibility and replicability, and hardware resources.

\end{abstract}




\maketitle

\section{Introduction}\label{sec:introduction}

In the last two decades, we have witnessed significant advancements in 
Machine Learning (\ml{}) and Deep Learning (\dl{}) techniques, specifically in the domain
of
image~\cite{Krizhevsky2012, Szegedy2015}, text~\cite{Lee2017, Abdeljaber2017}, 
and speech~\cite{Sainath2015, Greff2017, Graves2013} processing. 
These advancements, coupled with
a large amount of open-source code and associated artifacts, as well as the availability of
accelerated hardware, have encouraged researchers and practitioners to use \ml\ and \dl\ techniques
to address software engineering problems~\cite{Wan2019_396, Zhang2003_398, Allamanis2018_452, Le2020_297, Alsolai2020_186}.

The software engineering community has employed \ml\ and \dl\ techniques
for a variety of applications such as 
software testing \cite{Lima2020_74, Omri2020_83, Zhang2020_157}, source code representation \cite{Allamanis2018_452, Hellendoorn2017_455}, 
source code quality analysis \cite{Alsolai2020_186, Azeem2019_240},
program synthesis \cite{Le2020_297, Yahav2018_470},
code completion \cite{Liu2020_501},
refactoring \cite{Aniche2020_173},
code summarization \cite{Liu2018_430, LeClair2019_407, Allamanis2015_392},
and vulnerability analysis \cite{Shen2020_9, Shabtai2009_25, Ucci2019_54}
that involve source code analysis.

As the field of \textit{Machine Learning  for Software Engineering} ({\sc ml4se})  is expanding, the number of available resources, methods, and techniques as well as tools and datasets, is also increasing. 
This poses a challenge, to both researchers and practitioners,
to fully comprehend the landscape of the available resources and infer
the potential directions that the field is taking.
In fact, there have been numerous recent attempts to summarize the application-specific knowledge in the form of surveys.
For example, \citet{Allamanis2018_452} present key methods to model source code using \ml\ techniques.
\citet{Shen2020_9} provide a summary of research methods associated with
software vulnerability detection, software program repair, and software defect prediction.
\citet{Durelli2019_103} collect 48 primary studies focusing
on software testing using machine learning.
\citet{Alsolai2020_186} present a systematic review of 56 studies related to maintainability prediction using \ml\ techniques.
Recent surveys \cite{Tsintzira2020_194, AL-Shaaby2020_198, Azeem2019_240} summarize application of \ml\ techniques on software code smells and technical debt identification.
Similarly, literature reviews on program synthesis \cite{Le2020_297} and code summarization \cite{Nazar2016_439} have been attempted.
We compare in Table~\ref{tab:studies} the aspects investigated in our survey with respect to existing surveys that review \ml{} techniques for topics such as testing, vulnerabilities, and program comprehension with our survey. We can observe that our survey makes the following additions to the state-of-the-art surveys: it covers a wide range of software engineering activities; it summarizes a significantly large number of primary studies; it systematically examines available tools and datasets for \ml{} that would support researchers in their studies in this field; it identifies perceived challenges in the field to encourage the community to explore ways to overcome them.

% Even though related work also identifies challenges,
% those challenges are not systematically found and reported as in our case.
% Fnally, our survey is, to the best of our knowledge,
% the first that refers to \ml{}/\dl{} techniques used for source-code analysis.

\begin{table*}
    \caption{Comparison Among Surveys. The ``Category'' column refers to the software engineering task-based category of the survey where \ml{} is used, column ``Data\&Tools'' means that a survey reviews available datasets and tools for \ml{}-based applications, column ``Challenges'' shows whether the study identifies challenges related to \ml{} applications, column ``Type'' refers to the type of literature survey, and column ``\#Studies'' refers to the number of primary studies included in a considered study. We tag a study with ``--'' to indicate that the field is not applicable for the study and \textit{NA} for the number of studies column where the study does not explicitly mention the selection criteria and the number of selected studies.}
    \label{tab:studies}
    \centering
    \scalebox{0.97}{
    \begin{tabular}{llllllll}
    \toprule
      Category & Article &  Data \& Tools & Challenges & Type & \#Studies\\
      \midrule
        \multirow{2}{*}{\shortstack{Program \\Comprehension}} & \citet{738_Abbas2020} &  Yes & No & Meta-analysis & --\\
         &\citet{836_Uchoa2021} &  Yes & No & Meta-analysis & --\\
        \midrule
            \multirow{6}{*}{\shortstack{Testing}} &\citet{Omri2020_83} &  No & No & Lit. survey & NA\\
            &\citet{Durelli2019_103} &  No & Yes & Mapping study & 48\\
            &\citet{Hall2012_145} &  Yes & Yes & Meta-analysis & 21\\
            &\citet{Zhang2020_157} &  No & Yes & Lit. survey & 46\\
            &\citet{702_Pandey2021} &  No & Yes & Lit. survey & 154\\
            &\citet{710_Singh2018} &  No & No & Lit. survey & 13\\
        \midrule
            \multirow{5}{*}{\shortstack{Vulnerability\\ analysis}}&\citet{Li2019_4} & Yes & Yes & Meta-analysis & --\\
            &\citet{Shen2020_9}  & No & Yes & Meta-analysis & --\\
            &\citet{Ucci2019_54} &  No & Yes & Lit. survey & 64\\
            &\citet{537_Jie2016} &  No & No & Lit. survey & 19\\
            &\citet{545_Hanif2021} &  No & Yes & Lit. survey & 90\\
        \midrule
            \multirow{5}{*}{\shortstack{Quality \\assessment}}
            &\citet{Alsolai2020_186} &  No & No & Lit. survey & 56\\
            &\citet{Tsintzira2020_194} &  Yes & Yes & Lit. survey & 90\\
            &\citet{Azeem2019_240} &  Yes & No & Lit. survey & 15\\
            &\citet{751_Caram2019} &  No & No & Mapping study & 25\\
            &\citet{828_Lewowski2022} &  Yes & No & Lit. survey & 45\\
        \midrule
            \multirow{2}{*}{\shortstack{Program synthesis}} &\citet{Goues2019_281}  & No & Yes & Lit. survey & NA\\
            &\citet{Le2020_297} & Yes & Yes & Lit. survey & NA \\
        \midrule
            \shortstack{Program synthesis \&\\ code representation} &\citet{Allamanis2018_452}  & Yes & Yes & Lit. survey & 39+48\\
        \midrule
            \shortstack{Source-code\\ analysis}
            & Our study &  Yes & Yes & Lit. survey & 479\\
      \bottomrule
    \end{tabular}
    }
\end{table*}

In this paper, we focus on the usage of {\sc ml} and {\sc dl} techniques for source code analysis.
Source code analysis involves tasks that take the source code as input, process it, and/or produces source code as output.
Source code representation, code quality analysis, testing, code summarization,
and program synthesis are applications that involve source code analysis.
To the best of our knowledge, the software engineering literature
lacks a survey covering a wide range of source code analysis applications
using machine learning;
this work is an attempt to fill this gap.

In this survey, we aim to give a comprehensive, yet concise, overview of current knowledge on applied machine learning for source code analysis.
We also aim to collate and consolidate available resources (in the form of datasets and tools)
that researchers have used in previous studies on this topic.
Additionally, we aim to identify challenges in this domain and present them in a synthesized form.
We believe that our efforts to consolidate and summarize the techniques, resources, and challenges
will help the community to not only understand the state-of-the-art better, but also to focus their efforts on tackling the identified challenges.

This survey makes the following contributions to the field:

\begin{itemize}
	\item It presents a 
	summary of the applied machine learning studies attempted in source code analysis domain.
	\item It consolidates resources (such as datasets and tools) relevant for future studies in this domain.
	\item It provides a synthesized summary of the open challenges that requires the attention of the researchers.
\end{itemize}

In this paper, for the sake of simplicity,
we use \ml\ techniques to refer to both \ml\ and \dl\ techniques and models, unless explicitly specified.

The rest of the paper is organized as follows.
We present the followed methodology, including the literature search protocol and research questions,
 in Section~\ref{methodology}.
Section~\ref{results_rq1}, Section~\ref{results_rq2}, Section~\ref{results_rq3}, and Section~\ref{results_rq4}
provide the detailed results of our documented findings corresponding to each research question.
We present discussion in Section~\ref{discussion}, threats to validity in Section~\ref{threats}, and conclude the paper  in Section~\ref{conclusions}.

\section{Methodology} \label{methodology}
First, we present the objectives of this study and the research questions derived from such objectives.
Second, we describe the search protocol that we followed to identify relevant studies.
The protocol identifies  detailed steps to collect the initial set of articles as well as the inclusion and exclusion criteria to obtain a filtered set of studies.

\subsection{Research objectives}
This study aims to achieve the following objectives.

\begin{itemize}
    \item \textit{Identifying specific tasks involving source code analysis that has been attempted using machine learning.} \\
    We would like to investigate different types of code analysis tasks that have been attempted using
    \ml{} techniques.
    We aim to summarize how \ml{} methods and techniques are helping specific software
 	engineering tasks.
 	\item \textit{Summarizing the machine learning techniques used for source code analysis tasks.}\\
	This objective explores different \ml{} techniques commonly used for source code analysis.
	We attempt to synthesize a mapping of code analysis tasks along with sub-tasks and steps,
	and corresponding \ml{} techniques.
	\item \textit{Providing a list of available datasets and tools.}\\
	With this goal, we aim to provide a consolidated summary of available datasets and tools along with their purpose.
	\item \textit{Discussing the challenges and perceived deficiencies in \ml{}-enabled source code analysis.}\\
	With this concrete objective, we aim to present the perceived deficiencies, challenges, and opportunities in the
	software engineering field specifically in the context of applying \ml{} techniques
	observed from the collected articles.
\end{itemize}

\subsection{Literature search protocol}
\begin{figure}[!ht]
\centering
\includegraphics[width=0.6\columnwidth]{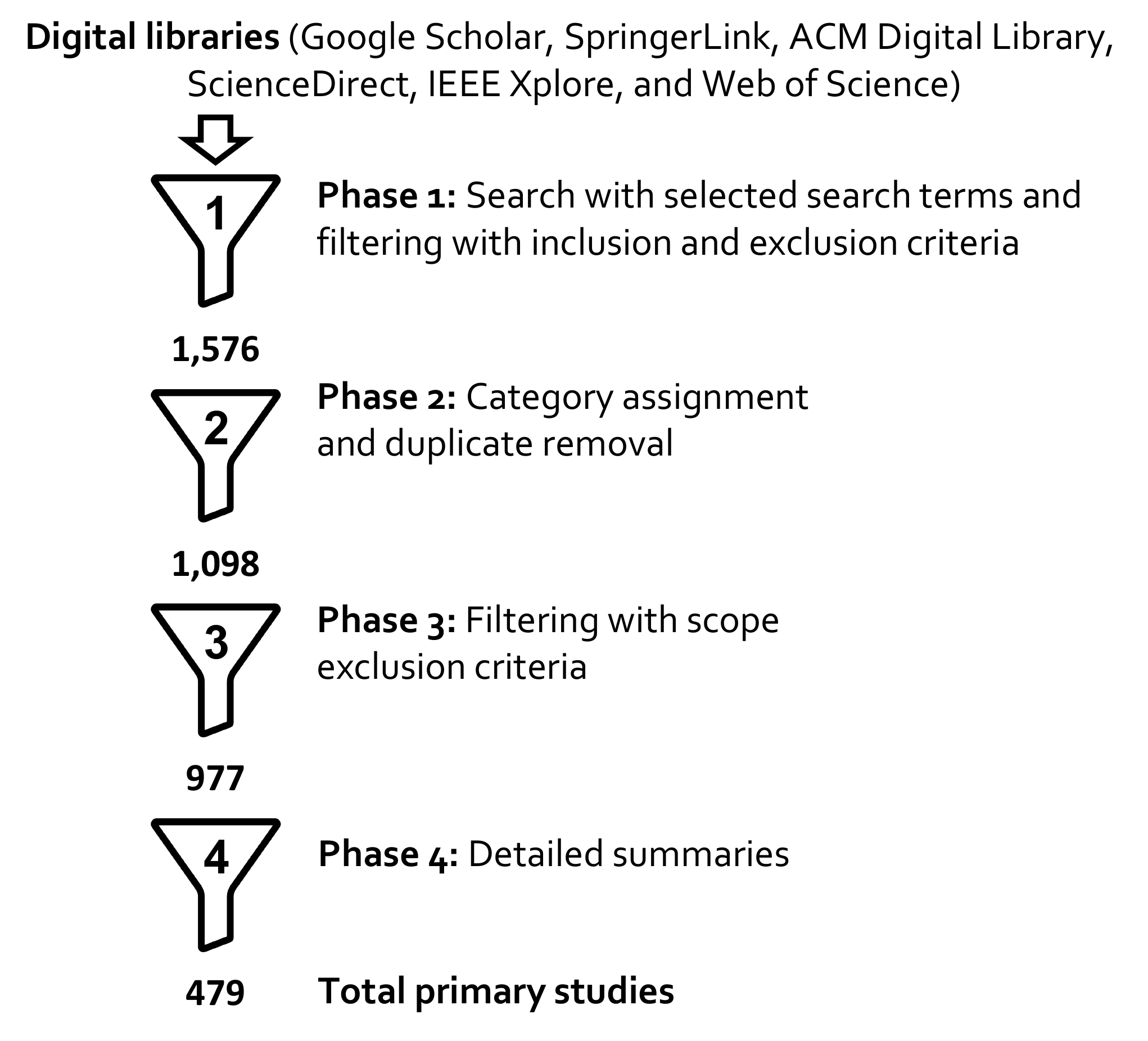}
\caption{Overview of the search process}
\label{fig:search}
\end{figure}

We identified $479$ relevant studies through a four step literature search.
Figure \ref{fig:search} summarizes the search process.
We elaborate on each of these phases in the rest of the section.

\subsubsection{Literature search --- Phase 1}
We split the phase 1 literature search into two rounds.
In the first round, we carried out an extensive initial search on six well-known digital libraries---Google Scholar,
SpringerLink, ACM Digital Library, ScienceDirect, IEEE Xplore, and Web of Science
during Feb-Mar 2021.
We formulated a set of search terms based on common tasks and software engineering activities
related to source code analysis.
Specifically, we used the following terms for the search:
\textit{machine learning code}, 
\textit{machine learning code representation,} 
\textit{machine learning testing}, 
\textit{machine learning code synthesis}, 
\textit{machine learning smell identification},
\textit{machine learning security source code analysis},
\textit{machine learning software quality assessment},
\textit{machine learning code summarization},
\textit{machine learning program repair},
\textit{machine learning code completion}, 
and \textit{machine learning refactoring}.
We searched minimum seven pages of search results for each search term manually;
beyond seven pages, we continued the search unless we get two continuous search pages without any new and relevant articles.
We adopted this mechanism to avoid missing any relevant articles in the context of our study.

In the second round of phase 1,
we identified a set of frequently occurring keywords in the articles obtained from the first round for each category individually.
To do that, we manually scanned the keywords mentioned in the articles belonging to each category, and noted the keywords that appeared at least three times.
If the selected keywords are too generic, we combined the keyword with other keywords or with other relevant words.
For example, \textit{machine learning} and \textit{program generation} occurred multiple times in the \textit{program synthesis} category; we combined both of these terms to make one search string \ie{} \textit{program generation using machine learning}.
We carried out this additional round of literature search to augment our initial search terms and reduce the risk of missing relevant articles in our search.
The search terms used in the second round of phase 1 can be found in our replication package~\cite{Replication_ML4SCA}.
Next, we defined inclusion and exclusion criteria to filter out irrelevant studies.

\task{Inclusion criteria}
\begin{itemize}
	\item Studies that discuss source code analysis using a \ml\ technique (including \dl{}).
	\item Surveys discussing source code analysis using \ml\ techniques.
	\item Resources revealing the deficiencies or challenges in the current set of methods, tools, and practices.
\end{itemize}

\task{Exclusion criteria}
\begin{itemize}
	\item Studies focusing on techniques other than \ml\  applied on source code analysis \textit{e.g.,} code smell detection using metrics.
	\item Articles that are not peer-reviewed (such as articles available only on arXiv.org).
	\item Articles constituting a keynote, extended abstract, editorial, tutorial, poster, or panel discussion (due to insufficient details and small size).
	\item Studies whose full text is not available, or be written in any other language than English.
\end{itemize}

During the search, we documented studies that satisfy our search protocol in a spreadsheet
including the required meta-data (such as title, bibtex record, and link of the source).
The spreadsheet with all the articles from each phase can be found in our replication package online~\cite{Replication_ML4SCA}.
Each selected article went through a manual inspection of title, keywords, and abstract.
The inspection applied the inclusion and exclusion criteria leading to inclusion or exclusion of the articles.
In the end, we obtained $1,576$ articles after completing {\it Phase 1} of the search process.

\subsubsection{Literature search --- Phase 2}
In {\it Phase 2}, we first identified a set of categories and sub-categories for common software engineering tasks.
These tasks are commonly referred in recent publications~\cite{Ferreira2021, Allamanis2018_452, Shen2020_9, Azeem2019_240}.
These categories and sub-categories of common software engineering tasks can be found in
Figure~\ref{fig:subcategory}.
Then, we manually assigned a category and sub-category, if applicable, to each selected article
based on the (sub-)category to which the article contributes the most.
The assignment is carried out by one of the authors and verified by two other authors;
disagreements were discussed and resolved to reach a consensus. 
In this phase,
we also discarded duplicates or irrelevant studies not meeting our inclusion criteria after reading their title and abstract.
After this phase, we were left with $1,098$ studies.

\subsubsection{Literature search --- Phase 3}
To keep our focus on the most recent studies,
we marked the studies published before 2011 as out of scope.
Also, we discarded papers that had not received enough attention from the community by filtering out all those having a `citation count $<$ (2021 – publication year)'.
We chose $2021$ as the base year to not penalize studies that came out recently; hence, the studies that are published in 2021 do not need to have any citation to be included in this search. 
We obtain the citation count from digital libraries manually during Mar-May 2022.
After applying this filter, we obtained $977$ studies.

\subsubsection{Literature search --- Phase 4}
In this phase, we discarded studies that do not satisfy our 
inclusion criteria 
(such as when the article is too small or do not employ any \ml\ technique for source code analysis and processing tasks) 
after reading the whole article.
The remaining $480$ articles are the primary studies that we examine in detail.
For each study, we extracted the core idea and contribution, the \ml\ techniques and tools used as well as challenges and findings unveiled. 
Next, we present our observations corresponding to
each research goal we pose.

\section{Categorizing ML-enabled Source code analysis tasks}
\label{results_rq1}

\begin{figure*}[!h]
	\centering
	\includegraphics[width=\textwidth]{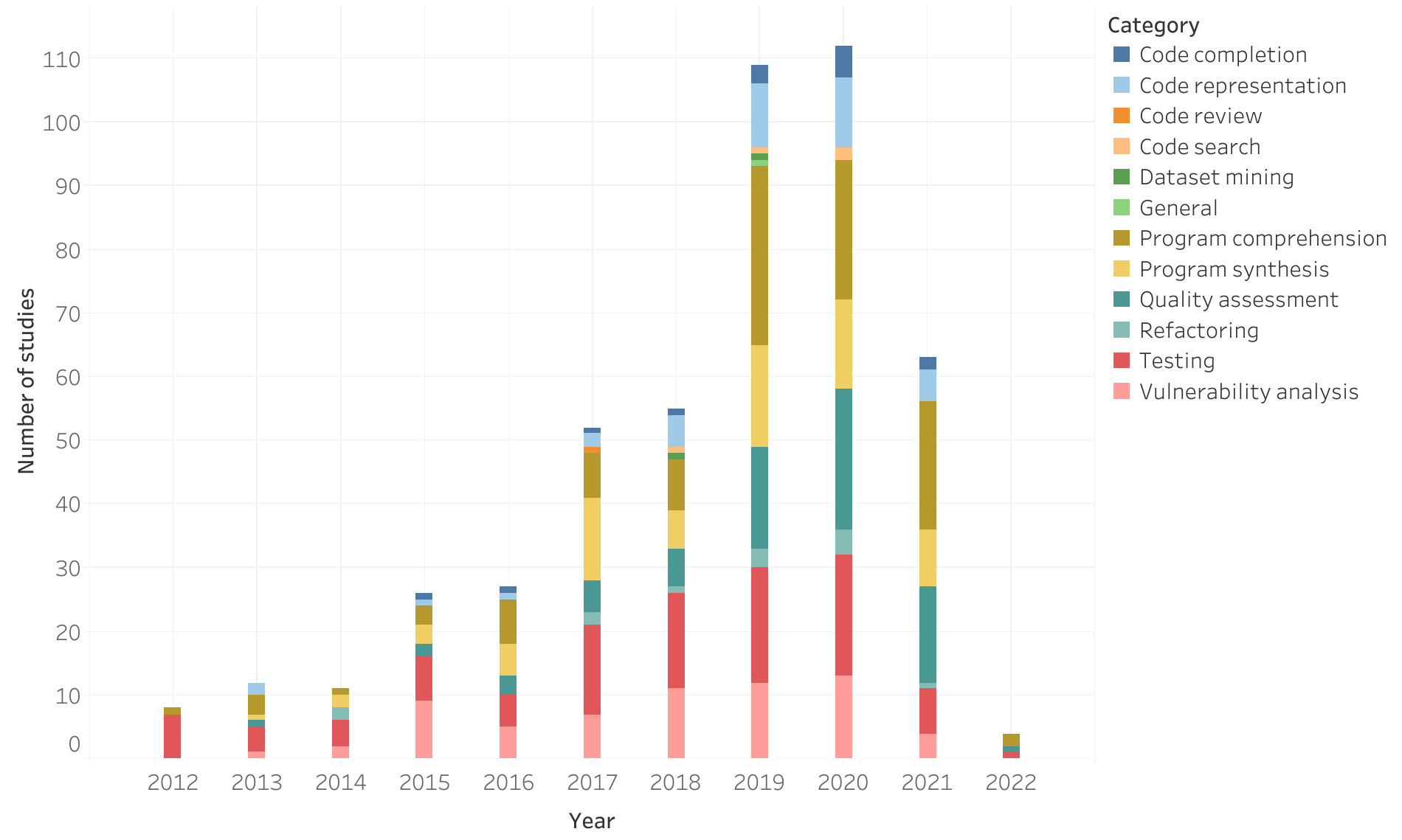}
	\caption{Category-wise distribution of studies}
	\label{fig:category}
\end{figure*}

We tagged each selected article with one of the task categories based on the primary focus of the study.
The categories represent common software engineering tasks that involve source code analysis.
These categories are \textit{code completion}, \textit{code representation},
\textit{code review}, \textit{code search},  \textit{dataset mining},
\textit{program comprehension}, \textit{program synthesis}, \textit{quality assessment},
\textit{refactoring}, \textit{testing},
and \textit{vulnerability analysis}.
If a given article does not fall in any of these categories but it is still relevant to our discussion 
as it offers overarching discussion on the topic; 
we put the study in the \textit{general} category.
Figure~\ref{fig:category} presents a category-wise distribution of studies per year.
It is evident that the topic is attracting the research community more and more and we observe, in general, a healthy upward trend.
Interestingly, the number of studies in the scope dropped significantly in the year $2021$.

\begin{figure*}[!h]
	\centering
	\includegraphics[width=\textwidth]{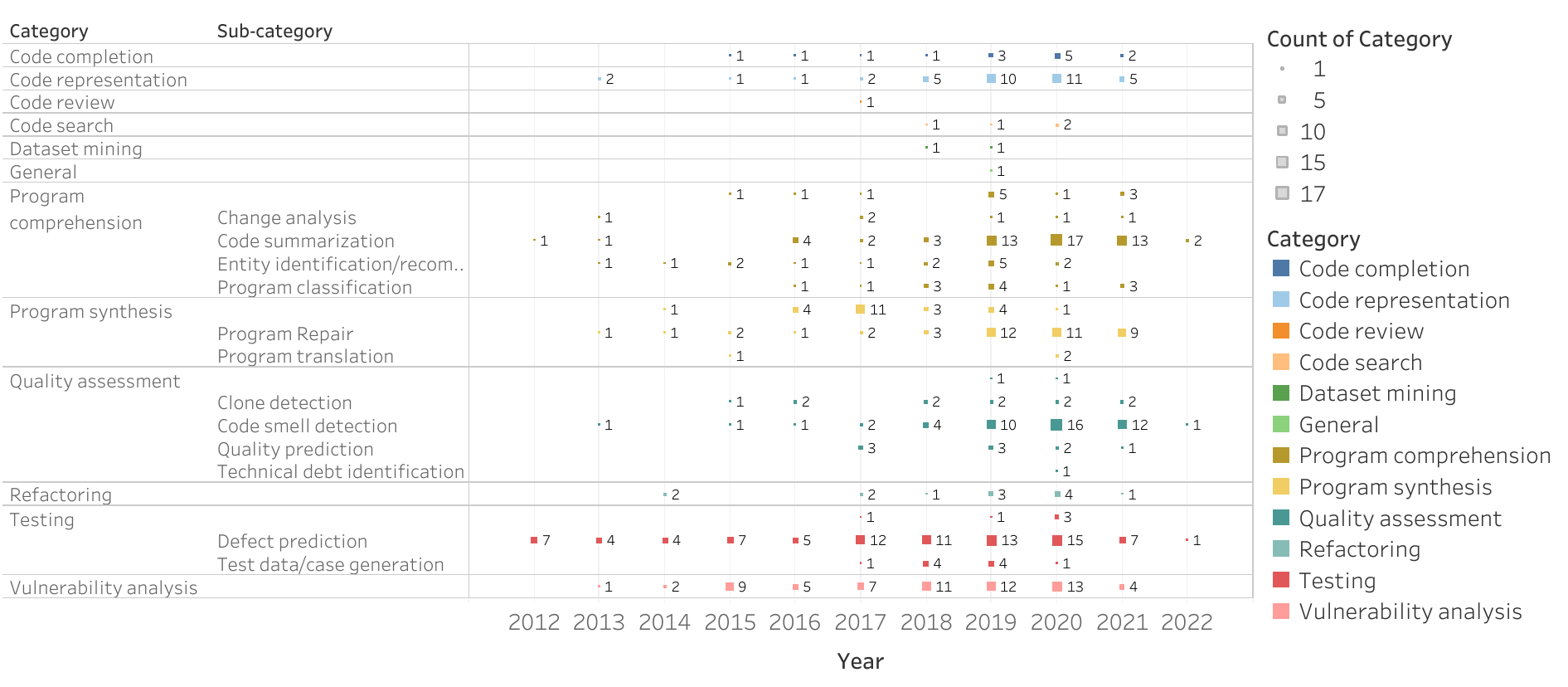}
	\caption{Category- and sub-categories-wise distribution of studies}
	\label{fig:subcategory}
\end{figure*}

Some of the categories are quite generic and hence further categorization is possible
based on specific tasks.
For example, category \textit{testing} is further divided into \textit{defect prediction},
and \textit{test data/case generation}.
We assigned sub-categories to the studies wherever applicable;
if none of the sub-categories is appropriate for a study,
we assigned it to the parent category.
Figure~\ref{fig:subcategory} presents the distribution of studies per year \textit{w.r.t. }
each category and corresponding sub-categories.

\begin{figure}[!ht]
	\centering
	\includegraphics[height=0.93\textheight]{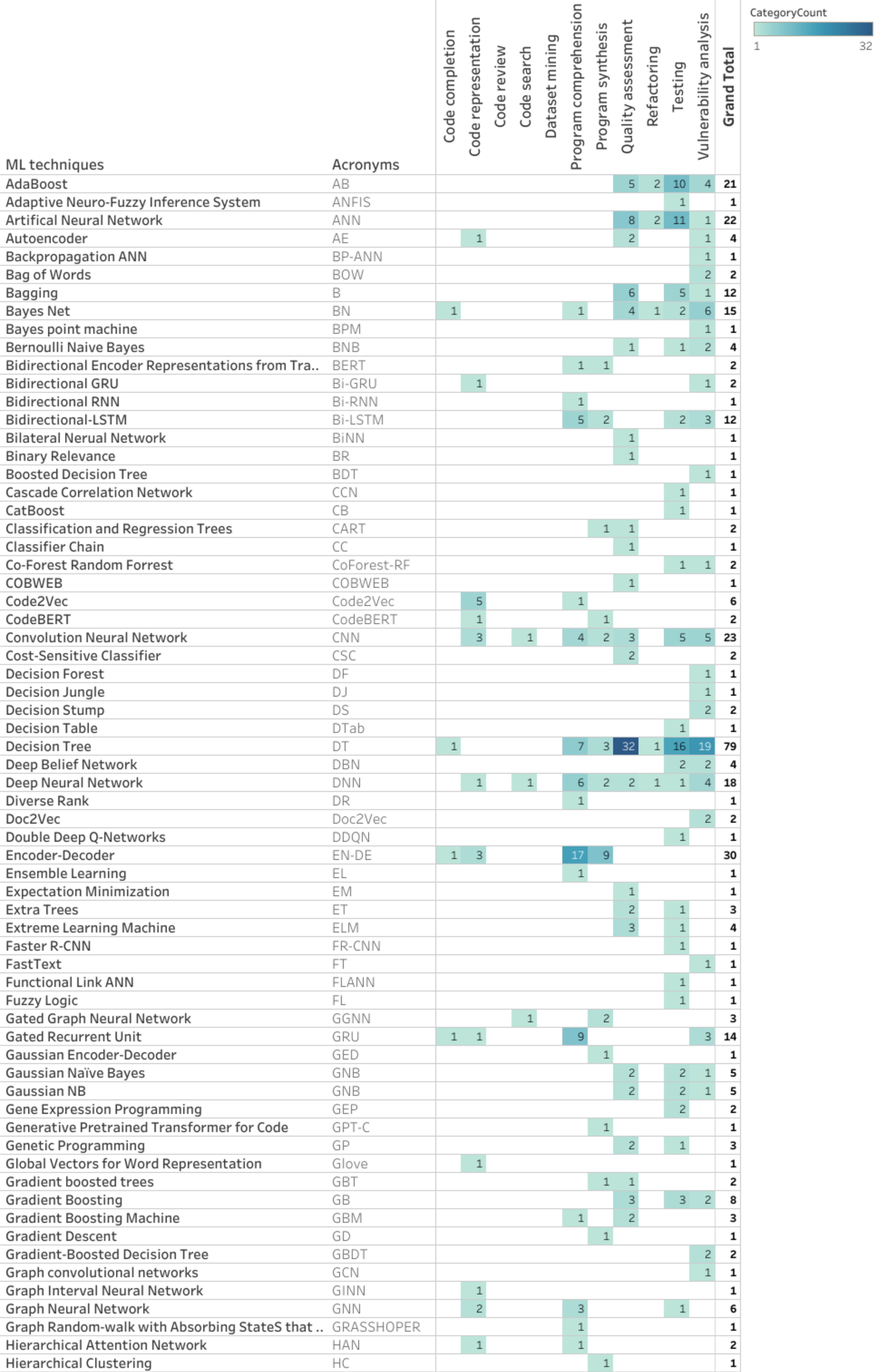}
	\caption{Usage of ML techniques in the primary studies}
	\label{fig:ml}
\end{figure}

\begin{figure}[!ht]
	\centering
	\includegraphics[height=0.8\textheight]{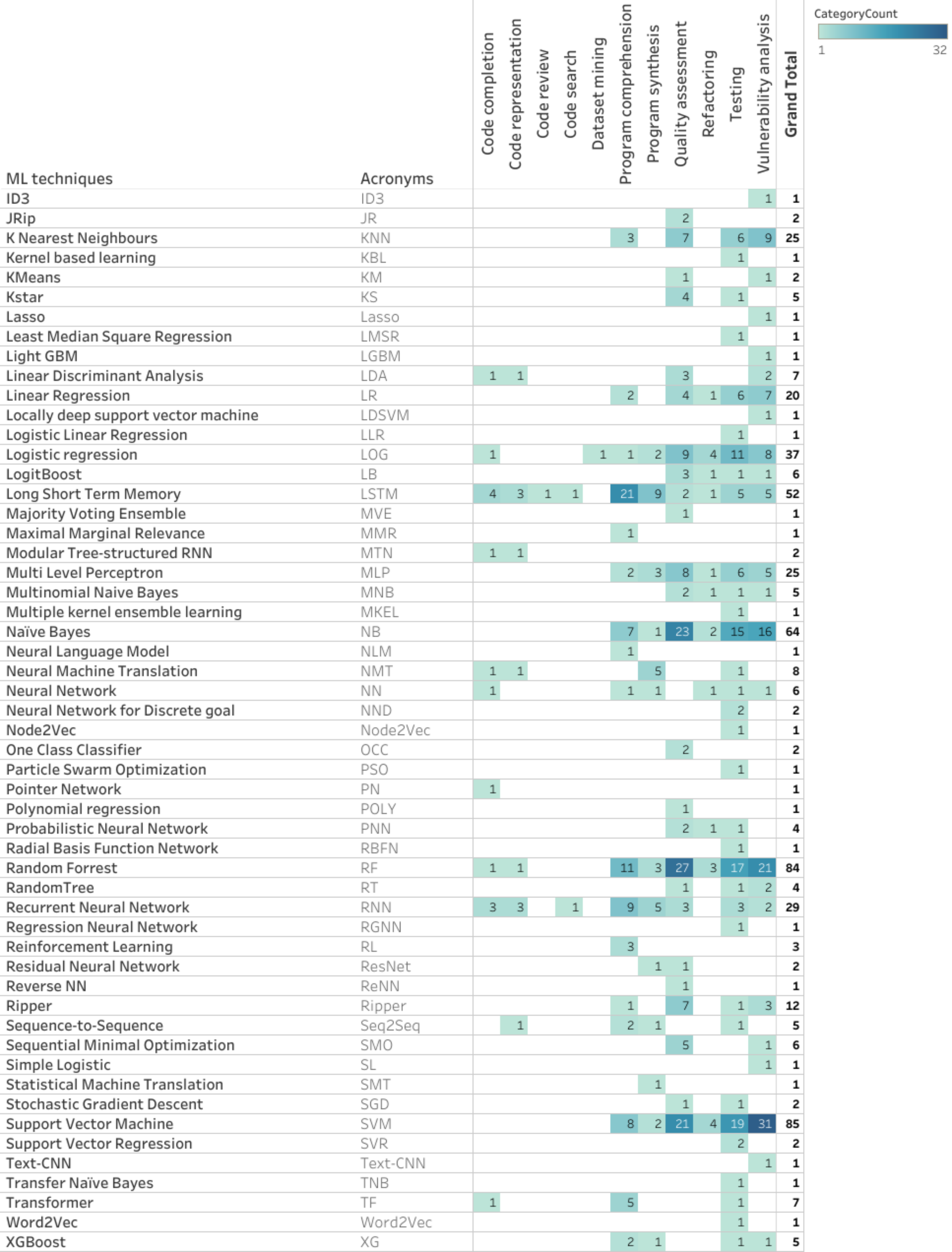}
	\caption{Usage of ML techniques in the primary studies}
	\label{fig:ml2}
\end{figure}

To quantify the growth of each category, we compute the average increase in the number of articles
from the last year for each category between the years 2012 and 2021.
We did not include the year 2022 because the year has not completed at the time of writing this survey and the obviously our collection includes the partial set of articles published in 2022.
We observed that the \textit{program comprehension} and \textit{vulnerability analysis} categories grew most 
with approximately $63.9\%$ and $64.2\%$ average growth each year, respectively.

\section{Machine learning techniques used for source code analysis}
\label{results_rq2}

We document our observations per category and subcategory
by providing a summary of the existing efforts.
Figure~\ref{fig:ml} and Figure~\ref{fig:ml2}
show the frequency of various \ml\ techniques per category
used in the primary studies.
The figures use commonly used acronyms for \ml\ techniques along with their corresponding expanded form;
we utilize these acronyms throughout the paper.
It is evident from the figures that \SVM{} and \RF{} are the most frequently employed \ml\ techniques.
From the \dl\ side, the \RNN\ family (including \LSTM{} and \GRU{}) is the most commonly used in this context.

In the rest of the section, we delve into each category and sub-category at a time,
break down the entire workflow of a code analysis task into fine-grained steps,
and summarize the method and \ml\ techniques used.
It is worth emphasizing that we structure the discussion around the crucial steps for each category (\textit{e.g.,} model generation, data sampling, feature extraction, and model training).

\subsection{Code representation}
Raw source code cannot be fed directly to a \dl\ model.
Code representation is the fundamental activity to make source code compatible with
\dl\  models by preparing a numerical representation of the code  to further solve a specific software engineering task.
Studies in this category emphasize that source code is a richer construct and hence should not
be treated simply as a collection of tokens or text~\cite{Nguyen2018_448, Allamanis2018_452};
the proposed techniques extensively utilize the syntax, structure, and semantics (such as type information from an \abst{}).
The activity transforms source code into a numerical representation making it easier to further
use the code by \ml\ models to solve specific tasks such as 
code pattern identification~\cite{Mou2016_466, Thaller2019_468}, 
method name prediction~\cite{Alon2019_464},
and comment classification~\cite{Wang2020_450}.

\begin{figure*}[!h]
	\centering
	\includegraphics[width=\textwidth]{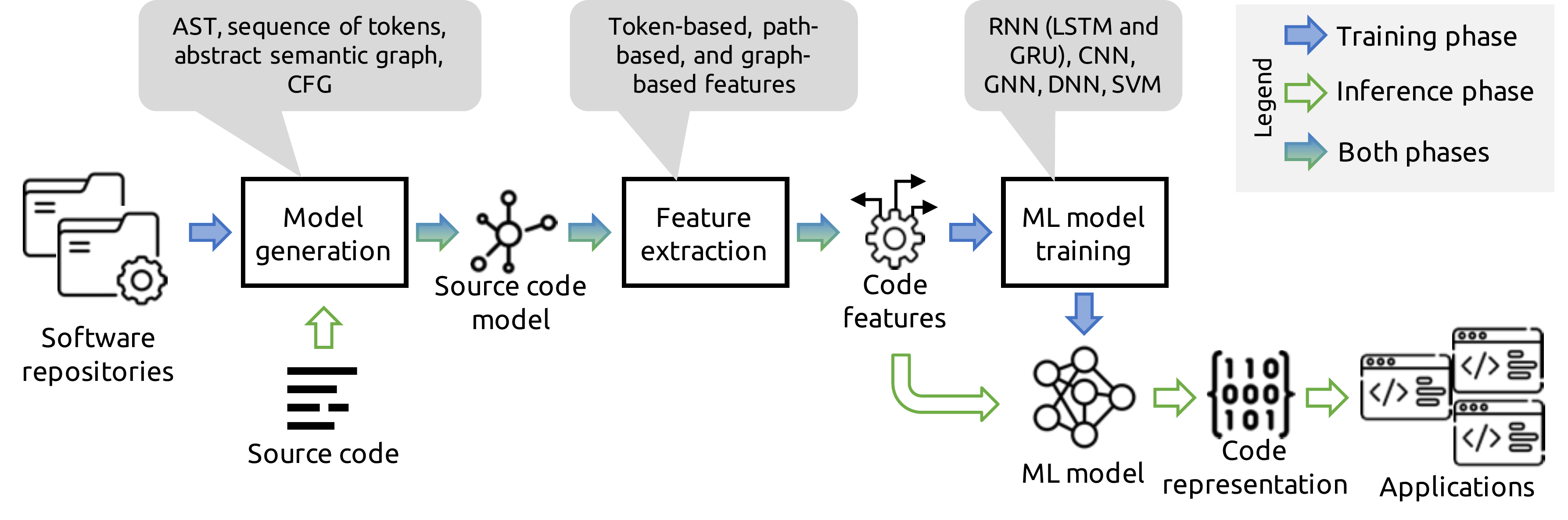}
	\caption{Overview of the code representation process pipeline}
	\label{fig:code-repr-overview}
\end{figure*}

Figure~\ref{fig:code-repr-overview} provides an overview of a typical pipeline associated
with code representation.
In the training phase, a large number of repositories are processed to train a model which is then used in the inference phase.
Source code is pre-processed to extract a
source code model (such as an \abst\ or a sequence of tokens)
which is fed into a feature extractor responsible to mine the necessary features 
(for instance, \abst\ paths and tree-based embeddings).
Then, an \ml\ model is trained using the extracted features.
The model produces a numerical (\ie{} a vector) representation that can be used further
for specific software engineering applications such as defect prediction, vulnerability detection, and code smells detection.

\task{Model generation}
Code representation efforts start with preparing a source code model.
The majority of the studies generate \abst\  \cite{Nguyen2018_448, Alon2018_449, 
Zhang2019_451, Allamanis2015_457, Chen2019_460, Alon2019_463, Alon2019_464, Yahav2018_470, Brockschmidt2019_471, Wang2020_475, 937_Chakraborty2022}.
Some studies \cite{Shedko2020_454, Allamanis2013_474, Azcona2019_482, 1003_Chakraborty2021, 1032_Zheng2019, 1087_Kanade2020, 1086_Nguyen2013, 1073_Movshovitz-Attias2013, 1068_Efstathiou2019}
parsed source code as tokens and prepared a sequence of tokens in this step.
\citet{Hoang2020_461} generated tokens representing only the code changes. 
Furthermore, \citet{Sui2020_469} compiled a program  into {\sc llvm-ir}. 
The inter-procedural value-flow graph ({\sc ivfg}) is built on top of the intermediate representation.
\citet{Thaller2019_468} used abstract semantic graph as their code model.
\citet{1009_Nie2021} used dataset offered by \citet{Jiang2018_351} that offers a large number code snippets and comment pairs.
Finally, \citet{Brauckmann2020_465} and \citet{Tufano2018_467} generated
 multiple source code models (\abst{}, {\sc cfg}, and byte code).

\task{Feature extraction}
Relevant features need to be extracted from the prepared source code model for further processing.
The first category of studies, based on applied feature extraction mechanism, uses token-based features.
\citet{Nguyen2018_448} prepared vectors of syntactic context (referred to as \textit{syntaxeme}),
type context (\textit{sememes}), and lexical tokens.
\citet{Shedko2020_454} generated a stream of tokens corresponding to function calls and control flow
expressions.
\citet{Karampatsis2020_456} split tokens as subwords to enable subwords prediction.
Path-based abstractions is the basis of the second category
where the studies extract a path typically from an \abst{}.
\citet{Alon2018_449} used paths between \abst\ nodes.
\citet{Kovalenko2020_458} extracted path context representing two tokens in code and a structural connection
along with paths between \abst\ nodes.
\citet{Alon2019_463} encoded each \abst\ path with its values as a vector and used the average of all of the
 \texttt{k} paths as the decoder's initial state
where the value of \texttt{k} depends on the number of leaf nodes in the \abst{}.
The decoder then generated an output sequence while attending over the \texttt{k} encoded paths.
Finally, \citet{Alon2019_464} also used path-based features along with distributed representation of context
where each of the path and leaf-values of a path-context is mapped to its corresponding real-valued vector representation.
Another set of studies belong to the category that used graph-based features.
\citet{Chen2019_460} created \abst\ node identified by an {\sc api} name and attached each node
to the corresponding \abst\ node belonging to the identifier.
\citet{Thaller2019_468} proposed feature maps; feature maps are human-interpretation, stacked,
named subtrees extracted from abstract semantic graph.
\citet{Brauckmann2020_465} created a dataflow-enriched \abst\ graph, 
where nodes are labeled as declarations, statements, and types as found in the Clang\footnote{\url{https://clang.llvm.org/}} \abst{}.
\citet{Cvitkovic2019_476} augmented \abst\ with semantic information by adding a graph-structured
vocabulary cache.
Finally, \citet{Zhang2019_451} extracted small statement trees along with multi-way statement
trees to capture the statement-level lexical and syntactical information.
The final category of studies used \dl\ \cite{Hoang2020_461, Tufano2018_467} to learn features automatically.

\task{ML model training}
The majority of the studies rely on the \RNN{}-based \dl\ model.
Among them, some of the studies \cite{Wang2020_450, Hellendoorn2017_455, Wang2020_475, Brauckmann2020_465, Alon2019_463} 
employed \LSTM{}-based models;
while others \cite{Zhang2019_451, Hoang2020_461, Karampatsis2020_456, Yahav2018_470, Brockschmidt2019_471}
used \GRU{}-based models.
Among the other kinds of \ml\ models, studies employed \GNN{}-based \cite{Cvitkovic2019_476, Wang2020_473},
{\sc dnn} \cite{Nguyen2018_448}, conditional random fields \cite{Alon2018_449},
\SVM{} \cite{Lim2018_453, Rabin2020_481}, and \CNN{}-based models \cite{Chen2019_460, Mou2016_466, Thaller2019_468}.
Some of the studies rely on the combination of different \dl\ models.
For example, \citet{Tufano2018_467} employed 
\RNN{}-based model for learning embedding in the first stage which is given to an Autoencoder-based
model to encode arbitrarily long streams of embeddings.

A typical output of a code representation technique is the vector representation of the source code.
The exact form of the output vector may differ based on the adopted mechanism.
Often, the code vectors are application specific depending upon the nature of features extracted and training mechanism.
For example, Code2Vec produces code vectors trained for method name prediction; however,
the same mechanism can be used for other applications after tuning and selecting appropriate features. 
\citet{1084_Kang2019} carried out an empirical study to observe whether the embeddings generated by Code2Vec can be used in other contexts.
Similarly, \citet{780_Pour2021} used Code2Vec, Code2Seq, and CodeBERT to explore the robustness of code embedding models by retraining the models using the generated adversarial examples.

The semantics of the produced embeddings depends significantly on the selected features. 
Studies in this domain identify this aspect and hence swiftly focused to extract features that capture the relevant semantics; for example, path-based features encode the order among the tokens. 
The chosen \ml{} model plays another important role to generate effective embeddings.
Given the success of \RNN{} with text processing tasks, due to its capability to identify sequence and pattern, \RNN{}-based models dominate this category.

\subsection{Testing}
In this section,
we point out the state-of-the-art regarding \ml\ techniques applied to software testing.
Testing is the process of identifying functional or non-functional bugs to improve the accuracy and reliability of a software.
Following the definition, we include defect prediction studies in this category
where authors extract features to train \ml\ models
to find bugs in software applications.
Then,
we offer a discussion on effort prediction models used to identify
the time needed to test an application.
Finally,
we present studies associated with test cases generation by employing
\ml\ techniques.

\subsubsection{Defect prediction\\}
\begin{figure*}[!h]
	\centering
	\includegraphics[width=\textwidth]{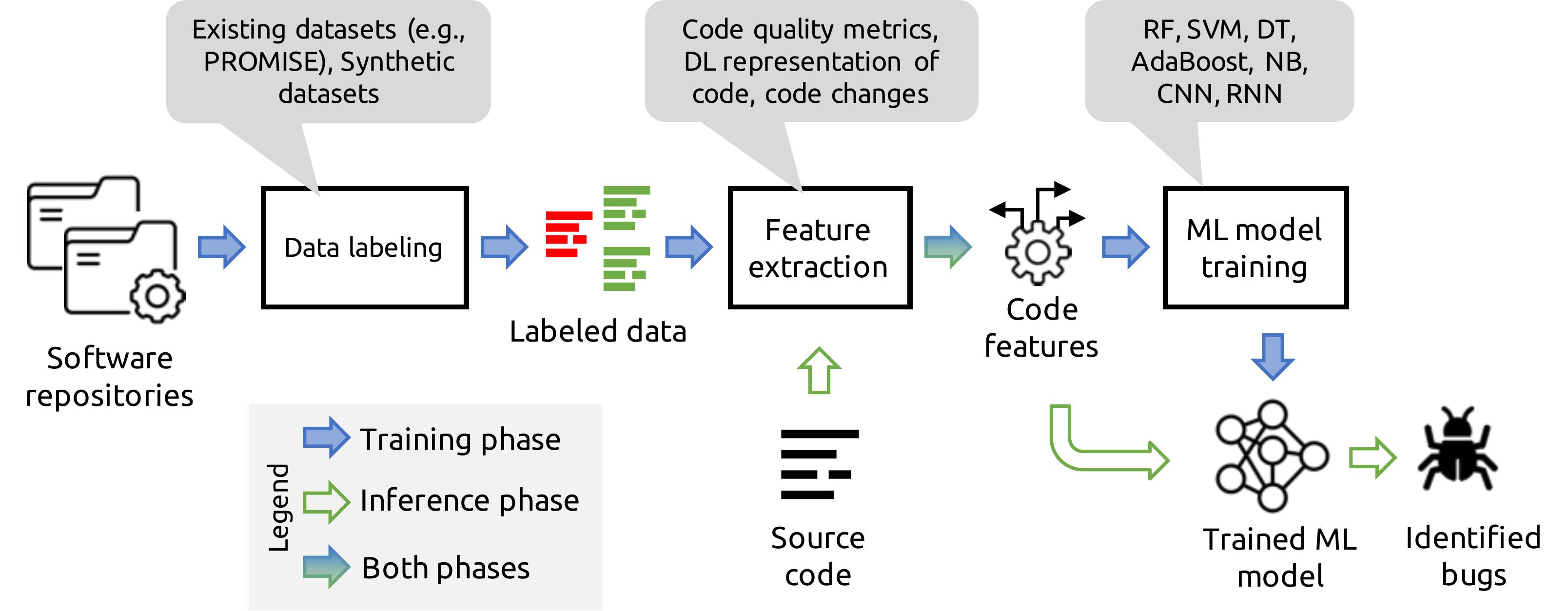}
	\caption{Overview of the defect prediction process pipeline}
	\label{fig:defect_predition}
\end{figure*}

To pinpoint bugs in software,
researchers used various \ml\ approaches.
Figure~\ref{fig:defect_predition} depicts a common pipeline
used to train a defect prediction model.
The first step of this process is to identify the
positive and negative samples from a dataset
where samples
could be a type of source code entity such as classes, modules, files, and methods.
Next,
features are extracted from the source code
and fed into an \ml\ model for training.
Finally,
the trained model can classify different code snippets
as buggy or benign based on the encoded knowledge.
To this end,
we discuss the collected studies based on (1) data labeling,
(2) features extract,
and (3) \ml\ model training.

\task{Data labeling}
To train an \ml{} model for predicting defects in source code
a labeled dataset is required.
For this purpose,
researchers have used some well-known and publicly available datasets.
For instance,
a large number of studies
\cite{Cetiner2020_65, Gondra2008_73, Malhotra2014_81,
	Singh2017_76, Challagulla2008_87, Bhandari2018_105,
	Malhotra2011_114, Singh2017_130, Ceylan2006_129,
	Wang2013_150, Chug2013_134, Li2011_124,
	Dhamayanthi2019_94, Prabha2020_133, Ma2012_147,
	Khan2020_82, Chen2020_140, Dam2019_100,
	Wang2016_78, Shi2020_116, Santos2020_148,
	Singh2020_146, Zhang2020_135, Butgereit2019_154, DiMartino2011, Sarro2012,
	526_Wang2018, 527_Lin2021, 637_Ren2014, 652_Li2017, 653_Kaur2015, 661_Okutan2014,
	663_Laradji2015, 676_Song2019, 680_Pan2019, 681_Malhotra2019, 682_Qiao2020, 684_Manjula2019, 717_Suresh2014, 722_Erturk2015, 809_Palomba2017, 705_Yohannese2017, 690_Sun2012, 718_Dejaeger2012, 729_Al2020, 1062_Aleem2015}
used the {\sc promise} dataset~\cite{Promise_dataset_2005}.
Some studies used other datasets in addition to {\sc promise} dataset.
For example, \citet{1094_Liang2019} used Apache projects and \citet{682_Qiao2020} used {\sc mis} dataset \cite{MIS1996}.
\citet{Xiao2020_102} utilized
a Continuous Integration ({\sc ci}) dataset and
\citet{Pradel2018_84} generated a synthetic dataset.
Apart from using the existing datasets, some other studies
prepared their own datasets by utilizing 
various {\sc GitHub} projects~\cite{Malhotra2017_121, Harman2014, Singh2020_146,
	Aggarwal2019_128, Malhotra2017_88, 1065_Pascarella2018, 1080_Tufano2019} including Apache \cite{Li2019_96,Bowes2016,Dambros2012, 529_Fan2019, 808_Palomba2016, 829_Sotto-Mayor2021, 925_Malhotra2012, 720_Choudhary2018, 725_Rathore2021}, Eclipse \cite{Zimmermann2007,Dambros2012}
and Mozilla \cite{Madhavan2007_117, Knab2006_118} projects, or industrial data\cite{Bowes2016}.

\task{Feature extraction}
The most common features to train a defect prediction model
are the source code metrics introduced by~\citet{Halstead_1977},
\citet{Chidamber1994},
and~\citet{McCabe1976}.
Most of the examined studies~\cite{Cetiner2020_65,
Gondra2008_73, Malhotra2014_81, Singh2017_76,
Challagulla2008_87, Malhotra2011_114,
Wang2013_150, Chug2013_134, Malhotra2017_121,
Malhotra2017_88, Ma2012_147, Khan2020_82,
Butgereit2019_154, Chappelly2017_107, Knab2006_118,
Sethi2016_93, 529_Fan2019, 653_Kaur2015, 656_Jing2014, 661_Okutan2014, 663_Laradji2015, 674_Arar2015, 675_Aljamaan2020, 676_Song2019, 678_Wang2016, 681_Malhotra2019, 682_Qiao2020, 684_Manjula2019, 690_Sun2012, 705_Yohannese2017, 717_Suresh2014, 718_Dejaeger2012, 720_Choudhary2018, 722_Erturk2015, 725_Rathore2021, 729_Al2020, 808_Palomba2016, 829_Sotto-Mayor2021, 866_Premalatha2017, 925_Malhotra2012, 1062_Aleem2015, 1065_Pascarella2018}
used a large number of metrics such as Lines of Code,
Number of Children,
Coupling Between Objects, and
Cyclomatic Complexity.
Some authors \cite{809_Palomba2017, 833_Soltanifar2016} combined detected code smells with code quality metrics.
Furthermore, \citet{666_Felix2017} used defect metrics such as defect density and defect velocity along with traditional code smells.

In addition to the above,
some authors~\cite{Ceylan2006_129, Dhamayanthi2019_94, Bhandari2018_105, Prabha2020_133}
suggested the use of dimensional space reduction
techniques---such as Principal Component Analysis ({\sc pca})---to limit
the number of features.
\citet{Pandey2018_137} used Sequential Forward Search ({\sc sfs})
to extract relevant source code metrics.
\citet{Santos2020_148} suggested a sampling-based approach to extract
source code metrics to train defect prediction models.
\citet{Kaur2017_71} suggested an approach
to fetch entropy of change metrics. Bowes et al. \cite{Bowes2016} introduced a novel set of metrics constructed in terms of mutants and the test cases that cover and detect them.

%embedding
Other authors~\cite{Pradel2018_84, Zhang2020_135} used embeddings
as features to train models.
Such studies, first generate \abst{} \cite{Li2019_96, 529_Fan2019, 652_Li2017, 680_Pan2019, 1094_Liang2019}, a variation of \abst{} such as Simplified \abst{} \cite{527_Lin2021, 1083_Chen2019},
or \abst{}-\textit{diff} \cite{526_Wang2018, 1080_Tufano2019}
for a selected method or file.
Then, embeddings are generated either using the token vector corresponding to each node in the generated tree or extracting a set of paths from \abst{}.
\citet{Singh2020_146} proposed a method
named \textit{Transfer Learning Code Vectorizer}
that generates features from source code by using a pre-trained code representation \dl\ model.
Another approach for detecting defects
is capturing the syntax and multiple levels
of semantics in the source code as suggested
by~\citet{Dam2019_100}.
To do so,
the authors trained a tree-base
{\sc lstm} model by using source code files as feature vectors.
Subsequently,
the trained model receives an \abst\
as input and predicts if the file is clear from bugs or not.

\citet{Wang2016_78} employed the Deep Belief Network algorithm ({\sc dbn})
to learn semantic features from token vectors,
which are fetched from applications' \abst{}s.
\citet{Shi2020_116} used a {\sc dnn} model
to automate the features extraction from the source code.
\citet{Xiao2020_102} collected the testing history information
of all previous {\sc ci} cycles,
within a {\sc ci} environment,
to train defect predict models.
Likewise to the above study,
\citet{Madhavan2007_117} and \citet{Aggarwal2019_128}
used the changes among various versions of a software
as features to train defect prediction models.

In contrast  to the above studies,
\citet{Chen2020_140} suggested the {\sc dtl-dp},
a framework to predict defects without the need of features extraction tools.
Specifically,
{\sc dtl-dp} visualizes the programs as images
and extracts features out of them
by using a self-attention mechanism \cite{Vaswani2017}.
Afterwards,
it utilizes transfer learning to reduce
the sample distribution differences
between the projects by feeding them to a model.

\task{ML model training}
In the following, we present the main categories
of \ml\ techniques found
in the examined papers.

\subtask{Traditional \ml\ models}
To train models,
most of the studies \cite{Cetiner2020_65, Gondra2008_73, Malhotra2014_81,
	Singh2017_76, Challagulla2008_87, Bhandari2018_105,
	Malhotra2011_114, Singh2017_130, Ceylan2006_129,
	Chug2013_134, Dhamayanthi2019_94, Prabha2020_133,
	Malhotra2017_121, Malhotra2017_88, Hammouri2018_127,
	Pandey2018_137, Santos2020_148, Singh2020_146,
	Khan2020_82, Kaur2017_71, Butgereit2019_154,
	Wang2016_78, 682_Qiao2020, 684_Manjula2019, 690_Sun2012, 705_Yohannese2017, 717_Suresh2014, 718_Dejaeger2012, 720_Choudhary2018, 725_Rathore2021, 808_Palomba2016, 829_Sotto-Mayor2021, 833_Soltanifar2016, 866_Premalatha2017, 925_Malhotra2012, 1062_Aleem2015, 1065_Pascarella2018, 653_Kaur2015, 661_Okutan2014, 663_Laradji2015, 666_Felix2017, 681_Malhotra2019, 676_Song2019, 675_Aljamaan2020, 637_Ren2014} 
	used  
traditional \ml\ algorithms such as \dt{}, \rf{}, \svm{},
 and \textit{AdaBoost}.
Similarly, \citet{656_Jing2014, 678_Wang2016} used \textit{Cost Sensitive Discriminative Learning}.
In addition,
authors~\cite{Li2011_124, Wang2013_150, Ma2012_147}
proposed changes
to traditional \ml\ algorithms
to train their models.
Specifically,
\citet{Wang2013_150} suggested
a dynamic version of \textit{AdaBoost.NC } that adjusts
its parameters automatically during training.
Similarly,
\citet{Li2011_124} proposed ACoForest,
an active semi-supervised learning method
to sample the most useful modules
to train defect prediction models.
\citet{Ma2012_147} introduced \textit{Transfer Naive Bayes},
an approach to facilitate transfer learning
from cross-company data information
and weighting training data.

\subtask{\dl{}-based models}
In contrast to the above studies,
researchers~\cite{Chen2020_140, Dam2019_100,
	Pradel2018_84, Li2019_96, Sethi2016_93}
used \dl\ models such as \CNN{} and \RNN{}-based models
for defect prediction.
Specifically, \citet{Chen2020_140, 729_Al2020, 652_Li2017, 680_Pan2019}
used \CNN{}-based models to predict bugs.
\RNN{}-based methods \cite{Dam2019_100, 1080_Tufano2019, 1083_Chen2019, 1094_Liang2019, 529_Fan2019, 527_Lin2021} are also frequently used where variations of \LSTM{} are used to for defect prediction.
Moreover,
by using \dl\ approaches,
authors achieved improved accuracy for defect prediction
and they pointed out bugs in real-world
applications~\cite{Pradel2018_84, Li2019_96}.

\subsubsection{Test data and test cases generation\\}
A usual approach to have a \ml\  model for generating
test oracles involves
capturing data from an application under test,
pre-processing the captured data,
extracting relevant features, using an \ml{} algorithm, and
evaluating the model.

\task{Data generation and pre-processing}
Researchers developed a number of ways for capturing
data from applications under test and pre-process them
before feeding them to an \ml\ model.
\citet{Braga2018_67} recorded traces
for applications to capture usage data.
They sanitized any irrelevant
information collected from the programs recording
components.
AppFlow ~\cite{Hu2018_72}
captures human-event sequences from a smart-phone
screen in order to identify tests.
Similarly,
\citet{Nguyen2019_126} suggested Shinobi,
a framework that uses a fast {\sc r-cnn} model
to identify input data fields from multiple web-sites.
\citet{Utting2020_92} captured user
and system execution traces to help
 generating missing {\sc api} tests.
To automatically identify metamorphic relations,
\citet{Nair2019_101} suggested an approach that
leverages \ml\ techniques and test mutants. 
By using a variety of code transformation techniques,
the authors' approach can generate a synthetic dataset
for training models to predict metamorphic relations.

\task{Feature extraction}
Some authors \cite{Braga2018_67, Utting2020_92} used execution traces as features.
\citet{Kim2018_90} suggested an approach
that replaces {\sc sbst}'s meta-heuristic algorithms
with deep reinforcement learning to generate test cases
based on branch coverage information.
\cite{Grano2018_91} used code quality metrics such as coupling, {\sc dit}, and {\sc nof} to generate test data; they use the generate test data to predict the code coverage in a continuous integration pipeline.

\task{Train ML model}
Researchers used supervised
and unsupervised \ml{} algorithms to generate test data and cases.
In some of the studies,
the authors utilized more than one
\ml{} algorithm to achieve their goal.
Specifically,
several studies~\cite{Braga2018_67, 
Kim2018_90, Utting2020_92, Nair2019_101}
used traditional \ml{} algorithms,
such as \svm{}, \nb{}, \dt{}, \mlp{}, \rf{}, \ada{}, \lr{}.
\citet{Nguyen2019_126} used the \dl{}
algorithm Fast {\sc r-cnn}.
Similarly, \cite{Godefroid2017_99} used \LSTM{} to automate generating the input grammar data for fuzzing.

\subsection{Program synthesis}

This section summarizes the \ml\ techniques used by automated program synthesis tools and techniques
in the examined software engineering literature.
Apart from a major sub-category \textit{program repair}, we also discuss state-of-the-art
corresponds to \textit{refactoring} and \textit{program translation} sub-categories in this section.

\subsubsection{Program repair\\}

Automated Program Repair ({\sc apr}) refers to techniques
that attempt to automatically identify patches for a given bug
(\ie{} programming mistakes that can cause unintended run-time behavior),
which can be applied to software
with a little or without human intervention~\cite{Goues2019_281}.
Program repair typically consists of two phases.
Initially, the repair tool uses fault localization to detect a bug
in the software under examination,
then, it generates patches using techniques such as
search-based software engineering and logic rules
that can possibly fix a given bug.
To validate the generated patch,
the (usually manual) evaluation
of the semantic correctness\footnote{The term semantic
	correctness is a criterion for evaluating whether a generated patch is similar to the human fix
	for a given bug~\cite{Liu2020}.}
of that patch follows.

According to~\citet{Goues2019_281},
the techniques for constructing repair patches
can be divided into three categories
(heuristic repair, constraint-based repair, and learning-aided repair)
if we consider the following two criteria:
what types of patches are constructed and how the search is conducted.
Here, we are interested in learning-aided repair,
which leverages the availability of previously generated patches and
bug fixes to generate patches.
In particular, learning-aided-based repair tools use \ml\
to learn patterns for patch generation.

\begin{figure*}
	\centering
	\includegraphics[width=\textwidth]{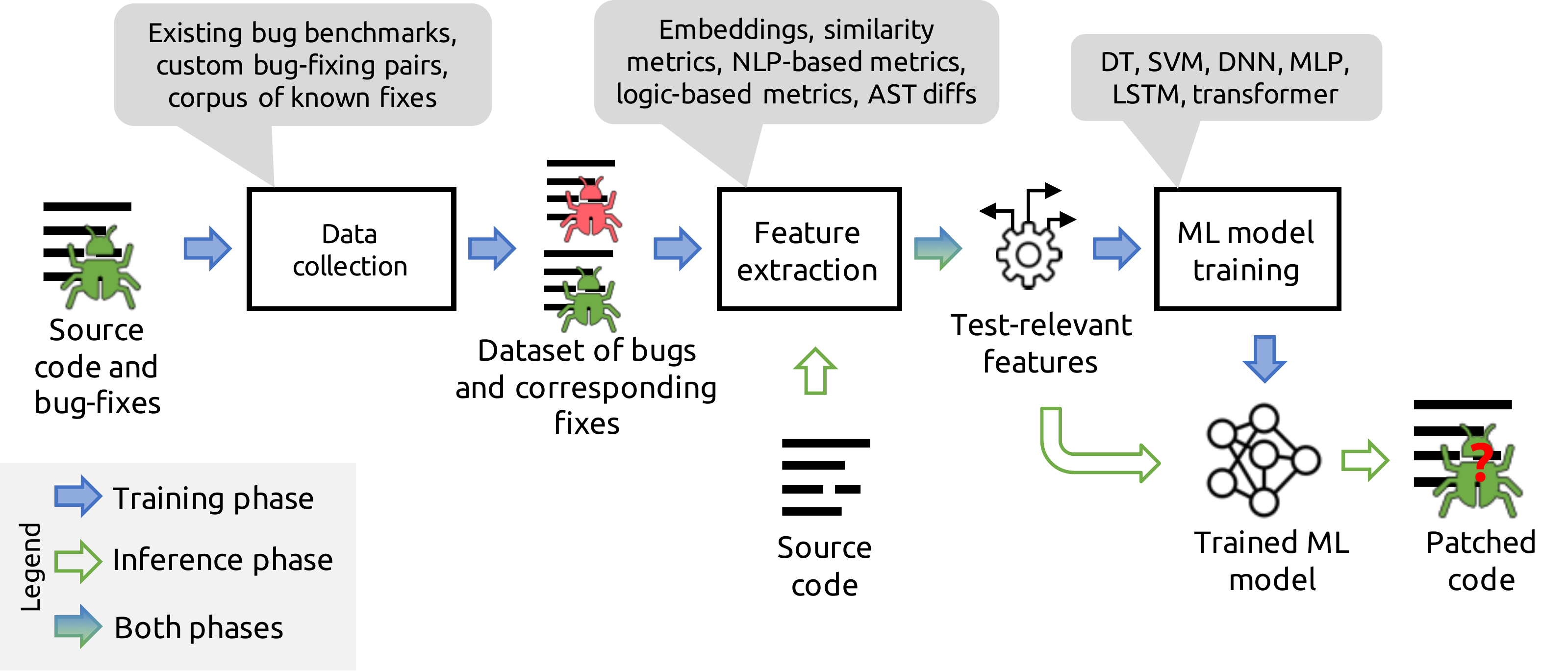}
	\caption{Overview of the program repair process pipeline.}
	\label{fig:approach_repair}
\end{figure*}

Figure~\ref{fig:approach_repair}
depicts the typical process performed by learning-aided-based repair tools.
Typically, at the pre-processing step, such methods
take source code of the buggy revision as an input,
and those revisions that fixes the buggy revision.
The revision with the fixes includes
a  patch carried out manually that corrects the buggy revision
and a test case that checks whether the bug has been fixed.
Learning-aided-based repair is mainly based on the hypothesis
that similar bugs will have similar fixes.
Therefore, during the training phase,
such techniques can use features such as similarity metrics
to match bug patterns to similar fixes.
Then, the generated patches rely on
those learnt patterns.
Next, we elaborate upon the individual steps involved in the process of
program repair using \ml{} techniques.

\task{Data collection}
The majority of the studies
extract buggy project revisions and manual fixes
from buggy software projects.
Most studies leverage source-code naturalness.
For instance, \citet{Tufano2019_279}
extracted millions of bug-fixing pairs from {\sc GitHub},
\citet{Amorim2018_272} leveraged the
naturalness obtained from a corpus of known fixes,
and \citet{Chen2016_323} used
natural language structures from source code.
Furthermore, many studies develop their own large-scale bug benchmarks.
\citet{Ahmed2018_293}
leveraged 4,500 erroneous C programs,
\citet{Gopinath2014_296}
used a suite of programs
and datasets
stemmed from real-world applications,
\citet{Long2016_285}
used a set of successful manual patches
from open-source software repositories, and
\citet{888_Mashhadi2021} used ManySStuBs4J dataset containing natural language description and code snippets to automatically generate code fixes.
\citet{Le2015_352} created an oracle for
predicting which bugs should be delegated to developers for fixing and
which should be fixed by repair tools.
\citet{890_Jiang2021} used a dataset containing more than 4 million methods extracted.
\citet{889_White2019} used Spoon, an open-source library for analyzing and transforming Java source code, to build a model for each buggy program revision.
\citet{915_Pinconschi2021} constructed a dataset containing vulnerability-fix pairs by aggregating five existing dataset (Mozilla Foundation Security Advisories, SecretPatch, NVD, Secbench, and Big-Vul). The dataset \ie{} \textit{PatchBundle} is publicly available on GitHub.
\citet{943_Cambronero2019} proposed a method to generate new supervised machine learning pipelines. To achieve the goal, the study trained using a collection of 500 supervised learning programs and their associated target datasets from Kaggle.
\citet{975_Liu2013} prepared their dataset by selecting 636 closed bug reports from the Linux kernel and Mozilla databases.
\citet{1092_Svyatkovskiy2021} constructed their experimental dataset from the 2700 top-starred Python source code repositories on GitHub.

Other studies use existing bug benchmarks,
such as {\sc Defects4J}~\cite{Defects4J} and {\sc IntroClass}~\cite{LeGoues15},
which already include buggy revisions and human fixes,
to evaluate their approaches.
For instance,
\citet{Saha2019_315},
\citet{896_Lou2020},
\citet{934_Zhu2021},
\citet{939_Renzullo2021},
\citet{Wang2019_316}, and
\citet{Chen2019_350}
leveraged {\sc Defects4J} for the evaluations of their approaches.
Additionally, \citet{Dantas2019_289} used the
{\sc IntroClass} benchmark and
\citet{Majd2020_353} conducted experiments
using 119,989 C/C++ programs
within {\sc Code4Bench}.
\citet{Wu2020_519} used the {\sc DeepFix} dataset
that contains 46,500 correct C programs and 6,975 programs with errors 
for their graph-based \dl\ approach for syntax error correction.

Some studies examine bugs in different programming languages.
For instance, \citet{Svyatkovskiy2020_321} used
1.2 billion lines of source code
in Python, C\#, JavaScript, and TypeScript programming languages.
Also, \citet{Lutellier2020_287}
used six popular benchmarks of four programming
languages (Java, C, Python, and JavaScript).

There are also studies that
mostly focus on syntax errors.
In particular,
\citet{Gupta2019_298}
used 6,975 erroneous C programs with typographic errors,
\citet{Santos2018_358}
used source code files with syntax errors, and
\citet{Sakkas2020_366}
used a corpus of 4,500 ill-typed {\sc OCaml} programs
that lead to compile-time errors.
\citet{Bhatia2018_337} examined
a corpus of syntactically correct submissions for a programming assignment.
They used a dataset comprising of
over 14,500 student submissions with syntax errors.

Finally, there is a number of studies that use
programming assignment from students.
For instance,
\citet{Bhatia2018_337},
\citet{Gupta2019_298},
and \citet{Sakkas2020_366}
used a corpus of 4,500 ill-typed {\sc OCaml} student programs.

\task{Feature extraction}
The majority of studies utilize similarity metrics
to extract similar bug patterns
and, respectively, correct bug fixes.
These studies mostly employ word embeddings
for code representation and abstraction.
In particular,
\citet{Amorim2018_272, Svyatkovskiy2020_321, Santos2018_358, 890_Jiang2021},
and \citet{Chen2016_323},
 leveraged source-code naturalness
and applied {\sc nlp}-based metrics.
\citet{Tian2020_309} employed
different representation learning approaches
for code changes to derive embeddings
for similarity computations.
Similarly, \citet{889_White2019} used Word2Vec to learn embeddings for each buggy program revision.
\citet{Ahmed2018_293}
used similar metrics for fixing compile-time errors.
Additionally, \citet{Saha2019_315} leveraged
a code similarity analysis,
which compares both syntactic and semantic features,
and the revision history of a software project under examination,
from {\sc Defects4J},
for fixing multi-hunk bugs,
\ie{} bugs that require applying a substantially similar patch
to different locations.
Furthermore, \citet{Wang2019_316}
investigated, using similarity metrics,
how these machine-generated  correct  patches
can be semantically equivalent
to human patches,
and how bug characteristics affect patch generation.
\citet{Sakkas2020_366} also applied similarity metrics.
\citet{1092_Svyatkovskiy2021} extracted structured representation of code (for example, lexemes, \abst{}s, and dataflow) and learn directly a task over those representations.

There are several approaches
that use logic-based metrics
based on the relationships of the features used.
Specifically, \citet{Hoang2018_280}
extracted twelve relations of statements and blocks
for Bi-gram model using Big code
to prune the search space,
and make the patches generated
by {\sc Prophet}~\cite{Long2016_285} more efficient and precise.
\citet{Alrajeh2015_283}
identified counterexamples and witness traces
using model checking for logic-based
learning to perform repair process automatically.
\citet{Cai2019_284} used publicly available examples of faulty models
written in the \texttt{B} formal specification language,
and proposed B-repair,
an approach that supports automated repair of
such a formal specification.
\citet{943_Cambronero2019} extracted dynamic program traces through identification of relevant APIs of the target library; the extracted traces help the employed machine learning model to generate pipelines for new datasets.

Many studies also extract and consider
the context where the bugs are related to. 
For instance, \citet{Tufano2019_279}
extracted Bug-Fixing Pairs ({\sc bfp}s)
from millions of bug fixes mined from {\sc GitHub}
(used as meaningful examples of such bug-fixes),
where such a pair consists of a buggy code component and
the corresponding fixed code.
Then, they used those pairs as input to
an Encoder-Decoder Natural Machine Translation ({\sc nmt}) model.
For the extraction of the pair,
they used the {\sc GumTree} {\sc Spoon} {\sc ast} Diff tool~\cite{Falleri2014}.
Additionally, \citet{Soto2018_292}
constructed a corpus
by delimiting debugging regions 
in a provided dataset.
Then, they recursively analyzed
the differences between
the Simplified Syntax Trees associated with EditEvent’s.
\citet{Mesbah2019_300} also generated {\sc ast} diffs
from the textual code changes
and transformed them into a domain-specific language called Delta
that encodes the changes that must be made to make the code compile.
Then, they fed the compiler diagnostic information (as source) and the Delta changes 
that resolved the diagnostic (as target) into a Neural Machine Translation network for training.
Furthermore, \citet{Li2020_303} used the prior bug fixes
and the surrounding code contexts of the fixes
for code transformation learning.
\citet{Saha2017_306} developed a \ml\ model
that relies on four features derived from
a program's context,
\ie{} the source-code surrounding the potential
repair location, and the bug report.
Similarly, \citet{888_Mashhadi2021} used a combination of natural language text and corresponding code snippet to generated an aggregated sequence representation for the downstream task.
Finally, \citet{Bader2019_312} utilized a ranking technique that
also considers the context of a code change, and
selects the most appropriate fix for a given bug.
\citet{Vasic2019_333} used results from
localization of variable-misuse bugs.
\citet{Wu2020_519}
developed an approach, {\sc ggf},
for syntax-error correction that treats the code as a mixture of the token sequences and graphs.
\citet{933_LIN2021} and \citet{934_Zhu2021} utilized \abst{} paths to generate code embeddings to predict the correctness of a patch.

\task{ML model training}
In the following, we present the main categories
of \ml\ techniques found
in the examined papers.

\subtask{Neural Machine Translation}
This category includes papers that
apply neural machine translation (\NMT{})
for enhancing automated program repair.
Such approaches can, for instance, include techniques that
use examples of bug fixing for one programming language
to fix similar bugs for other programming language.
\citet{Lutellier2020_287}
developed the repair tool called {\sc CoCoNuT}
that uses ensemble learning on the combination of \CNN{}s
and a new context-aware \NMT{}.
Additionally, \citet{Tufano2019_279} used \NMT{} techniques
 (Encoder-Decoder model) for learning bug-fixing patches for real defects, and generated repair patches.
\citet{Mesbah2019_300} introduced {\sc DeepDelta},
which used \NMT{}
for learning to repair compilation errors.
\citet{890_Jiang2021} proposed {\sc cure}, a \NMT{}-based approach to
automatically fix bugs.
\citet{915_Pinconschi2021} used SequenceR, a sequence-to-sequence model,
to patch security faults in C programs.
\citet{934_Zhu2021} proposed a tool Recoder, a syntax-guided edit decoder that takes encoded information and produces placeholders by selecting non-terminal nodes based on their probabilities.

\subtask{Natural Language Processing}
In this category, we include papers that combine
natural language processing ({\sc nlp}) techniques, embeddings,
similarity scores, and \ml\
for automated program repair.
\citet{Tian2020_309}
introduced an empirical work that investigates different representation learning approaches
for code changes to derive embeddings,
which are amendable to similarity computations.
This study uses  {\sc bert} transformer-based embeddings.
Furthermore, \citet{Amorim2018_272}
applied, a word embedding model ({\sc Word2Vec}),
to facilitate the evaluation of repair processes,
by considering the naturalness obtained from known bug fixes.
\citet{Hoang2018_280} have also applied word representations,
and extracted relations of statements and blocks for a Bi-gram model using Big code,
to improve the existing learning-aid-based repair tool {\sc Prophet}~\cite{Long2016_285}.
\citet{Gupta2019_298} used word embeddings and reinforcement learning
to fix erroneous C student programs with typographic errors.
\citet{Tian2020_309} applied a \ml\ predictor
with {\sc bert} transformer-based embeddings associated with logistic regression
to learn code representations in order to learn deep features
that can encode the properties of patch correctness.
\citet{Saha2019_315} used similarity analysis for repairing bugs
that may require applying a substantially similar patch at a number of locations.
Additionally, \citet{Wang2019_316} used also similarity metrics to compare the differences
among machine-generated and human patches.
\citet{Santos2018_358} used n-grams and {\sc nn}s
to detect and correct syntax errors.

\subtask{Logic-based rules}
\citet{Alrajeh2015_283}
combined model checking and logic-based learning to support automated program repair.
\citet{Cai2019_284} also combined model-checking and \ml\ for program repair.
\citet{Shim2020_301} used inductive program synthesis ({\sc DeeperCoder}), 
by creating a simple Domain Specific Language ({\sc dsl}), 
and \ml\ to generate computer programs that satisfies user requirements and specification.
\citet{Sakkas2020_366} combined type rules and \ml\
(\ie{} multi-class classification, \DNN{}s, and \MLP{})
for repairing compile errors.

\subtask{Probabilistic predictions}
Here, we list papers that use probabilistic learning
and \ml\ approaches
such as
association rules, \dt{}, and \svm{}
to predict bug locations and fixes for automated program repair.
\citet{Long2016_285}
introduced a repair tool called {\sc Prophet},
which uses a set of successful manual patches from open-source software repositories,
to learn a probabilistic model of correct code, and generate patches.
\citet{Soto2018_292}
conducted a granular analysis using different statement kinds to
identify those statements that are more likely to be modified than others
during bug fixing. For this, they used simplified syntax trees and association rules.
\citet{Gopinath2014_296} presented a data-driven approach for fixing of bugs in
database statements.
For  predicting the correct behavior for defect-inducing data,
this study uses \svm{} and \dt{}.
\citet{Saha2017_306} developed  {\sc Elixir} repair approach
that uses \logr\ models and similarity-score metrics.
\citet{Bader2019_312} developed a repair approach called {\sc Getafix}
that uses hierarchical clustering to summarize fix patterns
into a hierarchy ranging from general to specific patterns.
\citet{Xiong2018_327} introduced {\sc L}2{\sc S}
that uses \ml\ to estimate conditional probabilities
for the candidates at each search step,
and search algorithms to find the best possible solutions.
\citet{Gopinath2016_348} used \svm{} and {\sc ID}3 with path exploration
to repair bugs in complex data structures.
\citet{Le2015_352} conducted an empirical study on the capabilities of program repair tools,
and applied \rf{} to predict whether using genetic programming search in {\sc apr}
can lead to a repair within a desired time limit.
\citet{893_Aleti2021} used the most significant features as inputs to
\rf{}, \svm{}, \dt{}, and \textit{multi-layer perceptron} models.

\subtask{Recurrent neural networks}
\dl{}
approaches such as \RNN{}s (e.g., \LSTM{} and Transformer)
have been used
for synthesizing new code statements by learning
patterns from a previous list of code statement,
\ie{} this techniques can be used to mainly predict the next statement.
Such approaches often leverage word embeddings.
\citet{Dantas2019_289}
combined {\sc Doc2Vec} and \LSTM{},
to capture dependencies between source code statements,
and improve the fault-localization step of program repair.
\citet{Ahmed2018_293} developed a repair approach ({\sc Tracer})
for fixing compilation errors using \RNN{}s.
Recently, \citet{Li2020_303} introduced {\sc DLFix},
which is a context-based code transformation learning for automated program repair.
{\sc DLFix} uses \RNN{}s and treats automated program repair as code transformation learning,
by learning patterns from prior bug fixes and the surrounding code contexts of those fixes.
\citet{Svyatkovskiy2020_321} presented {\sc IntelliCode}
that uses a Transformer model that predicts sequences of code tokens of arbitrary types,
and generates entire lines of syntactically correct code.
\citet{Chen2016_323} used the \LSTM{}
for synthesizing {\tt if}--{\tt then} constructs.
Similarly, \citet{Vasic2019_333} applied the \LSTM{}
in multi-headed pointer networks for jointly learning to
localize and repair variable misuse bugs.
\citet{Bhatia2018_337} combined neural networks, and in particular \RNN{}s, with constraint-based reasoning to repair syntax errors in buggy programs.
\citet{Chen2019_350} applied \LSTM{} for sequence-to-sequence learning
achieving end-to-end program repair through the {\sc SequenceR} repair tool they developed.
\citet{Majd2020_353} developed {\sc SLDeep},
statement-level software defect prediction,
which uses \LSTM{} on static code features.

Apart from above-mentioned techniques,
\citet{889_White2019} developed DeepRepair, a recursive unsupervised deep learning-based approach, that automatically creates a representation of source code that accounts for the structure and semantics of lexical elements. The neural network language model is trained from the file-level corpus using embeddings.

\subsubsection{Program translation\\}
In this section,
we list studies that use \ml\
that can be used, for instance,
for translating source code from one programming language
to another by learning source-code patterns.
\citet{Le2020_297} presented a survey on \dl\ techniques
including machine translation algorithms and applications.
\citet{Chakraborty2020_291} developed a technique called {\sc codit}
that automates code changes for bug fixing using tree-based neural machine translation.
In particular, they proposed a tree-based neural machine translation model to learn the probability distribution of changes in code.
They evaluate {\sc codit} on a dataset of 30k real-world changes and 6k patches.
The evaluation reveals that {\sc codit} can effectively learn and suggest patches, as well as learn specific bug fix patterns on {\sc Defects4J}.
\citet{Oda2015_325} used statistical machine translation ({\sc smt})
and proposed a method to automatically generate pseudo-code from source code
for source-code comprehension.
To evaluate their approach
they conducted experiments,
and generated English or Japanese pseudo-code from Python statements using {\sc smt}.
Then, they found that the generated pseudo-code is mostly accurate, and it can facilitate code understanding.\subsection{Quality assessment}
The \textit{quality assessment} category has sub-categories \textit{code smell detection},
\textit{clone detection}, and
\textit{quality assessment/prediction}.
In this section, we elaborate upon the state-of-the-art related to each of these
categories within our scope.

\subsubsection{Code smell detection\\}
\begin{figure*}[!h]
	\centering
	\includegraphics[width=\textwidth]{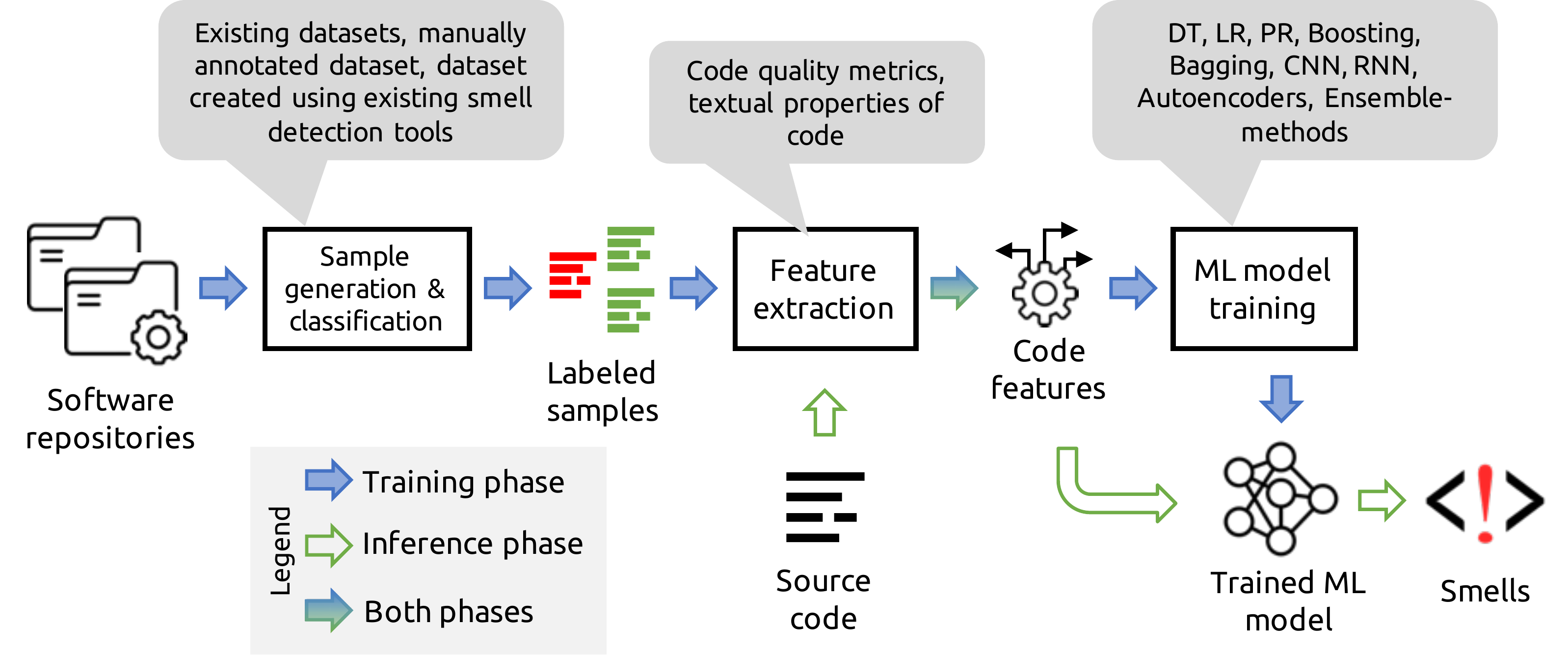}
	\caption{Overview of the code smell detection process pipeline}
	\label{fig:code-smells-overview}
\end{figure*}

Code smells impair the code quality and make the software difficult to extend and maintain \cite{Sharma2018}.
Extensive literature is available on detecting smells automatically \cite{Sharma2018}; \ml\ techniques have been used
to classify smelly snippets from non-smelly code. 
Figure \ref{fig:code-smells-overview} presents a common workflow for code smells detection using \ml{}.
First, source code is pre-processed to extract individual samples (such as a class, file, or method).
These samples are classified into positive and negative samples.
Afterwards, relevant features are identified from the source code and those features are then fed into an \ml{} model
for training.
The trained model classifies a source code sample into a smelly or non-smelly code.

\task{Sample generation and classification}
The process of identifying code smells requires a dataset as a ground truth for training an \ml\ model.
Each sample of the training dataset must be tagged appropriately as smelly sample (along with target smell types)
or non-smelly sample. 
Many authors built their datasets tagged manually with annotations.
For example, \citet{Fakhoury2018_232} developed a manually validated oracle containing
$1,700$ instances of linguistic smells.
\citet{Pecorelli2019_250} created a dataset of 8.5 thousand samples of smells
from $13$ open-source projects.
Some authors \cite{Al-Jamimi2013_244, Mhawish2020_252, 767_Cruz2020, 824_Jain2021, 840_Hadj-Kacem2019} employed existing datasets (Landfill and Qualitas) in their studies.
\citet{784_Tummalapalli2022, 793_Tummalapalli2020, 796_Tummalapalli2021} used 226 WSDL files from the tera-PROMISE dataset.  
\citet{Oliveira2020_195} relied on historical data and mined smell instances from history where
the smells are refactored.

Some efforts such as one by \citet{Sharma2021_510} used CodeSplit \cite{CodeSplitJava, CodeSplitCS} first
to split source code files into individual classes and methods.
Then, they used existing smell detection tools \cite{Designite, DesigniteJava}
to identify smells in the subject systems.
They used the output of both of these tasks to identify and segregate positive and negative samples.
Similarly, \citet{814_Kaur2021} used smells identified by \textit{Dr Java}, \textit{EMMA}, and \textit{FindBugs} as their gold-set.
\citet{820_Alazba2021} and \citet{816_Dewangan2021} used the dataset manually labelled instances detected by four code smell detector tools (\ie{} iPlasma, PMD, Fluid Tool, Anti-Pattern Scanner, and Marinescu's detection rule). The dataset labelled six code smells collected from 74 software systems.
\citet{835_Zhang2021} proposed a large dataset BrainCode consisting $270,000$ samples from 20 real-world applications. The study used iPlasma to identify smells in the subject systems.

\citet{Liu2019_512} adopted an usual mechanism to identify their positive and negative samples.
They assumed that popular well-known open-source projects are well-written and hence
all of the classes/methods of these projects are by default considered free from smells.
To obtain positive samples, they carried out \textit{reverse refactoring} \textit{e.g.,}
moving a method from a class to another class to create an instance of feature envy smell.

\task{Feature extraction}
The majority of the articles \cite{Barbez2020_180, Kaur2017_184, Kumar2018_189, Gupta2019_190, Agnihotri2020_193, Oliveira2020_195, Pritam2019_196, Fontana2013_207, ArcelliFontana2017_209, Fontana2015_211, Thongkum2020_217, Cruz2020_218, Amorim2015_223, Cunha2020_231, Mhawish2020_252, Liu2019_512, Hadj-Kacem2018_511, 749_Tummalapalli2019, 767_Cruz2020, 784_Tummalapalli2022, 787_Saidani2020, 793_Tummalapalli2020, 796_Tummalapalli2021, 814_Kaur2021, 815_Gupta2021, 816_Dewangan2021, 820_Alazba2021, 824_Jain2021, 835_Zhang2021, 847_Gupta2021} 
in this category use object-oriented metrics as features.
These metrics include class-level metrics 
(such as \textit{lines of code, lack of cohesion among methods, number of methods,
fan-in} and \textit{fan-out}) and method-level metrics (such as \textit{parameter count, lines of code, cyclomatic complexity,} and
\textit{depth of nested conditional}).
We observed that some of the attempts use a relatively small number of metrics
(\citet{Thongkum2020_217} and \citet{Agnihotri2020_193} used 10 and 16 metrics, respectively).
However, some of the authors chose to experiment with a large number of metrics.
For example, \citet{Amorim2015_223} employed 62, \citet{Mhawish2020_252} utilized 82,
and \citet{ArcelliFontana2017_209} used 63 class-level metrics and 84 method-level metrics.

Some efforts diverge from the mainstream usage of using metrics as features and used alternative features.
\citet{Lujan2020_181} used warnings generated from existing static analysis tools as features.
Similarly, \citet{Ochodek2019_257} analyzed individual lines in source code to extract textual properties
such as regex and keywords to formulate a set of vocabulary based features (such as bag of words).
\citet{795_Tummalapalli2021} and \citet{823_Gupta2021} used distributed word representation techniques such as Term frequency-inverse Document Frequency (TFIDF), 
Continuous Bag Of Words (CBW), 
Global Vectors for Word Representation (GloVe), and Skip Gram.
Similarly, \citet{840_Hadj-Kacem2019} generated \abst{} first and obtain the corresponding vector representation to train a model for smell detection.
Furthermore, \citet{Sharma2021_510} hypothesized that \dl\ methods can infer the features
by themselves and hence explicit feature extraction is not required.
They did not process the source code to extract features and feed the tokenized code to \ml\ models.

\task{ML model training}
The type of \ml\ models usage can be divided into three categories.

\subtask{Traditional \textsc{ml} models}
	In the first category, we can put studies that use one or more traditional \ml\ models.
	These models include \dt{}, \svm{}, \rf{}, \nb{}, \logr{}, \lr{}, \poly{}, \bg{}, and \mlp{}.
	The majority of studies~\cite{Lujan2020_181, Kumar2018_189, Gupta2019_190, Agnihotri2020_193, Oliveira2020_195, Pritam2019_196, Fontana2013_207, Fontana2015_211, Pecorelli2019_212, Thongkum2020_217, Cruz2020_218, DiNucci2018_220, Cunha2020_231, 749_Tummalapalli2019, 767_Cruz2020, 795_Tummalapalli2021, 796_Tummalapalli2021, 814_Kaur2021, 816_Dewangan2021, 820_Alazba2021, 823_Gupta2021, 824_Jain2021, 840_Hadj-Kacem2019, 847_Gupta2021}
	in this category compared the performance of various \ml\ models.
	Some of the authors experimented with individual \ml\ models;
	for example, \citet{Kaur2017_184} and \citet{Amorim2015_223} used \svm{} and
	\dt{}, respectively, for smell detection.
	
\subtask{Ensemble methods}
	The second category of studies employed ensemble methods to detect smells.
	\citet{Barbez2020_180} and \citet{Tummalapalli2020_254} experimented with ensemble
	techniques such as \textit{majority training ensemble} and \textit{best training ensemble}.
	\citet{787_Saidani2020} used the Ensemble Classifier Chain (ECC) model that transforms multi-label problems into several single-label problems to find the optimal detection rules for each anti-pattern type.
	
\subtask{\textsc{dl}-based models}
	Studies that use \dl\ form the third category.
	\citet{Sharma2021_510} used {\sc cnn, rnn (lstm)}, and Autoencoders-based \dl{} models.
	\citet{Hadj-Kacem2018_511} employed Autoencoder-based \dl\ model to first reduce the dimensionality of data
	and Artificial Neural Network to classify the samples into smelly and non-smelly instances.
	\citet{Liu2019_512} deployed four different \dl\ models based on \CNN{} and \RNN{}.
	It is common to use other kinds of layers (such as embeddings, dense, and dropout) along with \CNN{} and \RNN{}.
	\citet{815_Gupta2021} used eight \dl\ models and \citet{835_Zhang2021}
	proposed Metric–Attention-based Residual network (MARS) to detect brain class/method. MARS used metric–attention mechanism to calculate the weight of code metrics and detect code smells. 

\subtask{Discussion}
A typical \ml\ model trained to classify samples into either smelly or non-smelly samples.
The majority of the studies focused on a relatively small set of known code smells---
\textit{god class }\cite{Barbez2020_180, Lujan2020_181, Kaur2017_184, Gupta2019_190, Agnihotri2020_193, Oliveira2020_195, Fontana2013_207, Grodzicka2020_208, ArcelliFontana2017_209, Cruz2020_218, Caram2019_241, Hadj-Kacem2018_511},
\textit{feature envy} \cite{Barbez2020_180, Kaur2017_184, Agnihotri2020_193, Fontana2013_207, ArcelliFontana2017_209, Fontana2015_211, Cruz2020_218, Sharma2021_510, 
	Hadj-Kacem2018_511},
\textit{long method }\cite{Kaur2017_184, Gupta2019_190, Fontana2013_207, Grodzicka2020_208, ArcelliFontana2017_209, Fontana2015_211, Cruz2020_218, Azeem2019_240, Hadj-Kacem2018_511},
\textit{data class }\cite{Kaur2017_184, Oliveira2020_195, Fontana2013_207, Grodzicka2020_208, ArcelliFontana2017_209, Fontana2015_211},
and \textit{complex class }\cite{Lujan2020_181, Gupta2019_190, Oliveira2020_195}.
Results of these efforts vary significantly; F1 score of the \ml\ models vary between 0.3 to 0.99.
Among the investigated  \ml\ models, authors widely report that \dt{} \cite{Azeem2019_240, Fontana2015_211, AL-Shaaby2020_198, Gupta2019_190}
and \rf{} \cite{Azeem2019_240, Fontana2015_211, Kumar2018_189, ArcelliFontana2017_209, Mhawish2020_252}
perform the best.
Other methods that have been reported better than other \ml\ models in their respective studies are
\svm{} \cite{Tummalapalli2020_254}, \textit{Boosting} \cite{Luiz2019_242}, and \textit{Autoencoders} \cite{Sharma2021_510}.

Traditional \ml{} techniques are the prominent choice in this category because these techniques works well with fixed size, fixed column meaning vectors. 
Code quality metrics capture the features relevant to identify smells and they are fixed size, fixed column meaning vectors.
However, such vectors do not capture subjectivity inherent in the context and hence some studies rely on alternative features such as embeddings generated from \abst{} to feed to \dl{} models such as \RNN{}.

\subsubsection{Code clone detection\\}
\begin{figure*}[!h]
	\centering
	\includegraphics[width=\textwidth]{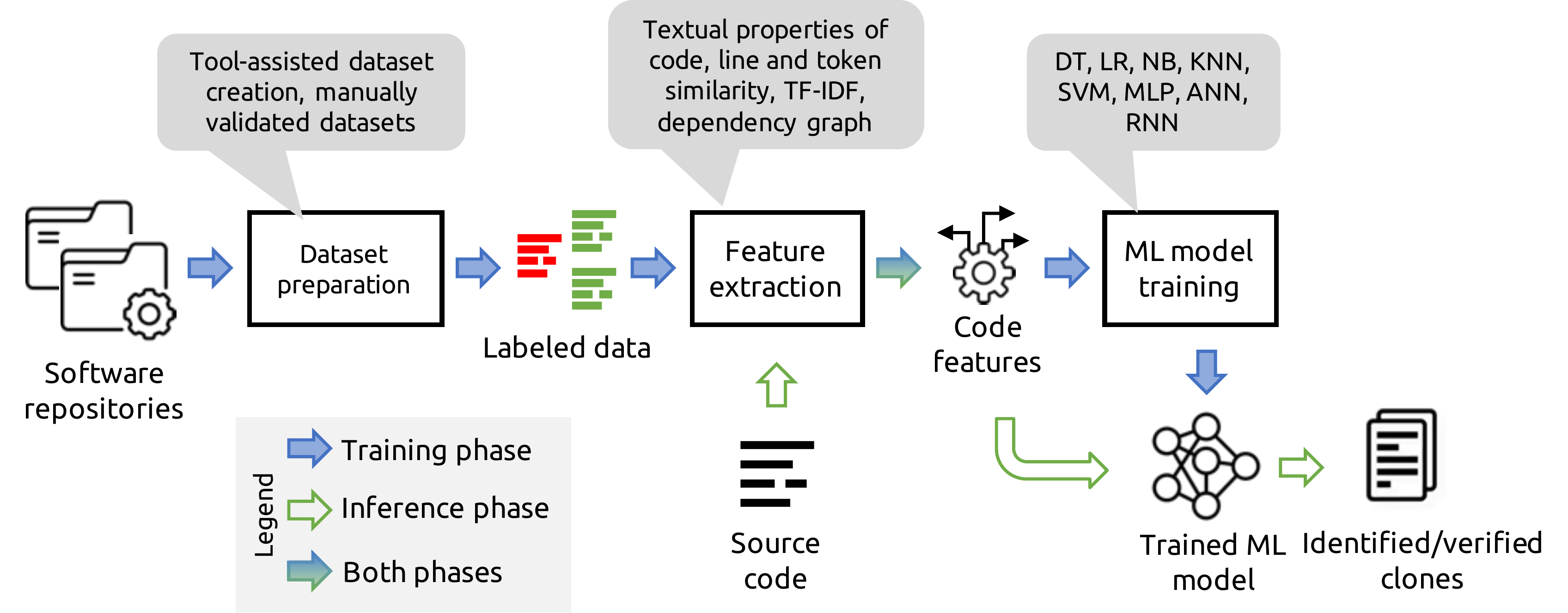}
	\caption{Overview of the code clone detection/validation process pipeline}
	\label{fig:code-clone-overview}
\end{figure*}

Code clone detection is the process of identifying duplicate code blocks in a given software system.
Software engineering researchers have proposed not only methods to detect code clones automatically,
but, also verify whether the reported clones from existing tools are false-positives or not using \ml\ techniques.
Figure~\ref{fig:code-clone-overview} provides an overview of techniques that detect code clones
using \ml{} techniques.
Studies in this category prepare a dataset containing source code samples classified as clones or non-clones.
Then, they apply feature extraction techniques to identify relevant features that are fed into
\ml{} models for training and evaluation.
The trained models identify clones among the sample pairs.

\task{Dataset preparation}
Manual annotation is a common way to prepare a dataset for applying \ml{} to identify
code clones \cite{Mostaeen2019_203, Mostaeen2020_178, White2016_215}.
\citet{Mostaeen2019_203} used a set of tools (NiCad, Deckard, iClones, CCFinderX and SourcererCC) to first
identify a list of code clones; they then manually validated each of the identified clone set.
\citet{Yang2014_202} used existing code clone detection tools to generate their training set.
Some authors (such as \citet{Bandara2011_179} and \citet{758_Hammad2021}) relied on existing code-clone datasets.
\citet{753_Zhang2021} used NiCad to detect all clone groups from each version of the software. The study mapped the clones from consecutive version and used the mapping to predict clone consistency at both the clone-creating and the clone-changing time.
\citet{Bui2018_214} deployed an interesting mechanism to prepare their code-clone dataset.
They crawled through {\sc GitHub} repositories to find different implementations of sorting algorithms;
they collected 3,500 samples from this process.

\task{Feature extraction}
The majority of the studies relied on the textual properties of the source code as features.
\citet{Bandara2011_179} identified features such as the number of characters and words, identifier count,
identifier character count, and underscore count using {\sc antlr} tool.
Some studies~\cite{Mostaeen2019_203, Mostaeen2020_178, Mostaeen2018_251} utilized line similarity
and token similarity.
\citet{Yang2014_202} and \citet{758_Hammad2021} computed \TFIDF{} along with other metrics such as position of clones in the file.
\citet{Cesare2013_204} extracted 30 package-level features including
the number of files, hashes of the files, and common filenames as they detected code clones at the package level.
\citet{753_Zhang2021} obtained code attribute set (\textit{e.g.,} lines of code and the number of parameters), context attribute set (\textit{e.g.,} method name similarity, and sum of parameter similarity).
Similarly, \citet{Sheneamer2016_260} obtained metrics such as the number of constructors,
number of field access, and super-constructor invocation from the program \abst{}.
They also employed program dependence graph features such as \textit{decl\_assign} and \textit{control\_decl}.
Along the similar lines, \citet{Zhao2018_216} used {\sc cfg} and {\sc dfg} (Data Flow Graph) for clone detection.
Some of the studies~\cite{Bui2018_214, White2016_215, Fang2020_225}
relied on \dl{} methods to encode the required features automatically without specifying an explicit set of features.

\task{ML model training}

\subtask{Traditional \ml{} models}
The majority of studies \cite{Mostaeen2020_178, Bandara2011_179, Mostaeen2018_251, Sheneamer2016_260, 753_Zhang2021}
experimented with a number of \ml{} approaches.
For example, \citet{Mostaeen2020_178} used \bn{}, \logr{}, and \dt{};
\citet{Bandara2011_179} employed \nb{}, \knn{}, \ada{}.
Similarly, \citet{Sheneamer2016_260} compared the performance of \svm{}, \lda{},
\ibk{}, \lkm{}, \dt{}, \nb{}, \mlp{}, and \lb{}.

\subtask{\dl{}-based models}
\dl{} models such as \ANN{} \cite{Mostaeen2019_203, Mostaeen2018_251}, \DNN{} \cite{Fang2020_225, Zhao2018_216},
and \RNN{} with \rtnn{} \cite{White2016_215} are also employed extensively.
\citet{Bui2019_200} and \citet{Bui2018_214} combined neural networks for \ml\ models training.
Specifically, \citet{Bui2019_200} built a \textit{Bilateral neural network} on top of two underlying sub-networks,
each of which encodes syntax and semantics of code in one language.
\citet{Bui2018_214} constructed  BiTBCNNs---a combination layer of sub-networks to encode
similarities and differences among code structures
in different languages. 
\citet{758_Hammad2021} proposed a Clone-Advisor, a \DNN{} model trained by fine-tuning GPT-2 over the BigCloneBench code clone dataset, for predicting code tokens and clone methods.

\subsubsection{Quality assessment/prediction\\}
Studies in this category assess or predict issues related to various quality attributes such as reliability,
maintainability, and run-time performance.
The process starts with dataset pre-processing and labeling
to obtain labeled data samples.
Feature extraction techniques are applied on the processed samples.
The extracted features are then fed into an \ml{} model for training.
The trained model assesses or predicts the quality issues in the analyzed source code.

\task{Dataset preprocessing and labeling}
\citet{Heo2017_246} generated data to train an \ml{} model
in pursuit to balance soundness and relevance in static analysis by selectively
allowing unsoundness only when it is likely to reduce false alarms.
\citet{Ribeiro2019_256}  used ensemble learning to learn from
multiple static analyzers and show that ensemble learning improves the accuracy. 
Specifically, they took three static analyzers (Clang-analyzer, CppCheck, and Frama-C)
and detected issues in Juliet dataset. 
Once the report is generated from all three tools, the authors combined the reports by converting them to a uniform format. 
Then, they tagged the samples.
Similarly, \citet{Alikhashashneh2018_268} used the Understand tool to detect various metrics,
and employed them on the Juliet test suite for C++.
\citet{920_Reddivari2019} extracted a subset of data belonging to open source projects such as Ant, Tomcat, and Jedit to predict reliability and maintainability using \ml{} techniques.
\citet{924_Malhotra$^1$2012} also prepared a custom dataset using two proprietary software systems as their subjects to predict maintainability of a class.

\task{Feature extraction}
\citet{Heo2017_246} extracted 37 low-level code features for loop (such as 
number of \texttt{Null}, array accesses, and number of exits) and library call constructs
(such as parameter count and whether the call is within a loop).
\citet{Ribeiro2019_256}  generated features only from the warnings (such as redundancy level and
number of warnings in the same file).
Some studies~\cite{Alikhashashneh2018_268, Kim2009_259, 920_Reddivari2019, 924_Malhotra$^1$2012} used source code metrics as features.

\task{ML model training}
\citet{Kim2009_259} used \svm{} to identify risky modules from a software system.
\citet{Alikhashashneh2018_268} employed \rf{}, \svm{}, \knn{}, and \dt{} to classify static code analysis tool warnings
as true positives, false positives, or false negatives.
\citet{920_Reddivari2019} predicted reliability and maintainability using the similar set of \ml{} techniques.
The study by~\citet{Ribeiro2019_256} claimed that ensemble methods such as \ada{}
works superior than standalone \ml{} methods.
Anomaly-detection techniques such as \textit{One-class} \svm{} have been used by~\citet{Heo2017_246}.
They applied their method on taint analysis and buffer overflow detection
to improve the recall of static analysis.
Whereas, some other studies~\cite{Ribeiro2019_256, Alikhashashneh2018_268} aimed to rank and
classify static analysis warnings.
\citet{Kim2009_259} estimated risky modules in the subject system.\subsection{Code completion}
Code auto-completion
is a state-of-the-art integral feature of modern source-code editors and {\sc ide}s~\cite{Bruch2009_500}.
The latest generation of auto-completion methods uses {\sc nlp} and advanced \ml\ models,
trained on publicly available software repositories,
to suggest source-code completions,
given the current context of the software-projects under examination.

\task{Data collection}
The majority of the studies mined a large number of repositories to construct their own dataset.
Specifically,
\citet{Gopalakrishnan2017_459} examined 116,000 open-source systems
to identify correlations between the latent topics in source code and the usage of 
architectural developer tactics (such as authentication and load-balancing).
\citet{Han2009_488, Han2011_489}
trained and tested their system by sampling
4,919 source code lines from open-source projects.
\citet{Raychev2016_504} used
large codebases from {\sc GitHub} to make 
predictions for JavaScript and Python code completion.
\citet{Svyatkovskiy2019_505} used 2,700 Python open-source software {\sc GitHub} repositories
for the evaluation of their novel approach, Pythia.

The rest of the approaches employed existing benchmarks and datasets.
\citet{Rahman2020_484} trained their proposed model using
the data extracted from Aizu Online Judge ({\sc aoj}) system.
\citet{Liu2020_485,Liu2020_501} performed experiments
on three real-world datasets
to evaluate the effectiveness of their model when compared with the state-of-the-art approaches.
\citet{Li2018_491} conducted experiments on two  datasets
to demonstrate the effectiveness of their approach consisting of an attention mechanism
and a pointer mixture network on code completion tasks.
\citet{Phan2020_506} used  three corpus for their experiments---a
large-scale corpus of English-German translation in 
{\sc nlp}~\cite{luong17},
the Conala corpus~\cite{YDC18}, which contains
Python software documentation as 116,000
English sentences, and
the {\sc msr} 2013 corpus~\cite{AS13}.
\citet{Schuster2021_508}
used a public archive of {\sc GitHub} from 2020~\cite{misc20}.

\task{Feature extraction}
Studies in this category extract source code information in variety of forms.
\citet{Gopalakrishnan2017_459} extracted relationships
between topical concepts in the source code
and the use of specific architectural developer tactics in that code.
~\citet{Phan2020_506} used machine translation
to learn the mapping from prefixes to code tokens for code suggestion.
They extracted the tokens from the documentation of the source code.
\citet{Liu2020_485,Liu2020_501} introduced a self-attentional neural architecture
for code completion with multi-task learning.
To achieve this, they extracted the hierarchical source code structural information from the programs considered.
Also, they captured the long-term dependency in the input programs,
and derived knowledge sharing between related tasks.
\citet{Li2018_491} used locally repeated terms in program source code
to predict out-of-vocabulary (OoV) words that restrict the code completion.
Chen and Wan~\cite{Chen2019_503} proposed a tree-to-sequence (Tree2Seq) model that
captures the structure information of source code
to generate comments for source code.
\citet{Raychev2016_504}
used {\sc ast}s and performed prediction
of a program element on a dynamically computed context.
\citet{Svyatkovskiy2019_505}
introduced a novel approach for code completion called Pythia,
which exploits state-of-the-art large-scale \dl\ models trained
on code contexts extracted from {\sc ast}s.

\task{ML model training}
The studies can be classified based on the used \ml\ technique for code completion.

\subtask{Recurrent Neural Networks}
 For code completion,
 researchers mainly try to predict the next token.
 Therefore, most approaches use \RNN{}s.
 In particular, \citet{Terada2019_487} used \LSTM{}
 for code completion to facilitate programming education.
 \citet{Rahman2020_484}
 also used \LSTM{}.
 \citet{Wang2019_494} used
 \LSTM{}-based neural network combined with several techniques such as
\textit{Word Embedding} models and \textit{Multi-head Attention Mechanism}
 to complete programming code.
 \citet{Zhong2019_499}
 applied several \dl\ techniques, including \LSTM{}, \textit{Attention Mechanism} (AM),
 and \textit{Sparse Point Network} ({\sc spn})
 for JavaScript code suggestions.
 
 Apart from \LSTM{},
 researchers have used \RNN{}
 with different approaches
 to perform code suggestions.
 \citet{Li2018_491}
 applied neural language models, which involve attention mechanism for \RNN{},
 by learning from large codebases
 to facilitate effective code completion for dynamically-typed programming languages.
 \citet{Hussain2020_492} presented {\sc CodeGRU}
 that uses  \GRU{}
 for capturing source codes contextual,
 syntactical, and structural dependencies.
 \citet{Yang2019_496} presented {\sc rep}
 to improve language modeling for code completion.
 Their approach uses learning of general token repetition
 of source code with optimized memory, and it outperforms \LSTM{}.
 \citet{Schumacher2020_497}
 combined neural and classical \ml\,
 including \RNN{}s,
 to improve code recommendations.
 
\subtask{Probabilistic Models}
 Earlier approaches for code completion
 used statistical learning for recommending code elements.
 In particular, \citet{Gopalakrishnan2017_459}
 developed a recommender system using prediction models
 including neural networks for latent topics.
 \citet{Han2009_488, Han2011_489} applied \textit{Hidden Markov Models}
 to improve the efficiency of code-writing
 by supporting code completion of multiple keywords
 based on non-predefined abbreviated input.
 \citet{Proksch2015_498} used \textit{Bayesian Networks}
 for intelligent code completion.
 \citet{Raychev2016_504}
 utilized a probabilistic model for code in any programming language with \dt{}.
 \citet{Svyatkovskiy2019_505}
 proposed {\sc Pythia}
 that employs a \textit{Markov Chain} language model.
 Their approach can generate ranked lists of
 methods and {\sc api} recommendations,
 which can be used by developers while writing programs.
 
\subtask{Other techniques}
 Recently, new approaches have been developed for code completion
 based on multi-task learning, code representations,
 and {\sc nmt}. 
 For instance, \citet{Liu2020_485,Liu2020_501}
 applied Multi-Task Learning ({\sc mtl})
 for suggesting code elements.
 \citet{Lee2021_495} developed {\sc MergeLogging},
 a \dl\-based merged network
 that uses code representations for automated logging decisions.
\citet{Chen2019_503} applied {\sc Tree2Seq} model
 with {\sc nmt} techniques
 for code comment generation.
 \citet{Phan2020_506} proposed {\sc PrefixMap},
 a code suggestion tool for all types of code tokens in the Java programming language.
 Their approach uses statistical machine translation
 that outperforms {\sc nmt}.

Program comprehension techniques attempt to understand the theory of
comprehension process of developers as well as the tools, techniques, and processes that
influence the comprehension activity  \cite{Storey2005}.
We summarized, in the rest of the section, program comprehension
studies into four sub-categories \ie\ code summarization, program classification, change analysis, and
entity identification/recommendation.

\subsubsection{Code summarization\\}
\begin{figure*}[!h]
	\centering
	\includegraphics[width=\textwidth]{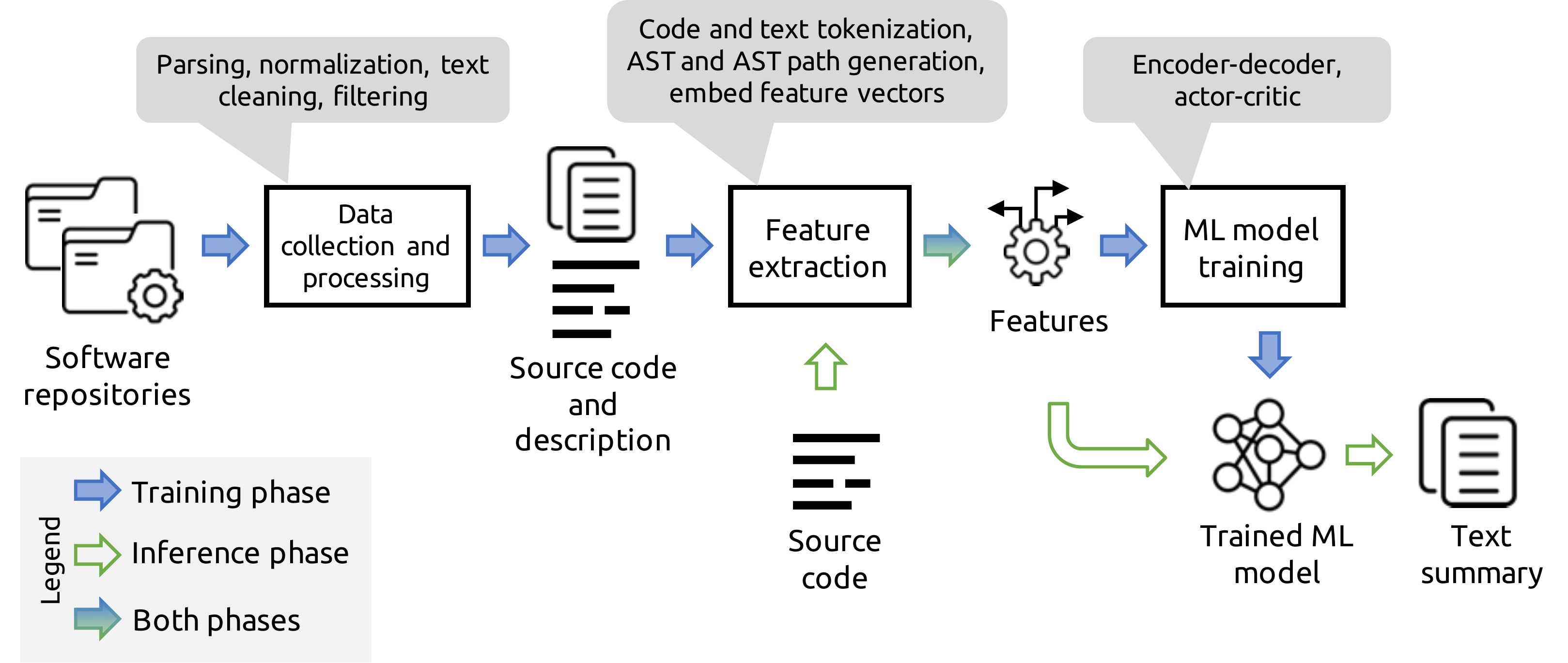}
	\caption{Overview of the code summarization process pipeline}
	\label{fig:summarization-overview}
\end{figure*}

Code summarization techniques attempt to provide a consolidated summary of the source code entity (typically a method).
A variety of attempts has been made in this direction.
The majority of the studies~\cite{Chen2018_406, LeClair2019_407, Liu2019_408, Ahmad2020_410, Shido2019_415, Yao2019_418, Hu2018_421, Li2020_422, Wang2020_424, LeClair2020_426, Ye2020_428, Wang2020_434, Zhang2020_435, Iyer2016_440, 762_Li2021, 982_Zhou2022, 989_Haque2020, 993_Zhou2021}
produces a summary for a small block (such as a method). 
This category also includes studies that summarize small code fragments~\cite{Nazar2015_437}, 
code folding within {\sc ide}s~\cite{Viuginov2019_405},
commit message generation \cite{Jiang2017_416, Liu2018_430, 952_Jiang2017, 955_Jiang2019, 988_Chen2021, 992_Wang2020},
and title generation for online posts from code \cite{Gao2020_441}.
Figure~\ref{fig:summarization-overview} provides an overview of the mechanism used by code summarization techniques.

\task{Data collection and processing}
The majority of the studies~\cite{Allamanis2016_403, Chen2018_406, LeClair2019_407, Liu2019_408, Ahmad2020_410, Hu2018_421, Chen2019_420, Li2020_422, Wang2020_424, Wan2018_427, Wang2020_434, 988_Chen2021, 993_Zhou2021}
in this category prepares pairs of code snippets and their corresponding natural language description.
Specifically, \citet{Chen2018_406} used more than 66 thousand pairs of C\# code and natural language description where
source code is tokenized using a modified version of the {\sc antlr} parser.
\citet{Ahmad2020_410} conducted their experiments on a dataset containing Java and Python snippets;
sequences of both the code and summary tokens are represented by a sequence of vectors.
\citet{Hu2018_421} and \citet{Li2020_422} prepared a large dataset from 9,714 {\sc GitHub} projects.
Similarly, \citet{Wang2020_424} mined code snippets and corresponding \texttt{javadoc} comments
for their experiment.
\citet{Chen2019_420} created their dataset from 12 popular open-source Java libraries with more than 10 thousand stars.
They considered method bodies as their inputs and method names along with method comments
as prediction targets.
\citet{998_Psarras2019} prepared their dataset by using Weka, SystemML, DL4J, Mahout, Neuroph, and Spark as their subject systems. The authors retained names and types of methods, and local and class variables.
\citet{Choi2020_438} collected and refined more than 114 thousand pairs of methods and corresponding code annotations from 100 open-source Java projects.
\citet{Iyer2016_440} mined StackOverflow and extracted title and code snippet from posts that contain exactly one code snippet.
Similarly, \citet{Gao2020_441} used a dump of StackOverflow dataset. 
They tokenized code snippets with respect to each programming language for pre-processing.
The common steps in preprocessing identifiers include making them lower case, splitting
the camel-cased and underline identifiers into sub-tokens, and normalizing the code with special tokens such as "VAR" and "NUMBER".
\citet{Nazar2015_437} used human annotators to summarize 127 code fragments retrieved from Eclipse and NetBeans official frequently asked questions.
\citet{Yang2021_522} built
a dataset with over 300K 
pairs of method and comment to evaluate their approach.
\citet{988_Chen2021} used dataset provided by \citet{Hu2018_421} and manually categorize comments into six intention categories for 20,000 code-comment pairs.
\citet{992_Wang2020} created a Python dataset that contains 128 thousand code-comment pairs.
\citet{1060_Lal2017} crawled over 6700 Java projects from Github to extract their methods and the corresponding Javadoc comments to create their dataset.

\citet{955_Jiang2019} used 18 popular Java projects from GitHub to prepare a dataset with approximately 50 thousand commits to generate commit messages automatically.
\citet{1010_Liu2020} processed 56 popular open-source projects and selected approximately 160K commits after filtering out the irrelevant commits.
\citet{1014_Liu2019} used RepoRepears to identify Java repositories to process. They collected pull-request meta data by using GitHub APIs. After preprocessing the collected information, they trained a model to generate pull request description automatically.
\citet{1015_Wang2021} prepared a dataset of 107K commits by mining 10K open-source repositories to generate context-aware commit messages.

Apart from source code, some of the studies used additional information generated from source code.
For example, \citet{LeClair2019_407} used \abst\ along with code and their corresponding summaries 
belonging to  more than 2 million Java methods.
Likewise, \citet{Shido2019_415} and \citet{Zhang2020_435} also generated \abst{}s of the collected code samples.
\citet{Liu2019_408} utilized call dependencies along with source code and corresponding comments from more than a thousand GitHub repositories.
\citet{LeClair2020_426} employed \abst\ along with adjacency matrix of \abst\ edges.

Some of the studies used existing datasets such as StaQC \cite{StaQC} and the dataset created by~\citet{Jiang2017_416}. 
Specifically, \citet{Liu2018_430, 952_Jiang2017} utilized a dataset of commits provided by~\citet{Jiang2017_416} that contains two million commits from one thousand popular Java projects.
\citet{Yao2019_418} and~\citet{Ye2020_428} used StaQC dataset~\cite{StaQC};
it contains more than 119 thousand pairs of question title and code snippet related to {\sc sql} mined from StackOverflow.
\citet{Xie2021_523} utilized two existing datasets---one each for Java~\cite{Leclair2019} and Python~\cite{Barone2017}.
\citet{Bansal2021_525} evaluated their code summarization technique
using a Java dataset of 2.1M Java methods from 28K projects created by~\citet{Leclair2019}.
\citet{762_Li2021} also used the Java dataset of 2.1M methods \citet{Leclair2019} to predict the inconsistent names from the implementation of the methods.
Simiarly, \citet{989_Haque2020, 994_LeClair2021, 995_Haque2021} relied on the Java dataset by \citet{Leclair2019} for summarizing methods.
\citet{982_Zhou2022} combined multiple datasets for their experiment. The first dataset \cite{Hu2018_421} contains over 87 thousand Java methods. The other datasets contained 2.1M Java methods \cite{Leclair2019} and 500 thousand Java methods respectively.

Efforts in the direction of automatic code folding also utilize
techniques similar to code summarization.
\citet{Viuginov2019_405} collected projects developed using IntelliJ platform.
They identified \texttt{foldable} and \texttt{FoldingDescription} elements from
\texttt{workspace.xml} belonging to 335 JavaScript and 304 Python repositories.

% -----
\task{Feature extraction}
Studies investigated different techniques for code and feature representations.
In the simplest form, \citet{Jiang2017_416} tokenized their code and text.
\citet{952_Jiang2017} extracted commit messages starting from ``verb + object'' and computed TFIDF for each word. 
\citet{995_Haque2021} extracted top-40 most-common action words from the dataset of 2.1m Java methods provided by \citet{Leclair2019}.
\citet{998_Psarras2019} used comments as well as source code elements such as method name, variables, and method definition to prepare bag-of-words representation for each class.
\citet{Liu2019_408}
represented the extracted call dependency features as a  sequence of tokens.

Some of the studies extracted explicit features from code or \abst{}.
For example, \citet{Viuginov2019_405} used 17 languages as independent and 8 languages as dependent features.
These features include \abst\ features such as \textit{depth of code blocks' root node}, \textit{number of \abst\ nodes}, 
and \textit{number of lines in the block}.
\citet{Hu2018_421} and \citet{Li2020_422} transformed \abst\ into Structure-Based Traversal ({\sc sbt}).
\citet{Yang2021_522} developed a \dl\ approach,
{\sc MMTrans}, for code summarization
that learns the representation of source code
from the two heterogeneous modalities of the \abst{},
\ie\ {\sc sbt} sequences and 
graphs. 
\citet{982_Zhou2022} extracted \abst{} and prepared tokenized code sequences and tokenized \abst{} to feed to semantic and structural encoders respectively.
\citet{993_Zhou2021, 1060_Lal2017} tokenized source code and parse them into \abst{}.
\citet{1006_Lin2021} proposed block-wise \abst{} splitting method; they split the code of a method based on the blocks in the dominator tree of the Control Flow Graph, and generated a split \abst{} for each block.
\citet{1010_Liu2020} worked with \abst{} \textit{diff} between commits as input to generate a commit summary.
\citet{1031_Lu2017} used Eclipse JDT to parse code snippets at method-level into \abst{} and extracted API sequences and corresponding comments to generate comments for API-based snippets.
\citet{1033_Huang2020} proposed a statement-based \abst{} traversal algorithm to generate the code token sequence preserving the semantic, syntactic and structural information in the code snippet.

The most common way of representing features in this category is to encode the features in the form of embeddings or feature vectors.
Specifically, \citet{LeClair2019_407} used embeddings layer for code, text, as well as for \abst{}. 
Similarly, \citet{Choi2020_438} transformed each of the tokenized source code into a vector of fixed length through an embedding layer.
\citet{Wang2020_424} extracted the functional keyword from the code and perform positional encoding.
\citet{Yao2019_418} used a code retrieval pre-trained model with natural language query and code snippet and annotated
each code snippet with the help of a trained model.
\citet{Ye2020_428} utilized two separate embedding layers to convert input sequences, belonging to both text and code, into high-dimensional vectors.
Furthermore, some authors encode source code models using various techniques.
For instance, \citet{Chen2019_420} represented every input code snippet as a series of \abst\ paths where
each path is seen as a sequence of embedding vectors associated with all the path nodes.
\citet{LeClair2020_426} used  a single embedding layer for both the source code and \abst\ 
node inputs to exploit a large overlap in vocabulary.
\citet{Wang2020_434} prepared a large-scale corpus of training data where each code sample is represented by
three sequences---code (in text form), \abst{}, and {\sc cfg}.
These sequences are encoded into vector forms using work2vec.
Studies also explored other mechanisms to encode features.
For example,
\citet{Liu2018_430} extracted commit \textit{diffs} and represented them as bag of words.
The corresponding model ignores grammar and word order, but keeps term frequencies.
The vector obtained from the model is referred to as \textit{diff vector}.
\citet{Zhang2020_435} parsed code snippets into \abst{}s and calculated their similarity using \abst{}s. 
\citet{Allamanis2016_403} and \citet{Ahmad2020_410} employed attention-based mechanism to encode tokens.
\citet{762_Li2021} used GloVe, a word embedding technique, to obtain the vector representation of the context; the study included method callers and callee as well as other methods in the enclosing class as the context for a method.
Similarly, \citet{999_Li2021} calculated edit vectors based on the lexical and semantic differences between input code and the similar code.

\task{ML model training} 
The \ml\ techniques used by the studies in this category can be divided into the following four categories.

\subtask{Encoder-decoder  models}
	The majority of the studies used attention-based \ENDE{} models to generate code summaries for code snippets.
	For instance,
	\citet{Gao2020_441} proposed an end-to-end sequence-to-sequence system enhanced with an attention mechanism to perform better content selection.
	A code snippet is transformed by a source-code encoder into a vector representation;
	the decoder reads the code embeddings to generate the target question titles.
	\citet{Jiang2017_416} trained an {\sc ntm} algorithm to ``translate''
	from diffs to commit messages.
	Similarly, \citet{Chen2019_420, Hu2018_421, 955_Jiang2019, 989_Haque2020, 999_Li2021, 1014_Liu2019, 1031_Lu2017, 1034_Takahashi2019} employed \LSTM{}-based \ENDE{}
	model to generate summaries. 
	\citet{Zhang2020_435} proposed \textit{Rencos} in which they first trained an attentional \ENDE{} 
	model to obtain an encoder for all code samples and a decoder for generating natural language summaries.
	Second, the approach retrieves the most similar code snippets from the training set for each input code snippet.
	Rencos uses the trained model to encode the input and retrieves two code snippets as context vectors. 
	It then decodes them simultaneously to adjust the conditional probability of the next word using the similarity values
	from the retrieved two code snippets. 
	\citet{Iyer2016_440} used an attention-based neural network to model the conditional distribution of a natural language
	summary.
	Their approach uses an {\sc lstm} model guided by attention on the source code snippet to generate a summary of one word at a time.
	\citet{Choi2020_438} transformed input source code into a context vector by detecting local structural features
	with \CNN{}s. 
	Also, attention mechanism is used with encoder \CNN{}s to identify interesting locations within the source code. 
	Their last module decoder generates source code summary.
	\citet{Ahmad2020_410} proposed to use Transformer to generate a natural language summary given a piece of source code.
	For both encoder and decoder, 
	the Transformer consists of stacked multi-head attention and parameterized linear transformation  layers. 
	\citet{LeClair2019_407} used attention mechanism to not only attend words in the output summary to words in the code word representation but also to attend the summary words to parts of the \abst{}. 
	The concatenated context vector is used to predict the summary of one word at a time.
	\citet{Yang2021_522}  developed a
	multi-modal transformer-based code summarization approach for smart contracts.
	\citet{Xie2021_523} designed a novel multi-task learning 
	({\sc mlt})  approach  for  code summarization
	through mining the relationship between method-code summaries and method names.
	\citet{Bansal2021_525}
	introduced a project-level encoder \dl\ model
	for code summarization.
	\citet{762_Li2021} used \RNN{}-based encoder-decoder model to generate a code representation of a method and check whether the current method name is inconsistent with the predicted name based on the semantic representation.
    \citet{995_Haque2021} compared five seq2seq-like approaches (\textit{attendgru}, \textit{ast-attendgru}, \textit{ast-attendgru-fc}, \textit{graph2seq}, and \textit{code2seq}) to explore the role of action word identification in code summarization.
	\citet{1015_Wang2021} proposed a new approach, named CoRec, to translate git diffs, using attentional Encoder-Decoder model, that include both code changes and non-code changes into commit messages.
	\citet{1060_Lal2017} presented ContextCC that uses a Seq2Seq Neural Network model with an attention mechanism to generate comments for Java methods.

\subtask{Extended encoder-decoder models}
Many studies extended the traditional \ENDE{} mechanism in a variety of ways.
\citet{Liu2019_408} proposed CallNN that utilizes call dependency information.
They employed two encoders, one for the source code and another for the call dependency sequence.
The generated output from the two encoders are integrated and used in a decoder for the target natural language summarization.
Similarly, \citet{Li2020_422} presented Hybrid-DeepCon model containing two encoders for code and \abst\ along with
a decoder to generate sequences of natural language annotations.
\citet{Shido2019_415} extended {\sc Tree-lstm} and proposed Multi-way {\sc Tree-lstm} as their encoder.
The rational behind the extension is that the proposed approach not only can handle an arbitrary
number of ordered children, but also factor-in interactions among children.
\citet{Wang2020_424} implemented a three step approach. 
In the first step, functional reinforcer extracts the most critical function-indicated tokens from source code
which are fed into the second module code encoder along with source code.
The output of the code encoder is given to a decoder that generates the target sequence by sequentially predicting
the probability of words one by one.
\citet{LeClair2020_426} proposed to use {\sc gnn}-based encoder to encode \abst\ of each method and
\RNN{}-based encoder to model the method as a sequence.
They used an attention mechanism to learn important tokens in the code and corresponding \abst{}.
Finally, the decoder generates a sequence of tokens based on the encoder output.
\citet{Ye2020_428} employed dual learning mechanism by using {\sc Bi-lstm}.
In one direction, the model is trained for code summarization task that takes code sequence as input and summarized into a sequence of text.
On the other hand, the code generation task takes the text sequence and generate code sequence. 
They reused the outcome of both tasks to improve performance of the other task.
	\citet{982_Zhou2022} used two encoders, semantic and structural, to generate summaries for Java methods. Their method combined text features with structure information of code snippets to train encoders with multiple graph attention layers.
\citet{993_Zhou2021} trained two separate \ENDE{} models, one for source code sequence and another for \abst{} via adversarial training, where each model is guided by a well-designed discriminator that learns to evaluate its outputs. 
\citet{1006_Lin2021} used a transformer to generate high-quality code summaries. The learned syntax encoding is combined with code encoding, and fed into the transformer.
\citet{1010_Liu2020} proposed a new approach ATOM that uses the diff between commits as input. The approach used BiLSTM module to generate a new message by using \textit{diff-diff} to retrieve the most relevant commit message. 
	
\subtask{Reinforcement learning models}
Some of the studies exploited reinforcement learning techniques for code summary generation.
In particular, \citet{Yao2019_418} proposed code annotation for code retrieval method that generates an natural language
annotation for a code snippet so that the generated annotation can be used for code retrieval.
They used \textit{Advanced Actor-Critic} model for annotation mechanism and \LSTM\ based model for code retrieval.
\citet{Wan2018_427} and \citet{Wang2020_434} used deep reinforcement learning model for training using annotated code samples.
The trained model is an \textit{Actor} network that generates comments for input code snippets.
The \textit{Critic} module evaluates whether the generated word is a good fit or not.
\citet{992_Wang2020} used a hierarchical attention network for comment generation. The study incorporated multiple code features, including type-augmented abstract syntax trees and program control ﬂows, along with plain code sequences. The extracted features are injected into an actor-critic network.
\citet{1033_Huang2020} proposed a composite learning model, which combines the actor-critic algorithm of reinforcement learning with the encoder-decoder algorithm, to generate block comments.

\subtask{Other techniques}
\citet{952_Jiang2017} used \nb\ to classify the diff files into the verb groups.
For automated code folding, \citet{Viuginov2019_405} used \rf\ and \dt\ to classify whether
a code block needs to be folded.
Similarly, \citet{Nazar2015_437} used \svm\ and \nb\ classifiers to generate summaries from the extracted features.
\citet{988_Chen2021} compared six \ml{} techniques to demonstrate that comment category prediction can boost code summarization to reach better results.
\citet{1013_Etemadi2020} compared NNGen, SimpleNNGen, and EXC-NNGen to explore the origin of nearest diffs selected by the neural network.

%----------------------------------
\subsubsection{Program classification\\}
Studies targeting this category classify software artifacts 
based on programming language \cite{Ugurel2002_394}, application domain \cite{Ugurel2002_394}, and
type of commits (such as buggy and adaptive) \cite{Ji2018_382, Meqdadi2019_387}.
We summarize these efforts below from dataset preparation, feature extraction, and \ml\ model training perspective.

\task{Dataset and benchmarks}
\citet{Ma2018_373} identified more than 91 thousand open-source repositories from {\sc GitHub} as subject systems.
They created an oracle  by manually classifying software artifacts from 383 sample projects.
\citet{Shimonaka2016_381} 
conducted experiments on source code generated by four kinds of code generators
to evaluate their technique that
identify auto-generated code automatically by using \ml\ techniques.
\citet{Ji2018_382} and \citet{Meqdadi2019_387} analyzed the {\sc GitHub} commit history.
\citet{Ugurel2002_394} relied on C and C++ projects from Ibiblio and the Sourceforge archives.
\citet{965_Levin2017} used eleven popular open-source projects and annotated 1151 commits manually to train a model that can classify commits into maintenance activities.
Similarly, \citet{970_Mariano2021} and \citet{971_Mariano2019} classify commits by maintenance activities; they identify a large number of open-source GitHub repositories.
Along the similar lines, \citet{964_Meng2021} classified commits messages into categories such as bug fix and feature addition and \citet{967_Li2019} predicted the impact of single commit on the program.
They used popular a small set (specifically, 5 and 10 respectively) of Java projects as their dataset.
Furthermore, \citet{949_Sabetta2018} proposed an approach to classify security-related commits. To achieve the goal, they used 660 such commits from 152 open-source Java projects that are used in SAP software.
\citet{1022_Gharbi2019} created a dataset containing 29K commits from 12 open source projects.
\citet{947_Abdalkareem2020} built a dataset to improve the detection CI skip commits i.e., commits where `[ci skip]' or `[skip ci]' is used to skip continuous integration pipeline to execute on the pushed commit. To build the dataset, the authors used BigQuery GitHub dataset to identify repositories where at least 10\% of commits skipped the CI pipeline.

\task{Feature extraction}
Features in this category of studies belong to either source code features category or repository features.
A subset of studies \cite{Shimonaka2016_381, Ma2018_373, Ugurel2002_394} relies on features extracted
from source code token including language specific keywords and other syntactic information.
Other studies \cite{Ji2018_382, Meqdadi2019_387} collect repository metrics (such as number of changed statements, methods, hunks, and files)
to classify commits.
\citet{Ben-Nun2018_388} leveraged both the underlying data- and control-flow of a program
to learn code semantics performance prediction.
\citet{1022_Gharbi2019} used {\sc tf-idf} to weight the tokens extracted from change messages.
\citet{1023_Ghadhab2021} curated a set of 768 BERT-generated features, a set of 70 code change-based features and a set of 20 keyword-based features for training a model to classify commits.
Similarly, \citet{970_Mariano2021} and \citet{971_Mariano2019} extracted a 71 features majorly belonging to source code changes and keyword occurrences categories.
\citet{964_Meng2021} and \citet{967_Li2019} computed change metrics (such as number lines added and removed) as well as natural language metrics extracted from commit messages.
\citet{947_Abdalkareem2020} employed 23 commit-level repository metrics.
\citet{949_Sabetta2018} analyzed changes in source code associated with each commit and extracted the terms that the developer used to name entities in the source code (\textit{e.g.,} names of classes).

\task{ML model training}
A variety of \ml\ approaches have been applied.
Specifically, \citet{Ma2018_373} used \svm{}, \dt{}, and \bn{} for artifact classification.
\citet{Meqdadi2019_387} employed \nb{}, \textit{Ripper}, as well as  \dt{} and \citet{Ugurel2002_394}  used \svm{} to classify specific commits.
\citet{Ben-Nun2018_388} proposed an approach based on an \RNN{} architecture and fixed {\sc inst2vec} embeddings
for code analysis tasks.
\citet{965_Levin2017, 970_Mariano2021, 971_Mariano2019} used \dt{} and \rf{} for commits classification into maintenance activities.
\citet{1022_Gharbi2019} applied \logr{} model to determine the commit classes for each new commit message.
\citet{1023_Ghadhab2021} trained a \DNN{} classifier to fine-tune the BERT model on the task of commit classification.
\citet{964_Meng2021} used a \CNN{}-based model to classify code commits.
\citet{949_Sabetta2018} trained \rf{}, \nb{}, and \svm{} to identify security-relevant commits.

%----------------------------------
\subsubsection{Change analysis\\}
Researchers have explored applications of \ml\ techniques to identify or predict  relevant code changes \cite{Tollin2017_374, Tufano2019_390}.

We briefly describe the efforts in this domain \textit{w.r.t.} three major steps---dataset preparation, feature extraction, and \ml\ model training.

\task{Dataset preparation}
\citet{Tollin2017_374} performed their study on two industrial projects.
\citet{Tufano2019_390} extracted 236K pairs of code snippets
identified before and after the implementation of the changes provided in the pull requests.
\citet{742_Kumar2017} used eBay web-services as their subject systems.
\citet{836_Uchoa2021} used the data provided by the Code Review Open Platform (CROP), an open-source dataset that links code review data to software changes, to predict impactful changes in code review.
\citet{918_Malhotra2013} considered three open-source projects to investigate the relationship between code quality metrics and change proneness.

\task{Feature extraction}
\citet{Tollin2017_374}
extracted features related to the code quality
from the issues of two industrial projects.
\citet{Tufano2019_390} used features from pull requests
to investigate the ability of a {\sc nmt} modes.
\citet{738_Abbas2020} and \citet{918_Malhotra2013} computed well-known C\&K metrics to investigate the relationship between change proneness and object-oriented metrics. 
Similarly, \citet{742_Kumar2017} computed 21 code quality metrics to predict change-prone web-services.
\citet{836_Uchoa2021} combines metrics from different sources --- 21 features related to source code, modification history of the files, and the textual description of the change, 20 features that characterize the developer’s experience, and 27 code smells detected by DesigniteJava\cite{DesigniteJava}.

\task{ML model training}
\citet{Tollin2017_374} employed \dt{}, \rf{}, and \nb{} {\sc ml} algorithms for their prediction task.
\citet{Tufano2019_390}
used \ENDE{} architecture of a typical \NMT{} model
to learn the changes introduced in pull requests.
\citet{918_Malhotra2013} experimented with \b{}, \mlp{}, and \rf{} to observe relationship between code metrics and change proneness.
\citet{738_Abbas2020} compared ten \ml{} models including \rf{}, \dt{}, \mlp{}, and \bn{}.
Similarly, \citet{742_Kumar2017} used \svm{} to the predict change proneness in web-services.
\citet{836_Uchoa2021} used six \ml{} models such as \svm{}, \dt{}, and \rf{} to investigate whether predicted impactful changes are helpful for code reviewers.

% ---------------
\subsubsection{Entity identification/recommendation\\}
This category represents studies that recommend source code entities (such as method and class names) \cite{Allamanis2015_392, Malik2019_389, Xu2019_386, Jiang2019_384, Hellendoorn2018_378}
or identify entities such as design patterns~\cite{gamma_design_1994} in code using \ml{}~\cite{Uchiyama2014_379, 927_Alhusain2013, 1049_Zanoni2015, 1051_Dwivedi2016, 1052_Chaturvedi2018}.
Specifically, \citet{Linstead2008_371} proposed a method to identify functional components in source code and to understand code evolution to analyze emergence of functional topics with time.
\citet{Huang2020_377} found commenting position in code using \ml\ techniques.
\citet{Uchiyama2014_379} identified design patterns and \citet{Abuhamad2018_383} recommended code authorship.
Similar  approaches include recommending method name  \cite{Allamanis2015_392, Jiang2019_384, Xu2019_386}, method signature \cite{Malik2019_389}, class name \cite{Allamanis2015_392}, and type inference \cite{Hellendoorn2018_378}.
We summarize these efforts classified in three steps of applying \ml\ techniques below.

\task{Dataset preparation}
The majority of the studies employed {\sc GitHub} projects for their experiments.
Specifically, \citet{Linstead2008_371} used two large, open source Java projects, Eclipse and Argo{\sc UML}
in their experiments to apply unsupervised statistical topic models.
Similarly, \citet{Hellendoorn2018_378} downloaded 1,000 open-source TypeScript projects and extracted identifiers with corresponding type information.
\citet{Abuhamad2018_383}
evaluated their approach over the entire Google Code Jam ({\sc gcj}) dataset
(from 2008 to 2016) and over real-world code samples (from 1987) extracted from public repositories on {\sc GitHub}.
\citet{Allamanis2015_392} mined 20 software projects
from {\sc GitHub} to predict method and class names.
\citet{Jiang2019_384} used the Code2Seq dataset containing 3.8 million methods as their experimental data.

A subset of studies focused on identifying design patterns using \ml{} techniques.
\citet{Uchiyama2014_379} performed experimental evaluations with five programs
to evaluate their approach on predicting design patterns.
\citet{927_Alhusain2013} applied a set of design patterns detection tools on 400 open source repositories; they selected all identified instances where at least two tools report a design pattern instance.
\citet{1049_Zanoni2015} manually identified 2,794 design patterns instances from ten open-source repositories.
\citet{1051_Dwivedi2016} analyzed JHotDraw and identified 59 instances of abstract factory and 160 instances of adapter pattern for their experiment.

\task{Feature extraction}
Several studies generated embeddings from their feature set.
Specifically, \citet{Huang2020_377} used embeddings generated from \textit{Word2vec} capturing code semantics.
Similarly, \citet{Jiang2019_384} employed \textit{Code2vec} embeddings and \citet{Allamanis2015_392} used embeddings 
that contain semantic information about sub-tokens of a method name 
to identify similar embeddings utilized in similar contexts.
\citet{Zhang2020_518}
utilized knowledge graph embeddings
to extract interrelations of code
for bug localization.
\citet{Abuhamad2018_383}
extracted code authorship attributes from samples of code.
\citet{Malik2019_389} used function names, formal parameters, and corresponding comments as features.

In addition, 
\citet{Uchiyama2014_379, 1052_Chaturvedi2018, 1051_Dwivedi2016, 927_Alhusain2013}  used several source-code metrics as features
to detect design patterns in software programs.

\task{ML model training}
The majority of studies in this category use \RNN{}-based \dl\ models.
In particular, \citet{Huang2020_377} and \citet{Hellendoorn2018_378}
used bidirectional \RNN{} models.
Similarly, \citet{Abuhamad2018_383} and \citet{Malik2019_389} also employed \RNN{} models
to identify code authorship and function signatures respectively.
\citet{Zhang2020_518} created a bug-localization tool, {\sc KGBugLocator} utilizing 
knowledge graph embeddings and bi-directional 
attention models.
\citet{Xu2019_386} employed the \GRU{}-based \ENDE{} model for method name prediction.
\citet{Uchiyama2014_379} used a hierarchical neural network as their classifier.
\citet{Allamanis2015_392} utilized neural language models for predicting method and class names.

Other studies used traditional \ml{} techniques.
Specifically, \citet{1052_Chaturvedi2018} compared four \ml{} techniques (\lr{}, \poly{}, \textit{support vector regression}, and \textit{neural network}).
\citet{1051_Dwivedi2016} used \dt{} and \citet{1049_Zanoni2015} trained \nb{}, \dt{}, \rf{}, and \svm{} to detect design patterns using \ml{}.
\subsection{Code review}
Code Review is the process of systematically check the code written by a developer performed by one or more different developers.
A very small set of studies explore the role of \ml\ in the process of code review.
Specifically, \citet{Lal2017_446} labeled check-in code samples as \textit{clean} and \textit{buggy}. 
On code samples, they carried out extensive pre-processing such as normalization and label encoding
before using \TFIDF\ to convert the samples into vectors.
They used a \nb\ model to classify samples into buggy or clean.
Similarly, \citet{Axelsson2009_447} developed a tool referred to as `Code Distance Visualiser'
to help reviewers  find problematic sections of code. 
The authors experimented on two subject systems---one open-source and another proprietary.
The authors provide an interactive, supervised self-learning static analysis tool based on 
\textit{Normalised Compression Distance} ({\sc ncd}) metric that relies on \knn{}.
\subsection{Code search}
Code search is an activity of searching a code snippet based on individual's need typically in Q\&A sites
such as StackOverflow  \cite{Sachdev2018_444, Shuai2020_442, Wan2019_443}.
The studies in this category define the following coarse-grained steps.
In the first step, the techniques prepare a training set by collecting source code and often corresponding description or query.
A feature extraction step then identifies and extracts relevant features from the input code and text.
Next, these features are fed into \ml{} models for training which is later used to execute test queries.

\task{Dataset preparation}
\citet{Shuai2020_442} utilized commented code as input.
\citet{Wan2019_443} used source code in the the form of tokens, \abst{}, and \CFG{}.
\citet{Sachdev2018_444} employed a simple tokenizer to extract
all tokens from source code by removing non–alphanumeric tokens. 
\citet{Ling2020_515} mined software projects from {\sc GitHub} for the training of their approach.

\task{Feature extraction}
Code search studies typically use embeddings representing the input code.
\citet{Shuai2020_442}  performed embeddings on code, where source code elements
(method name, {\sc api} sequence, and tokens) are processed separately. 
They generated embeddings for code comments independently.
\citet{Wan2019_443} employed a multi-modal code representation, where
they learnt the representation of each modality via \LSTM{}, {\sc Tree-lstm} and 
{\sc ggnn}, respectively. 
\citet{Sachdev2018_444} identified words from source code and transformed the extracted
tokens into a natural language documents.
Similarly, \citet{Ling2020_515} used an unsupervised word embedding technique
to construct a matching matrix to represent lexical similarities in software projects
and used an \RNN\ model to capture latent syntactic patterns for adaptive code search.

\task{ML model training}
\citet{Shuai2020_442} used a \CNN{}-based \ml{} model named {\sc carlcs-cnn}. 
The corresponding model learns interdependent representations for embedded code and query by a co-attention mechanism. 
Based on the embedded code and query, the co-attention mechanism learns a correlation matrix
and leverages row/column-wise max-pooling on the matrix.
\citet{Wan2019_443} employed a multi-modal attention fusion. 
The model learns representations of different modality and assigns weights using an attention layer.
Next, the attention vectors are fused into a single vector.
\citet{Sachdev2018_444} utilized word and documentation embeddings and performed code search using the learned embeddings.
Similarly, \citet{Ling2020_515} used an \textit{Autoencoder} network and a metric
(believability) to measure the degree to which a sentence is approved or disapproved within a discussion in a issue-tracking system.

Once an \ml{} model is trained, code search can be initiated using a query and a code snippet.
\citet{Shuai2020_442} used the given query and code sample to measure the semantic similarity using cosine 
similarity.
\citet{Wan2019_443} ranked all the code snippets by their similarities with the input query. 
Similarly, \citet{Sachdev2018_444} were able to answer almost 43\% of the collected
StackOverflow questions directly from code.

\subsection{Refactoring}
Refactoring transformations are intended to improve code quality (specifically maintainability),
while preserving the program behavior (functional requirements) from users' perspective~\cite{Suryanarayana2014}.
This section summarizes the studies that identify
refactoring candidates or predict refactoring commits by analyzing source code and by applying \ml\ techniques on code.
A process pipeline
typically adopted by the studies in this category can be viewed as a three step process. 
In the first step, the source code of the projects is used to prepare a dataset for training.
Then, individual samples (\ie\ either a method, class, or a file) is processed to extract relevant features.
The extracted features are then fed to an \ml\ model for training. 
Once trained, the model is used to predict whether an input sample is a candidate for refactoring or not.

\task{Dataset preparation}
The first set of studies created their own dataset for model training.
For instance,
\citet{Rodriguez2019_159} and \citet{Amal2014_170} created datasets where each sample is reviewed by a human to identify
an applicable refactoring operation; the identified operation is carried out by automated means.
\citet{Kosker2009_161} employed four versions of the same repository, computed their complexity metrics,
and classified their classes as refactored
if their complexity metric values are reduced from the previous version.
\citet{Nyamawe2019_163} analyzed 43 open-source repositories with 13.5 thousand commits to prepare their dataset.
Similarly, \citet{Aniche2020_173} created a dataset comprising over two million refactorings 
from more than 11 thousand open-source repositories.
\citet{733_Sagar2021} identified 5004 commits randomly selected from all the commits obtained from 800 open-source repositories where RefactoringMiner \cite{RefactoringMiner2.0} identified at least one refactoring.
Along the similar lines, \citet{762_Li2021} used RefactoringMiner and RefDiff tools to identify refactoring operations in the selected commits.
\citet{745_Xu2017, 770_Krasniqi2020} used manual analysis and tagging for identifying refactoring operations.
Finally, \citet{Kurbatova2020_172} generated synthetic data by moving methods to other classes to prepare a dataset for feature envy smell.
The rest of the studies in this category \cite{Kumar2017_162, Kumar2019_169, Aribandi2019_171}, 
used the tera-{\sc promise} dataset containing various metrics for open-source projects where the classes that need refactoring are tagged.

\task{Feature extraction}
A variety of features, belonging to product as well as process metrics, has been employed by the studies in this category.
Some of the studies rely on code quality metrics.
Specifically, \citet{Kosker2009_161} computed cyclomatic complexity along with 25 other code quality metrics.
Similarly, \citet{Kumar2019_169} computed 25 different code quality metrics using the SourceMeter tool;
these metrics include cyclomatic complexity, class class and clone complexity, {\sc loc}, outgoing method invocations,
and so on. 
Some of the studies~\cite{Kumar2017_162, Aribandi2019_171, 734_Sidhu2022, 757_Wang2014} calculated a large number of metrics.
Specifically, \citet{Kumar2017_162} computed 102 metrics and then applied {\sc pca} to reduce the number of features to 31,
while~\citet{Aribandi2019_171} used 125 metrics.
\citet{734_Sidhu2022} used metrics capturing design characteristics of a model including inheritance, coupling and modularity, and size.
\citet{757_Wang2014} computed a wide range of metrics related to clones such as number of clone fragements in a class, clone type (type1, type2, or type3), and lines of code in the cloned method.

Some other studies did not limit themselves to only code quality metrics.
Particularly, \citet{Yue2018_164} collected 34 features belonging to code, evolution history, \textit{diff} between commits, and co-change.
Similarly, \citet{Aniche2020_173} extracted code quality metrics, process metrics, and code ownership metrics.

In addition, \citet{Nyamawe2019_163, 760_Nyamawe_2020} carried out standard \NLP\ preprocessing and generated \TFIDF\ embeddings for each sample.
Along the similar lines, \citet{Kurbatova2020_172} used {\it code2vec} to generate embeddings for each method.
\citet{733_Sagar2021} extracted keywords from commit messages and used GloVe to obtain the corresponding embedding.
\citet{770_Krasniqi2020} tagged each commit message with their parts-of-speech and prepared a language model dependency tree to detect refactoring operations from commit messages.

\task{ML model training}
Majority of the studies in this category utilized traditional \ml\ techniques.
\citet{Rodriguez2019_159} proposed a method to identify  web-service groups for refactoring using \textit{K-means}, {\sc cobweb}, and expectation maximization.
\citet{Kosker2009_161} trained a \nb{}-based classifier to identify classes that need refactoring.
\citet{Kumar2017_162}  used \textit{Least Square}-\svm{}  ({\sc ls-svm}) along with {\sc smote} as classifier.
 They found that {\sc ls-svm} with \textit{Radial Basis Function} ({\sc rbf}) kernel gives the best results.
\citet{Nyamawe2019_163}  recommended refactorings based on the history of requested features and applied refactorings. 
Their approach involves two classification tasks;
first, a binary classification that suggests whether refactoring is needed or not and second,
a multi-label classification that suggests the type of refactoring. 
The authors used \lr{}, \textit{Multinomial} \nb{} ({\sc mnb}), \svm{}, and \rf{} classifiers. 
\citet{Yue2018_164} presented {\sc crec}---a learning-based approach that automatically extracts refactored
and non-refactored clones groups from software repositories, and trains an \ada\ model to recommend clones for refactoring.
\citet{Kumar2019_169} employed a set of \ml\ models such as \lr{}, \nb{},  \bn{}, \rf{}, \ada{}, and \lb{}
to develop a  recommendation system to suggest the need of refactoring for a method. 
\citet{Amal2014_170} proposed the use of \ANN\ to generate a sequence of refactoring.
\citet{Aribandi2019_171}  predicted the classes that are likely to be refactored in the future iterations.
To achieve their aim, the authors used various variants of \ANN{}, \svm{},
as well as \textit{Best-in-training based Ensemble} ({\sc bte}) and
\textit{Majority Voting Ensemble} ({\sc mve}) as ensemble techniques.
\citet{Kurbatova2020_172}
proposed an approach to recommend move method refactoring  based on a path-based presentation of code using \svm{}.
Similarly, \citet{Aniche2020_173} used \lr{}, \nb{}, \svm{}, \dt{}, \rf{}, and \textit{Neural Network}
to predict applicable refactoring operations.
\citet{734_Sidhu2022, 745_Xu2017, 757_Wang2014} used \DNN{}, \textit{gradient boosting}, and \dt{} respectively to identify refactoring candidate.
\citet{733_Sagar2021, 760_Nyamawe_2020} employed various classifiers such as \svm{}, \lr{}, and \rf{} to predict commits with refactoring operations.
\subsection{Vulnerability analysis}
\begin{figure*}[!h]
	\centering
	\includegraphics[width=\textwidth]{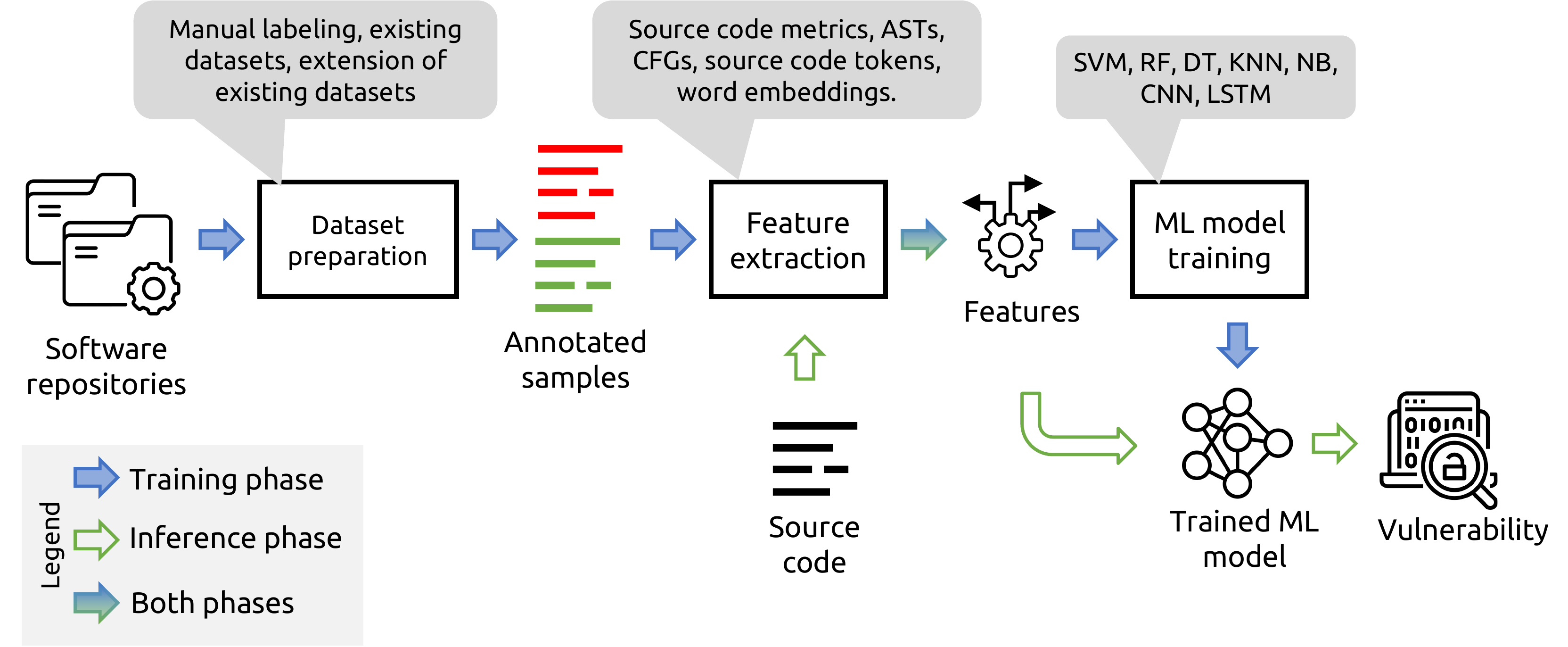}
	\caption{Overview of vulnerability detection  process pipeline}
	\label{fig:vulnerabilities_detection}
\end{figure*}

The studies in this domain analyze source code to identify potential security vulnerabilities.
In this section,
we point out the state-of-the-art in software vulnerability detection using \ml\ techniques.
Figure~\ref{fig:vulnerabilities_detection} presents an overview of a typical process
to detect vulnerabilities with the help of \ml\ techniques.
First, the studies prepare a dataset or identify an existing dataset for \ml\ training.
Next, the studies extract relevant features from the identified subject systems.
Then, the features are fed into a \ml\ model for training. 
The trained model is then used to predict vulnerabilities in the source code.

\task{Dataset preparation}
Authors used existing labeled datasets as well as created their own datasets to train \ml{} models.
Specifically, a set of studies
\cite{Pereira2019_12, Milosevic2017_36, Rahman2017_48,
Saccente2019_50, Kim2019_59, Bilgin2020_63, Spreitzenbarth2014_248, lin_deep_2020, Yosifova_2021, sultana_using_2021, law_is_2010, Pang_2017_vul, Abunadi_Alenezi_2015, younis_malaiya_2014, vishnu_jevitha_2014, khalid_predicting_2019, kim_hong_oh_lee_2018, zhang_li_2020, mateless_decompiled_2020, du_chen_li_guo_zhou_liu_jiang_2019, shiqi_android_2018, du_wang_want_2015, yang_li_wu_lu_han_2019, batur_canan_abualigah_2021, narayanan_chandramohan_chen_liu_2018, wang_want_sun_batcheller_jajodia_2020, chen_santosa_yi_sharma_sharma_mo_2020, Li2015AHM, ren_buffer_2019, ban_performance_2019} used available labeled datasets
for {\sc php}, Java, C, C++, and Android applications
to train vulnerability detection models.
In other cases,
\citet{Russell2018_16} extended an existing dataset
with millions of C and C++ functions
and then labeled it based on the output of three static analyzers
(\ie\ Clang, CppCheck, and Flawfinder).

Many studies~\cite{Ma2019_3, Ali2016_14, Cui2020_60, Ndichu2019_5, Elovici2007_15, Medeiros2014_17,
Ferenc2019_18, Piskachev2019_21, Kronjee2018_26, Pang2016_27, Alves2016_29, Gupta2021_30,
Clemente2018_34, Chernis2018_39, Moskovitch2009_43, Hou2010_44, Santos2013_46, Yang2018_47,
Zheng2020_56, Perl2015_61, Shar2015_64, Jimenez2019, lin_deep_2020, lin_zhang_luo_pan_xiang_de_montague_2018} created their own datasets.
\citet{Ma2019_3, Ali2016_14, Cui2020_60}, and~\citet{Gupta2021_30} created datasets
to train vulnerability detectors for Android applications.
In particular,
\citet{Ma2019_3} decompiled and generated  {\sc cfg}s of approximately 10 thousand, both benign and
vulnerable, Android applications from {\it AndroZoo} and {\it Android Malware} datasets;
\citet{Ali2016_14} collected 5,063 Android applications where 1,000 of them were marked
as benign and the remaining as malware;
\citet{Cui2020_60} selected an open-source dataset comprised of 1,179 Android applications
that have 4,416 different version (of the 1,179 applications)
and labeled the selected dataset by using the Androrisk tool;
and~\citet{Gupta2021_30} used two Android applications (Android-universal-image-loader and JHotDraw)
which they have manually labeled based on the projects {\sc pmd} reports (true if a vulnerability was reported
in a {\sc pmd} file and false otherwise).
To create datasets of {\sc php} projects,
\citet{Medeiros2014_17} collected 35 open-source {\sc php} projects
and intentionally injected  76 vulnerabilities
in their dataset.
\citet{Shar2015_64} used {\it phpminer} to extract 15 datasets
that include {\sc sql} injections, cross-site scripting,
remote code execution, and file inclusion vulnerabilities,
and labeled only 20\% of their dataset to point out
the precision of their approach.
\citet{Ndichu2019_5} collected 5,024 JavaScript code snippets
from {\sc d3m, jsunpack}, and 100 top websites
where the half of the code snippets were benign and the other half malicious.
In other cases,
authors~\cite{Yang2018_47, Rahman2017_48, Perl2015_61} collected large number of commit messages
and mapped them to known vulnerabilities by using Google's Play Store,
National Vulnerability Database ({\sc nvd}),
Synx, Node Security Project, and so on,
while in limited cases authors~\cite{Piskachev2019_21} manually label their dataset.
\citet{Hou2010_44, Moskovitch2009_43} and \citet{Santos2013_46} created their datasets
by collecting web-page samples from StopBadWare and VxHeavens.
\citet{lin_deep_2020} constructed a dataset and manually
labeled 1,471 vulnerable functions and 1,320 vulnerable files from nine open-source applications,
named Asterisk, FFmpag, {\sc httpd}, LibPNG, LibTIFF, OpenSSL, Pidgin, {\sc vlc} Player, and Xen.
\citet{lin_zhang_luo_pan_xiang_de_montague_2018} have used more then 30,000 non-vulnerable functions and manually labeled 475 vulnerable functions for their experiments.

\task{Feature extraction}
Authors used static source code metrics,
{\sc cfg}s, \abst{}s, source code tokens,
and word embeddings as features.

\subtask{Source code metrics}
	A set of studies
	\cite{Medeiros2014_17, Ferenc2019_18, Alves2016_29, Gupta2021_30,
		Clemente2018_34, Rahman2017_48, Cui2020_60, Piskachev2019_21, ren_buffer_2019, du_chen_li_guo_zhou_liu_jiang_2019, kim_hong_oh_lee_2018, medeiros_neves_correia_2016, Abunadi_Alenezi_2015, law_is_2010, sultana_using_2021}
	used more than 20 static source code metrics 
	(such as \textit{cyclomatic complexity}, \textit{maximum depth of class in inheritance tree},
	\textit{number of statements,} and \textit{number of blank lines}).
	
\subtask{Data/control flow and \abst{}}
	\citet{Ma2012_147, Kim2019_59, Bilgin2020_63, Kronjee2018_26, wang_want_sun_batcheller_jajodia_2020, du_wang_want_2015, medeiros_neves_correia_2016} used {\sc cfg}s,
	{\sc ast}s, or data flow analysis as features.
	More specifically,
	\citet{Ma2019_3} extracted the {\sc api} calls from the {\sc cfg}s
	of their dataset and collected information such as
	the usage of {\sc api}s (which {\sc api}s the application uses),
	the {\sc api} frequencies (how many times the application uses {\sc api}s)
	and {\sc api} sequence (the order the application uses {\sc api}s).
	\citet{Kim2019_59} extracted {\sc ast}s and {\sc gfc}s which they tokenized and fed into \ml\ models,
	while \citet{Bilgin2020_63} extracted {\sc ast}s and translated their
	representation of source code into a one-dimensional numerical array
	to fed them to a model.
	\citet{Kronjee2018_26} used data-flow analysis
	to extract features,
	while~\citet{Spreitzenbarth2014_248} used static, dynamic analysis,
	and information collected from {\tt ltrace}
	to collect features and train a linear vulnerability detection model.
	\citet{lin_zhang_luo_pan_xiang_de_montague_2018} created {\sc ast}s and from there they extracted code semantics as features.
	
\subtask{Repository and file metrics}
	\citet{Perl2015_61} collected {\sc GitHub} repository meta-data
	(\ie{} \textit{programming language, star count, fork count,} and \textit{number of commits}) in addition to source code metrics.
	Other authors~\cite{Pereira2019_12, Elovici2007_15} used file meta-data such as 
	\textit{files' creation and modification
	time, machine type, file size,} and \textit{linker version}.
	
\subtask{Code and Text tokens}
	\citet{Chernis2018_39} used simple token features (\textit{character
	count, character diversity, entropy, maximum nesting depth, arrow
	count, ``if'' count, ``if'' complexity, ``while'' count,} and \textit{``for'' count})
	and complex features (\textit{character n-grams, word n-grams,} and \textit{suffix trees}).
	\citet{Hou2010_44} collected 10 features such as \textit{length of the document,
average length of word,  word count,  word count in a line,} and
\textit{ number of NULL characters}.
	The remaining studies~\cite{Russell2018_16, Pang2016_27, Moskovitch2009_43, Santos2013_46,
		Yang2018_47, Saccente2019_50, Zheng2020_56, Shar2015_64, chen_santosa_yi_sharma_sharma_mo_2020, narayanan_chandramohan_chen_liu_2018, Russell2018_16, mateless_decompiled_2020, fang_fastembed_2020, zhang_li_2020, Pang_2017_vul, ban_performance_2019, Yosifova_2021, lin_deep_2020} tokenized parts of the source
	code or text-based information with various techniques such as the most frequent occurrences of operational codes,
	capture the meaning of critical tokens,
	or applied techniques to reduce the vocabulary size
	in order to retrieve the most important tokens.
	In some other cases, authors~\cite{Li2015AHM} used statistical techniques to reduce the feature space to reduce the number of code tokens.
	
\subtask{Other features}
	\citet{Ali2016_14, Ndichu2019_5} and \citet{Milosevic2017_36} extracted permission-related features.
	In other cases, authors~\cite{yang_li_wu_lu_han_2019} combined software metrics and N-grams as features to train models and others~\cite{shiqi_android_2018} created text-based images to extract features.
	Likewise, \citet{sultana_kazi_2017} extracted traceable patterns such as CompoundBox, Immutable, Implementor, Overrider, Sink,
    Stateless, FunctionObject, and LimitSel and used Understand tool to extract various software metrics.
	\citet{wei_luo_weng_zhong_zhang_yan_2017} extracted system calls and function call-related information to use as features,
	while \citet{vishnu_jevitha_2014} extracted {\sc url}-based features
	like number of chars, duplicated characters, special characters, script tags, cookies, and re-directions.
	\citet{padmanabhuni_tan_2015} extracted buffer usage patterns and defensive mechanisms statements constructs by analyzing files.

\task{Model training}
To train models,
the selected studies used a variety of traditional \ml{}
and \dl{} algorithms.

	\subtask{Traditional ML techniques}
	One set of studies
	\cite{Ali2016_14, Ndichu2019_5, Pereira2019_12,  Russell2018_16, Pang2016_27, Moskovitch2009_43, Perl2015_61, Shar2015_64, Yosifova_2021, sultana_using_2021, padmanabhuni_tan_2015, law_is_2010, Abunadi_Alenezi_2015, younis_malaiya_2014, sultana_kazi_2017, vishnu_jevitha_2014, wei_luo_weng_zhong_zhang_yan_2017, du_chen_li_guo_zhou_liu_jiang_2019, fang_fastembed_2020, medeiros_neves_correia_2016, du_wang_want_2015, narayanan_chandramohan_chen_liu_2018, wang_want_sun_batcheller_jajodia_2020, chen_santosa_yi_sharma_sharma_mo_2020, ren_buffer_2019}
	used traditional \ml{} algorithms such as
	\nb{}, \dt{}, \svm{}, \lr{}, \dt{}, and \rf{} to train their models.
	Specifically,
	\citet{Ali2016_14, Russell2018_16, Perl2015_61} selected \svm{} because it is not affected
	by over-fitting when having very high dimensional variable spaces.
	Along the similar lines,
	\citet{Ndichu2019_5} used \svm{} to train their model
	with linear kernel.
	\citet{Pereira2019_12} used \dt{}, \lr{}, and {\it Lasso}
	to train their models,
	while \cite{Abunadi_Alenezi_2015} found that \rf{} is the best
	model for predicting cross-project vulnerabilities.
	Compared to the above studies,
	\citet{Shar2015_64} used both supervised (\ie{} \lr{} and \rf{})
	and semi-supervised (\ie{} \textit{Co-trained} \rf{}) algorithms to train
	their models since most of that datasets were not labeled.
	\citet{Yosifova_2021} used text-based features to train \nb, \svm{}, and \rf{} models.
	\citet{du_chen_li_guo_zhou_liu_jiang_2019} created the {\sc leopard} framework that does not require prior knowledge about known vulnerabilities and used \rf{}, \nb{}, \svm{}, and \dt{} to point
	them out.

	Other studies~\cite{Medeiros2014_17, Ferenc2019_18, Piskachev2019_21, Kronjee2018_26,
		Alves2016_29, Gupta2021_30, Clemente2018_34, Milosevic2017_36, Chernis2018_39,
		Hou2010_44, Santos2013_46, Rahman2017_48, Cui2020_60} used up to 32
	different \ml{} algorithms to train models and compared their performance.
	Specifically,
	~\citet{Medeiros2014_17} experimented with multiple variants of  \dt{}, \rf{}, \nb{}, \knn{}, \lr{}, \mlp{},
	and \svm{} models and identified  \svm{} as the best performing classifier
	for their experiment.
	Likewise,
	\citet{Milosevic2017_36} and \citet{Rahman2017_48} employed multiple \ml{} algorithms,
	respectively, and found that \svm{} offers the highest accuracy rate
	for training vulnerability detectors.
	In contrast to the above  studies,
	\citet{Ferenc2019_18} showed that \knn{}  offers the best performance for their dataset
	after experimenting with
	\DNN{}, \knn{}, \svm{}, \lr{}, \dt{}, \rf{}, and \nb{}.
	In order to find out which is the best model
	for the {\sc swan} tool,
	\citet{Piskachev2019_21} evaluated the 
	\svm{}, \nb{}, \bn{}, \dt{}, \textit{Stump}, and \textit{Ripper}.
	Their results pointed out the \svm{} as the best performing model
	to detect vulnerabilities.
	Similarly,
	\citet{Kronjee2018_26}, \citet{Cui2020_60}, and  \citet{Gupta2021_30}
	compared different \ml{} algorithms
	and found  \dt{} and  \rf{} as the best performing algorithms.
	
	\subtask{DL techniques}
	A large number of studies~\cite{Yang2018_47, Saccente2019_50, Kim2019_59, lin_deep_2020, ban_performance_2019, kim_hong_oh_lee_2018, mateless_decompiled_2020, lin_zhang_luo_pan_xiang_de_montague_2018, shiqi_android_2018, batur_canan_abualigah_2021}
	used \dl{} methods such as 
	\CNN{}, \RNN{}, and \ANN{} to train models.
	In more details,
	\citet{Yang2018_47} utilized the {\sc bp-ann} algorithm to train vulnerability detectors.
	For the project \textit{Achilles},
	\citet{Saccente2019_50} used an array of \LSTM{} models
    to train on data containing Java code snippets
	for a specific set of vulnerability types.
	In another study,
	\citet{Kim2019_59} suggested a \dl{} framework that makes use of \RNN{} models
	to train vulnerability detectors.
	Specifically,
	the authors framework first feeds the code embeddings into a {\sc b}i-\LSTM{} model
	to capture the feature semantics,
	then an attention layer is used to get the vector weights,
	and, finally, passed into a dense layer to output if a code is safe or vulnerable.
	Compared to the studies that examined traditional \ml{} or \dl{} algorithms,
	\citet{Zheng2020_56} examined both of them.
	They used
	\rf{}, \knn{}, \svm{}, \lr{} among the traditional \ml{} algorithms
	along with {\sc b}i-\LSTM{}, \GRU{}, and \CNN{}.
	There results indicate {\sc b}i-\LSTM{} as the best performing model.
    \cite{lin_deep_2020} developed a benchmarking framework that can use
    {\sc b}i-\LSTM{}, \LSTM{}, {\sc b}i-\GRU{}, \GRU{}, \DNN{} and Text-{\sc cnn}, but can be extended to use more deep learning models.
    \citet{kim_hong_oh_lee_2018} generating graphical semantics that reflect on code semantic features and use them for Graph Convulutional Network to automatically identify and learn semantic and extract features for vulnerability detection,
    while \citet{shiqi_android_2018} created textual images and fed them to Deep Belief Networks to classify malware.

\section{Datasets and tools}
\label{results_rq3}

This section provides a consolidated summary of available datasets and tools
that are used by the studies considered in the survey.
We carefully examined each primary study and noted the used resources (\ie\ datasets and tools).
We define the following criteria to include a resource in our catalog.

\begin{itemize}
	\item The referenced resource must have been used by at least one primary study.
	\item The referenced resource must be publicly available at the time of writing this paper (Jul 2022).
	\item The resource provides bare-minimum usage instructions to build and execute (wherever applicable) and to use the artifact. 
	\item The resource is useful either by providing an implementation of a \ml\ technique, 
	helping the user to generate information/data which is further used by a \ml\ technique, 
	or by providing a processed dataset that can be directly employed in a \ml\ study.
\end{itemize}

Table~\ref{tab:tools} lists all the tools  that we found in this exploration.
Each resource is listed with it's category, name and link to access the resource, number of citations (as of Jul 2022), and the time when it was first introduced along with the time when the resource was last updated. We collected the metadata about the resources manually by searching the digital libraries, repositories, and authors' websites. The cases where we could not find the required information, we mark the entry with "--". We also provide a short description of the resource.

\newpage

\topcaption{A list of tools useful for analyzing source code and applying machine learning techniques}
%\rowcolors{2}{gray!25}{white}
	\label{tab:tools}
\begin{supertabular}{p{.14\textwidth}p{.20\textwidth}p{.08\textwidth}p{.08\textwidth}p{.08\textwidth}p{.29\textwidth}}
	\rowcolor{gray!50}
	\textbf{Category} & \textbf{Name} & \textbf{\#Citation}& \textbf{Introd.} & \textbf{Updated} & \textbf{Description} \\
	
	 \multirow{15}{*}{\shortstack{Code\\ Representation}}&
	 \href{https://github.com/spcl/ncc}{ncc}~\cite{Ben-Nun2018_388}
	& 3 & Dec 2018 & Aug 2021
	& Learns representations of code semantics\\
	& \cellcolor{gray!25} \href{https://github.com/tech-srl/code2vec/}{Code2vec}~\cite{Alon2019_464}
	& \cellcolor{gray!25} 271 & \cellcolor{gray!25} Jan 2019 & \cellcolor{gray!25} Feb 2022
	& \cellcolor{gray!25} Generates distributed representation of code \\
	& \href{https://github.com/tech-srl/code2seq}{Code2seq}~\cite{Alon2019_463}
	& 418 & May 2019 & Jul 2022
	& Generates sequences  from structured representation of code\\
	& \cellcolor{gray!25} \href{http://zenodo.org/record/3647645} {Vector representation for coding style} \cite{Kovalenko2020_458}
	& \cellcolor{gray!25} 1 & \cellcolor{gray!25} Sep 2020 & \cellcolor{gray!25} Jul 2022 
	& \cellcolor{gray!25} Implements vector representation of individual coding style \\
	& \href{https://github.com/CC2Vec/CC2Vec} {CC2Vec} \cite{Hoang2020_461}
	& 23 & Oct 2020 & --
	& Implements distributed representation of code changes \\
	&\cellcolor{gray!25}  \href{https://github.com/micheletufano/AutoenCODE} {AutoenCODE} \cite{Tufano2018_467}
	& \cellcolor{gray!25} 1 & \cellcolor{gray!25} -- & \cellcolor{gray!25} -- 
	&  \cellcolor{gray!25} Encodes source code fragments into vector representations \\
	& \href{https://github.com/Microsoft/graph-based-code-modelling} {Graph-based code modeling} \cite{Allamanis2018_326}
	& 544 & May 2018 & May 2021
	& Generates code modeling with graphs\\
	&\cellcolor{gray!25}  \href{https://github.com/mwcvitkovic/Open-Vocabulary-Learning-on-Source-Code-with-a-Graph-Structured-Cache--Code-Preprocessor} {Vocabulary learning on code} \cite{Cvitkovic2019_476}
	& \cellcolor{gray!25} 34 & \cellcolor{gray!25} Jan 2019 & \cellcolor{gray!25} -- 
	& \cellcolor{gray!25}  Generates an augmented \abst\ from Java source code \\
	 & \href{https://github.com/dazcona/user2code2vec} {User2code2vec} \cite{Azcona2019_482}
	 & 14 & Mar 2019 & May 2019
	 &  Generates embeddings for developers based on distributed representation of code \\\hline
	
    \multirow{2}{*}{\shortstack{Code Search}} &\cellcolor{gray!25} 
    \href{https://github.com/guxd/deep-code-search/}{Deep Code Search}~\cite{Gu2018}
    & \cellcolor{gray!25} 160 & \cellcolor{gray!25} May 2018 & \cellcolor{gray!25} May 2022
    & \cellcolor{gray!25} Searches code by using code embeddings\\ \hline

	\multirow{25}{*}{\shortstack{Program\\ Comprehension}} & \href{https://github.com/basedrhys/obfuscated-code2vec}{Obfuscated-code2vec}~\cite{Compton2020_380} 
	& 14 & Oct 2022 & --
	& Embeds Java Classes with Code2vec  \\
    & \cellcolor{gray!25} \href{https://github.com/DeepTyper/DeepTyper}{{\sc DeepTyper}} \cite{Hellendoorn2018_378} 
    & \cellcolor{gray!25} 54 & \cellcolor{gray!25} Oct 2018 & \cellcolor{gray!25} Feb 2020
    & \cellcolor{gray!25} Annotates types for JavaScript and TypeScript \\
	& \href{https://github.com/yorhaz40/CallNN} {CallNN} \cite{Liu2019_408} 
	& 6 & Oct 2019 & --
	&  Implements a code summarization approach by using call dependencies \\
	&\cellcolor{gray!25}  \href{https://github.com/wasiahmad/NeuralCodeSum} {NeuralCodeSum} \cite{Ahmad2020_410}
	& \cellcolor{gray!25} 147 & \cellcolor{gray!25} May 2020 & \cellcolor{gray!25} Oct 2021
	& \cellcolor{gray!25} Implements a code summarization method by using transformers \\
	& \href{https://github.com/sh1doy/summarization_tf} {Summarization\_tf} \cite{Shido2019_415}  
	& 9 & Jul 2019 & --
	&  Summarizes code with Extended {\sc Tree-lstm} \\
	&\cellcolor{gray!25} \href{https://github.com/LittleYUYU/CoaCor} {CoaCor} \cite{Yao2019_418}  & \cellcolor{gray!25} 16 & \cellcolor{gray!25} Jul 2019 & \cellcolor{gray!25} May 2020 
	&\cellcolor{gray!25}  Explores the role of rich annotation for code retrieval \\
	& \href{https://github.com/xing-hu/EMSE-DeepCom} {DeepCom} \cite{Li2020_422}
	& 12 & Nov 2020 & May 2021
	& Generates code comments \\
	
	&\cellcolor{gray!25} \href{https://github.com/zhangj111/rencos} {Rencos} \cite{Zhang2020_435} 
	& \cellcolor{gray!25}79 & \cellcolor{gray!25} Oct 2020 & \cellcolor{gray!25}--
	& \cellcolor{gray!25} Generates code summary by using both neural and retrieval-based techniques \\
	
	& \href{http://www.ing.unisannio.it/spanichella/pages/tools/CODES/} {{\sc codes}} \cite{Panichella2012}
	& 107 & Jul 2012 & Jul 2016
	& Extracts method description from StackOverflow discussions\\
	
	&\cellcolor{gray!25} \href{http://oscar-lab.org/CFS/} {{\sc cfs}}
	& \cellcolor{gray!25} -- & \cellcolor{gray!25} -- & \cellcolor{gray!25}--
	&\cellcolor{gray!25} Summarizes code fragments using \SVM\ and \NB\\\
	
	& \href{https://github.com/mast-group/tassal} {{\sc tassal}}
	& -- & -- & --
	& Summarizes code using autofolding\\
	
	&\cellcolor{gray!25} \href{https://github.com/SEMERU-WM/ChangeScribe}{ChangeScribe} \cite{CortesCoy2014OnAG}
	& \cellcolor{gray!25} 158 & \cellcolor{gray!25} Dec 2014 &\cellcolor{gray!25} Dec 2015 
	&\cellcolor{gray!25} Generates commit messages\\
	
	& \href{https://github.com/masud-technope/CodeInsight-Replication-Package-SCAM2015}{CodeInsight} \cite{scam2015masud}
	& 59 & Nov 2015 & May 2019
	& Recommends insightful comments for source code\\
	
	&\cellcolor{gray!25} \href{https://github.com/sriniiyer/codenn}{CodeNN} \cite{Iyer2016_440}  
	& \cellcolor{gray!25} 479 & \cellcolor{gray!25} Aug 2016 & \cellcolor{gray!25} May 2017  
	&\cellcolor{gray!25} Summarizes code using neural attention model\\
	
	& \href{https://github.com/beyondacm/Code2Que}{Code2Que} \cite{Gao2020_441}
	& 3 & Jul 2020 & Aug 2021
	& Suggests improvements in question titles from mined code in StackOverflow\\
	
	&\cellcolor{gray!25} \href{https://github.com/bdqnghi/bi-tbcnn}{{\sc bi-tbcnn}} \cite{Bui2018_214}
	& \cellcolor{gray!25} 30 & \cellcolor{gray!25} Mar 2019 & \cellcolor{gray!25} May 2019
	&\cellcolor{gray!25} Implements a {\sc b}i-{\sc tbcnn} model to classify algorithms\\
	
	& \href{https://github.com/parasol-aser/deepsim}{DeepSim} \cite{Zhao2018_216}
	& 150 & Oct 2018 & --
	& Implements a \dl\ approach to measure code functional similarity \\
	
	&\cellcolor{gray!25} \href{https://github.com/shiyy123/FCDetector}{FCDetector} \cite{Fang2020_225}
	& \cellcolor{gray!25} 22 & \cellcolor{gray!25} Jul 2020 & \cellcolor{gray!25} -- 
	&\cellcolor{gray!25} Proposes a fine-grained granularity of source code for functionality identification\\\hline

	\multirow{20}{*}{\shortstack{Quality\\Assessment}} & \href{http://www.sonarqube.org/}{{\sc SonarQube}}
	& -- & -- & --
	&  Analyzes code quality \\
	
	&\cellcolor{gray!25} \href{https://github.com/SVF-tools/SVF} {{\sc svf}} \cite{sui2016svf}
	& \cellcolor{gray!25} 216 & \cellcolor{gray!25} Mar 2016 & \cellcolor{gray!25} Jul 2022
	& \cellcolor{gray!25} Enables inter-procedural dependency analysis for {\sc llvm}-based languages\\
	
	& \href{http://www.designite-tools.com} {Designite} \cite{Designite}
	& 18 & Mar 2016 & Jul 2022
	& Detects code smells and computes quality metrics in Java and C\# code\\
	
	&\cellcolor{gray!25} \href{https://github.com/pseudoPixels/CloneCognition}{CloneCognition} \cite{Mostaeen2018}
	& \cellcolor{gray!25} 4 & \cellcolor{gray!25} Nov 2018 & \cellcolor{gray!25} May 2019
	&\cellcolor{gray!25} Proposes a \ml\ framework to validate code clones\\
	
	& \href{https://github.com/antoineBarbez/SMAD}{{\sc smad}} \cite{Barbez2020_180}
	& 25 & Mar 2020 & Feb 2021
	& Implements smell detection (God class and Feature envy) using \ml\ \\
	
	&\cellcolor{gray!25} \href{https://checkstyle.sourceforge.io/}{Checkstyle}
	& \cellcolor{gray!25} -- & \cellcolor{gray!25} -- & \cellcolor{gray!25}--
	&\cellcolor{gray!25} Checks for coding convention in Java code \\
	
	& \href{http://findbugs.sourceforge.net/findbugs2.htm}{FindBugs}
	& -- & -- & --
	&Implements a static analysis tool for Java\\
	
	& \cellcolor{gray!25} \href{https://pmd.github.io/latest/}{{\sc pmd}}
	& \cellcolor{gray!25} -- & \cellcolor{gray!25} -- & \cellcolor{gray!25} --
	& \cellcolor{gray!25} Finds common programming flaws in Java and six other languages\\
	
	& \href{https://github.com/pseudoPixels/ML_CloneValidationFramework}{ML Clone Validation Framework} \cite{Mostaeen2018_251}
	& 11 & Aug 2019 & Aug 2019
	& Implements a \ml\ framework for automatic code clone validation\\
	
	& \cellcolor{gray!25} \href{https://github.com/mochodek/py-ccflex}{py-ccflex} \cite{Ochodek2019_257}
	& \cellcolor{gray!25} 6 & \cellcolor{gray!25} Mar 2017 & \cellcolor{gray!25} Oct 2020
	& \cellcolor{gray!25} Mimics code metrics by using \ml\ \\
	
	& \href{https://github.com/tushartushar/DeepLearningSmells}{Deep learning smells} \cite{Sharma2021_510}
	& 18 & Jul 2021 & Nov 2020
	& Implements \dl\ (\CNN{}, \RNN{}, and Autoencoder-based models) to identify four smells \\
	
	& \cellcolor{gray!25} \href{https://github.com/soniapku/CREC}{{\sc crec}} \cite{Yue2018_164}
	& \cellcolor{gray!25} 26 & \cellcolor{gray!25} Nov 2018 & \cellcolor{gray!25} --
	& \cellcolor{gray!25} Recommends clones for refactoring \\
	
	& \href{https://github.com/refactoring-ai/predicting-refactoring-ml}{\ml\ for software refactoring} \cite{Aniche2020_173}
	& 31 & Sep 2020 & --
	& Recommends refactoring by using \ml\ \\\hline
	
	\multirow{2}{*}{\shortstack{Program \\Synthesis}} & \cellcolor{gray!25} \href{https://github.com/lin-tan/CoCoNut-Artifact}{{\sc CoCoNuT}} \cite{Lutellier2020_287}
	& \cellcolor{gray!25} 86 &\cellcolor{gray!25} Jul 2020 & \cellcolor{gray!25} Sep 2021
	& \cellcolor{gray!25}  Repairs Java programs\\
	
    & \href{https://bitbucket.org/iiscseal/deepfix/src/master/}{DeepFix} \cite{Gupta2017}
    & 359 & Feb 2017 & Dec 2017
    & Fixes common C errors\\\hline
		
	\multirow{15}{*}{Testing} & \cellcolor{gray!25} \href{https://github.com/columbia/appflow} {AppFlow} \cite{Hu2018_72}
	& \cellcolor{gray!25} 47 & \cellcolor{gray!25} Oct 2018 & \cellcolor{gray!25} --
	& \cellcolor{gray!25}   Automates {\sc ui} tests generation \\
	
	 & \href{https://github.com/s3team/DeepFuzz} {DeepFuzz} \cite{Liu2019_85}
	 & 42 & Jul 2019 & Mar 2020 
	 & Grammar fuzzer that generates C programs \\
	 
	 & \cellcolor{gray!25} \href{https://github.com/utting/agilkia} {Agilika} \cite{Utting2020_92} & \cellcolor{gray!25} 7 & \cellcolor{gray!25} Aug 2020 & \cellcolor{gray!25} Mar 2022   
	 & \cellcolor{gray!25}  Generates tests from execution traces \\
	 
	 & \href{https://github.com/OOPSLA-2019-BugDetection} {BugDetection} \cite{Li2019_96}
	 & 66 & Oct 2019 & May 2021
	 & Trains models for defect prediction \\
	 
	 & \cellcolor{gray!25} \href{https://zenodo.org/record/3373409} {{\sc dtldp}} \cite{Chen2020_140}
	 & \cellcolor{gray!25} 21 & \cellcolor{gray!25} Aug 2019 & \cellcolor{gray!25} --
	 & \cellcolor{gray!25}  Implements a deep transfer learning framework \\
	 
	 & \href{https://github.com/michaelpradel/DeepBugs} {DeepBugs} \cite{Pradel2018_84}
	 & 210 & Nov 2018 & May 2021
	 & Implements a framework for learning name-based bug detectors\\
	 
	 & \cellcolor{gray!25} \href{https://github.com/randoop/randoop}{Randoop}
	 & \cellcolor{gray!25} -- & \cellcolor{gray!25} -- & \cellcolor{gray!25} Jul 2022
	 & \cellcolor{gray!25}	Generates  tests automatic for Java code \\
	 
	 & \href{https://www.ifi.uzh.ch/en/seal/people/panichella/tools/TestDescriber.html}{TestDescriber}
	 & -- & -- & --
	 & Implements test case summary generator and evaluator\\\hline
	
	\multirow{6}{*}{\shortstack{Vulnerability\\ Analysis}}  & \cellcolor{gray!25} \href{http://awap.sourceforge.net/download.html} {{\sc wap}} \cite{wap_2013}
	& \cellcolor{gray!25} 3 & \cellcolor{gray!25} Oct 2013 & \cellcolor{gray!25}  Nov 2015 
	& \cellcolor{gray!25}  Detects and corrects input validation vulnerabilities\\
	
	& \href{https://github.com/secure-software-engineering/swan} {{\sc swan}\cite{Piskachev2019_21} } 
    & 6 & Oct 2019 & May 2022
    &  Identifies  vulnerabilities \\
	
	& \cellcolor{gray!25} \href{https://github.com/hperl/vccfinder} {{\sc vccf}inder} \cite{Perl2015_61}
	& \cellcolor{gray!25} 174 & \cellcolor{gray!25} Oct 2015
	& \cellcolor{gray!25}  May 2017 
	& \cellcolor{gray!25}  Finds potentially dangerous code in repositories \\\hline
	
	\multirow{10}{*}{General} & \href{https://github.com/google-research/bert} {{\sc bert} }
	& 43462 & Oct 2018 & Mar 2020
	& {\sc nlp} pre-trained models \\
	
	 & \cellcolor{gray!25} \href{http://www.cs.ubc.ca/nest/lci/bc3/framework.html} {{\sc bc3} Annotation Framework }
	 & \cellcolor{gray!25} -- & \cellcolor{gray!25}  -- & \cellcolor{gray!25} --
	 & \cellcolor{gray!25}  Annotates emails/conversations easily \\
	 
	 & \href{http://jgibblda.sourceforge.net/} {JGibLDA}
	 & -- & -- & --
	 & Implements Latent Dirichlet Allocation\\
	 
	 & \cellcolor{gray!25} \href{http://nlp.stanford.edu/software/lex-parser.shtml} {Stanford NLP Parser }
	 & \cellcolor{gray!25} -- & \cellcolor{gray!25} -- & \cellcolor{gray!25} --
	 & \cellcolor{gray!25}  A statistical NLP parser \\
	 
	 & \href{http://www.srcml.org} {srcML}
	 & -- & -- & May 2022
	 & Generates {\sc xml} representation of sourcecode\\
	 
	 & \cellcolor{gray!25} \href{https://github.com/gousiosg/java-callgraph} {CallGraph}
	 & \cellcolor{gray!25}  8 & \cellcolor{gray!25}  Oct 2017 & \cellcolor{gray!25}  Oct 2018
	 & \cellcolor{gray!25} Generates static and dynamic call graphs for Java code\\
	 
	 & \href{https://www.sri.inf.ethz.ch/research/plml}{ML for programming}
	 & -- & -- & --
	 & Offers various tools such as JSNice, Nice2Predict, and {\sc debin}\\\hline
\end{supertabular}

\vspace{5mm}
The list of datasets found in this exploration are presented in 
Table~\ref{tab:datasets}.
Similar to Tools' table, Table~\ref{tab:datasets} lists each resource with it's category, name and link to access the resource, number of citations (as of Jul 2022), the time when it was first introduced along with the time when the resource was last updated, and a short description of the resource.

\vspace{5mm}
\topcaption{A list of datasets useful for analyzing source code and applying machine learning techniques}
%\rowcolors{2}{gray!25}{white}
\label{tab:datasets}
\begin{supertabular}{p{.14\textwidth}p{.20\textwidth}p{.08\textwidth}p{.08\textwidth}p{.08\textwidth}p{.29\textwidth}}
	\rowcolor{gray!50}
	\textbf{Category} & \textbf{Name} & \textbf{\#Citation}& \textbf{Introd.} & \textbf{Updated} & \textbf{Description} \\
	
	\multirow{4}{*}{\shortstack{Code\\ Representation}} & \href{https://github.com/tech-srl/code2seq} {Code2seq} \cite{Alon2019_464}
	& 271 & Jan 2019 & Feb 2022
	& Sequences generated from structured representation of code\\
	
    &\cellcolor{gray!25}  \href{https://ghtorrent.org/} {{GHTorrent}} \cite{Gousi13}
    & \cellcolor{gray!25} 645 & \cellcolor{gray!25}Oct 2013 &\cellcolor{gray!25} Sep 2020
    & \cellcolor{gray!25}  Meta-data from {\sc GitHub} repositories \\\hline

    \multirow{4}{*}{\shortstack{Code\\ Completion}} & \href{https://github.com/jack57lee/neuralCodeCompletion}{Neural Code Completion} & 148 & Nov 2017 & Sep 2019
    & Dataset and code for code completion with neural attention and pointer networks\\

    & \cellcolor{gray!25} \href{https://github.com/xing-hu/TL-CodeSum}{TL-CodeSum}~\cite{Xing18} 
    & \cellcolor{gray!25}150 & \cellcolor{gray!25} Feb 2019 & \cellcolor{gray!25} Sep 2020
    & \cellcolor{gray!25} Dataset for code summarization\\\hline

    \multirow{4}{*}{\shortstack{Program\\ Synthesis}} & \href{https://conala-corpus.github.io/} {{\sc CoNaLa} corpus} \cite{YDC18} 
    & 130 & Dec 2018 & Oct 2021
    & Python snippets and corresponding natural language description \\
    
    & \cellcolor{gray!25} \href{https://github.com/ProgramRepair/IntroClass} {IntroClass} \cite{LeGoues15}
    & \cellcolor{gray!25} 144 & \cellcolor{gray!25} Jul 2015 & \cellcolor{gray!25} Feb 2016
    & \cellcolor{gray!25} Program repair dataset of C programs \\\hline

    \multirow{5}{*}{\shortstack{Program \\Comprehension}}  & \href{https://dijkstra.eecs.umich.edu/code-summary/} {Program comprehension dataset} \cite{Stapleton2020_404}
    & 61 & May 2018 & Aug 2021
    & Contains code for a program comprehension user survey\\
    
    & \cellcolor{gray!25} \href{https://sjiang1.github.io/commitgen/} {CommitGen} \cite{Jiang2017_416}
    & \cellcolor{gray!25} -- & \cellcolor{gray!25} -- & \cellcolor{gray!25} --
    & \cellcolor{gray!25}  Commit messages and the diffs from 1,006 Java projects\\
    
     & \href{https://github.com/LittleYUYU/StackOverflow-Question-Code-Dataset} {StaQC} \cite{StaQC}
     & -- & -- & Nov 2019
     & 148K Python and 120K {\sc sql} question-code pairs from StackOverflow\\\hline

    \multirow{12}{*}{\shortstack{Quality\\ Assessment}} & \cellcolor{gray!25} \href{https://github.com/src-d/datasets} {src-d datasets}
    & \cellcolor{gray!25}-- & \cellcolor{gray!25}-- & \cellcolor{gray!25}--
    & \cellcolor{gray!25}  Various labeled datasets (commit messages, duplicates, DockerHub, and Nuget)\\
    
    & \href{https://github.com/clonebench/BigCloneBench} {BigCloneBench \cite{Svajlenko2014}}
    & 187 & Dec 2014 & Mar 2021
    & Known clones in the IJaDataset source repository \\
    
    & \cellcolor{gray!25} \href{https://github.com/thiru578/Multilabel-Dataset}{Multi-label smells} \cite{Guggulothu2020_206}
    & \cellcolor{gray!25} 28 & \cellcolor{gray!25} May 2020 & \cellcolor{gray!25} --
    & \cellcolor{gray!25} A dataset of 445 instances of two code smells and 82 metrics\\
    
    & \href{https://github.com/tushartushar/DeepLearningSmells}{Deep learning smells} \cite{Sharma2021_510} 
    & 15 & Jul 2021 & Nov 2020
    & A dataset of four smells in tokenized form from 1,072 C\# and 100 Java repositories \\
    
    & \cellcolor{gray!25} \href{https://zenodo.org/record/3547639}{\ml\ for software refactoring} \cite{Aniche2020_173}
    & 6 & Nov 2019 & --
    & \cellcolor{gray!25} Dataset for applying 
    \ml\ to recommend refactoring \\
    
    & \href{https://zenodo.org/record/4468361}{QScored} \cite{QScored}
    & 0 & Aug 2021 & --
    & Code smell and metrics dataset for more than 86 thousand open-source repositories \\\hline

    \multirow{10}{*}{Testing} & \href{https://github.com/rjust/defects4j} {Defects4J}~\cite{Defects4J}
    & 858 & Jul 2014 & Jul 2022
    & Java reproducible bugs \\
    
    & \cellcolor{gray!25} \href{https://promise.site.uottawa.ca/SERepository/datasets-page.html} {{\sc promise}}~\cite{Promise_dataset_2005}
    & \cellcolor{gray!25} 41 & \cellcolor{gray!25} -- & \cellcolor{gray!25} Jan 2021
    & \cellcolor{gray!25} Various datasets including defect prediction and cost estimation \\
    
    & \href{https://github.com/OOPSLA-2019-BugDetection/OOPSLA-2019-BugDetection} {BugDetection}~\cite{Li2019_96}
    & 66 & Oct 2019 & May 2021
    &  A bug prediction dataset containing 4.973M methods belonging to 92 different Java project versions \\
    
    & \cellcolor{gray!25} \href{https://github.com/aravi11/data-augmented-metamorphic-testing} {{\sc damt}}~\cite{Nair2019_101}
    & \cellcolor{gray!25} 15 & \cellcolor{gray!25} Aug 2019 & \cellcolor{gray!25} Dec 2019
    & \cellcolor{gray!25}  Metamorphic testing dataset \\
    
    & \href{https://zenodo.org/record/3373409} {{\sc dtldp}}~\cite{Chen2020_140}
    & 21 & Oct 2020 & -- 
    & Dataset for deep transfer learning for defect prediction\\
    
    & \cellcolor{gray!25} \href{https://www.sri.inf.ethz.ch/js150} {\sc DeepBugs}~\cite{Pradel2018_84}
    & \cellcolor{gray!25} 210 & \cellcolor{gray!25} Oct 2018 & \cellcolor{gray!25} Apr 2021 
    & \cellcolor{gray!25} A JavaScript code corpus with 150K code snippets\\\hline

    \multirow{18}{*}{\shortstack{Vulnerability\\ Analysis}} & \cellcolor{gray!25} \href{https://wpscan.com/wordpresses} {{\sc wp}scan }
    & \cellcolor{gray!25} -- & \cellcolor{gray!25}-- & \cellcolor{gray!25}--
    & \cellcolor{gray!25}  a {\sc php} dataset for WordPress plugin vulnerabilities \\
    
    & \href{https://www.malgenomeproject.org/} {Genome}~\cite{genome_2012}
    & 1139 & Jul 2012 & Dec 2015
    & 1,200 malware samples covering the majority of existing malware families \\

    & \cellcolor{gray!25}  \href{https://www.nist.gov/publications/juliet-11-cc-and-java-test-suite} {Juliet}~\cite{juliet_test_suite_2012}
    & \cellcolor{gray!25} -- & \cellcolor{gray!25} -- & \cellcolor{gray!25} -- 
    & \cellcolor{gray!25} 81K synthetic C/C++ and Java programs with known flaws \\

    & \href{http://androzoo.uni.lu/lists} {AndroZoo}~\cite{AndroZoo_2016}
    & -- & -- & --
    & 15.7M {\sc apk}s from Google's Play Store \\
    
    & \cellcolor{gray!25} \href{https://github.com/DanielLin1986/TransferRepresentationLearning} {{\sc trl}}~\cite{trl_2018}
    &  \cellcolor{gray!25} 108 & \cellcolor{gray!25} Apr 2018 & \cellcolor{gray!25} Jan 2019
    & \cellcolor{gray!25} Vulnerabilities in six C programs \\

    & \href{https://osf.io/d45bw/} {Draper {\sc vdisc}}~\cite{vdisc_2018}
    & 247 & Jul 2018 & Nov 2018
    & 1.27 million functions mined from {\sc c} and {\sc c++} applications\\

    & \cellcolor{gray!25} \href{samate.nist.gov/SRD/view.php} {{\sc samate}}~\cite{samate_2007}
    & \cellcolor{gray!25} -- & \cellcolor{gray!25} -- & \cellcolor{gray!25} -- 
    & \cellcolor{gray!25} A set of known security flaws from {\sc nist} for {\sc c, c++,} and Java programs\\

    & \href{https://www.inf.u-szeged.hu/~ferenc/papers/JSVulnerability AnalysisDataSet/} {{\sc js}Vulner}~\cite{Ferenc2019_18}
    & -- & -- & --
    & JavaScript Vulnerability Analysis dataset\\

    & \cellcolor{gray!25} \href{https://github.com/secure-software-engineering/swan} {{\sc swan}}~\cite{Piskachev2019_21}
    & \cellcolor{gray!25}6 & \cellcolor{gray!25} Jul 2019 & \cellcolor{gray!25} Jul 2022
    & \cellcolor{gray!25} A Vulnerability Analysis collection of 12 Java applications \\

    & \href{https://github.com/SAP/project-kb/tree/master/MSR2019}{Project-KB}~\cite{Ponta2019_400}
    & 49 & Aug 2019 & --
    & A Manually-Curated dataset of fixes to vulnerabilities of open-source software\\\hline

    \multirow{6}{*}{General} & \cellcolor{gray!25} \href{https://groups.inf.ed.ac.uk/cup/javaGithub/} {GitHub Java Corpus}~\cite{Allamanis2013_474}
    & \cellcolor{gray!25} 333 & \cellcolor{gray!25} -- & \cellcolor{gray!25}--
    & \cellcolor{gray!25} A large collection of Java repositories\\
    
    & \href{https://www.sri.inf.ethz.ch/py150} {150k Python dataset}~\cite{Raychev2016_504}
    & -- & -- & -- 
    & Contains parsed \abst\ for 150K Python files\\

    & \cellcolor{gray!25} \href{https://www.ics.uci.edu/~lopes/datasets/} {{\sc uci} source code dataset}~\cite{Lopes2010}
    & \cellcolor{gray!25} 38 & \cellcolor{gray!25} Apr 2010 & \cellcolor{gray!25} Nov 2013
    & \cellcolor{gray!25}  Various large scale source code analysis datasets\\\hline
\end{supertabular}

\section{Challenges and perceived deficiencies}
\label{results_rq4}

The aim of this section is to focus on the perceived deficiencies, challenges, and opportunities
 in applying \ml\ techniques in the context of 
source code analysis
observed from the primary studies.
We document challenges or deficiencies mentioned in the considered primary studies
while studying and summarizing them. 
After the summarization phase was over, we consolidated all the documented notes and
synthesized a summary that we present below.

\begin{itemize}
	\item{\textit{\textbf{Standard datasets:}}}
	\ml\ is by nature data hungry;
	specifically, supervised learning methods need a considerably large, cleaned, and
	annotated dataset.
	Though the size of available open software engineering artifacts is increasing day by day,
	lack of high-quality datasets (\ie\ clean and reliably annotated)
	are one of the biggest challenges in the domain~\cite{Ghaffarian2017_52, Ucci2019_54, Gondra2008_73, Kumar2012_77, Durelli2019_103, Chen2020_140, Barbez2020_180, Alsolai2020_186, Tsintzira2020_194, Soto2018_292, Tian2020_309, Svyatkovskiy2020_321, Gopinath2016_348, Sakkas2020_366, Liu2019_512, Wan2019_396, Shen2020_9,Jimenez2019}.
	Therefore, there is a need for defining standardized datasets.
	Authors have cited low performance, poor generalizability, and over-fitting due to poor dataset quality 	as the results of the lack of standard validated high-quality datasets.
	
	\item {\textit{\textbf{Reproducibility and replicability:}}} 
	Reproducibility and replicability of any \ml\ implementation can be compromised by
	factors discussed below.
	\begin{itemize}
		\item \textit{Insufficient information:}  Aspects such as \ml\ model, their hyper-parameters,
		data size and ratio (of benign and faulty samples, for instance) are needed to understand
		and replicate the study.
		During our exploration, we found numerous studies that do not present even the bare-minimum 
		pieces of information to replicate and reproduce their results.
		Likewise, \citet{DiNucci2018_220} carried out a detailed replication study and reported that
		the replicated results were lower by up to 90\% compared to what was reported in the original study.
		\item \textit{Handling of data imbalance:}
		It is very common to have imbalanced datasets in software engineering applications.
		Authors use techniques such as under-sampling and over-sampling to overcome the challenge for training.
		However, test datasets must retain the original sample ratio as found in the real world~\cite{DiNucci2018_220};
		carrying out a performance evaluation based on a balanced dataset is flawed. Obviously, the model will perform
		significantly inferior when it is put at work in a real-world context.
		We noted many studies~\cite{Agnihotri2020_193, Oliveira2020_195, Guggulothu2020_206, Fontana2013_207, Fontana2015_211, Thongkum2020_217, Cunha2020_231} that used balanced samples and often did not provide the size and ratio
		of the training and testing dataset.
		Such improper handling of data imbalance contributes to poor reproducibility.
	\end{itemize}
	\item {\textbf{\textit{Maturity in \textsc{ml} development:}}} 
	Development of \ml\ systems are inherently different from traditional software development~\cite{Wan2019_396}. 
	Phases of \ml\ development are very exploratory in nature and highly domain and problem dependent~\cite{Wan2019_396}.
	Identifying the most appropriate \ml\ model, their appropriate parameters, and configuration is  largely driven by \textit{trial and error} manner~\cite{Wan2019_396, Azeem2019_240, Shen2020_9}.
	Such an \textit{ad hoc} and immature software development environment poses a huge challenge to the community.

	A related challenge is lack of tools and techniques for \ml\ software development.
	It includes effective tools for testing \ml\ programs, ensuring that the dataset are pre-processed adequately, debugging, and
	effective data management~\cite{Wan2019_396, Patel2008_397, Giray2021}.
	In addition, quality aspects such as explainability and trust-worthiness are new desired quality aspects especially applicable for
	\ml\ code where current practices and knowledge is inadequate~\cite{Giray2021}.
	
	\item \textit{\textbf{Data privacy and bias:}}
	Data hungry \ml\ models are considered as good as the data they are consuming.
	Data collection and preparation without data diversity leads to bias and unfairness.
	Although we are witnessing more efforts to understand these sensitive aspects \cite{Zhang2021, Brun2018}, 
the present set of methods and practices lack the support to deal with data privacy issues at large as well as data diversity and fairness \cite{Brun2018, Giray2021}. 
	
\item \textit{\textbf{Effective feature engineering:}}
Features represent the problem-specific knowledge in pieces extracted from the data; the effectiveness of any \ml\ model
depends on the features fed into it.
Many studies identified the importance of effective feature engineering 
and the challenges in gathering the same~\cite{Tsintzira2020_194, Shen2020_9, Patel2008_397, Wan2019_396, Ivers2019_395}.
Specifically, software engineering researchers have notified that identifying and extracting relevant features beyond code quality metrics
is non-trivial.
For example, \citet{Ivers2019_395} discusses that identifying features that establishes a relationship among different code elements
is a significant challenge for \ml\  implementations applied on source code analysis.
\citet{Sharma2021_510} have shown in their study that smell detection using \ml\ techniques
perform poorly especially for design smells where multiple code elements and their properties has to be observed.

\item \textbf{\textit{Skill gap:}}
\citet{Wan2019_396} identified that \ml\ software development requires an extended set of skills beyond software development
including \ml\ techniques, statistics, and mathematics apart from the application domain.
Similarly, \citet{Hall2012_145} also reports a serious lack of \ml\ expertise in academic software engineering efforts.
Other authors~\cite{Patel2008_397} have emphasized the importance of domain knowledge to design effective \ml\ models.

\item \textbf{\textit{Hardware resources:}}
Given the need of large training dataset and many hidden layers, often \ml\ training requires high-end processing units (such as {\sc gpu}s and memory) \cite{Wan2019_396, Giray2021}.
A user-survey study~\cite{Wan2019_396} highlights the need to special hardware for \ml\ training. 
Such requirements poses a challenge to researchers constrained with limited hardware resources.
\end{itemize}

\section{Discussion} \label{discussion}
This section provides a discussion on the top venues for articles belonging to each selected category in our scope
as well as on potential mitigations for the challenges we identified in the previous section.

\subsection{Venue and article categories}
\begin{figure*}[!h]
	\centering
	\includegraphics[width=0.8\textwidth]{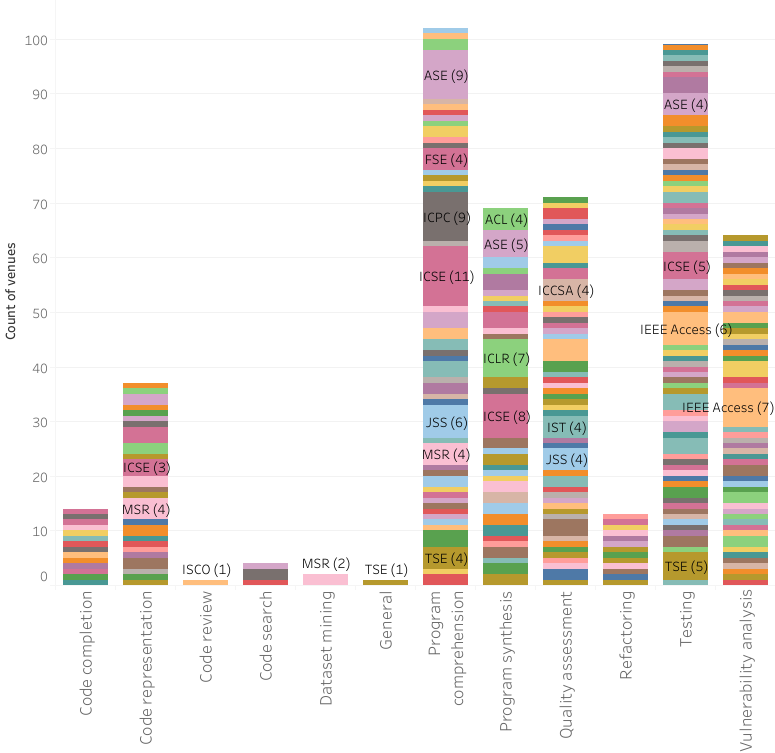}
	\caption{Top venues for each considered category}
	\label{fig:venue}
\end{figure*}

The goal of the exploration is to understand the top venues for each considered category.
We identified and manually curated the software engineering venue for each primary study discussed in our literature review.
Figure ~\ref{fig:venue} shows the venues for the considered categories.
We show the most prominent venues per category. 
Each label includes a number indicating  the number of articles published at the same venue in that category.

We observe that {\sc icse} and {\sc ase} are among the top venues, appearing in four and three categories respectively.
{\sc tse}, {\sc jss}, and {\sc ieee} \textit{Access} are the top journals for the considered categories.
Machine learning conferences such as {\sc iclr} also appear as the one of top venues for \textit{program synthesis} category).

The categories of \textit{program comprehension} and \textit{program synthesis} exhibit the highest concentration of articles to a relatively small list of top venues where 45\% and 35\% of articles, respectively, come from the top venues.
On the other hand, researchers publish articles related to \textit{testing} and \textit{vulnerability} in a rather large number of venues. 

\subsection{Model selection}
Selecting a \ml{} model for a given task depends on many factors such as nature of the problem, properties of training and input samples, and expected output.
Below, we provide an analysis of employed \ml{} models based on these factors.
\begin{itemize}
    \item 
    One of the factors that influence the choice of \ml{} models is the chosen features and their properties. Studies in the \textit{quality assessment} category majorly relied on token-based features and code quality metrics. Such features allowed studies in this categories to use traditional \ml{} models. Some authors applied \dl{} models such as \DNN{} when higher-granularity constructs such as {\sc cfg} and {\sc dfg} are used as features.
    \item 
    Similarly, the majority of the studies in \textit{testing} category relied on code quality metrics. Therefore, they have fixed size, fixed meaning (for each column) vectors to feed to a \ml{} model. With such inputs, traditional \ml{} approaches, such as \rf{} and \svm{}, work well. Other studies used a variation of \abst{} or \abst{} of the changes to generate the embeddings. \dl{} models including \DNN{} and \RNN{}-based models are used to first train a model for embeddings. A typical \ml{} classifier use the embeddings to classify samples in buggy or benign.
    \item
    Typical output of a \textit{code representation} study is embedding representing code in the vector form. The semantics of the produced embeddings depends significantly on the selected features.  Studies in this domain identify this aspect and hence swiftly focused to extract features that capture the relevant semantics; for example, path-based features encode the order among the tokens. The chosen \ml{} model plays another important role to generate effective embeddings. Given the success of \RNN{} with text processing tasks, due to its capability to identify sequence and pattern, \RNN{}-based models dominate this category.
    \item
    \textit{Program repair} is typically a sequence to sequence transformation \ie{} a sequence of buggy code is the input and a sequence of fixed code is the output. Given the nature of the problem, it is not surprising to observe that the majority of the studies in this category used Encoder-Decoder-based models. \RNN{} are considered a popular choice to realize Encoder-Decoder models due to its capability to remember long sequences.
\end{itemize}

\subsection{Mitigating the challenges}
\subsubsection{Availability of standard datasets:}
Although available datasets have increased, given a wide number of software engineering tasks and variations in these tasks as well as the need of application-specific datasets,
the community still looks for application-specific, large, and high-quality datasets.
To mitigate the issue, the community has focused on developing new datasets and making them publicly available 
by organizing a dedicated track, for example, the {\sc msr} data showcase track.
Dataset search engines such as Google dataset search\footnote{\url{https://datasetsearch.research.google.com/ }}
could be used to search available datasets. 
Researchers may also propose generic datasets that can serve multiple application domains or at least different variations of a software engineering task.
In addition, recent advancements in \ml\ techniques such as active learning \cite{Prince2004, Settles2009, Ren2020survey} may reduce the need of large datasets.
Besides, the way the data is used for model validation must be improved. For example, Jimenez et al. \cite{Jimenez2019} showed that previous studies on vulnerability prediction trained predictive models by using perfect labelling information (i.e., including future labels, as yet undiscovered vulnerabilities) and showed that such an unrealistic labelling assumption can profoundly affect the scientific conclusions of a study as the prediction performance worsen dramatically when one fully accounts for realistically available labelling.

\subsubsection{Reproducibility and replicability:}
The importance of reproducibility and replicability has been emphasized and understood by the software engineering community~\cite{ChaoLiu2020}.
It has lead to a concrete artifact evaluation mechanism adopted by leading software engineering conferences. 
For example, {\sc fse} artifact evaluation divides artifacts into five categories---\textit{functional, reusable, available, results reproduced,} and \textit{results replicated}.\footnote{\url{https://2021.esec-fse.org/track/fse-2021-artifacts}}
Such thorough evaluation encouraging software engineering authors to produce high-quality documentation along with easily replicate experiment results using their developed artifacts. 
In addition, efforts (such as model engineering process~\cite{Banna2021}) are being made to support \ml\ research reproducible and replicable.

\subsubsection{Maturity in ML development:}
The ad-hoc trial and error \ml\ development can be addressed by improved tools and techniques.
Even though the variety of \ml\ development environments including managed services such as {\sc aws} Sagemaker and Google Notebooks
attempt to make \ml\ development easier, they essentially do not offer much help in reducing the ad-hoc nature of the development.
A significant research push from the community would make \ml\ development relatively  systematic and organized.

Recent advancements in the form of available tools not only help a developer to comprehend the process but also let them effectively manage code, data, and experimental results. 
Examples of such tools and methods include {\sc darviz}~\cite{Sankaran2017} for \dl\ model visualization, {\sc mlf}low\footnote{\url{https://mlflow.org/}} for managing the \ml\ lifecycle, and DeepFault~\cite{Eniser2019} for identifying faults in \dl\ programs. 
Such efforts are expected to address the challenge.

Software Engineering for Machine Learning ({\sc se4ml}) brings another perspective to this issue by bringing best practices from software engineering to
\ml\ development. 
Efforts in this direction not only can make \ml\ specific code maintainable and reliable but also can contribute back to reproducibility and replicability.

\subsubsection{Hardware resources:}
\ml\ development is resource hungry. 
Certain \dl\ models (such as models based on \RNN{}) consume excessive hardware resources.
The need for a large-scale hardware infrastructure is increasing with the increase in size of the captured features and the training samples.
To address the challenge, infrastructure at institution and country level are maintained in some countries;
however, a generic and widely-applicable solution is needed for more globally-inclusive research.

The first internal threats to validity relates to the concern of covering all the relevant articles in the selected domain.
To mitigate the concern, we defined our scope \ie{} studies that use \ml\ techniques to solve a software engineering
problem by analyzing source code.
We also carefully defined inclusion and exclusion criteria for selecting relevant studies.
We carry out extensive manual search process on commonly used digital libraries with the help of a comprehensive set of search terms;
we augment the search terms that are used in related articles to maximize the chances of identifying the relevant articles.

Another threat to validity is the validity of data extraction and their interpretation applicable to the generated summary and metadata for each primary study.
We mitigated this threat by dividing the task of summarization to all the authors and cross verifying the generated information.
During the manual summarization phase, metadata of each paper was reviewed by, at least, two authors.

External validity concerns  the generalizability and reproducibility of the produced results and observations.
We provide a spreadsheet~\cite{Replication_ML4SCA} containing all the metadata for all the articles selected in each of the phases of article selection.
In addition, inspired by previous surveys \cite{Allamanis2018_452, hort21tse}, we have developed a website\footnote{\url{http://www.tusharma.in/ML4SCA}} as a \textit{living documentation and literature survey} to facilitate easy navigation, exploration, and extension. 
The website can be easily extended as the new studies emerge in the domain; we have made the repository\footnote{\url{https://github.com/tushartushar/ML4SCA}} open-source to allow the community to extend the living literature survey.\section{Conclusions}\label{conclusions}
With the increasing presence of \ml\ techniques in software engineering research,
it has become challenging to have a comprehensive overview of its advancements.
This survey aims to provide a detailed overview of the studies at the intersection of source code analysis
and \ml{}.
We have selected 479 primary studies spanning from 2011 to 2021 (and to some extent 2022) covering 12 software engineering categories.
We present a synthesized summary of the selected studies arranged in categories, subcategories, and their corresponding involved steps.
Also, the survey consolidates useful resources (datasets and tools) that could ease the task for future studies. 
Finally, we present perceived challenges and opportunities in the field.
The presented opportunities invite practitioners as well as researchers to propose new methods, tools, and techniques to make the integration of \ml\ techniques for software engineering applications easy, flexible, and maintainable.

\begin{acks}
%\section*{Acknowledgments}
This work is supported by the {\sc erc} Advanced fellowship grant  no. 741278 ({\sc epic}).
\end{acks}

\bibliographystyle{ACM-Reference-Format}
\bibliography{ref_curated}


\begin{thebibliography}{521}

%%% the Web address.

\ifx \showCODEN    \undefined \def \showCODEN     #1{\unskip}     \fi
\ifx \showDOI      \undefined \def \showDOI       #1{#1}\fi
\ifx \showISBNx    \undefined \def \showISBNx     #1{\unskip}     \fi
\ifx \showISBNxiii \undefined \def \showISBNxiii  #1{\unskip}     \fi
\ifx \showISSN     \undefined \def \showISSN      #1{\unskip}     \fi
\ifx \showLCCN     \undefined \def \showLCCN      #1{\unskip}     \fi
\ifx \shownote     \undefined \def \shownote      #1{#1}          \fi
\ifx \showarticletitle \undefined \def \showarticletitle #1{#1}   \fi
\ifx \showURL      \undefined \def \showURL       {\relax}        \fi
% invisible to TeX
\providecommand\bibfield[2]{#2}
\providecommand\bibinfo[2]{#2}
\providecommand\natexlab[1]{#1}
\providecommand\showeprint[2][]{arXiv:#2}

\bibitem[\protect\citeauthoryear{??}{mis}{2020}]%
        {misc20}
 \bibinfo{year}{2020}\natexlab{}.
\newblock \bibinfo{title}{GitHub archive}.
\newblock
\newblock
\urldef\tempurl%
\url{https://www.gharchive.org/}
\showURL{%
\tempurl}


\bibitem[\protect\citeauthoryear{Abbas, Albalooshi, and Hammad}{Abbas
  et~al\mbox{.}}{2020}]%
        {738_Abbas2020}
\bibfield{author}{\bibinfo{person}{Raja Abbas},
  \bibinfo{person}{Fawzi~Abdulaziz Albalooshi}, {and} \bibinfo{person}{Mustafa
  Hammad}.} \bibinfo{year}{2020}\natexlab{}.
\newblock \showarticletitle{Software change proneness prediction using machine
  learning}. In \bibinfo{booktitle}{\emph{2020 International Conference on
  Innovation and Intelligence for Informatics, Computing and Technologies
  (3ICT)}}. IEEE, \bibinfo{pages}{1--7}.
\newblock


\bibitem[\protect\citeauthoryear{Abdalkareem, Mujahid, and Shihab}{Abdalkareem
  et~al\mbox{.}}{2020}]%
        {947_Abdalkareem2020}
\bibfield{author}{\bibinfo{person}{Rabe Abdalkareem}, \bibinfo{person}{Suhaib
  Mujahid}, {and} \bibinfo{person}{Emad Shihab}.}
  \bibinfo{year}{2020}\natexlab{}.
\newblock \showarticletitle{A machine learning approach to improve the
  detection of ci skip commits}.
\newblock \bibinfo{journal}{\emph{IEEE Transactions on Software Engineering}}
  (\bibinfo{year}{2020}).
\newblock


\bibitem[\protect\citeauthoryear{Abdeljaber, Avci, Kiranyaz, Gabbouj, and
  Inman}{Abdeljaber et~al\mbox{.}}{2017}]%
        {Abdeljaber2017}
\bibfield{author}{\bibinfo{person}{Osama Abdeljaber}, \bibinfo{person}{Onur
  Avci}, \bibinfo{person}{Serkan Kiranyaz}, \bibinfo{person}{Moncef Gabbouj},
  {and} \bibinfo{person}{Daniel~J Inman}.} \bibinfo{year}{2017}\natexlab{}.
\newblock \showarticletitle{Real-time vibration-based structural damage
  detection using one-dimensional convolutional neural networks}.
\newblock \bibinfo{journal}{\emph{Journal of Sound and Vibration}}
  \bibinfo{volume}{388} (\bibinfo{year}{2017}), \bibinfo{pages}{154--170}.
\newblock


\bibitem[\protect\citeauthoryear{Abuhamad, AbuHmed, Mohaisen, and
  Nyang}{Abuhamad et~al\mbox{.}}{2018}]%
        {Abuhamad2018_383}
\bibfield{author}{\bibinfo{person}{Mohammed Abuhamad}, \bibinfo{person}{Tamer
  AbuHmed}, \bibinfo{person}{Aziz Mohaisen}, {and} \bibinfo{person}{DaeHun
  Nyang}.} \bibinfo{year}{2018}\natexlab{}.
\newblock \showarticletitle{Large-Scale and Language-Oblivious Code Authorship
  Identification}. In \bibinfo{booktitle}{\emph{Proceedings of the 2018 ACM
  SIGSAC Conference on Computer and Communications Security}} (Toronto, Canada)
  \emph{(\bibinfo{series}{CCS '18})}. \bibinfo{pages}{101–114}.
\newblock
\showISBNx{9781450356930}
\urldef\tempurl%
\url{https://doi.org/10.1145/3243734.3243738}
\showDOI{\tempurl}


\bibitem[\protect\citeauthoryear{Abunadi and Alenezi}{Abunadi and
  Alenezi}{2015}]%
        {Abunadi_Alenezi_2015}
\bibfield{author}{\bibinfo{person}{Ibrahim Abunadi} {and}
  \bibinfo{person}{Mamdouh Alenezi}.} \bibinfo{year}{2015}\natexlab{}.
\newblock \showarticletitle{Towards Cross Project Vulnerability Prediction in
  Open Source Web Applications}. In \bibinfo{booktitle}{\emph{Proceedings of
  the The International Conference on Engineering \& MIS 2015}} (Istanbul,
  Turkey) \emph{(\bibinfo{series}{ICEMIS '15})}.
  \bibinfo{publisher}{Association for Computing Machinery},
  \bibinfo{address}{New York, NY, USA}, Article \bibinfo{articleno}{42},
  \bibinfo{numpages}{5}~pages.
\newblock
\showISBNx{9781450334181}
\urldef\tempurl%
\url{https://doi.org/10.1145/2832987.2833051}
\showDOI{\tempurl}


\bibitem[\protect\citeauthoryear{Aggarwal}{Aggarwal}{2019}]%
        {Aggarwal2019_128}
\bibfield{author}{\bibinfo{person}{Simran Aggarwal}.}
  \bibinfo{year}{2019}\natexlab{}.
\newblock \showarticletitle{Software Code Analysis Using Ensemble Learning
  Techniques}. In \bibinfo{booktitle}{\emph{Proceedings of the International
  Conference on Advanced Information Science and System}} (Singapore,
  Singapore) \emph{(\bibinfo{series}{AISS '19})}. Article
  \bibinfo{articleno}{9}, \bibinfo{numpages}{7}~pages.
\newblock
\showISBNx{9781450372916}
\urldef\tempurl%
\url{https://doi.org/10.1145/3373477.3373486}
\showDOI{\tempurl}


\bibitem[\protect\citeauthoryear{Agnihotri and Chug}{Agnihotri and
  Chug}{2020}]%
        {Agnihotri2020_193}
\bibfield{author}{\bibinfo{person}{Mansi Agnihotri} {and}
  \bibinfo{person}{Anuradha Chug}.} \bibinfo{year}{2020}\natexlab{}.
\newblock \showarticletitle{Application of machine learning algorithms for code
  smell prediction using object-oriented software metrics}.
\newblock \bibinfo{journal}{\emph{Journal of Statistics and Management
  Systems}} \bibinfo{volume}{23}, \bibinfo{number}{7} (\bibinfo{year}{2020}),
  \bibinfo{pages}{1159--1171}.
\newblock
\urldef\tempurl%
\url{https://doi.org/10.1080/09720510.2020.1799576}
\showDOI{\tempurl}
\showeprint{https://doi.org/10.1080/09720510.2020.1799576}


\bibitem[\protect\citeauthoryear{Ahmad, Chakraborty, Ray, and Chang}{Ahmad
  et~al\mbox{.}}{2020}]%
        {Ahmad2020_410}
\bibfield{author}{\bibinfo{person}{Wasi Ahmad}, \bibinfo{person}{Saikat
  Chakraborty}, \bibinfo{person}{Baishakhi Ray}, {and} \bibinfo{person}{Kai-Wei
  Chang}.} \bibinfo{year}{2020}\natexlab{}.
\newblock \showarticletitle{A Transformer-based Approach for Source Code
  Summarization}. In \bibinfo{booktitle}{\emph{Proceedings of the 58th Annual
  Meeting of the Association for Computational Linguistics}}.
  \bibinfo{pages}{4998--5007}.
\newblock
\urldef\tempurl%
\url{https://doi.org/10.18653/v1/2020.acl-main.449}
\showDOI{\tempurl}


\bibitem[\protect\citeauthoryear{Ahmed, Kumar, Karkare, Kar, and Gulwani}{Ahmed
  et~al\mbox{.}}{2018}]%
        {Ahmed2018_293}
\bibfield{author}{\bibinfo{person}{Umair~Z. Ahmed}, \bibinfo{person}{Pawan
  Kumar}, \bibinfo{person}{Amey Karkare}, \bibinfo{person}{Purushottam Kar},
  {and} \bibinfo{person}{Sumit Gulwani}.} \bibinfo{year}{2018}\natexlab{}.
\newblock \showarticletitle{Compilation Error Repair: For the Student Programs,
  from the Student Programs}. In \bibinfo{booktitle}{\emph{Proceedings of the
  40th International Conference on Software Engineering: Software Engineering
  Education and Training}} (Gothenburg, Sweden)
  \emph{(\bibinfo{series}{ICSE-SEET '18})}. \bibinfo{pages}{78–87}.
\newblock
\showISBNx{9781450356602}
\urldef\tempurl%
\url{https://doi.org/10.1145/3183377.3183383}
\showDOI{\tempurl}


\bibitem[\protect\citeauthoryear{{Al-Jamimi} and {Ahmed}}{{Al-Jamimi} and
  {Ahmed}}{2013}]%
        {Al-Jamimi2013_244}
\bibfield{author}{\bibinfo{person}{H.~A. {Al-Jamimi}} {and} \bibinfo{person}{M.
  {Ahmed}}.} \bibinfo{year}{2013}\natexlab{}.
\newblock \showarticletitle{Machine Learning-Based Software Quality Prediction
  Models: State of the Art}. In \bibinfo{booktitle}{\emph{2013 International
  Conference on Information Science and Applications (ICISA)}}.
  \bibinfo{pages}{1--4}.
\newblock
\urldef\tempurl%
\url{https://doi.org/10.1109/ICISA.2013.6579473}
\showDOI{\tempurl}


\bibitem[\protect\citeauthoryear{Al~Qasem, Akour, and Alenezi}{Al~Qasem
  et~al\mbox{.}}{2020}]%
        {729_Al2020}
\bibfield{author}{\bibinfo{person}{Osama Al~Qasem}, \bibinfo{person}{Mohammed
  Akour}, {and} \bibinfo{person}{Mamdouh Alenezi}.}
  \bibinfo{year}{2020}\natexlab{}.
\newblock \showarticletitle{The influence of deep learning algorithms factors
  in software fault prediction}.
\newblock \bibinfo{journal}{\emph{IEEE Access}}  \bibinfo{volume}{8}
  (\bibinfo{year}{2020}), \bibinfo{pages}{63945--63960}.
\newblock


\bibitem[\protect\citeauthoryear{AL-Shaaby, Aljamaan, and Alshayeb}{AL-Shaaby
  et~al\mbox{.}}{2020}]%
        {AL-Shaaby2020_198}
\bibfield{author}{\bibinfo{person}{A. AL-Shaaby}, \bibinfo{person}{Hamoud~I.
  Aljamaan}, {and} \bibinfo{person}{M. Alshayeb}.}
  \bibinfo{year}{2020}\natexlab{}.
\newblock \showarticletitle{Bad Smell Detection Using Machine Learning
  Techniques: A Systematic Literature Review}.
\newblock \bibinfo{journal}{\emph{Arabian Journal for Science and Engineering}}
   \bibinfo{volume}{45} (\bibinfo{year}{2020}), \bibinfo{pages}{2341--2369}.
\newblock


\bibitem[\protect\citeauthoryear{Alazba and Aljamaan}{Alazba and
  Aljamaan}{2021}]%
        {820_Alazba2021}
\bibfield{author}{\bibinfo{person}{Amal Alazba} {and} \bibinfo{person}{Hamoud
  Aljamaan}.} \bibinfo{year}{2021}\natexlab{}.
\newblock \showarticletitle{Code smell detection using feature selection and
  stacking ensemble: An empirical investigation}.
\newblock \bibinfo{journal}{\emph{Information and Software Technology}}
  \bibinfo{volume}{138} (\bibinfo{year}{2021}), \bibinfo{pages}{106648}.
\newblock


\bibitem[\protect\citeauthoryear{Aleem, Capretz, Ahmed, et~al\mbox{.}}{Aleem
  et~al\mbox{.}}{2015}]%
        {1062_Aleem2015}
\bibfield{author}{\bibinfo{person}{Saiqa Aleem}, \bibinfo{person}{Luiz~Fernando
  Capretz}, \bibinfo{person}{Faheem Ahmed}, {et~al\mbox{.}}}
  \bibinfo{year}{2015}\natexlab{}.
\newblock \showarticletitle{Comparative performance analysis of machine
  learning techniques for software bug detection}. In
  \bibinfo{booktitle}{\emph{Proceedings of the 4th International Conference on
  Software Engineering and Applications}}. AIRCC Press Chennai, Tamil Nadu,
  India, \bibinfo{pages}{71--79}.
\newblock


\bibitem[\protect\citeauthoryear{Aleti and Martinez}{Aleti and
  Martinez}{2021}]%
        {893_Aleti2021}
\bibfield{author}{\bibinfo{person}{Aldeida Aleti} {and} \bibinfo{person}{Matias
  Martinez}.} \bibinfo{year}{2021}\natexlab{}.
\newblock \showarticletitle{E-APR: mapping the effectiveness of automated
  program repair techniques}.
\newblock \bibinfo{journal}{\emph{Empirical Software Engineering}}
  \bibinfo{volume}{26}, \bibinfo{number}{5} (\bibinfo{year}{2021}),
  \bibinfo{pages}{1--30}.
\newblock


\bibitem[\protect\citeauthoryear{Alhusain, Coupland, John, and
  Kavanagh}{Alhusain et~al\mbox{.}}{2013}]%
        {927_Alhusain2013}
\bibfield{author}{\bibinfo{person}{Sultan Alhusain}, \bibinfo{person}{Simon
  Coupland}, \bibinfo{person}{Robert John}, {and} \bibinfo{person}{Maria
  Kavanagh}.} \bibinfo{year}{2013}\natexlab{}.
\newblock \showarticletitle{Towards machine learning based design pattern
  recognition}. In \bibinfo{booktitle}{\emph{2013 13th UK Workshop on
  Computational Intelligence (UKCI)}}. IEEE, \bibinfo{pages}{244--251}.
\newblock


\bibitem[\protect\citeauthoryear{Ali~Alatwi, Oh, Fokoue, and
  Stackpole}{Ali~Alatwi et~al\mbox{.}}{2016}]%
        {Ali2016_14}
\bibfield{author}{\bibinfo{person}{Huda Ali~Alatwi}, \bibinfo{person}{Tae Oh},
  \bibinfo{person}{Ernest Fokoue}, {and} \bibinfo{person}{Bill Stackpole}.}
  \bibinfo{year}{2016}\natexlab{}.
\newblock \showarticletitle{Android Malware Detection Using Category-Based
  Machine Learning Classifiers}. In \bibinfo{booktitle}{\emph{Proceedings of
  the 17th Annual Conference on Information Technology Education}} (Boston,
  Massachusetts, USA) \emph{(\bibinfo{series}{SIGITE '16})}.
  \bibinfo{pages}{54–59}.
\newblock
\showISBNx{9781450344524}
\urldef\tempurl%
\url{https://doi.org/10.1145/2978192.2978218}
\showDOI{\tempurl}


\bibitem[\protect\citeauthoryear{{Alikhashashneh}, {Raje}, and
  {Hill}}{{Alikhashashneh} et~al\mbox{.}}{2018}]%
        {Alikhashashneh2018_268}
\bibfield{author}{\bibinfo{person}{E.~A. {Alikhashashneh}},
  \bibinfo{person}{R.~R. {Raje}}, {and} \bibinfo{person}{J.~H. {Hill}}.}
  \bibinfo{year}{2018}\natexlab{}.
\newblock \showarticletitle{Using Machine Learning Techniques to Classify and
  Predict Static Code Analysis Tool Warnings}. In
  \bibinfo{booktitle}{\emph{2018 IEEE/ACS 15th International Conference on
  Computer Systems and Applications (AICCSA)}}. \bibinfo{pages}{1--8}.
\newblock
\urldef\tempurl%
\url{https://doi.org/10.1109/AICCSA.2018.8612819}
\showDOI{\tempurl}


\bibitem[\protect\citeauthoryear{Aljamaan and Alazba}{Aljamaan and
  Alazba}{2020}]%
        {675_Aljamaan2020}
\bibfield{author}{\bibinfo{person}{Hamoud Aljamaan} {and} \bibinfo{person}{Amal
  Alazba}.} \bibinfo{year}{2020}\natexlab{}.
\newblock \showarticletitle{Software defect prediction using tree-based
  ensembles}. In \bibinfo{booktitle}{\emph{Proceedings of the 16th ACM
  international conference on predictive models and data analytics in software
  engineering}}. \bibinfo{pages}{1--10}.
\newblock


\bibitem[\protect\citeauthoryear{Allamanis, Barr, Bird, and Sutton}{Allamanis
  et~al\mbox{.}}{2015a}]%
        {Allamanis2015_392}
\bibfield{author}{\bibinfo{person}{Miltiadis Allamanis},
  \bibinfo{person}{Earl~T. Barr}, \bibinfo{person}{Christian Bird}, {and}
  \bibinfo{person}{Charles Sutton}.} \bibinfo{year}{2015}\natexlab{a}.
\newblock \showarticletitle{Suggesting Accurate Method and Class Names}. In
  \bibinfo{booktitle}{\emph{Proceedings of the 2015 10th Joint Meeting on
  Foundations of Software Engineering}} (Bergamo, Italy)
  \emph{(\bibinfo{series}{ESEC/FSE 2015})}. \bibinfo{pages}{38–49}.
\newblock
\showISBNx{9781450336758}
\urldef\tempurl%
\url{https://doi.org/10.1145/2786805.2786849}
\showDOI{\tempurl}


\bibitem[\protect\citeauthoryear{Allamanis, Barr, Devanbu, and
  Sutton}{Allamanis et~al\mbox{.}}{2018a}]%
        {Allamanis2018_452}
\bibfield{author}{\bibinfo{person}{Miltiadis Allamanis},
  \bibinfo{person}{Earl~T. Barr}, \bibinfo{person}{Premkumar Devanbu}, {and}
  \bibinfo{person}{Charles Sutton}.} \bibinfo{year}{2018}\natexlab{a}.
\newblock \showarticletitle{A Survey of Machine Learning for Big Code and
  Naturalness}.
\newblock \bibinfo{journal}{\emph{ACM Comput. Surv.}} \bibinfo{volume}{51},
  \bibinfo{number}{4}, Article \bibinfo{articleno}{81} (\bibinfo{date}{July}
  \bibinfo{year}{2018}), \bibinfo{numpages}{37}~pages.
\newblock
\showISSN{0360-0300}
\urldef\tempurl%
\url{https://doi.org/10.1145/3212695}
\showDOI{\tempurl}


\bibitem[\protect\citeauthoryear{Allamanis, Brockschmidt, and
  Khademi}{Allamanis et~al\mbox{.}}{2018b}]%
        {Allamanis2018_326}
\bibfield{author}{\bibinfo{person}{Miltiadis Allamanis}, \bibinfo{person}{Marc
  Brockschmidt}, {and} \bibinfo{person}{Mahmoud Khademi}.}
  \bibinfo{year}{2018}\natexlab{b}.
\newblock \showarticletitle{Learning to Represent Programs with Graphs}. In
  \bibinfo{booktitle}{\emph{International Conference on Learning
  Representations}}.
\newblock


\bibitem[\protect\citeauthoryear{Allamanis, Peng, and Sutton}{Allamanis
  et~al\mbox{.}}{2016}]%
        {Allamanis2016_403}
\bibfield{author}{\bibinfo{person}{Miltiadis Allamanis}, \bibinfo{person}{Hao
  Peng}, {and} \bibinfo{person}{Charles Sutton}.}
  \bibinfo{year}{2016}\natexlab{}.
\newblock \bibinfo{title}{A Convolutional Attention Network for Extreme
  Summarization of Source Code}.
\newblock
\newblock
\showeprint[arxiv]{1602.03001}~[cs.LG]


\bibitem[\protect\citeauthoryear{{Allamanis} and {Sutton}}{{Allamanis} and
  {Sutton}}{2013}]%
        {Allamanis2013_474}
\bibfield{author}{\bibinfo{person}{M. {Allamanis}} {and} \bibinfo{person}{C.
  {Sutton}}.} \bibinfo{year}{2013}\natexlab{}.
\newblock \showarticletitle{Mining source code repositories at massive scale
  using language modeling}. In \bibinfo{booktitle}{\emph{2013 10th Working
  Conference on Mining Software Repositories (MSR)}}.
  \bibinfo{pages}{207--216}.
\newblock
\urldef\tempurl%
\url{https://doi.org/10.1109/MSR.2013.6624029}
\showDOI{\tempurl}


\bibitem[\protect\citeauthoryear{Allamanis and Sutton}{Allamanis and
  Sutton}{2013}]%
        {AS13}
\bibfield{author}{\bibinfo{person}{Miltiadis Allamanis} {and}
  \bibinfo{person}{Charles Sutton}.} \bibinfo{year}{2013}\natexlab{}.
\newblock \showarticletitle{Mining source code repositories at massive scale
  using language modeling}. In \bibinfo{booktitle}{\emph{10th Working
  Conference on Mining Software Repositories (MSR)}}.
  \bibinfo{pages}{207--216}.
\newblock
\urldef\tempurl%
\url{https://doi.org/10.1109/MSR.2013.6624029}
\showDOI{\tempurl}


\bibitem[\protect\citeauthoryear{Allamanis, Tarlow, Gordon, and Wei}{Allamanis
  et~al\mbox{.}}{2015b}]%
        {Allamanis2015_457}
\bibfield{author}{\bibinfo{person}{Miltiadis Allamanis},
  \bibinfo{person}{Daniel Tarlow}, \bibinfo{person}{Andrew~D. Gordon}, {and}
  \bibinfo{person}{Yi Wei}.} \bibinfo{year}{2015}\natexlab{b}.
\newblock \showarticletitle{Bimodal Modelling of Source Code and Natural
  Language}. In \bibinfo{booktitle}{\emph{Proceedings of the 32nd International
  Conference on International Conference on Machine Learning - Volume 37}}
  (Lille, France) \emph{(\bibinfo{series}{ICML'15})}.
  \bibinfo{pages}{2123–2132}.
\newblock


\bibitem[\protect\citeauthoryear{Allix, Bissyand{\'e}, Klein, and
  Le~Traon}{Allix et~al\mbox{.}}{2016}]%
        {AndroZoo_2016}
\bibfield{author}{\bibinfo{person}{Kevin Allix},
  \bibinfo{person}{Tegawend{\'e}~F. Bissyand{\'e}}, \bibinfo{person}{Jacques
  Klein}, {and} \bibinfo{person}{Yves Le~Traon}.}
  \bibinfo{year}{2016}\natexlab{}.
\newblock \showarticletitle{AndroZoo: Collecting Millions of Android Apps for
  the Research Community}. In \bibinfo{booktitle}{\emph{Proceedings of the 13th
  International Conference on Mining Software Repositories}} (Austin, Texas)
  \emph{(\bibinfo{series}{MSR '16})}. \bibinfo{pages}{468--471}.
\newblock
\showISBNx{978-1-4503-4186-8}
\urldef\tempurl%
\url{https://doi.org/10.1145/2901739.2903508}
\showDOI{\tempurl}


\bibitem[\protect\citeauthoryear{Alon, Brody, Levy, and Yahav}{Alon
  et~al\mbox{.}}{2019a}]%
        {Alon2019_463}
\bibfield{author}{\bibinfo{person}{Uri Alon}, \bibinfo{person}{Shaked Brody},
  \bibinfo{person}{Omer Levy}, {and} \bibinfo{person}{Eran Yahav}.}
  \bibinfo{year}{2019}\natexlab{a}.
\newblock \bibinfo{title}{code2seq: Generating Sequences from Structured
  Representations of Code}.
\newblock
\newblock
\showeprint[arxiv]{1808.01400}~[cs.LG]


\bibitem[\protect\citeauthoryear{Alon, Zilberstein, Levy, and Yahav}{Alon
  et~al\mbox{.}}{2018}]%
        {Alon2018_449}
\bibfield{author}{\bibinfo{person}{Uri Alon}, \bibinfo{person}{Meital
  Zilberstein}, \bibinfo{person}{Omer Levy}, {and} \bibinfo{person}{Eran
  Yahav}.} \bibinfo{year}{2018}\natexlab{}.
\newblock \showarticletitle{A General Path-Based Representation for Predicting
  Program Properties}.
\newblock \bibinfo{journal}{\emph{SIGPLAN Not.}} \bibinfo{volume}{53},
  \bibinfo{number}{4} (\bibinfo{date}{June} \bibinfo{year}{2018}),
  \bibinfo{pages}{404–419}.
\newblock
\showISSN{0362-1340}
\urldef\tempurl%
\url{https://doi.org/10.1145/3296979.3192412}
\showDOI{\tempurl}


\bibitem[\protect\citeauthoryear{Alon, Zilberstein, Levy, and Yahav}{Alon
  et~al\mbox{.}}{2019b}]%
        {Alon2019_464}
\bibfield{author}{\bibinfo{person}{Uri Alon}, \bibinfo{person}{Meital
  Zilberstein}, \bibinfo{person}{Omer Levy}, {and} \bibinfo{person}{Eran
  Yahav}.} \bibinfo{year}{2019}\natexlab{b}.
\newblock \showarticletitle{Code2vec: Learning Distributed Representations of
  Code}.
\newblock \bibinfo{journal}{\emph{Proc. ACM Program. Lang.}}
  \bibinfo{volume}{3}, \bibinfo{number}{POPL}, Article \bibinfo{articleno}{40}
  (\bibinfo{date}{January} \bibinfo{year}{2019}), \bibinfo{numpages}{29}~pages.
\newblock
\urldef\tempurl%
\url{https://doi.org/10.1145/3290353}
\showDOI{\tempurl}


\bibitem[\protect\citeauthoryear{Alrajeh, Kramer, Russo, and Uchitel}{Alrajeh
  et~al\mbox{.}}{2015}]%
        {Alrajeh2015_283}
\bibfield{author}{\bibinfo{person}{Dalal Alrajeh}, \bibinfo{person}{Jeff
  Kramer}, \bibinfo{person}{Alessandra Russo}, {and} \bibinfo{person}{Sebastian
  Uchitel}.} \bibinfo{year}{2015}\natexlab{}.
\newblock \showarticletitle{Automated Support for Diagnosis and Repair}.
\newblock \bibinfo{journal}{\emph{Commun. ACM}} \bibinfo{volume}{58},
  \bibinfo{number}{2} (\bibinfo{date}{January} \bibinfo{year}{2015}),
  \bibinfo{pages}{65–72}.
\newblock
\showISSN{0001-0782}
\urldef\tempurl%
\url{https://doi.org/10.1145/2658986}
\showDOI{\tempurl}


\bibitem[\protect\citeauthoryear{Alsolai and Roper}{Alsolai and Roper}{2020}]%
        {Alsolai2020_186}
\bibfield{author}{\bibinfo{person}{Hadeel Alsolai} {and} \bibinfo{person}{Marc
  Roper}.} \bibinfo{year}{2020}\natexlab{}.
\newblock \showarticletitle{A systematic literature review of machine learning
  techniques for software maintainability prediction}.
\newblock \bibinfo{journal}{\emph{Information and Software Technology}}
  \bibinfo{volume}{119} (\bibinfo{year}{2020}), \bibinfo{pages}{106214}.
\newblock
\showISSN{0950-5849}
\urldef\tempurl%
\url{https://doi.org/10.1016/j.infsof.2019.106214}
\showDOI{\tempurl}


\bibitem[\protect\citeauthoryear{{Alves}, {Fonseca}, and {Antunes}}{{Alves}
  et~al\mbox{.}}{2016}]%
        {Alves2016_29}
\bibfield{author}{\bibinfo{person}{H. {Alves}}, \bibinfo{person}{B. {Fonseca}},
  {and} \bibinfo{person}{N. {Antunes}}.} \bibinfo{year}{2016}\natexlab{}.
\newblock \showarticletitle{Experimenting Machine Learning Techniques to
  Predict Vulnerabilities}. In \bibinfo{booktitle}{\emph{2016 Seventh
  Latin-American Symposium on Dependable Computing (LADC)}}.
  \bibinfo{pages}{151--156}.
\newblock
\urldef\tempurl%
\url{https://doi.org/10.1109/LADC.2016.32}
\showDOI{\tempurl}


\bibitem[\protect\citeauthoryear{Amal, Kessentini, Bechikh, Dea, and Said}{Amal
  et~al\mbox{.}}{2014}]%
        {Amal2014_170}
\bibfield{author}{\bibinfo{person}{Boukhdhir Amal}, \bibinfo{person}{Marouane
  Kessentini}, \bibinfo{person}{Slim Bechikh}, \bibinfo{person}{Josselin Dea},
  {and} \bibinfo{person}{Lamjed~Ben Said}.} \bibinfo{year}{2014}\natexlab{}.
\newblock \showarticletitle{On the Use of Machine Learning and Search-Based
  Software Engineering for Ill-Defined Fitness Function: A Case Study on
  Software Refactoring}. In \bibinfo{booktitle}{\emph{Search-Based Software
  Engineering}}, \bibfield{editor}{\bibinfo{person}{Claire Le~Goues} {and}
  \bibinfo{person}{Shin Yoo}} (Eds.). \bibinfo{pages}{31--45}.
\newblock
\showISBNx{978-3-319-09940-8}


\bibitem[\protect\citeauthoryear{{Amorim}, {Costa}, {Antunes}, {Fonseca}, and
  {Ribeiro}}{{Amorim} et~al\mbox{.}}{2015}]%
        {Amorim2015_223}
\bibfield{author}{\bibinfo{person}{L. {Amorim}}, \bibinfo{person}{E. {Costa}},
  \bibinfo{person}{N. {Antunes}}, \bibinfo{person}{B. {Fonseca}}, {and}
  \bibinfo{person}{M. {Ribeiro}}.} \bibinfo{year}{2015}\natexlab{}.
\newblock \showarticletitle{Experience report: Evaluating the effectiveness of
  decision trees for detecting code smells}. In \bibinfo{booktitle}{\emph{2015
  IEEE 26th International Symposium on Software Reliability Engineering
  (ISSRE)}}. \bibinfo{pages}{261--269}.
\newblock
\urldef\tempurl%
\url{https://doi.org/10.1109/ISSRE.2015.7381819}
\showDOI{\tempurl}


\bibitem[\protect\citeauthoryear{{Amorim}, {Freitas}, {Dantas}, {de Souza},
  {Camilo-Junior}, and {Martins}}{{Amorim} et~al\mbox{.}}{2018}]%
        {Amorim2018_272}
\bibfield{author}{\bibinfo{person}{L.~A. {Amorim}}, \bibinfo{person}{M.~F.
  {Freitas}}, \bibinfo{person}{A. {Dantas}}, \bibinfo{person}{E.~F. {de
  Souza}}, \bibinfo{person}{C.~G. {Camilo-Junior}}, {and}
  \bibinfo{person}{W.~S. {Martins}}.} \bibinfo{year}{2018}\natexlab{}.
\newblock \showarticletitle{A New Word Embedding Approach to Evaluate Potential
  Fixes for Automated Program Repair}. In \bibinfo{booktitle}{\emph{2018
  International Joint Conference on Neural Networks (IJCNN)}}.
  \bibinfo{pages}{1--8}.
\newblock
\urldef\tempurl%
\url{https://doi.org/10.1109/IJCNN.2018.8489079}
\showDOI{\tempurl}


\bibitem[\protect\citeauthoryear{{Aniche}, {Maziero}, {Durelli}, and
  {Durelli}}{{Aniche} et~al\mbox{.}}{2020}]%
        {Aniche2020_173}
\bibfield{author}{\bibinfo{person}{M. {Aniche}}, \bibinfo{person}{E.
  {Maziero}}, \bibinfo{person}{R. {Durelli}}, {and} \bibinfo{person}{V.
  {Durelli}}.} \bibinfo{year}{2020}\natexlab{}.
\newblock \showarticletitle{The Effectiveness of Supervised Machine Learning
  Algorithms in Predicting Software Refactoring}.
\newblock \bibinfo{journal}{\emph{IEEE Transactions on Software Engineering}}
  (\bibinfo{year}{2020}), \bibinfo{pages}{1--1}.
\newblock
\urldef\tempurl%
\url{https://doi.org/10.1109/TSE.2020.3021736}
\showDOI{\tempurl}


\bibitem[\protect\citeauthoryear{Arar and Ayan}{Arar and Ayan}{2015}]%
        {674_Arar2015}
\bibfield{author}{\bibinfo{person}{{\""O}mer~Faruk Arar} {and}
  \bibinfo{person}{K{\""u}r{\c{s}}at Ayan}.} \bibinfo{year}{2015}\natexlab{}.
\newblock \showarticletitle{Software defect prediction using cost-sensitive
  neural network}.
\newblock \bibinfo{journal}{\emph{Applied Soft Computing}}
  \bibinfo{volume}{33} (\bibinfo{year}{2015}), \bibinfo{pages}{263--277}.
\newblock


\bibitem[\protect\citeauthoryear{{Arcelli Fontana} and Zanoni}{{Arcelli
  Fontana} and Zanoni}{2017}]%
        {ArcelliFontana2017_209}
\bibfield{author}{\bibinfo{person}{Francesca {Arcelli Fontana}} {and}
  \bibinfo{person}{Marco Zanoni}.} \bibinfo{year}{2017}\natexlab{}.
\newblock \showarticletitle{Code smell severity classification using machine
  learning techniques}.
\newblock \bibinfo{journal}{\emph{Knowledge-Based Systems}}
  \bibinfo{volume}{128} (\bibinfo{year}{2017}), \bibinfo{pages}{43 -- 58}.
\newblock
\showISSN{0950-7051}
\urldef\tempurl%
\url{https://doi.org/10.1016/j.knosys.2017.04.014}
\showDOI{\tempurl}


\bibitem[\protect\citeauthoryear{Aribandi, Kumar, Bhanu Murthy~Neti, and
  Krishna}{Aribandi et~al\mbox{.}}{2019}]%
        {Aribandi2019_171}
\bibfield{author}{\bibinfo{person}{Vamsi~Krishna Aribandi},
  \bibinfo{person}{Lov Kumar}, \bibinfo{person}{Lalita Bhanu Murthy~Neti},
  {and} \bibinfo{person}{Aneesh Krishna}.} \bibinfo{year}{2019}\natexlab{}.
\newblock \showarticletitle{Prediction of Refactoring-Prone Classes Using
  Ensemble Learning}. In \bibinfo{booktitle}{\emph{Neural Information
  Processing}}, \bibfield{editor}{\bibinfo{person}{Tom Gedeon},
  \bibinfo{person}{Kok~Wai Wong}, {and} \bibinfo{person}{Minho Lee}} (Eds.).
  \bibinfo{pages}{242--250}.
\newblock
\showISBNx{978-3-030-36802-9}


\bibitem[\protect\citeauthoryear{Axelsson, Baca, Feldt, Sidlauskas, and
  Kacan}{Axelsson et~al\mbox{.}}{2009}]%
        {Axelsson2009_447}
\bibfield{author}{\bibinfo{person}{S. Axelsson}, \bibinfo{person}{D. Baca},
  \bibinfo{person}{Robert Feldt}, \bibinfo{person}{Darius Sidlauskas}, {and}
  \bibinfo{person}{Denis Kacan}.} \bibinfo{year}{2009}\natexlab{}.
\newblock \showarticletitle{Detecting Defects with an Interactive Code Review
  Tool Based on Visualisation and Machine Learning}. In
  \bibinfo{booktitle}{\emph{SEKE}}.
\newblock


\bibitem[\protect\citeauthoryear{Azcona, Arora, Hsiao, and Smeaton}{Azcona
  et~al\mbox{.}}{2019}]%
        {Azcona2019_482}
\bibfield{author}{\bibinfo{person}{David Azcona}, \bibinfo{person}{Piyush
  Arora}, \bibinfo{person}{I-Han Hsiao}, {and} \bibinfo{person}{Alan Smeaton}.}
  \bibinfo{year}{2019}\natexlab{}.
\newblock \showarticletitle{User2code2vec: Embeddings for Profiling Students
  Based on Distributional Representations of Source Code}. In
  \bibinfo{booktitle}{\emph{Proceedings of the 9th International Conference on
  Learning Analytics \&amp; Knowledge}} (Tempe, AZ, USA)
  \emph{(\bibinfo{series}{LAK19})}. \bibinfo{pages}{86–95}.
\newblock
\showISBNx{9781450362566}
\urldef\tempurl%
\url{https://doi.org/10.1145/3303772.3303813}
\showDOI{\tempurl}


\bibitem[\protect\citeauthoryear{Azeem, Palomba, Shi, and Wang}{Azeem
  et~al\mbox{.}}{2019}]%
        {Azeem2019_240}
\bibfield{author}{\bibinfo{person}{Muhammad~Ilyas Azeem},
  \bibinfo{person}{Fabio Palomba}, \bibinfo{person}{Lin Shi}, {and}
  \bibinfo{person}{Qing Wang}.} \bibinfo{year}{2019}\natexlab{}.
\newblock \showarticletitle{Machine learning techniques for code smell
  detection: A systematic literature review and meta-analysis}.
\newblock \bibinfo{journal}{\emph{Information and Software Technology}}
  \bibinfo{volume}{108} (\bibinfo{year}{2019}), \bibinfo{pages}{115 -- 138}.
\newblock
\showISSN{0950-5849}
\urldef\tempurl%
\url{https://doi.org/10.1016/j.infsof.2018.12.009}
\showDOI{\tempurl}


\bibitem[\protect\citeauthoryear{Bader, Scott, Pradel, and Chandra}{Bader
  et~al\mbox{.}}{2019}]%
        {Bader2019_312}
\bibfield{author}{\bibinfo{person}{Johannes Bader}, \bibinfo{person}{Andrew
  Scott}, \bibinfo{person}{Michael Pradel}, {and} \bibinfo{person}{Satish
  Chandra}.} \bibinfo{year}{2019}\natexlab{}.
\newblock \showarticletitle{Getafix: Learning to Fix Bugs Automatically}.
\newblock \bibinfo{journal}{\emph{Proc. ACM Program. Lang.}}
  \bibinfo{volume}{3}, \bibinfo{number}{OOPSLA}, Article
  \bibinfo{articleno}{159} (\bibinfo{date}{October} \bibinfo{year}{2019}),
  \bibinfo{numpages}{27}~pages.
\newblock
\urldef\tempurl%
\url{https://doi.org/10.1145/3360585}
\showDOI{\tempurl}


\bibitem[\protect\citeauthoryear{Ban, Liu, Chen, and Chua}{Ban
  et~al\mbox{.}}{2019}]%
        {ban_performance_2019}
\bibfield{author}{\bibinfo{person}{Xinbo Ban}, \bibinfo{person}{Shigang Liu},
  \bibinfo{person}{Chao Chen}, {and} \bibinfo{person}{Caslon Chua}.}
  \bibinfo{year}{2019}\natexlab{}.
\newblock \showarticletitle{A performance evaluation of deep-learnt features
  for software vulnerability detection}.
\newblock \bibinfo{journal}{\emph{Concurrency and Computation: Practice and
  Experience}} \bibinfo{volume}{31}, \bibinfo{number}{19}
  (\bibinfo{year}{2019}), \bibinfo{pages}{e5103}.
\newblock
\showISSN{1532-0634}
\urldef\tempurl%
\url{https://doi.org/10.1002/cpe.5103}
\showDOI{\tempurl}
\newblock
\shownote{\_eprint: https://onlinelibrary.wiley.com/doi/pdf/10.1002/cpe.5103.}


\bibitem[\protect\citeauthoryear{Bandara and Wijayarathna}{Bandara and
  Wijayarathna}{2011}]%
        {Bandara2011_179}
\bibfield{author}{\bibinfo{person}{U. Bandara} {and} \bibinfo{person}{G.
  Wijayarathna}.} \bibinfo{year}{2011}\natexlab{}.
\newblock \showarticletitle{A Machine Learning Based Tool for Source Code
  Plagiarism Detection}.
\newblock \bibinfo{journal}{\emph{International Journal of Machine Learning and
  Computing}} (\bibinfo{year}{2011}), \bibinfo{pages}{337--343}.
\newblock


\bibitem[\protect\citeauthoryear{Banna, Chinnakotla, Yan, Vegesana, Vivek,
  Krishnappa, Jiang, Lu, Thiruvathukal, and Davis}{Banna et~al\mbox{.}}{2021}]%
        {Banna2021}
\bibfield{author}{\bibinfo{person}{Vishnu Banna}, \bibinfo{person}{Akhil
  Chinnakotla}, \bibinfo{person}{Zhengxin Yan}, \bibinfo{person}{Anirudh
  Vegesana}, \bibinfo{person}{Naveen Vivek}, \bibinfo{person}{Kruthi
  Krishnappa}, \bibinfo{person}{Wenxin Jiang}, \bibinfo{person}{Yung{-}Hsiang
  Lu}, \bibinfo{person}{George~K. Thiruvathukal}, {and}
  \bibinfo{person}{James~C. Davis}.} \bibinfo{year}{2021}\natexlab{}.
\newblock \showarticletitle{An Experience Report on Machine Learning
  Reproducibility: Guidance for Practitioners and TensorFlow Model Garden
  Contributors}.
\newblock \bibinfo{journal}{\emph{CoRR}}  \bibinfo{volume}{abs/2107.00821}
  (\bibinfo{year}{2021}).
\newblock
\showeprint[arxiv]{2107.00821}
\urldef\tempurl%
\url{https://arxiv.org/abs/2107.00821}
\showURL{%
\tempurl}


\bibitem[\protect\citeauthoryear{Bansal, Haque, and McMillan}{Bansal
  et~al\mbox{.}}{2021}]%
        {Bansal2021_525}
\bibfield{author}{\bibinfo{person}{A. Bansal}, \bibinfo{person}{S. Haque},
  {and} \bibinfo{person}{C. McMillan}.} \bibinfo{year}{2021}\natexlab{}.
\newblock \showarticletitle{Project-Level Encoding for Neural Source Code
  Summarization of Subroutines}. In \bibinfo{booktitle}{\emph{2021 2021
  IEEE/ACM 29th International Conference on Program Comprehension (ICPC)
  (ICPC)}}. \bibinfo{publisher}{IEEE Computer Society},
  \bibinfo{pages}{253--264}.
\newblock
\urldef\tempurl%
\url{https://doi.org/10.1109/ICPC52881.2021.00032}
\showDOI{\tempurl}


\bibitem[\protect\citeauthoryear{Barbez, Khomh, and Guéhéneuc}{Barbez
  et~al\mbox{.}}{2020}]%
        {Barbez2020_180}
\bibfield{author}{\bibinfo{person}{Antoine Barbez}, \bibinfo{person}{Foutse
  Khomh}, {and} \bibinfo{person}{Yann-Gaël Guéhéneuc}.}
  \bibinfo{year}{2020}\natexlab{}.
\newblock \showarticletitle{A machine-learning based ensemble method for
  anti-patterns detection}.
\newblock \bibinfo{journal}{\emph{Journal of Systems and Software}}
  \bibinfo{volume}{161} (\bibinfo{year}{2020}), \bibinfo{pages}{110486}.
\newblock
\showISSN{0164-1212}
\urldef\tempurl%
\url{https://doi.org/10.1016/j.jss.2019.110486}
\showDOI{\tempurl}


\bibitem[\protect\citeauthoryear{Barone and Sennrich}{Barone and
  Sennrich}{2017}]%
        {Barone2017}
\bibfield{author}{\bibinfo{person}{Antonio Valerio~Miceli Barone} {and}
  \bibinfo{person}{Rico Sennrich}.} \bibinfo{year}{2017}\natexlab{}.
\newblock \bibinfo{title}{A parallel corpus of Python functions and
  documentation strings for automated code documentation and code generation}.
\newblock
\newblock


\bibitem[\protect\citeauthoryear{Batur~\c{S}ahin and Abualigah}{Batur~\c{S}ahin
  and Abualigah}{2021}]%
        {batur_canan_abualigah_2021}
\bibfield{author}{\bibinfo{person}{Canan Batur~\c{S}ahin} {and}
  \bibinfo{person}{Laith Abualigah}.} \bibinfo{year}{2021}\natexlab{}.
\newblock \showarticletitle{A Novel Deep Learning-Based Feature Selection Model
  for Improving the Static Analysis of Vulnerability Detection}.
\newblock \bibinfo{journal}{\emph{Neural Comput. Appl.}} \bibinfo{volume}{33},
  \bibinfo{number}{20} (\bibinfo{date}{oct} \bibinfo{year}{2021}),
  \bibinfo{pages}{14049–14067}.
\newblock
\showISSN{0941-0643}
\urldef\tempurl%
\url{https://doi.org/10.1007/s00521-021-06047-x}
\showDOI{\tempurl}


\bibitem[\protect\citeauthoryear{Ben-Nun, Jakobovits, and Hoefler}{Ben-Nun
  et~al\mbox{.}}{2018}]%
        {Ben-Nun2018_388}
\bibfield{author}{\bibinfo{person}{Tal Ben-Nun},
  \bibinfo{person}{Alice~Shoshana Jakobovits}, {and} \bibinfo{person}{Torsten
  Hoefler}.} \bibinfo{year}{2018}\natexlab{}.
\newblock \showarticletitle{Neural Code Comprehension: A Learnable
  Representation of Code Semantics}. In \bibinfo{booktitle}{\emph{Proceedings
  of the 32nd International Conference on Neural Information Processing
  Systems}} (Montr\'{e}al, Canada) \emph{(\bibinfo{series}{NIPS'18})}.
  \bibinfo{pages}{3589–3601}.
\newblock


\bibitem[\protect\citeauthoryear{{Bhandari} and {Gupta}}{{Bhandari} and
  {Gupta}}{2018}]%
        {Bhandari2018_105}
\bibfield{author}{\bibinfo{person}{G.~P. {Bhandari}} {and} \bibinfo{person}{R.
  {Gupta}}.} \bibinfo{year}{2018}\natexlab{}.
\newblock \showarticletitle{Machine learning based software fault prediction
  utilizing source code metrics}. In \bibinfo{booktitle}{\emph{2018 IEEE 3rd
  International Conference on Computing, Communication and Security (ICCCS)}}.
  \bibinfo{pages}{40--45}.
\newblock
\urldef\tempurl%
\url{https://doi.org/10.1109/CCCS.2018.8586805}
\showDOI{\tempurl}


\bibitem[\protect\citeauthoryear{Bhatia, Kohli, and Singh}{Bhatia
  et~al\mbox{.}}{2018}]%
        {Bhatia2018_337}
\bibfield{author}{\bibinfo{person}{Sahil Bhatia}, \bibinfo{person}{Pushmeet
  Kohli}, {and} \bibinfo{person}{Rishabh Singh}.}
  \bibinfo{year}{2018}\natexlab{}.
\newblock \showarticletitle{Neuro-Symbolic Program Corrector for Introductory
  Programming Assignments}. In \bibinfo{booktitle}{\emph{Proceedings of the
  40th International Conference on Software Engineering}} (Gothenburg, Sweden)
  \emph{(\bibinfo{series}{ICSE '18})}. \bibinfo{pages}{60–70}.
\newblock
\showISBNx{9781450356381}
\urldef\tempurl%
\url{https://doi.org/10.1145/3180155.3180219}
\showDOI{\tempurl}


\bibitem[\protect\citeauthoryear{{Bilgin}, {Ersoy}, {Soykan}, {Tomur},
  {Çomak}, and {Karaçay}}{{Bilgin} et~al\mbox{.}}{2020}]%
        {Bilgin2020_63}
\bibfield{author}{\bibinfo{person}{Z. {Bilgin}}, \bibinfo{person}{M.~A.
  {Ersoy}}, \bibinfo{person}{E.~U. {Soykan}}, \bibinfo{person}{E. {Tomur}},
  \bibinfo{person}{P. {Çomak}}, {and} \bibinfo{person}{L. {Karaçay}}.}
  \bibinfo{year}{2020}\natexlab{}.
\newblock \showarticletitle{Vulnerability Prediction From Source Code Using
  Machine Learning}.
\newblock \bibinfo{journal}{\emph{IEEE Access}}  \bibinfo{volume}{8}
  (\bibinfo{year}{2020}), \bibinfo{pages}{150672--150684}.
\newblock
\urldef\tempurl%
\url{https://doi.org/10.1109/ACCESS.2020.3016774}
\showDOI{\tempurl}


\bibitem[\protect\citeauthoryear{Black}{Black}{2007}]%
        {samate_2007}
\bibfield{author}{\bibinfo{person}{Paul~E. Black}.}
  \bibinfo{year}{2007}\natexlab{}.
\newblock \showarticletitle{Software {Assurance} with {SAMATE} {Reference}
  {Dataset}, {Tool} {Standards}, and {Studies}}.
\newblock  (\bibinfo{date}{Oct.} \bibinfo{year}{2007}).
\newblock


\bibitem[\protect\citeauthoryear{Boland and Black}{Boland and Black}{2012}]%
        {juliet_test_suite_2012}
\bibfield{author}{\bibinfo{person}{Frederick Boland} {and}
  \bibinfo{person}{Paul Black}.} \bibinfo{year}{2012}\natexlab{}.
\newblock \showarticletitle{The Juliet 1.1 C/C++ and Java Test Suite}.
\newblock  \bibinfo{number}{45} (\bibinfo{date}{2012-10-01}
  \bibinfo{year}{2012}).
\newblock
\urldef\tempurl%
\url{https://doi.org/10.1109/MC.2012.345}
\showDOI{\tempurl}


\bibitem[\protect\citeauthoryear{Bowes, Hall, Harman, Jia, Sarro, and Wu}{Bowes
  et~al\mbox{.}}{2016}]%
        {Bowes2016}
\bibfield{author}{\bibinfo{person}{David Bowes}, \bibinfo{person}{Tracy Hall},
  \bibinfo{person}{Mark Harman}, \bibinfo{person}{Yue Jia},
  \bibinfo{person}{Federica Sarro}, {and} \bibinfo{person}{Fan Wu}.}
  \bibinfo{year}{2016}\natexlab{}.
\newblock \showarticletitle{Mutation-Aware Fault Prediction}. In
  \bibinfo{booktitle}{\emph{Proceedings of the 25th International Symposium on
  Software Testing and Analysis}} (Saarbr\"{u}cken, Germany)
  \emph{(\bibinfo{series}{ISSTA 2016})}. \bibinfo{publisher}{Association for
  Computing Machinery}, \bibinfo{address}{New York, NY, USA},
  \bibinfo{pages}{330–341}.
\newblock
\showISBNx{9781450343909}
\urldef\tempurl%
\url{https://doi.org/10.1145/2931037.2931039}
\showDOI{\tempurl}


\bibitem[\protect\citeauthoryear{Braga, Neto, Rab\^{e}lo, Santiago, and
  Souza}{Braga et~al\mbox{.}}{2018}]%
        {Braga2018_67}
\bibfield{author}{\bibinfo{person}{Rony\'{e}rison Braga},
  \bibinfo{person}{Pedro~Santos Neto}, \bibinfo{person}{Ricardo Rab\^{e}lo},
  \bibinfo{person}{Jos\'{e} Santiago}, {and} \bibinfo{person}{Matheus Souza}.}
  \bibinfo{year}{2018}\natexlab{}.
\newblock \showarticletitle{A Machine Learning Approach to Generate Test
  Oracles}. In \bibinfo{booktitle}{\emph{Proceedings of the XXXII Brazilian
  Symposium on Software Engineering}} (Sao Carlos, Brazil)
  \emph{(\bibinfo{series}{SBES '18})}. \bibinfo{pages}{142–151}.
\newblock
\showISBNx{9781450365031}
\urldef\tempurl%
\url{https://doi.org/10.1145/3266237.3266273}
\showDOI{\tempurl}


\bibitem[\protect\citeauthoryear{Brauckmann, Goens, Ertel, and
  Castrillon}{Brauckmann et~al\mbox{.}}{2020}]%
        {Brauckmann2020_465}
\bibfield{author}{\bibinfo{person}{Alexander Brauckmann},
  \bibinfo{person}{Andr\'{e}s Goens}, \bibinfo{person}{Sebastian Ertel}, {and}
  \bibinfo{person}{Jeronimo Castrillon}.} \bibinfo{year}{2020}\natexlab{}.
\newblock \showarticletitle{Compiler-Based Graph Representations for Deep
  Learning Models of Code}. In \bibinfo{booktitle}{\emph{Proceedings of the
  29th International Conference on Compiler Construction}} (San Diego, CA, USA)
  \emph{(\bibinfo{series}{CC 2020})}. \bibinfo{pages}{201–211}.
\newblock
\showISBNx{9781450371209}


\bibitem[\protect\citeauthoryear{Brockschmidt, Allamanis, Gaunt, and
  Polozov}{Brockschmidt et~al\mbox{.}}{2019}]%
        {Brockschmidt2019_471}
\bibfield{author}{\bibinfo{person}{Marc Brockschmidt},
  \bibinfo{person}{Miltiadis Allamanis}, \bibinfo{person}{Alexander~L. Gaunt},
  {and} \bibinfo{person}{Oleksandr Polozov}.} \bibinfo{year}{2019}\natexlab{}.
\newblock \showarticletitle{Generative Code Modeling with Graphs}. In
  \bibinfo{booktitle}{\emph{International Conference on Learning
  Representations}}.
\newblock


\bibitem[\protect\citeauthoryear{Bruch, Monperrus, and Mezini}{Bruch
  et~al\mbox{.}}{2009}]%
        {Bruch2009_500}
\bibfield{author}{\bibinfo{person}{Marcel Bruch}, \bibinfo{person}{Martin
  Monperrus}, {and} \bibinfo{person}{Mira Mezini}.}
  \bibinfo{year}{2009}\natexlab{}.
\newblock \showarticletitle{Learning from Examples to Improve Code Completion
  Systems}. In \bibinfo{booktitle}{\emph{Proceedings of the 7th Joint Meeting
  of the European Software Engineering Conference and the ACM SIGSOFT Symposium
  on The Foundations of Software Engineering}} (Amsterdam, The Netherlands)
  \emph{(\bibinfo{series}{ESEC/FSE '09})}. \bibinfo{pages}{213–222}.
\newblock
\showISBNx{9781605580012}
\urldef\tempurl%
\url{https://doi.org/10.1145/1595696.1595728}
\showDOI{\tempurl}


\bibitem[\protect\citeauthoryear{Brun and Meliou}{Brun and Meliou}{2018}]%
        {Brun2018}
\bibfield{author}{\bibinfo{person}{Yuriy Brun} {and} \bibinfo{person}{Alexandra
  Meliou}.} \bibinfo{year}{2018}\natexlab{}.
\newblock \showarticletitle{Software Fairness}. In
  \bibinfo{booktitle}{\emph{Proceedings of the 2018 26th ACM Joint Meeting on
  European Software Engineering Conference and Symposium on the Foundations of
  Software Engineering}} (Lake Buena Vista, FL, USA)
  \emph{(\bibinfo{series}{ESEC/FSE 2018})}. \bibinfo{publisher}{Association for
  Computing Machinery}, \bibinfo{address}{New York, NY, USA},
  \bibinfo{pages}{754–759}.
\newblock
\showISBNx{9781450355735}
\urldef\tempurl%
\url{https://doi.org/10.1145/3236024.3264838}
\showDOI{\tempurl}


\bibitem[\protect\citeauthoryear{Bui, Jiang, and Yu}{Bui et~al\mbox{.}}{2018}]%
        {Bui2018_214}
\bibfield{author}{\bibinfo{person}{Nghi D.~Q. Bui}, \bibinfo{person}{Lingixao
  Jiang}, {and} \bibinfo{person}{Y. Yu}.} \bibinfo{year}{2018}\natexlab{}.
\newblock \showarticletitle{Cross-Language Learning for Program Classification
  using Bilateral Tree-Based Convolutional Neural Networks}. In
  \bibinfo{booktitle}{\emph{AAAI Workshops}}.
\newblock


\bibitem[\protect\citeauthoryear{{Bui}, {Yu}, and {Jiang}}{{Bui}
  et~al\mbox{.}}{2019}]%
        {Bui2019_200}
\bibfield{author}{\bibinfo{person}{N.~D.~Q. {Bui}}, \bibinfo{person}{Y. {Yu}},
  {and} \bibinfo{person}{L. {Jiang}}.} \bibinfo{year}{2019}\natexlab{}.
\newblock \showarticletitle{Bilateral Dependency Neural Networks for
  Cross-Language Algorithm Classification}. In \bibinfo{booktitle}{\emph{2019
  IEEE 26th International Conference on Software Analysis, Evolution and
  Reengineering (SANER)}}. \bibinfo{pages}{422--433}.
\newblock
\urldef\tempurl%
\url{https://doi.org/10.1109/SANER.2019.8667995}
\showDOI{\tempurl}


\bibitem[\protect\citeauthoryear{{Butgereit}}{{Butgereit}}{2019}]%
        {Butgereit2019_154}
\bibfield{author}{\bibinfo{person}{L. {Butgereit}}.}
  \bibinfo{year}{2019}\natexlab{}.
\newblock \showarticletitle{Using Machine Learning to Prioritize Automated
  Testing in an Agile Environment}. In \bibinfo{booktitle}{\emph{2019
  Conference on Information Communications Technology and Society (ICTAS)}}.
  \bibinfo{pages}{1--6}.
\newblock
\urldef\tempurl%
\url{https://doi.org/10.1109/ICTAS.2019.8703639}
\showDOI{\tempurl}


\bibitem[\protect\citeauthoryear{Cai, Sun, and Dobbie}{Cai
  et~al\mbox{.}}{2019}]%
        {Cai2019_284}
\bibfield{author}{\bibinfo{person}{Cheng-Hao Cai}, \bibinfo{person}{Jing Sun},
  {and} \bibinfo{person}{Gillian Dobbie}.} \bibinfo{year}{2019}\natexlab{}.
\newblock \showarticletitle{Automatic B-model repair using model checking and
  machine learning}.
\newblock \bibinfo{journal}{\emph{Automated Software Engineering}}
  \bibinfo{volume}{26}, \bibinfo{number}{3} (\bibinfo{date}{Jan.}
  \bibinfo{year}{2019}).
\newblock
\showISSN{1573-7535}
\urldef\tempurl%
\url{https://doi.org/10.1007/s10515-019-00264-4}
\showDOI{\tempurl}


\bibitem[\protect\citeauthoryear{Cambronero and Rinard}{Cambronero and
  Rinard}{2019}]%
        {943_Cambronero2019}
\bibfield{author}{\bibinfo{person}{Jos{\'e}~P Cambronero} {and}
  \bibinfo{person}{Martin~C Rinard}.} \bibinfo{year}{2019}\natexlab{}.
\newblock \showarticletitle{AL: autogenerating supervised learning programs}.
\newblock \bibinfo{journal}{\emph{Proceedings of the ACM on Programming
  Languages}} \bibinfo{volume}{3}, \bibinfo{number}{OOPSLA}
  (\bibinfo{year}{2019}), \bibinfo{pages}{1--28}.
\newblock


\bibitem[\protect\citeauthoryear{Caram, Rodrigues, Campanelli, and
  Parreiras}{Caram et~al\mbox{.}}{2019a}]%
        {751_Caram2019}
\bibfield{author}{\bibinfo{person}{Frederico~Luiz Caram},
  \bibinfo{person}{Bruno Rafael De~Oliveira Rodrigues},
  \bibinfo{person}{Amadeu~Silveira Campanelli}, {and}
  \bibinfo{person}{Fernando~Silva Parreiras}.}
  \bibinfo{year}{2019}\natexlab{a}.
\newblock \showarticletitle{Machine learning techniques for code smells
  detection: a systematic mapping study}.
\newblock \bibinfo{journal}{\emph{International Journal of Software Engineering
  and Knowledge Engineering}} \bibinfo{volume}{29}, \bibinfo{number}{02}
  (\bibinfo{year}{2019}), \bibinfo{pages}{285--316}.
\newblock


\bibitem[\protect\citeauthoryear{Caram, Rodrigues, Campanelli, and
  Parreiras}{Caram et~al\mbox{.}}{2019b}]%
        {Caram2019_241}
\bibfield{author}{\bibinfo{person}{Frederico~Luiz Caram},
  \bibinfo{person}{Bruno Rafael De~Oliveira Rodrigues},
  \bibinfo{person}{Amadeu~Silveira Campanelli}, {and}
  \bibinfo{person}{Fernando~Silva Parreiras}.}
  \bibinfo{year}{2019}\natexlab{b}.
\newblock \showarticletitle{Machine Learning Techniques for Code Smells
  Detection: A Systematic Mapping Study}.
\newblock \bibinfo{journal}{\emph{International Journal of Software Engineering
  and Knowledge Engineering}} \bibinfo{volume}{29}, \bibinfo{number}{02}
  (\bibinfo{year}{2019}), \bibinfo{pages}{285--316}.
\newblock
\urldef\tempurl%
\url{https://doi.org/10.1142/S021819401950013X}
\showDOI{\tempurl}
\showeprint{https://doi.org/10.1142/S021819401950013X}


\bibitem[\protect\citeauthoryear{Cesare, Xiang, and Zhang}{Cesare
  et~al\mbox{.}}{2013}]%
        {Cesare2013_204}
\bibfield{author}{\bibinfo{person}{Silvio Cesare}, \bibinfo{person}{Yang
  Xiang}, {and} \bibinfo{person}{Jun Zhang}.} \bibinfo{year}{2013}\natexlab{}.
\newblock \showarticletitle{Clonewise -- Detecting Package-Level Clones Using
  Machine Learning}. In \bibinfo{booktitle}{\emph{Security and Privacy in
  Communication Networks}}, \bibfield{editor}{\bibinfo{person}{Tanveer Zia},
  \bibinfo{person}{Albert Zomaya}, \bibinfo{person}{Vijay Varadharajan}, {and}
  \bibinfo{person}{Morley Mao}} (Eds.). \bibinfo{pages}{197--215}.
\newblock
\showISBNx{978-3-319-04283-1}


\bibitem[\protect\citeauthoryear{{Cetiner} and {Sahingoz}}{{Cetiner} and
  {Sahingoz}}{2020}]%
        {Cetiner2020_65}
\bibfield{author}{\bibinfo{person}{M. {Cetiner}} {and} \bibinfo{person}{O.~K.
  {Sahingoz}}.} \bibinfo{year}{2020}\natexlab{}.
\newblock \showarticletitle{A Comparative Analysis for Machine Learning based
  Software Defect Prediction Systems}. In \bibinfo{booktitle}{\emph{2020 11th
  International Conference on Computing, Communication and Networking
  Technologies (ICCCNT)}}. \bibinfo{pages}{1--7}.
\newblock
\urldef\tempurl%
\url{https://doi.org/10.1109/ICCCNT49239.2020.9225352}
\showDOI{\tempurl}


\bibitem[\protect\citeauthoryear{{Ceylan}, {Kutlubay}, and {Bener}}{{Ceylan}
  et~al\mbox{.}}{2006}]%
        {Ceylan2006_129}
\bibfield{author}{\bibinfo{person}{E. {Ceylan}}, \bibinfo{person}{F.~O.
  {Kutlubay}}, {and} \bibinfo{person}{A.~B. {Bener}}.}
  \bibinfo{year}{2006}\natexlab{}.
\newblock \showarticletitle{Software Defect Identification Using Machine
  Learning Techniques}. In \bibinfo{booktitle}{\emph{32nd EUROMICRO Conference
  on Software Engineering and Advanced Applications (EUROMICRO'06)}}.
  \bibinfo{pages}{240--247}.
\newblock
\urldef\tempurl%
\url{https://doi.org/10.1109/EUROMICRO.2006.56}
\showDOI{\tempurl}


\bibitem[\protect\citeauthoryear{{Chakraborty}, {Ding}, {Allamanis}, and
  {Ray}}{{Chakraborty} et~al\mbox{.}}{2020}]%
        {Chakraborty2020_291}
\bibfield{author}{\bibinfo{person}{S. {Chakraborty}}, \bibinfo{person}{Y.
  {Ding}}, \bibinfo{person}{M. {Allamanis}}, {and} \bibinfo{person}{B. {Ray}}.}
  \bibinfo{year}{2020}\natexlab{}.
\newblock \showarticletitle{CODIT: Code Editing with Tree-Based Neural Models}.
\newblock \bibinfo{journal}{\emph{IEEE Transactions on Software Engineering}}
  (\bibinfo{year}{2020}), \bibinfo{pages}{1--1}.
\newblock
\urldef\tempurl%
\url{https://doi.org/10.1109/TSE.2020.3020502}
\showDOI{\tempurl}


\bibitem[\protect\citeauthoryear{Chakraborty, Ding, Allamanis, and
  Ray}{Chakraborty et~al\mbox{.}}{2022}]%
        {937_Chakraborty2022}
\bibfield{author}{\bibinfo{person}{Saikat Chakraborty},
  \bibinfo{person}{Yangruibo Ding}, \bibinfo{person}{Miltiadis Allamanis},
  {and} \bibinfo{person}{Baishakhi Ray}.} \bibinfo{year}{2022}\natexlab{}.
\newblock \showarticletitle{CODIT: Code Editing With Tree-Based Neural Models}.
\newblock \bibinfo{journal}{\emph{IEEE Transactions on Software Engineering}}
  \bibinfo{volume}{48}, \bibinfo{number}{4} (\bibinfo{year}{2022}),
  \bibinfo{pages}{1385--1399}.
\newblock
\urldef\tempurl%
\url{https://doi.org/10.1109/TSE.2020.3020502}
\showDOI{\tempurl}


\bibitem[\protect\citeauthoryear{Chakraborty and Ray}{Chakraborty and
  Ray}{2021}]%
        {1003_Chakraborty2021}
\bibfield{author}{\bibinfo{person}{Saikat Chakraborty} {and}
  \bibinfo{person}{Baishakhi Ray}.} \bibinfo{year}{2021}\natexlab{}.
\newblock \showarticletitle{On Multi-Modal Learning of Editing Source Code}. In
  \bibinfo{booktitle}{\emph{2021 36th IEEE/ACM International Conference on
  Automated Software Engineering (ASE)}}. \bibinfo{pages}{443--455}.
\newblock
\urldef\tempurl%
\url{https://doi.org/10.1109/ASE51524.2021.1003_Chakraborty2021}
\showDOI{\tempurl}


\bibitem[\protect\citeauthoryear{CHALLAGULLA, BASTANI, YEN, and
  PAUL}{CHALLAGULLA et~al\mbox{.}}{2008}]%
        {Challagulla2008_87}
\bibfield{author}{\bibinfo{person}{VENKATA UDAYA~B. CHALLAGULLA},
  \bibinfo{person}{FAROKH~B. BASTANI}, \bibinfo{person}{I-LING YEN}, {and}
  \bibinfo{person}{RAYMOND~A. PAUL}.} \bibinfo{year}{2008}\natexlab{}.
\newblock \showarticletitle{EMPIRICAL ASSESSMENT OF MACHINE LEARNING BASED
  SOFTWARE DEFECT PREDICTION TECHNIQUES}.
\newblock \bibinfo{journal}{\emph{International Journal on Artificial
  Intelligence Tools}} \bibinfo{volume}{17}, \bibinfo{number}{02}
  (\bibinfo{year}{2008}), \bibinfo{pages}{389--400}.
\newblock
\urldef\tempurl%
\url{https://doi.org/10.1142/S0218213008003947}
\showDOI{\tempurl}
\showeprint{https://doi.org/10.1142/S0218213008003947}


\bibitem[\protect\citeauthoryear{{Chappelly}, {Cifuentes}, {Krishnan}, and
  {Gevay}}{{Chappelly} et~al\mbox{.}}{2017}]%
        {Chappelly2017_107}
\bibfield{author}{\bibinfo{person}{T. {Chappelly}}, \bibinfo{person}{C.
  {Cifuentes}}, \bibinfo{person}{P. {Krishnan}}, {and} \bibinfo{person}{S.
  {Gevay}}.} \bibinfo{year}{2017}\natexlab{}.
\newblock \showarticletitle{Machine learning for finding bugs: An initial
  report}. In \bibinfo{booktitle}{\emph{2017 IEEE Workshop on Machine Learning
  Techniques for Software Quality Evaluation (MaLTeSQuE)}}.
  \bibinfo{pages}{21--26}.
\newblock
\urldef\tempurl%
\url{https://doi.org/10.1109/MALTESQUE.2017.7882012}
\showDOI{\tempurl}


\bibitem[\protect\citeauthoryear{Chaturvedi, Chaturvedi, Tiwari, and
  Agarwal}{Chaturvedi et~al\mbox{.}}{2018}]%
        {1052_Chaturvedi2018}
\bibfield{author}{\bibinfo{person}{Shivam Chaturvedi}, \bibinfo{person}{Amrita
  Chaturvedi}, \bibinfo{person}{Anurag Tiwari}, {and} \bibinfo{person}{Shalini
  Agarwal}.} \bibinfo{year}{2018}\natexlab{}.
\newblock \showarticletitle{Design pattern detection using machine learning
  techniques}. In \bibinfo{booktitle}{\emph{2018 7th International Conference
  on Reliability, Infocom Technologies and Optimization (Trends and Future
  Directions)(ICRITO)}}. IEEE, \bibinfo{pages}{1--6}.
\newblock


\bibitem[\protect\citeauthoryear{Chen, Chen, Li, Xie, and Mu}{Chen
  et~al\mbox{.}}{2019a}]%
        {1083_Chen2019}
\bibfield{author}{\bibinfo{person}{Deyu Chen}, \bibinfo{person}{Xiang Chen},
  \bibinfo{person}{Hao Li}, \bibinfo{person}{Junfeng Xie}, {and}
  \bibinfo{person}{Yanzhou Mu}.} \bibinfo{year}{2019}\natexlab{a}.
\newblock \showarticletitle{Deepcpdp: Deep learning based cross-project defect
  prediction}.
\newblock \bibinfo{journal}{\emph{IEEE Access}}  \bibinfo{volume}{7}
  (\bibinfo{year}{2019}), \bibinfo{pages}{184832--184848}.
\newblock


\bibitem[\protect\citeauthoryear{Chen, Hu, Yu, Chen, Xuan, Liu, and
  Filkov}{Chen et~al\mbox{.}}{2020a}]%
        {Chen2020_140}
\bibfield{author}{\bibinfo{person}{Jinyin Chen}, \bibinfo{person}{Keke Hu},
  \bibinfo{person}{Yue Yu}, \bibinfo{person}{Zhuangzhi Chen},
  \bibinfo{person}{Qi Xuan}, \bibinfo{person}{Yi Liu}, {and}
  \bibinfo{person}{Vladimir Filkov}.} \bibinfo{year}{2020}\natexlab{a}.
\newblock \showarticletitle{Software Visualization and Deep Transfer Learning
  for Effective Software Defect Prediction}. In
  \bibinfo{booktitle}{\emph{Proceedings of the ACM/IEEE 42nd International
  Conference on Software Engineering}} (Seoul, South Korea)
  \emph{(\bibinfo{series}{ICSE '20})}. \bibinfo{pages}{578–589}.
\newblock
\showISBNx{9781450371216}
\urldef\tempurl%
\url{https://doi.org/10.1145/3377811.3380389}
\showDOI{\tempurl}


\bibitem[\protect\citeauthoryear{Chen, Ye, and Zhang}{Chen
  et~al\mbox{.}}{2019}]%
        {Chen2019_460}
\bibfield{author}{\bibinfo{person}{Long Chen}, \bibinfo{person}{Wei Ye}, {and}
  \bibinfo{person}{Shikun Zhang}.} \bibinfo{year}{2019}\natexlab{}.
\newblock \showarticletitle{Capturing Source Code Semantics via Tree-Based
  Convolution over API-Enhanced AST}. In \bibinfo{booktitle}{\emph{Proceedings
  of the 16th ACM International Conference on Computing Frontiers}} (Alghero,
  Italy) \emph{(\bibinfo{series}{CF '19})}. \bibinfo{pages}{174–182}.
\newblock
\showISBNx{9781450366854}
\urldef\tempurl%
\url{https://doi.org/10.1145/3310273.3321560}
\showDOI{\tempurl}


\bibitem[\protect\citeauthoryear{{Chen} and {Wan}}{{Chen} and {Wan}}{2019}]%
        {Chen2019_503}
\bibfield{author}{\bibinfo{person}{M. {Chen}} {and} \bibinfo{person}{X.
  {Wan}}.} \bibinfo{year}{2019}\natexlab{}.
\newblock \showarticletitle{Neural Comment Generation for Source Code with
  Auxiliary Code Classification Task}. In \bibinfo{booktitle}{\emph{2019 26th
  Asia-Pacific Software Engineering Conference (APSEC)}}.
  \bibinfo{pages}{522--529}.
\newblock
\urldef\tempurl%
\url{https://doi.org/10.1109/APSEC48747.2019.00076}
\showDOI{\tempurl}


\bibitem[\protect\citeauthoryear{Chen, Hu, and Liu}{Chen
  et~al\mbox{.}}{2019b}]%
        {Chen2019_420}
\bibfield{author}{\bibinfo{person}{Qiuyuan Chen}, \bibinfo{person}{Han Hu},
  {and} \bibinfo{person}{Zhaoyi Liu}.} \bibinfo{year}{2019}\natexlab{b}.
\newblock \showarticletitle{Code Summarization with Abstract Syntax Tree}. In
  \bibinfo{booktitle}{\emph{Neural Information Processing}},
  \bibfield{editor}{\bibinfo{person}{Tom Gedeon}, \bibinfo{person}{Kok~Wai
  Wong}, {and} \bibinfo{person}{Minho Lee}} (Eds.). \bibinfo{pages}{652--660}.
\newblock
\showISBNx{978-3-030-36802-9}


\bibitem[\protect\citeauthoryear{Chen, Xia, Hu, Lo, and Li}{Chen
  et~al\mbox{.}}{2021}]%
        {988_Chen2021}
\bibfield{author}{\bibinfo{person}{Qiuyuan Chen}, \bibinfo{person}{Xin Xia},
  \bibinfo{person}{Han Hu}, \bibinfo{person}{David Lo}, {and}
  \bibinfo{person}{Shanping Li}.} \bibinfo{year}{2021}\natexlab{}.
\newblock \showarticletitle{Why my code summarization model does not work: Code
  comment improvement with category prediction}.
\newblock \bibinfo{journal}{\emph{ACM Transactions on Software Engineering and
  Methodology (TOSEM)}} \bibinfo{volume}{30}, \bibinfo{number}{2}
  (\bibinfo{year}{2021}), \bibinfo{pages}{1--29}.
\newblock


\bibitem[\protect\citeauthoryear{{Chen} and {Zhou}}{{Chen} and {Zhou}}{2018}]%
        {Chen2018_406}
\bibfield{author}{\bibinfo{person}{Q. {Chen}} {and} \bibinfo{person}{M.
  {Zhou}}.} \bibinfo{year}{2018}\natexlab{}.
\newblock \showarticletitle{A Neural Framework for Retrieval and Summarization
  of Source Code}. In \bibinfo{booktitle}{\emph{2018 33rd IEEE/ACM
  International Conference on Automated Software Engineering (ASE)}}.
  \bibinfo{pages}{826--831}.
\newblock
\urldef\tempurl%
\url{https://doi.org/10.1145/3238147.3240471}
\showDOI{\tempurl}


\bibitem[\protect\citeauthoryear{Chen, Liu, Shin, Song, and Chen}{Chen
  et~al\mbox{.}}{2016}]%
        {Chen2016_323}
\bibfield{author}{\bibinfo{person}{Xinyun Chen}, \bibinfo{person}{Chang Liu},
  \bibinfo{person}{Richard Shin}, \bibinfo{person}{Dawn Song}, {and}
  \bibinfo{person}{Mingcheng Chen}.} \bibinfo{year}{2016}\natexlab{}.
\newblock \showarticletitle{Latent Attention for If-Then Program Synthesis}. In
  \bibinfo{booktitle}{\emph{Proceedings of the 30th International Conference on
  Neural Information Processing Systems}} (Barcelona, Spain)
  \emph{(\bibinfo{series}{NIPS'16})}. \bibinfo{pages}{4581–4589}.
\newblock
\showISBNx{9781510838819}


\bibitem[\protect\citeauthoryear{Chen, Santosa, Yi, Sharma, Sharma, and
  Lo}{Chen et~al\mbox{.}}{2020b}]%
        {chen_santosa_yi_sharma_sharma_mo_2020}
\bibfield{author}{\bibinfo{person}{Yang Chen}, \bibinfo{person}{Andrew~E.
  Santosa}, \bibinfo{person}{Ang~Ming Yi}, \bibinfo{person}{Abhishek Sharma},
  \bibinfo{person}{Asankhaya Sharma}, {and} \bibinfo{person}{David Lo}.}
  \bibinfo{year}{2020}\natexlab{b}.
\newblock \bibinfo{booktitle}{\emph{A Machine Learning Approach for
  Vulnerability Curation}}.
\newblock \bibinfo{publisher}{Association for Computing Machinery},
  \bibinfo{address}{New York, NY, USA}, \bibinfo{pages}{32–42}.
\newblock
\showISBNx{9781450375177}
\urldef\tempurl%
\url{https://doi.org/10.1145/3379597.3387461}
\showURL{%
\tempurl}


\bibitem[\protect\citeauthoryear{{Chen}, {Kommrusch}, {Tufano}, {Pouchet},
  {Poshyvanyk}, and {Monperrus}}{{Chen} et~al\mbox{.}}{2019}]%
        {Chen2019_350}
\bibfield{author}{\bibinfo{person}{Z. {Chen}}, \bibinfo{person}{S.~J.
  {Kommrusch}}, \bibinfo{person}{M. {Tufano}}, \bibinfo{person}{L. {Pouchet}},
  \bibinfo{person}{D. {Poshyvanyk}}, {and} \bibinfo{person}{M. {Monperrus}}.}
  \bibinfo{year}{2019}\natexlab{}.
\newblock \showarticletitle{SEQUENCER: Sequence-to-Sequence Learning for
  End-to-End Program Repair}.
\newblock \bibinfo{journal}{\emph{IEEE Transactions on Software Engineering}}
  (\bibinfo{year}{2019}), \bibinfo{pages}{1--1}.
\newblock
\urldef\tempurl%
\url{https://doi.org/10.1109/TSE.2019.2940179}
\showDOI{\tempurl}


\bibitem[\protect\citeauthoryear{Chernis and Verma}{Chernis and Verma}{2018}]%
        {Chernis2018_39}
\bibfield{author}{\bibinfo{person}{Boris Chernis} {and} \bibinfo{person}{Rakesh
  Verma}.} \bibinfo{year}{2018}\natexlab{}.
\newblock \showarticletitle{Machine Learning Methods for Software Vulnerability
  Detection}. In \bibinfo{booktitle}{\emph{Proceedings of the Fourth ACM
  International Workshop on Security and Privacy Analytics}} (Tempe, AZ, USA)
  \emph{(\bibinfo{series}{IWSPA '18})}. \bibinfo{pages}{31–39}.
\newblock
\showISBNx{9781450356343}
\urldef\tempurl%
\url{https://doi.org/10.1145/3180445.3180453}
\showDOI{\tempurl}


\bibitem[\protect\citeauthoryear{Chidamber and Kemerer}{Chidamber and
  Kemerer}{1994}]%
        {Chidamber1994}
\bibfield{author}{\bibinfo{person}{S.~R. Chidamber} {and}
  \bibinfo{person}{C.~F. Kemerer}.} \bibinfo{year}{1994}\natexlab{}.
\newblock \showarticletitle{A Metrics Suite for Object Oriented Design}.
\newblock \bibinfo{journal}{\emph{IEEE Transaction of Software Engineering}}
  \bibinfo{volume}{20}, \bibinfo{number}{6} (\bibinfo{date}{June}
  \bibinfo{year}{1994}), \bibinfo{pages}{476--493}.
\newblock
\showISSN{0098-5589}
\urldef\tempurl%
\url{https://doi.org/10.1109/32.295895}
\showDOI{\tempurl}


\bibitem[\protect\citeauthoryear{{Choi}, {Kim}, and {Lee}}{{Choi}
  et~al\mbox{.}}{2020}]%
        {Choi2020_438}
\bibfield{author}{\bibinfo{person}{Y. {Choi}}, \bibinfo{person}{S. {Kim}},
  {and} \bibinfo{person}{J. {Lee}}.} \bibinfo{year}{2020}\natexlab{}.
\newblock \showarticletitle{Source Code Summarization Using Attention-Based
  Keyword Memory Networks}. In \bibinfo{booktitle}{\emph{2020 IEEE
  International Conference on Big Data and Smart Computing (BigComp)}}.
  \bibinfo{pages}{564--570}.
\newblock
\urldef\tempurl%
\url{https://doi.org/10.1109/BigComp48618.2020.00011}
\showDOI{\tempurl}


\bibitem[\protect\citeauthoryear{Choudhary, Kumar, Kumar, Mishra, and
  Catal}{Choudhary et~al\mbox{.}}{2018}]%
        {720_Choudhary2018}
\bibfield{author}{\bibinfo{person}{Garvit~Rajesh Choudhary},
  \bibinfo{person}{Sandeep Kumar}, \bibinfo{person}{Kuldeep Kumar},
  \bibinfo{person}{Alok Mishra}, {and} \bibinfo{person}{Cagatay Catal}.}
  \bibinfo{year}{2018}\natexlab{}.
\newblock \showarticletitle{Empirical analysis of change metrics for software
  fault prediction}.
\newblock \bibinfo{journal}{\emph{Computers \& Electrical Engineering}}
  \bibinfo{volume}{67} (\bibinfo{year}{2018}), \bibinfo{pages}{15--24}.
\newblock


\bibitem[\protect\citeauthoryear{{Chug} and {Dhall}}{{Chug} and
  {Dhall}}{2013}]%
        {Chug2013_134}
\bibfield{author}{\bibinfo{person}{A. {Chug}} {and} \bibinfo{person}{S.
  {Dhall}}.} \bibinfo{year}{2013}\natexlab{}.
\newblock \showarticletitle{Software defect prediction using supervised
  learning algorithm and unsupervised learning algorithm}. In
  \bibinfo{booktitle}{\emph{Confluence 2013: The Next Generation Information
  Technology Summit (4th International Conference)}}.
  \bibinfo{pages}{173--179}.
\newblock
\urldef\tempurl%
\url{https://doi.org/10.1049/cp.2013.2313}
\showDOI{\tempurl}


\bibitem[\protect\citeauthoryear{{Clemente}, {Jaafar}, and {Malik}}{{Clemente}
  et~al\mbox{.}}{2018}]%
        {Clemente2018_34}
\bibfield{author}{\bibinfo{person}{C.~J. {Clemente}}, \bibinfo{person}{F.
  {Jaafar}}, {and} \bibinfo{person}{Y. {Malik}}.}
  \bibinfo{year}{2018}\natexlab{}.
\newblock \showarticletitle{Is Predicting Software Security Bugs Using Deep
  Learning Better Than the Traditional Machine Learning Algorithms?}. In
  \bibinfo{booktitle}{\emph{2018 IEEE International Conference on Software
  Quality, Reliability and Security (QRS)}}. \bibinfo{pages}{95--102}.
\newblock
\urldef\tempurl%
\url{https://doi.org/10.1109/QRS.2018.00023}
\showDOI{\tempurl}


\bibitem[\protect\citeauthoryear{Compton, Frank, Patros, and Koay}{Compton
  et~al\mbox{.}}{2020}]%
        {Compton2020_380}
\bibfield{author}{\bibinfo{person}{Rhys Compton}, \bibinfo{person}{Eibe Frank},
  \bibinfo{person}{Panos Patros}, {and} \bibinfo{person}{Abigail Koay}.}
  \bibinfo{year}{2020}\natexlab{}.
\newblock \showarticletitle{Embedding Java Classes with Code2vec: Improvements
  from Variable Obfuscation}. In \bibinfo{booktitle}{\emph{Proceedings of the
  17th International Conference on Mining Software Repositories}} (Seoul,
  Republic of Korea) \emph{(\bibinfo{series}{MSR '20})}.
  \bibinfo{pages}{243–253}.
\newblock
\showISBNx{9781450375177}
\urldef\tempurl%
\url{https://doi.org/10.1145/3379597.3387445}
\showDOI{\tempurl}


\bibitem[\protect\citeauthoryear{Cortes-Coy, V{\'a}squez, Aponte, and
  Poshyvanyk}{Cortes-Coy et~al\mbox{.}}{2014}]%
        {CortesCoy2014OnAG}
\bibfield{author}{\bibinfo{person}{Luis~Fernando Cortes-Coy},
  \bibinfo{person}{M. V{\'a}squez}, \bibinfo{person}{Jairo Aponte}, {and}
  \bibinfo{person}{D. Poshyvanyk}.} \bibinfo{year}{2014}\natexlab{}.
\newblock \showarticletitle{On Automatically Generating Commit Messages via
  Summarization of Source Code Changes}.
\newblock \bibinfo{journal}{\emph{2014 IEEE 14th International Working
  Conference on Source Code Analysis and Manipulation}} (\bibinfo{year}{2014}),
  \bibinfo{pages}{275--284}.
\newblock


\bibitem[\protect\citeauthoryear{Cruz, Santana, and Figueiredo}{Cruz
  et~al\mbox{.}}{2020a}]%
        {767_Cruz2020}
\bibfield{author}{\bibinfo{person}{Daniel Cruz}, \bibinfo{person}{Amanda
  Santana}, {and} \bibinfo{person}{Eduardo Figueiredo}.}
  \bibinfo{year}{2020}\natexlab{a}.
\newblock \showarticletitle{Detecting bad smells with machine learning
  algorithms: an empirical study}. In \bibinfo{booktitle}{\emph{Proceedings of
  the 3rd International Conference on Technical Debt}}.
  \bibinfo{pages}{31--40}.
\newblock


\bibitem[\protect\citeauthoryear{Cruz, Santana, and Figueiredo}{Cruz
  et~al\mbox{.}}{2020b}]%
        {Cruz2020_218}
\bibfield{author}{\bibinfo{person}{Daniel Cruz}, \bibinfo{person}{Amanda
  Santana}, {and} \bibinfo{person}{Eduardo Figueiredo}.}
  \bibinfo{year}{2020}\natexlab{b}.
\newblock \showarticletitle{Detecting Bad Smells with Machine Learning
  Algorithms: An Empirical Study}. In \bibinfo{booktitle}{\emph{Proceedings of
  the 3rd International Conference on Technical Debt}} (Seoul, Republic of
  Korea) \emph{(\bibinfo{series}{TechDebt '20})}. \bibinfo{pages}{31–40}.
\newblock
\showISBNx{9781450379601}
\urldef\tempurl%
\url{https://doi.org/10.1145/3387906.3388618}
\showDOI{\tempurl}


\bibitem[\protect\citeauthoryear{Cui, Wang, Zhao, and Zhang}{Cui
  et~al\mbox{.}}{2020}]%
        {Cui2020_60}
\bibfield{author}{\bibinfo{person}{Jianfeng Cui}, \bibinfo{person}{Lixin Wang},
  \bibinfo{person}{Xin Zhao}, {and} \bibinfo{person}{Hongyi Zhang}.}
  \bibinfo{year}{2020}\natexlab{}.
\newblock \showarticletitle{Towards predictive analysis of android
  vulnerability using statistical codes and machine learning for IoT
  applications}.
\newblock \bibinfo{journal}{\emph{Computer Communications}}
  \bibinfo{volume}{155} (\bibinfo{year}{2020}), \bibinfo{pages}{125 -- 131}.
\newblock
\showISSN{0140-3664}
\urldef\tempurl%
\url{https://doi.org/10.1016/j.comcom.2020.02.078}
\showDOI{\tempurl}


\bibitem[\protect\citeauthoryear{Cunha, Armijo, and de~Camargo}{Cunha
  et~al\mbox{.}}{2020}]%
        {Cunha2020_231}
\bibfield{author}{\bibinfo{person}{Warteruzannan~Soyer Cunha},
  \bibinfo{person}{Guisella~Angulo Armijo}, {and}
  \bibinfo{person}{Valter~Vieira de Camargo}.} \bibinfo{year}{2020}\natexlab{}.
\newblock \bibinfo{booktitle}{\emph{Investigating Non-Usually Employed Features
  in the Identification of Architectural Smells: A Machine Learning-Based
  Approach}}.
\newblock \bibinfo{pages}{21–30}.
\newblock
\showISBNx{9781450387545}


\bibitem[\protect\citeauthoryear{Cvitkovic, Singh, and Anandkumar}{Cvitkovic
  et~al\mbox{.}}{2019}]%
        {Cvitkovic2019_476}
\bibfield{author}{\bibinfo{person}{Milan Cvitkovic}, \bibinfo{person}{Badal
  Singh}, {and} \bibinfo{person}{Animashree Anandkumar}.}
  \bibinfo{year}{2019}\natexlab{}.
\newblock \showarticletitle{Open Vocabulary Learning on Source Code with a
  Graph-Structured Cache} \emph{(\bibinfo{series}{Proceedings of Machine
  Learning Research}, Vol.~\bibinfo{volume}{97})},
  \bibfield{editor}{\bibinfo{person}{Kamalika Chaudhuri} {and}
  \bibinfo{person}{Ruslan Salakhutdinov}} (Eds.). \bibinfo{pages}{1475--1485}.
\newblock


\bibitem[\protect\citeauthoryear{Dam, Pham, Ng, Tran, Grundy, Ghose, Kim, and
  Kim}{Dam et~al\mbox{.}}{2019}]%
        {Dam2019_100}
\bibfield{author}{\bibinfo{person}{Hoa~Khanh Dam}, \bibinfo{person}{Trang
  Pham}, \bibinfo{person}{Shien~Wee Ng}, \bibinfo{person}{Truyen Tran},
  \bibinfo{person}{John Grundy}, \bibinfo{person}{Aditya Ghose},
  \bibinfo{person}{Taeksu Kim}, {and} \bibinfo{person}{Chul-Joo Kim}.}
  \bibinfo{year}{2019}\natexlab{}.
\newblock \showarticletitle{Lessons Learned from Using a Deep Tree-Based Model
  for Software Defect Prediction in Practice}. In
  \bibinfo{booktitle}{\emph{Proceedings of the 16th International Conference on
  Mining Software Repositories}} (Montreal, Quebec, Canada)
  \emph{(\bibinfo{series}{MSR '19})}. \bibinfo{pages}{46–57}.
\newblock
\urldef\tempurl%
\url{https://doi.org/10.1109/MSR.2019.00017}
\showDOI{\tempurl}


\bibitem[\protect\citeauthoryear{D'Ambros, Lanza, and Robbes}{D'Ambros
  et~al\mbox{.}}{2012}]%
        {Dambros2012}
\bibfield{author}{\bibinfo{person}{Marco D'Ambros}, \bibinfo{person}{Michele
  Lanza}, {and} \bibinfo{person}{Romain Robbes}.}
  \bibinfo{year}{2012}\natexlab{}.
\newblock \showarticletitle{Evaluating Defect Prediction Approaches: A
  Benchmark and an Extensive Comparison}.
\newblock \bibinfo{journal}{\emph{Empirical Softw. Engg.}}
  \bibinfo{volume}{17}, \bibinfo{number}{4–5} (\bibinfo{date}{Aug.}
  \bibinfo{year}{2012}), \bibinfo{pages}{531–577}.
\newblock
\showISSN{1382-3256}
\urldef\tempurl%
\url{https://doi.org/10.1007/s10664-011-9173-9}
\showDOI{\tempurl}


\bibitem[\protect\citeauthoryear{Dantas, de~Souza, Souza, and
  Camilo-Junior}{Dantas et~al\mbox{.}}{2019}]%
        {Dantas2019_289}
\bibfield{author}{\bibinfo{person}{Altino Dantas}, \bibinfo{person}{Eduardo~F.
  de Souza}, \bibinfo{person}{Jerffeson Souza}, {and} \bibinfo{person}{Celso~G.
  Camilo-Junior}.} \bibinfo{year}{2019}\natexlab{}.
\newblock \showarticletitle{Code Naturalness to Assist Search Space Exploration
  in Search-Based Program Repair Methods}. In
  \bibinfo{booktitle}{\emph{Search-Based Software Engineering}},
  \bibfield{editor}{\bibinfo{person}{Shiva Nejati} {and}
  \bibinfo{person}{Gregory Gay}} (Eds.). \bibinfo{pages}{164--170}.
\newblock
\showISBNx{978-3-030-27455-9}


\bibitem[\protect\citeauthoryear{Dejaeger, Verbraken, and Baesens}{Dejaeger
  et~al\mbox{.}}{2012}]%
        {718_Dejaeger2012}
\bibfield{author}{\bibinfo{person}{Karel Dejaeger}, \bibinfo{person}{Thomas
  Verbraken}, {and} \bibinfo{person}{Bart Baesens}.}
  \bibinfo{year}{2012}\natexlab{}.
\newblock \showarticletitle{Toward comprehensible software fault prediction
  models using bayesian network classifiers}.
\newblock \bibinfo{journal}{\emph{IEEE Transactions on Software Engineering}}
  \bibinfo{volume}{39}, \bibinfo{number}{2} (\bibinfo{year}{2012}),
  \bibinfo{pages}{237--257}.
\newblock


\bibitem[\protect\citeauthoryear{Dewangan, Rao, Mishra, and Gupta}{Dewangan
  et~al\mbox{.}}{2021}]%
        {816_Dewangan2021}
\bibfield{author}{\bibinfo{person}{Seema Dewangan},
  \bibinfo{person}{Rajwant~Singh Rao}, \bibinfo{person}{Alok Mishra}, {and}
  \bibinfo{person}{Manjari Gupta}.} \bibinfo{year}{2021}\natexlab{}.
\newblock \showarticletitle{A Novel Approach for Code Smell Detection: An
  Empirical Study}.
\newblock \bibinfo{journal}{\emph{IEEE Access}}  \bibinfo{volume}{9}
  (\bibinfo{year}{2021}), \bibinfo{pages}{162869--162883}.
\newblock


\bibitem[\protect\citeauthoryear{Dhamayanthi and Lavanya}{Dhamayanthi and
  Lavanya}{2019}]%
        {Dhamayanthi2019_94}
\bibfield{author}{\bibinfo{person}{N. Dhamayanthi} {and} \bibinfo{person}{B.
  Lavanya}.} \bibinfo{year}{2019}\natexlab{}.
\newblock \showarticletitle{Improvement in Software Defect Prediction Outcome
  Using Principal Component Analysis and Ensemble Machine Learning Algorithms}.
  In \bibinfo{booktitle}{\emph{International Conference on Intelligent Data
  Communication Technologies and Internet of Things (ICICI) 2018}},
  \bibfield{editor}{\bibinfo{person}{Jude Hemanth}, \bibinfo{person}{Xavier
  Fernando}, \bibinfo{person}{Pavel Lafata}, {and} \bibinfo{person}{Zubair
  Baig}} (Eds.). \bibinfo{pages}{397--406}.
\newblock
\showISBNx{978-3-030-03146-6}


\bibitem[\protect\citeauthoryear{Di~Martino, Ferrucci, Gravino, and
  Sarro}{Di~Martino et~al\mbox{.}}{2011}]%
        {DiMartino2011}
\bibfield{author}{\bibinfo{person}{Sergio Di~Martino},
  \bibinfo{person}{Filomena Ferrucci}, \bibinfo{person}{Carmine Gravino}, {and}
  \bibinfo{person}{Federica Sarro}.} \bibinfo{year}{2011}\natexlab{}.
\newblock \showarticletitle{A Genetic Algorithm to Configure Support Vector
  Machines for Predicting Fault-Prone Components}. In
  \bibinfo{booktitle}{\emph{Product-Focused Software Process Improvement}},
  \bibfield{editor}{\bibinfo{person}{Danilo Caivano}, \bibinfo{person}{Markku
  Oivo}, \bibinfo{person}{Maria~Teresa Baldassarre}, {and}
  \bibinfo{person}{Giuseppe Visaggio}} (Eds.). \bibinfo{publisher}{Springer
  Berlin Heidelberg}, \bibinfo{address}{Berlin, Heidelberg},
  \bibinfo{pages}{247--261}.
\newblock
\showISBNx{978-3-642-21843-9}


\bibitem[\protect\citeauthoryear{{Di Nucci}, {Palomba}, {Tamburri},
  {Serebrenik}, and {De Lucia}}{{Di Nucci} et~al\mbox{.}}{2018}]%
        {DiNucci2018_220}
\bibfield{author}{\bibinfo{person}{D. {Di Nucci}}, \bibinfo{person}{F.
  {Palomba}}, \bibinfo{person}{D.~A. {Tamburri}}, \bibinfo{person}{A.
  {Serebrenik}}, {and} \bibinfo{person}{A. {De Lucia}}.}
  \bibinfo{year}{2018}\natexlab{}.
\newblock \showarticletitle{Detecting code smells using machine learning
  techniques: Are we there yet?}. In \bibinfo{booktitle}{\emph{2018 IEEE 25th
  International Conference on Software Analysis, Evolution and Reengineering
  (SANER)}}. \bibinfo{pages}{612--621}.
\newblock
\urldef\tempurl%
\url{https://doi.org/10.1109/SANER.2018.8330266}
\showDOI{\tempurl}


\bibitem[\protect\citeauthoryear{Dos~Santos, Figueiredo, Veloso, Viggiato, and
  Ziviani}{Dos~Santos et~al\mbox{.}}{2020}]%
        {Santos2020_148}
\bibfield{author}{\bibinfo{person}{Geanderson~Esteves Dos~Santos},
  \bibinfo{person}{E. Figueiredo}, \bibinfo{person}{Adriano Veloso},
  \bibinfo{person}{Markos Viggiato}, {and} \bibinfo{person}{N. Ziviani}.}
  \bibinfo{year}{2020}\natexlab{}.
\newblock \showarticletitle{Understanding machine learning software defect
  predictions}.
\newblock \bibinfo{journal}{\emph{Autom. Softw. Eng.}}  \bibinfo{volume}{27}
  (\bibinfo{year}{2020}), \bibinfo{pages}{369--392}.
\newblock


\bibitem[\protect\citeauthoryear{Du, Chen, Li, Guo, Zhou, Liu, and Jiang}{Du
  et~al\mbox{.}}{2019}]%
        {du_chen_li_guo_zhou_liu_jiang_2019}
\bibfield{author}{\bibinfo{person}{Xiaoning Du}, \bibinfo{person}{Bihuan Chen},
  \bibinfo{person}{Yuekang Li}, \bibinfo{person}{Jianmin Guo},
  \bibinfo{person}{Yaqin Zhou}, \bibinfo{person}{Yang Liu}, {and}
  \bibinfo{person}{Yu Jiang}.} \bibinfo{year}{2019}\natexlab{}.
\newblock \showarticletitle{LEOPARD: Identifying Vulnerable Code for
  Vulnerability Assessment Through Program Metrics}. In
  \bibinfo{booktitle}{\emph{2019 IEEE/ACM 41st International Conference on
  Software Engineering (ICSE)}}. \bibinfo{pages}{60--71}.
\newblock
\urldef\tempurl%
\url{https://doi.org/10.1109/ICSE.2019.00024}
\showDOI{\tempurl}


\bibitem[\protect\citeauthoryear{Du, Wang, and Wang}{Du et~al\mbox{.}}{2015}]%
        {du_wang_want_2015}
\bibfield{author}{\bibinfo{person}{Yao Du}, \bibinfo{person}{Xiaoqing Wang},
  {and} \bibinfo{person}{Junfeng Wang}.} \bibinfo{year}{2015}\natexlab{}.
\newblock \showarticletitle{A Static Android Malicious Code Detection Method
  Based on Multi-Source Fusion}.
\newblock \bibinfo{journal}{\emph{Sec. and Commun. Netw.}} \bibinfo{volume}{8},
  \bibinfo{number}{17} (\bibinfo{date}{nov} \bibinfo{year}{2015}),
  \bibinfo{pages}{3238–3246}.
\newblock
\showISSN{1939-0114}
\urldef\tempurl%
\url{https://doi.org/10.1002/sec.1248}
\showDOI{\tempurl}


\bibitem[\protect\citeauthoryear{{Durelli}, {Durelli}, {Borges}, {Endo},
  {Eler}, {Dias}, and {Guimarães}}{{Durelli} et~al\mbox{.}}{2019}]%
        {Durelli2019_103}
\bibfield{author}{\bibinfo{person}{V.~H.~S. {Durelli}}, \bibinfo{person}{R.~S.
  {Durelli}}, \bibinfo{person}{S.~S. {Borges}}, \bibinfo{person}{A.~T. {Endo}},
  \bibinfo{person}{M.~M. {Eler}}, \bibinfo{person}{D.~R.~C. {Dias}}, {and}
  \bibinfo{person}{M.~P. {Guimarães}}.} \bibinfo{year}{2019}\natexlab{}.
\newblock \showarticletitle{Machine Learning Applied to Software Testing: A
  Systematic Mapping Study}.
\newblock \bibinfo{journal}{\emph{IEEE Transactions on Reliability}}
  \bibinfo{volume}{68}, \bibinfo{number}{3} (\bibinfo{year}{2019}),
  \bibinfo{pages}{1189--1212}.
\newblock
\urldef\tempurl%
\url{https://doi.org/10.1109/TR.2019.2892517}
\showDOI{\tempurl}


\bibitem[\protect\citeauthoryear{Dwivedi, Tirkey, Ray, and Rath}{Dwivedi
  et~al\mbox{.}}{2016}]%
        {1051_Dwivedi2016}
\bibfield{author}{\bibinfo{person}{Ashish~Kumar Dwivedi},
  \bibinfo{person}{Anand Tirkey}, \bibinfo{person}{Ransingh~Biswajit Ray},
  {and} \bibinfo{person}{Santanu~Kumar Rath}.} \bibinfo{year}{2016}\natexlab{}.
\newblock \showarticletitle{Software design pattern recognition using machine
  learning techniques}. In \bibinfo{booktitle}{\emph{2016 ieee region 10
  conference (tencon)}}. IEEE, \bibinfo{pages}{222--227}.
\newblock


\bibitem[\protect\citeauthoryear{Efstathiou and Spinellis}{Efstathiou and
  Spinellis}{2019}]%
        {1068_Efstathiou2019}
\bibfield{author}{\bibinfo{person}{Vasiliki Efstathiou} {and}
  \bibinfo{person}{Diomidis Spinellis}.} \bibinfo{year}{2019}\natexlab{}.
\newblock \showarticletitle{Semantic Source Code Models Using Identifier
  Embeddings}. In \bibinfo{booktitle}{\emph{2019 IEEE/ACM 16th International
  Conference on Mining Software Repositories (MSR)}}. \bibinfo{pages}{29--33}.
\newblock
\urldef\tempurl%
\url{https://doi.org/10.1109/MSR.2019.00015}
\showDOI{\tempurl}


\bibitem[\protect\citeauthoryear{Elovici, Shabtai, Moskovitch, Tahan, and
  Glezer}{Elovici et~al\mbox{.}}{2007}]%
        {Elovici2007_15}
\bibfield{author}{\bibinfo{person}{Yuval Elovici}, \bibinfo{person}{Asaf
  Shabtai}, \bibinfo{person}{Robert Moskovitch}, \bibinfo{person}{Gil Tahan},
  {and} \bibinfo{person}{Chanan Glezer}.} \bibinfo{year}{2007}\natexlab{}.
\newblock \showarticletitle{Applying Machine Learning Techniques for Detection
  of Malicious Code in Network Traffic}. In \bibinfo{booktitle}{\emph{KI 2007:
  Advances in Artificial Intelligence}},
  \bibfield{editor}{\bibinfo{person}{Joachim Hertzberg},
  \bibinfo{person}{Michael Beetz}, {and} \bibinfo{person}{Roman Englert}}
  (Eds.). \bibinfo{pages}{44--50}.
\newblock
\showISBNx{978-3-540-74565-5}


\bibitem[\protect\citeauthoryear{Eniser, Gerasimou, and Sen}{Eniser
  et~al\mbox{.}}{2019}]%
        {Eniser2019}
\bibfield{author}{\bibinfo{person}{Hasan~Ferit Eniser}, \bibinfo{person}{Simos
  Gerasimou}, {and} \bibinfo{person}{Alper Sen}.}
  \bibinfo{year}{2019}\natexlab{}.
\newblock \showarticletitle{DeepFault: Fault Localization for Deep Neural
  Networks}. In \bibinfo{booktitle}{\emph{Fundamental Approaches to Software
  Engineering}}, \bibfield{editor}{\bibinfo{person}{Reiner H{\"a}hnle} {and}
  \bibinfo{person}{Wil van~der Aalst}} (Eds.). \bibinfo{publisher}{Springer
  International Publishing}, \bibinfo{address}{Cham},
  \bibinfo{pages}{171--191}.
\newblock
\showISBNx{978-3-030-16722-6}


\bibitem[\protect\citeauthoryear{Erturk and Sezer}{Erturk and Sezer}{2015}]%
        {722_Erturk2015}
\bibfield{author}{\bibinfo{person}{Ezgi Erturk} {and}
  \bibinfo{person}{Ebru~Akcapinar Sezer}.} \bibinfo{year}{2015}\natexlab{}.
\newblock \showarticletitle{A comparison of some soft computing methods for
  software fault prediction}.
\newblock \bibinfo{journal}{\emph{Expert systems with applications}}
  \bibinfo{volume}{42}, \bibinfo{number}{4} (\bibinfo{year}{2015}),
  \bibinfo{pages}{1872--1879}.
\newblock


\bibitem[\protect\citeauthoryear{Etemadi and Monperrus}{Etemadi and
  Monperrus}{2020}]%
        {1013_Etemadi2020}
\bibfield{author}{\bibinfo{person}{Khashayar Etemadi} {and}
  \bibinfo{person}{Martin Monperrus}.} \bibinfo{year}{2020}\natexlab{}.
\newblock \showarticletitle{On the Relevance of Cross-project Learning with
  Nearest Neighbours for Commit Message Generation}. In
  \bibinfo{booktitle}{\emph{Proceedings of the IEEE/ACM 42nd International
  Conference on Software Engineering Workshops}}. \bibinfo{pages}{470--475}.
\newblock


\bibitem[\protect\citeauthoryear{{Fakhoury}, {Arnaoudova}, {Noiseux}, {Khomh},
  and {Antoniol}}{{Fakhoury} et~al\mbox{.}}{2018}]%
        {Fakhoury2018_232}
\bibfield{author}{\bibinfo{person}{S. {Fakhoury}}, \bibinfo{person}{V.
  {Arnaoudova}}, \bibinfo{person}{C. {Noiseux}}, \bibinfo{person}{F. {Khomh}},
  {and} \bibinfo{person}{G. {Antoniol}}.} \bibinfo{year}{2018}\natexlab{}.
\newblock \showarticletitle{Keep it simple: Is deep learning good for
  linguistic smell detection?}. In \bibinfo{booktitle}{\emph{2018 IEEE 25th
  International Conference on Software Analysis, Evolution and Reengineering
  (SANER)}}. \bibinfo{pages}{602--611}.
\newblock
\urldef\tempurl%
\url{https://doi.org/10.1109/SANER.2018.8330265}
\showDOI{\tempurl}


\bibitem[\protect\citeauthoryear{Falleri, Morandat, Blanc, Martinez, and
  Monperrus}{Falleri et~al\mbox{.}}{2014}]%
        {Falleri2014}
\bibfield{author}{\bibinfo{person}{Jean-R\'{e}my Falleri},
  \bibinfo{person}{Flor\'{e}al Morandat}, \bibinfo{person}{Xavier Blanc},
  \bibinfo{person}{Matias Martinez}, {and} \bibinfo{person}{Martin Monperrus}.}
  \bibinfo{year}{2014}\natexlab{}.
\newblock \showarticletitle{Fine-Grained and Accurate Source Code
  Differencing}. In \bibinfo{booktitle}{\emph{Proceedings of the 29th ACM/IEEE
  International Conference on Automated Software Engineering}} (Vasteras,
  Sweden) \emph{(\bibinfo{series}{ASE '14})}. \bibinfo{pages}{313–324}.
\newblock
\showISBNx{9781450330138}
\urldef\tempurl%
\url{https://doi.org/10.1145/2642937.2642982}
\showDOI{\tempurl}


\bibitem[\protect\citeauthoryear{Fan, Diao, Yu, Yang, and Chen}{Fan
  et~al\mbox{.}}{2019}]%
        {529_Fan2019}
\bibfield{author}{\bibinfo{person}{Guisheng Fan}, \bibinfo{person}{Xuyang
  Diao}, \bibinfo{person}{Huiqun Yu}, \bibinfo{person}{Kang Yang}, {and}
  \bibinfo{person}{Liqiong Chen}.} \bibinfo{year}{2019}\natexlab{}.
\newblock \showarticletitle{Deep semantic feature learning with embedded static
  metrics for software defect prediction}. In \bibinfo{booktitle}{\emph{2019
  26th Asia-Pacific Software Engineering Conference (APSEC)}}. IEEE,
  \bibinfo{pages}{244--251}.
\newblock


\bibitem[\protect\citeauthoryear{Fang, Liu, Shi, Huang, and Shi}{Fang
  et~al\mbox{.}}{2020b}]%
        {Fang2020_225}
\bibfield{author}{\bibinfo{person}{Chunrong Fang}, \bibinfo{person}{Zixi Liu},
  \bibinfo{person}{Yangyang Shi}, \bibinfo{person}{Jeff Huang}, {and}
  \bibinfo{person}{Qingkai Shi}.} \bibinfo{year}{2020}\natexlab{b}.
\newblock \showarticletitle{Functional Code Clone Detection with Syntax and
  Semantics Fusion Learning}. In \bibinfo{booktitle}{\emph{Proceedings of the
  29th ACM SIGSOFT International Symposium on Software Testing and Analysis}}
  (Virtual Event, USA) \emph{(\bibinfo{series}{ISSTA 2020})}.
  \bibinfo{pages}{516–527}.
\newblock
\showISBNx{9781450380089}
\urldef\tempurl%
\url{https://doi.org/10.1145/3395363.3397362}
\showDOI{\tempurl}


\bibitem[\protect\citeauthoryear{Fang, Liu, Huang, and Liu}{Fang
  et~al\mbox{.}}{2020a}]%
        {fang_fastembed_2020}
\bibfield{author}{\bibinfo{person}{Yong Fang}, \bibinfo{person}{Yongcheng Liu},
  \bibinfo{person}{Cheng Huang}, {and} \bibinfo{person}{Liang Liu}.}
  \bibinfo{year}{2020}\natexlab{a}.
\newblock \showarticletitle{{FastEmbed}: {Predicting} vulnerability
  exploitation possibility based on ensemble machine learning algorithm}.
\newblock \bibinfo{journal}{\emph{PLoS ONE}}  \bibinfo{volume}{15}
  (\bibinfo{date}{Feb.} \bibinfo{year}{2020}), \bibinfo{pages}{e0228439}.
\newblock
\urldef\tempurl%
\url{https://doi.org/10.1371/journal.pone.0228439}
\showDOI{\tempurl}
\newblock
\shownote{ADS Bibcode: 2020PLoSO..1528439F.}


\bibitem[\protect\citeauthoryear{Felix and Lee}{Felix and Lee}{2017}]%
        {666_Felix2017}
\bibfield{author}{\bibinfo{person}{Ebubeogu~Amarachukwu Felix} {and}
  \bibinfo{person}{Sai~Peck Lee}.} \bibinfo{year}{2017}\natexlab{}.
\newblock \showarticletitle{Integrated approach to software defect prediction}.
\newblock \bibinfo{journal}{\emph{IEEE Access}}  \bibinfo{volume}{5}
  (\bibinfo{year}{2017}), \bibinfo{pages}{21524--21547}.
\newblock


\bibitem[\protect\citeauthoryear{Ferenc, Hegedundefineds, Gyimesi, Antal,
  B\'{a}n, and Gyim\'{o}thy}{Ferenc et~al\mbox{.}}{2019}]%
        {Ferenc2019_18}
\bibfield{author}{\bibinfo{person}{Rudolf Ferenc}, \bibinfo{person}{P\'{e}ter
  Hegedundefineds}, \bibinfo{person}{P\'{e}ter Gyimesi},
  \bibinfo{person}{G\'{a}bor Antal}, \bibinfo{person}{D\'{e}nes B\'{a}n}, {and}
  \bibinfo{person}{Tibor Gyim\'{o}thy}.} \bibinfo{year}{2019}\natexlab{}.
\newblock \showarticletitle{Challenging Machine Learning Algorithms in
  Predicting Vulnerable JavaScript Functions}. In
  \bibinfo{booktitle}{\emph{Proceedings of the 7th International Workshop on
  Realizing Artificial Intelligence Synergies in Software Engineering}}
  (Montreal, Quebec, Canada) \emph{(\bibinfo{series}{RAISE '19})}.
  \bibinfo{pages}{8–14}.
\newblock
\urldef\tempurl%
\url{https://doi.org/10.1109/RAISE.2019.00010}
\showDOI{\tempurl}


\bibitem[\protect\citeauthoryear{Ferreira, Silva, and Valente}{Ferreira
  et~al\mbox{.}}{2021}]%
        {Ferreira2021}
\bibfield{author}{\bibinfo{person}{Fabio Ferreira},
  \bibinfo{person}{Luciana~Lourdes Silva}, {and} \bibinfo{person}{Marco~Tulio
  Valente}.} \bibinfo{year}{2021}\natexlab{}.
\newblock \showarticletitle{Software Engineering Meets Deep Learning: A Mapping
  Study}. In \bibinfo{booktitle}{\emph{Proceedings of the 36th Annual ACM
  Symposium on Applied Computing}} (Virtual Event, Republic of Korea)
  \emph{(\bibinfo{series}{SAC '21})}. \bibinfo{publisher}{Association for
  Computing Machinery}, \bibinfo{address}{New York, NY, USA},
  \bibinfo{pages}{1542–1549}.
\newblock
\showISBNx{9781450381048}
\urldef\tempurl%
\url{https://doi.org/10.1145/3412841.3442029}
\showDOI{\tempurl}


\bibitem[\protect\citeauthoryear{Fontana, M{\"a}ntyl{\"a}, Zanoni, and
  Marino}{Fontana et~al\mbox{.}}{2015}]%
        {Fontana2015_211}
\bibfield{author}{\bibinfo{person}{F. Fontana}, \bibinfo{person}{M.
  M{\"a}ntyl{\"a}}, \bibinfo{person}{Marco Zanoni}, {and}
  \bibinfo{person}{Alessandro Marino}.} \bibinfo{year}{2015}\natexlab{}.
\newblock \showarticletitle{Comparing and experimenting machine learning
  techniques for code smell detection}.
\newblock \bibinfo{journal}{\emph{Empirical Software Engineering}}
  \bibinfo{volume}{21} (\bibinfo{year}{2015}), \bibinfo{pages}{1143--1191}.
\newblock


\bibitem[\protect\citeauthoryear{{Fontana}, {Zanoni}, {Marino}, and
  {Mäntylä}}{{Fontana} et~al\mbox{.}}{2013}]%
        {Fontana2013_207}
\bibfield{author}{\bibinfo{person}{F.~A. {Fontana}}, \bibinfo{person}{M.
  {Zanoni}}, \bibinfo{person}{A. {Marino}}, {and} \bibinfo{person}{M.~V.
  {Mäntylä}}.} \bibinfo{year}{2013}\natexlab{}.
\newblock \showarticletitle{Code Smell Detection: Towards a Machine
  Learning-Based Approach}. In \bibinfo{booktitle}{\emph{2013 IEEE
  International Conference on Software Maintenance}}.
  \bibinfo{pages}{396--399}.
\newblock


\bibitem[\protect\citeauthoryear{Gamma, Helm, Johnson, and Vlissides}{Gamma
  et~al\mbox{.}}{1994}]%
        {gamma_design_1994}
\bibfield{author}{\bibinfo{person}{Erich Gamma}, \bibinfo{person}{Richard
  Helm}, \bibinfo{person}{Ralph Johnson}, {and} \bibinfo{person}{John
  Vlissides}.} \bibinfo{year}{1994}\natexlab{}.
\newblock \bibinfo{booktitle}{\emph{Design {Patterns}: {Elements} of {Reusable}
  {Object}-{Oriented} {Software}} (\bibinfo{edition}{1st} ed.)}.
\newblock \bibinfo{publisher}{Addison-Wesley Professional. Part of the
  Addison-Wesley Professional Computing Series series.}
\newblock
\showISBNx{978-0-201-63361-0}
\urldef\tempurl%
\url{https://www.informit.com/store/design-patterns-elements-of-reusable-object-oriented-9780201633610?w_ptgrevartcl=Grady+Booch+on+Design+Patterns%2c+OOP%2c+and+Coffee_1405569}
\showURL{%
\tempurl}


\bibitem[\protect\citeauthoryear{Gao, Xia, Grundy, Lo, and Li}{Gao
  et~al\mbox{.}}{2020}]%
        {Gao2020_441}
\bibfield{author}{\bibinfo{person}{Zhipeng Gao}, \bibinfo{person}{Xin Xia},
  \bibinfo{person}{John Grundy}, \bibinfo{person}{David Lo}, {and}
  \bibinfo{person}{Yuan-Fang Li}.} \bibinfo{year}{2020}\natexlab{}.
\newblock \showarticletitle{Generating Question Titles for Stack Overflow from
  Mined Code Snippets}.
\newblock \bibinfo{journal}{\emph{ACM Trans. Softw. Eng. Methodol.}}
  \bibinfo{volume}{29}, \bibinfo{number}{4}, Article \bibinfo{articleno}{26}
  (\bibinfo{date}{September} \bibinfo{year}{2020}),
  \bibinfo{numpages}{37}~pages.
\newblock
\showISSN{1049-331X}
\urldef\tempurl%
\url{https://doi.org/10.1145/3401026}
\showDOI{\tempurl}


\bibitem[\protect\citeauthoryear{Ghadhab, Jenhani, Mkaouer, and
  Messaoud}{Ghadhab et~al\mbox{.}}{2021}]%
        {1023_Ghadhab2021}
\bibfield{author}{\bibinfo{person}{Lobna Ghadhab}, \bibinfo{person}{Ilyes
  Jenhani}, \bibinfo{person}{Mohamed~Wiem Mkaouer}, {and}
  \bibinfo{person}{Montassar~Ben Messaoud}.} \bibinfo{year}{2021}\natexlab{}.
\newblock \showarticletitle{Augmenting commit classification by using
  fine-grained source code changes and a pre-trained deep neural language
  model}.
\newblock \bibinfo{journal}{\emph{Information and Software Technology}}
  \bibinfo{volume}{135} (\bibinfo{year}{2021}), \bibinfo{pages}{106566}.
\newblock


\bibitem[\protect\citeauthoryear{Ghaffarian and Shahriari}{Ghaffarian and
  Shahriari}{2017}]%
        {Ghaffarian2017_52}
\bibfield{author}{\bibinfo{person}{Seyed~Mohammad Ghaffarian} {and}
  \bibinfo{person}{Hamid~Reza Shahriari}.} \bibinfo{year}{2017}\natexlab{}.
\newblock \showarticletitle{Software Vulnerability Analysis and Discovery Using
  Machine-Learning and Data-Mining Techniques: A Survey}.
\newblock \bibinfo{journal}{\emph{ACM Comput. Surv.}} \bibinfo{volume}{50},
  \bibinfo{number}{4}, Article \bibinfo{articleno}{56} (\bibinfo{date}{August}
  \bibinfo{year}{2017}), \bibinfo{numpages}{36}~pages.
\newblock
\showISSN{0360-0300}
\urldef\tempurl%
\url{https://doi.org/10.1145/3092566}
\showDOI{\tempurl}


\bibitem[\protect\citeauthoryear{Gharbi, Mkaouer, Jenhani, and Messaoud}{Gharbi
  et~al\mbox{.}}{2019}]%
        {1022_Gharbi2019}
\bibfield{author}{\bibinfo{person}{Sirine Gharbi},
  \bibinfo{person}{Mohamed~Wiem Mkaouer}, \bibinfo{person}{Ilyes Jenhani},
  {and} \bibinfo{person}{Montassar~Ben Messaoud}.}
  \bibinfo{year}{2019}\natexlab{}.
\newblock \showarticletitle{On the classification of software change messages
  using multi-label active learning}. In \bibinfo{booktitle}{\emph{Proceedings
  of the 34th ACM/SIGAPP Symposium on Applied Computing}}.
  \bibinfo{pages}{1760--1767}.
\newblock


\bibitem[\protect\citeauthoryear{Giray}{Giray}{2021}]%
        {Giray2021}
\bibfield{author}{\bibinfo{person}{Görkem Giray}.}
  \bibinfo{year}{2021}\natexlab{}.
\newblock \showarticletitle{A software engineering perspective on engineering
  machine learning systems: State of the art and challenges}.
\newblock \bibinfo{journal}{\emph{Journal of Systems and Software}}
  \bibinfo{volume}{180} (\bibinfo{year}{2021}), \bibinfo{pages}{111031}.
\newblock
\showISSN{0164-1212}
\urldef\tempurl%
\url{https://doi.org/10.1016/j.jss.2021.111031}
\showDOI{\tempurl}


\bibitem[\protect\citeauthoryear{{Godefroid}, {Peleg}, and {Singh}}{{Godefroid}
  et~al\mbox{.}}{2017}]%
        {Godefroid2017_99}
\bibfield{author}{\bibinfo{person}{P. {Godefroid}}, \bibinfo{person}{H.
  {Peleg}}, {and} \bibinfo{person}{R. {Singh}}.}
  \bibinfo{year}{2017}\natexlab{}.
\newblock \showarticletitle{Learn Fuzz: Machine learning for input fuzzing}. In
  \bibinfo{booktitle}{\emph{2017 32nd IEEE/ACM International Conference on
  Automated Software Engineering (ASE)}}. \bibinfo{pages}{50--59}.
\newblock
\urldef\tempurl%
\url{https://doi.org/10.1109/ASE.2017.8115618}
\showDOI{\tempurl}


\bibitem[\protect\citeauthoryear{Gondra}{Gondra}{2008}]%
        {Gondra2008_73}
\bibfield{author}{\bibinfo{person}{Iker Gondra}.}
  \bibinfo{year}{2008}\natexlab{}.
\newblock \showarticletitle{Applying machine learning to software
  fault-proneness prediction}.
\newblock \bibinfo{journal}{\emph{Journal of Systems and Software}}
  \bibinfo{volume}{81}, \bibinfo{number}{2} (\bibinfo{year}{2008}),
  \bibinfo{pages}{186 -- 195}.
\newblock
\showISSN{0164-1212}
\urldef\tempurl%
\url{https://doi.org/10.1016/j.jss.2007.05.035}
\showDOI{\tempurl}
\newblock
\shownote{Model-Based Software Testing.}


\bibitem[\protect\citeauthoryear{{Gopalakrishnan}, {Sharma}, {Mirakhorli}, and
  {Galster}}{{Gopalakrishnan} et~al\mbox{.}}{2017}]%
        {Gopalakrishnan2017_459}
\bibfield{author}{\bibinfo{person}{R. {Gopalakrishnan}}, \bibinfo{person}{P.
  {Sharma}}, \bibinfo{person}{M. {Mirakhorli}}, {and} \bibinfo{person}{M.
  {Galster}}.} \bibinfo{year}{2017}\natexlab{}.
\newblock \showarticletitle{Can Latent Topics in Source Code Predict Missing
  Architectural Tactics?}. In \bibinfo{booktitle}{\emph{2017 IEEE/ACM 39th
  International Conference on Software Engineering (ICSE)}}.
  \bibinfo{pages}{15--26}.
\newblock
\urldef\tempurl%
\url{https://doi.org/10.1109/ICSE.2017.10}
\showDOI{\tempurl}


\bibitem[\protect\citeauthoryear{Gopinath, Khurshid, Saha, and
  Chandra}{Gopinath et~al\mbox{.}}{2014}]%
        {Gopinath2014_296}
\bibfield{author}{\bibinfo{person}{Divya Gopinath}, \bibinfo{person}{Sarfraz
  Khurshid}, \bibinfo{person}{Diptikalyan Saha}, {and} \bibinfo{person}{Satish
  Chandra}.} \bibinfo{year}{2014}\natexlab{}.
\newblock \showarticletitle{Data-Guided Repair of Selection Statements}. In
  \bibinfo{booktitle}{\emph{Proceedings of the 36th International Conference on
  Software Engineering}} (Hyderabad, India) \emph{(\bibinfo{series}{ICSE
  2014})}. \bibinfo{pages}{243–253}.
\newblock
\showISBNx{9781450327565}
\urldef\tempurl%
\url{https://doi.org/10.1145/2568225.2568303}
\showDOI{\tempurl}


\bibitem[\protect\citeauthoryear{{Gopinath}, {Wang}, {Hua}, and
  {Khurshid}}{{Gopinath} et~al\mbox{.}}{2016}]%
        {Gopinath2016_348}
\bibfield{author}{\bibinfo{person}{D. {Gopinath}}, \bibinfo{person}{K. {Wang}},
  \bibinfo{person}{J. {Hua}}, {and} \bibinfo{person}{S. {Khurshid}}.}
  \bibinfo{year}{2016}\natexlab{}.
\newblock \showarticletitle{Repairing Intricate Faults in Code Using Machine
  Learning and Path Exploration}. In \bibinfo{booktitle}{\emph{2016 IEEE
  International Conference on Software Maintenance and Evolution (ICSME)}}.
  \bibinfo{pages}{453--457}.
\newblock
\urldef\tempurl%
\url{https://doi.org/10.1109/ICSME.2016.75}
\showDOI{\tempurl}


\bibitem[\protect\citeauthoryear{Goues, Pradel, and Roychoudhury}{Goues
  et~al\mbox{.}}{2019}]%
        {Goues2019_281}
\bibfield{author}{\bibinfo{person}{Claire~Le Goues}, \bibinfo{person}{Michael
  Pradel}, {and} \bibinfo{person}{Abhik Roychoudhury}.}
  \bibinfo{year}{2019}\natexlab{}.
\newblock \showarticletitle{Automated Program Repair}.
\newblock \bibinfo{journal}{\emph{Commun. ACM}} \bibinfo{volume}{62},
  \bibinfo{number}{12} (\bibinfo{date}{November} \bibinfo{year}{2019}),
  \bibinfo{pages}{56–65}.
\newblock
\showISSN{0001-0782}
\urldef\tempurl%
\url{https://doi.org/10.1145/3318162}
\showDOI{\tempurl}


\bibitem[\protect\citeauthoryear{Gousios}{Gousios}{2013}]%
        {Gousi13}
\bibfield{author}{\bibinfo{person}{Georgios Gousios}.}
  \bibinfo{year}{2013}\natexlab{}.
\newblock \showarticletitle{The {GHTorrent} dataset and tool suite}. In
  \bibinfo{booktitle}{\emph{Proceedings of the 10th Working Conference on
  Mining Software Repositories}} \emph{(\bibinfo{series}{MSR '13})}.
  \bibinfo{publisher}{IEEE Press}, \bibinfo{address}{Piscataway, NJ, USA},
  \bibinfo{pages}{233--236}.
\newblock
\showISBNx{978-1-4673-2936-1}
\urldef\tempurl%
\url{http://dl.acm.org/citation.cfm?id=2487085.2487132}
\showURL{%
\tempurl}


\bibitem[\protect\citeauthoryear{{Grano}, {Titov}, {Panichella}, and
  {Gall}}{{Grano} et~al\mbox{.}}{2018}]%
        {Grano2018_91}
\bibfield{author}{\bibinfo{person}{G. {Grano}}, \bibinfo{person}{T.~V.
  {Titov}}, \bibinfo{person}{S. {Panichella}}, {and} \bibinfo{person}{H.~C.
  {Gall}}.} \bibinfo{year}{2018}\natexlab{}.
\newblock \showarticletitle{How high will it be? Using machine learning models
  to predict branch coverage in automated testing}. In
  \bibinfo{booktitle}{\emph{2018 IEEE Workshop on Machine Learning Techniques
  for Software Quality Evaluation (MaLTeSQuE)}}. \bibinfo{pages}{19--24}.
\newblock
\urldef\tempurl%
\url{https://doi.org/10.1109/MALTESQUE.2018.8368454}
\showDOI{\tempurl}


\bibitem[\protect\citeauthoryear{Graves, Jaitly, and Mohamed}{Graves
  et~al\mbox{.}}{2013}]%
        {Graves2013}
\bibfield{author}{\bibinfo{person}{Alex Graves}, \bibinfo{person}{Navdeep
  Jaitly}, {and} \bibinfo{person}{Abdel-rahman Mohamed}.}
  \bibinfo{year}{2013}\natexlab{}.
\newblock \showarticletitle{Hybrid speech recognition with deep bidirectional
  LSTM}. In \bibinfo{booktitle}{\emph{Automatic Speech Recognition and
  Understanding (ASRU), 2013 IEEE Workshop on}}. IEEE,
  \bibinfo{pages}{273--278}.
\newblock


\bibitem[\protect\citeauthoryear{Greff, Srivastava, Koutn{\'\i}k, Steunebrink,
  and Schmidhuber}{Greff et~al\mbox{.}}{2017}]%
        {Greff2017}
\bibfield{author}{\bibinfo{person}{Klaus Greff}, \bibinfo{person}{Rupesh~K
  Srivastava}, \bibinfo{person}{Jan Koutn{\'\i}k}, \bibinfo{person}{Bas~R
  Steunebrink}, {and} \bibinfo{person}{J{\"u}rgen Schmidhuber}.}
  \bibinfo{year}{2017}\natexlab{}.
\newblock \showarticletitle{LSTM: A search space odyssey}.
\newblock \bibinfo{journal}{\emph{IEEE transactions on neural networks and
  learning systems}} \bibinfo{volume}{28}, \bibinfo{number}{10}
  (\bibinfo{year}{2017}), \bibinfo{pages}{2222--2232}.
\newblock


\bibitem[\protect\citeauthoryear{Grodzicka, Ziobrowski, {\L}akomiak, Kawa, and
  Madeyski}{Grodzicka et~al\mbox{.}}{2020}]%
        {Grodzicka2020_208}
\bibfield{author}{\bibinfo{person}{Hanna Grodzicka}, \bibinfo{person}{Arkadiusz
  Ziobrowski}, \bibinfo{person}{Zofia {\L}akomiak}, \bibinfo{person}{Micha{\l}
  Kawa}, {and} \bibinfo{person}{Lech Madeyski}.}
  \bibinfo{year}{2020}\natexlab{}.
\newblock \bibinfo{booktitle}{\emph{Code Smell Prediction Employing Machine
  Learning Meets Emerging Java Language Constructs}}.
\newblock \bibinfo{pages}{137--167}.
\newblock
\showISBNx{978-3-030-34706-2}
\urldef\tempurl%
\url{https://doi.org/10.1007/978-3-030-34706-2\_8}
\showDOI{\tempurl}


\bibitem[\protect\citeauthoryear{Gu, Zhang, and Kim}{Gu et~al\mbox{.}}{2018}]%
        {Gu2018}
\bibfield{author}{\bibinfo{person}{Xiaodong Gu}, \bibinfo{person}{Hongyu
  Zhang}, {and} \bibinfo{person}{Sunghun Kim}.}
  \bibinfo{year}{2018}\natexlab{}.
\newblock \showarticletitle{Deep Code Search}. In
  \bibinfo{booktitle}{\emph{2018 IEEE/ACM 40th International Conference on
  Software Engineering (ICSE)}}. \bibinfo{pages}{933--944}.
\newblock
\urldef\tempurl%
\url{https://doi.org/10.1145/3180155.3180167}
\showDOI{\tempurl}


\bibitem[\protect\citeauthoryear{Guggulothu and Moiz}{Guggulothu and
  Moiz}{2020}]%
        {Guggulothu2020_206}
\bibfield{author}{\bibinfo{person}{Thirupathi Guggulothu} {and}
  \bibinfo{person}{S.~A. Moiz}.} \bibinfo{year}{2020}\natexlab{}.
\newblock \showarticletitle{Code smell detection using multi-label
  classification approach}.
\newblock \bibinfo{journal}{\emph{Software Quality Journal}}
  (\bibinfo{year}{2020}), \bibinfo{pages}{1--24}.
\newblock


\bibitem[\protect\citeauthoryear{Gupta, Suri, Kumar, and Jain}{Gupta
  et~al\mbox{.}}{2021d}]%
        {Gupta2021_30}
\bibfield{author}{\bibinfo{person}{Aakanshi Gupta}, \bibinfo{person}{Bharti
  Suri}, \bibinfo{person}{Vijay Kumar}, {and} \bibinfo{person}{Pragyashree
  Jain}.} \bibinfo{year}{2021}\natexlab{d}.
\newblock \showarticletitle{Extracting rules for vulnerabilities detection with
  static metrics using machine learning}.
\newblock \bibinfo{journal}{\emph{International Journal of System Assurance
  Engineering and Management}}  \bibinfo{volume}{12} (\bibinfo{year}{2021}),
  \bibinfo{pages}{65--76}.
\newblock


\bibitem[\protect\citeauthoryear{Gupta, Suri, and Lamba}{Gupta
  et~al\mbox{.}}{2021c}]%
        {847_Gupta2021}
\bibfield{author}{\bibinfo{person}{Aakanshi Gupta}, \bibinfo{person}{Bharti
  Suri}, {and} \bibinfo{person}{Lakshay Lamba}.}
  \bibinfo{year}{2021}\natexlab{c}.
\newblock \showarticletitle{Tracing Bad Code Smells Behavior Using Machine
  Learning with Software Metrics}.
\newblock \bibinfo{journal}{\emph{Smart and Sustainable Intelligent Systems}}
  (\bibinfo{year}{2021}), \bibinfo{pages}{245--257}.
\newblock


\bibitem[\protect\citeauthoryear{Gupta, Gulanikar, Kumar, and Neti}{Gupta
  et~al\mbox{.}}{2021a}]%
        {823_Gupta2021}
\bibfield{author}{\bibinfo{person}{Himanshu Gupta},
  \bibinfo{person}{Abhiram~Anand Gulanikar}, \bibinfo{person}{Lov Kumar}, {and}
  \bibinfo{person}{Lalita Bhanu~Murthy Neti}.}
  \bibinfo{year}{2021}\natexlab{a}.
\newblock \showarticletitle{Empirical Analysis on Effectiveness of NLP Methods
  for Predicting Code Smell}. In \bibinfo{booktitle}{\emph{International
  Conference on Computational Science and Its Applications}}. Springer,
  \bibinfo{pages}{43--53}.
\newblock


\bibitem[\protect\citeauthoryear{Gupta, Kulkarni, Kumar, Neti, and
  Krishna}{Gupta et~al\mbox{.}}{2021b}]%
        {815_Gupta2021}
\bibfield{author}{\bibinfo{person}{Himanshu Gupta},
  \bibinfo{person}{Tanmay~Girish Kulkarni}, \bibinfo{person}{Lov Kumar},
  \bibinfo{person}{Lalita Bhanu~Murthy Neti}, {and} \bibinfo{person}{Aneesh
  Krishna}.} \bibinfo{year}{2021}\natexlab{b}.
\newblock \showarticletitle{An empirical study on predictability of software
  code smell using deep learning models}. In
  \bibinfo{booktitle}{\emph{International Conference on Advanced Information
  Networking and Applications}}. Springer, \bibinfo{pages}{120--132}.
\newblock


\bibitem[\protect\citeauthoryear{{Gupta}, {Kumar}, and {Neti}}{{Gupta}
  et~al\mbox{.}}{2019}]%
        {Gupta2019_190}
\bibfield{author}{\bibinfo{person}{H. {Gupta}}, \bibinfo{person}{L. {Kumar}},
  {and} \bibinfo{person}{L.~B.~M. {Neti}}.} \bibinfo{year}{2019}\natexlab{}.
\newblock \showarticletitle{An Empirical Framework for Code Smell Prediction
  using Extreme Learning Machine*}. In \bibinfo{booktitle}{\emph{2019 9th
  Annual Information Technology, Electromechanical Engineering and
  Microelectronics Conference (IEMECON)}}. \bibinfo{pages}{189--195}.
\newblock
\urldef\tempurl%
\url{https://doi.org/10.1109/IEMECONX.2019.8877082}
\showDOI{\tempurl}


\bibitem[\protect\citeauthoryear{Gupta, Kanade, and Shevade}{Gupta
  et~al\mbox{.}}{2019}]%
        {Gupta2019_298}
\bibfield{author}{\bibinfo{person}{Rahul Gupta}, \bibinfo{person}{Aditya
  Kanade}, {and} \bibinfo{person}{Shirish Shevade}.}
  \bibinfo{year}{2019}\natexlab{}.
\newblock \showarticletitle{Deep Reinforcement Learning for Syntactic Error
  Repair in Student Programs}.
\newblock \bibinfo{journal}{\emph{Proceedings of the AAAI Conference on
  Artificial Intelligence}}  \bibinfo{volume}{33} (\bibinfo{date}{07}
  \bibinfo{year}{2019}), \bibinfo{pages}{930--937}.
\newblock
\urldef\tempurl%
\url{https://doi.org/10.1609/aaai.v33i01.3301930}
\showDOI{\tempurl}


\bibitem[\protect\citeauthoryear{Gupta, Pal, Kanade, and Shevade}{Gupta
  et~al\mbox{.}}{2017}]%
        {Gupta2017}
\bibfield{author}{\bibinfo{person}{Rahul Gupta}, \bibinfo{person}{Soham Pal},
  \bibinfo{person}{Aditya Kanade}, {and} \bibinfo{person}{Shirish Shevade}.}
  \bibinfo{year}{2017}\natexlab{}.
\newblock \showarticletitle{{DeepFix}: Fixing Common C Language Errors by Deep
  Learning.}. In \bibinfo{booktitle}{\emph{AAAI}}. \bibinfo{pages}{1345--1351}.
\newblock


\bibitem[\protect\citeauthoryear{Hadj-Kacem and Bouassida}{Hadj-Kacem and
  Bouassida}{2018}]%
        {Hadj-Kacem2018_511}
\bibfield{author}{\bibinfo{person}{Mouna Hadj-Kacem} {and}
  \bibinfo{person}{Nadia Bouassida}.} \bibinfo{year}{2018}\natexlab{}.
\newblock \showarticletitle{A Hybrid Approach To Detect Code Smells using Deep
  Learning.}. In \bibinfo{booktitle}{\emph{ENASE}}. \bibinfo{pages}{137--146}.
\newblock


\bibitem[\protect\citeauthoryear{Hadj-Kacem and Bouassida}{Hadj-Kacem and
  Bouassida}{2019}]%
        {840_Hadj-Kacem2019}
\bibfield{author}{\bibinfo{person}{Mouna Hadj-Kacem} {and}
  \bibinfo{person}{Nadia Bouassida}.} \bibinfo{year}{2019}\natexlab{}.
\newblock \showarticletitle{Deep representation learning for code smells
  detection using variational auto-encoder}. In \bibinfo{booktitle}{\emph{2019
  International Joint Conference on Neural Networks (IJCNN)}}. IEEE,
  \bibinfo{pages}{1--8}.
\newblock


\bibitem[\protect\citeauthoryear{{Hall} and {Bowes}}{{Hall} and
  {Bowes}}{2012}]%
        {Hall2012_145}
\bibfield{author}{\bibinfo{person}{T. {Hall}} {and} \bibinfo{person}{D.
  {Bowes}}.} \bibinfo{year}{2012}\natexlab{}.
\newblock \showarticletitle{The State of Machine Learning Methodology in
  Software Fault Prediction}. In \bibinfo{booktitle}{\emph{2012 11th
  International Conference on Machine Learning and Applications}},
  Vol.~\bibinfo{volume}{2}. \bibinfo{pages}{308--313}.
\newblock
\urldef\tempurl%
\url{https://doi.org/10.1109/ICMLA.2012.226}
\showDOI{\tempurl}


\bibitem[\protect\citeauthoryear{Halstead}{Halstead}{1977}]%
        {Halstead_1977}
\bibfield{author}{\bibinfo{person}{Maurice~H. Halstead}.}
  \bibinfo{year}{1977}\natexlab{}.
\newblock \bibinfo{booktitle}{\emph{Elements of Software Science (Operating and
  Programming Systems Series)}}.
\newblock \bibinfo{address}{USA}.
\newblock
\showISBNx{0444002057}


\bibitem[\protect\citeauthoryear{Hammad, Babur, Basit, and van~den
  Brand}{Hammad et~al\mbox{.}}{2021}]%
        {758_Hammad2021}
\bibfield{author}{\bibinfo{person}{Muhammad Hammad},
  \bibinfo{person}{{\""O}nder Babur}, \bibinfo{person}{Hamid~Abdul Basit},
  {and} \bibinfo{person}{Mark van~den Brand}.} \bibinfo{year}{2021}\natexlab{}.
\newblock \showarticletitle{Clone-advisor: recommending code tokens and clone
  methods with deep learning and information retrieval}.
\newblock \bibinfo{journal}{\emph{PeerJ Computer Science}}  \bibinfo{volume}{7}
  (\bibinfo{year}{2021}), \bibinfo{pages}{e737}.
\newblock


\bibitem[\protect\citeauthoryear{Hammouri, Hammad, Alnabhan, and
  Alsarayrah}{Hammouri et~al\mbox{.}}{2018}]%
        {Hammouri2018_127}
\bibfield{author}{\bibinfo{person}{Awni Hammouri}, \bibinfo{person}{Mustafa
  Hammad}, \bibinfo{person}{Mohammad Alnabhan}, {and} \bibinfo{person}{Fatima
  Alsarayrah}.} \bibinfo{year}{2018}\natexlab{}.
\newblock \showarticletitle{Software Bug Prediction using Machine Learning
  Approach}.
\newblock \bibinfo{journal}{\emph{International Journal of Advanced Computer
  Science and Applications}}  \bibinfo{volume}{9} (\bibinfo{date}{01}
  \bibinfo{year}{2018}).
\newblock
\urldef\tempurl%
\url{https://doi.org/10.14569/IJACSA.2018.090212}
\showDOI{\tempurl}


\bibitem[\protect\citeauthoryear{{Han}, {Wallace}, and {Miller}}{{Han}
  et~al\mbox{.}}{2009}]%
        {Han2009_488}
\bibfield{author}{\bibinfo{person}{S. {Han}}, \bibinfo{person}{D.~R.
  {Wallace}}, {and} \bibinfo{person}{R.~C. {Miller}}.}
  \bibinfo{year}{2009}\natexlab{}.
\newblock \showarticletitle{Code Completion from Abbreviated Input}. In
  \bibinfo{booktitle}{\emph{2009 IEEE/ACM International Conference on Automated
  Software Engineering}}. \bibinfo{pages}{332--343}.
\newblock
\urldef\tempurl%
\url{https://doi.org/10.1109/ASE.2009.64}
\showDOI{\tempurl}


\bibitem[\protect\citeauthoryear{Han, Wallace, and Miller}{Han
  et~al\mbox{.}}{2011}]%
        {Han2011_489}
\bibfield{author}{\bibinfo{person}{Sangmok Han}, \bibinfo{person}{David~R.
  Wallace}, {and} \bibinfo{person}{Robert~C. Miller}.}
  \bibinfo{year}{2011}\natexlab{}.
\newblock \showarticletitle{Code Completion of Multiple Keywords from
  Abbreviated Input}.
\newblock \bibinfo{journal}{\emph{Automated Software Engg.}}
  \bibinfo{volume}{18}, \bibinfo{number}{3–4} (\bibinfo{date}{December}
  \bibinfo{year}{2011}), \bibinfo{pages}{363–398}.
\newblock
\showISSN{0928-8910}
\urldef\tempurl%
\url{https://doi.org/10.1007/s10515-011-0083-2}
\showDOI{\tempurl}


\bibitem[\protect\citeauthoryear{Hanif, Md~Nasir, Ab~Razak, Firdaus, and
  Anuar}{Hanif et~al\mbox{.}}{2021}]%
        {545_Hanif2021}
\bibfield{author}{\bibinfo{person}{Hazim Hanif}, \bibinfo{person}{Mohd
  Hairul~Nizam Md~Nasir}, \bibinfo{person}{Mohd~Faizal Ab~Razak},
  \bibinfo{person}{Ahmad Firdaus}, {and} \bibinfo{person}{Nor~Badrul Anuar}.}
  \bibinfo{year}{2021}\natexlab{}.
\newblock \showarticletitle{The rise of software vulnerability: {Taxonomy} of
  software vulnerabilities detection and machine learning approaches}.
\newblock \bibinfo{journal}{\emph{Journal of Network and Computer
  Applications}}  \bibinfo{volume}{179} (\bibinfo{date}{April}
  \bibinfo{year}{2021}), \bibinfo{pages}{103009}.
\newblock
\showISSN{1084-8045}
\urldef\tempurl%
\url{https://doi.org/10.1016/j.jnca.2021.103009}
\showDOI{\tempurl}


\bibitem[\protect\citeauthoryear{Haque, Bansal, Wu, and McMillan}{Haque
  et~al\mbox{.}}{2021}]%
        {995_Haque2021}
\bibfield{author}{\bibinfo{person}{Sakib Haque}, \bibinfo{person}{Aakash
  Bansal}, \bibinfo{person}{Lingfei Wu}, {and} \bibinfo{person}{Collin
  McMillan}.} \bibinfo{year}{2021}\natexlab{}.
\newblock \showarticletitle{Action word prediction for neural source code
  summarization}. In \bibinfo{booktitle}{\emph{2021 IEEE International
  Conference on Software Analysis, Evolution and Reengineering (SANER)}}. IEEE,
  \bibinfo{pages}{330--341}.
\newblock


\bibitem[\protect\citeauthoryear{Haque, LeClair, Wu, and McMillan}{Haque
  et~al\mbox{.}}{2020}]%
        {989_Haque2020}
\bibfield{author}{\bibinfo{person}{Sakib Haque}, \bibinfo{person}{Alexander
  LeClair}, \bibinfo{person}{Lingfei Wu}, {and} \bibinfo{person}{Collin
  McMillan}.} \bibinfo{year}{2020}\natexlab{}.
\newblock \showarticletitle{Improved automatic summarization of subroutines via
  attention to file context}. In \bibinfo{booktitle}{\emph{Proceedings of the
  17th International Conference on Mining Software Repositories}}.
  \bibinfo{pages}{300--310}.
\newblock


\bibitem[\protect\citeauthoryear{Harman, Islam, Jia, Minku, Sarro, and
  Srivisut}{Harman et~al\mbox{.}}{2014}]%
        {Harman2014}
\bibfield{author}{\bibinfo{person}{Mark Harman}, \bibinfo{person}{Syed Islam},
  \bibinfo{person}{Yue Jia}, \bibinfo{person}{Leandro~L. Minku},
  \bibinfo{person}{Federica Sarro}, {and} \bibinfo{person}{Komsan Srivisut}.}
  \bibinfo{year}{2014}\natexlab{}.
\newblock \showarticletitle{Less is More: Temporal Fault Predictive Performance
  over Multiple Hadoop Releases}. In \bibinfo{booktitle}{\emph{Search-Based
  Software Engineering}}, \bibfield{editor}{\bibinfo{person}{Claire Le~Goues}
  {and} \bibinfo{person}{Shin Yoo}} (Eds.). \bibinfo{publisher}{Springer
  International Publishing}, \bibinfo{address}{Cham},
  \bibinfo{pages}{240--246}.
\newblock
\showISBNx{978-3-319-09940-8}


\bibitem[\protect\citeauthoryear{Hellendoorn, Bird, Barr, and
  Allamanis}{Hellendoorn et~al\mbox{.}}{2018}]%
        {Hellendoorn2018_378}
\bibfield{author}{\bibinfo{person}{Vincent~J. Hellendoorn},
  \bibinfo{person}{Christian Bird}, \bibinfo{person}{Earl~T. Barr}, {and}
  \bibinfo{person}{Miltiadis Allamanis}.} \bibinfo{year}{2018}\natexlab{}.
\newblock \showarticletitle{Deep Learning Type Inference}
  \emph{(\bibinfo{series}{ESEC/FSE 2018})}. \bibinfo{pages}{152–162}.
\newblock
\showISBNx{9781450355735}
\urldef\tempurl%
\url{https://doi.org/10.1145/3236024.3236051}
\showDOI{\tempurl}


\bibitem[\protect\citeauthoryear{Hellendoorn and Devanbu}{Hellendoorn and
  Devanbu}{2017}]%
        {Hellendoorn2017_455}
\bibfield{author}{\bibinfo{person}{Vincent~J. Hellendoorn} {and}
  \bibinfo{person}{Premkumar Devanbu}.} \bibinfo{year}{2017}\natexlab{}.
\newblock \showarticletitle{Are Deep Neural Networks the Best Choice for
  Modeling Source Code?}. In \bibinfo{booktitle}{\emph{Proceedings of the 2017
  11th Joint Meeting on Foundations of Software Engineering}} (Paderborn,
  Germany) \emph{(\bibinfo{series}{ESEC/FSE 2017})}.
  \bibinfo{pages}{763–773}.
\newblock
\showISBNx{9781450351058}
\urldef\tempurl%
\url{https://doi.org/10.1145/3106237.3106290}
\showDOI{\tempurl}


\bibitem[\protect\citeauthoryear{Heo, Oh, and Yi}{Heo et~al\mbox{.}}{2017}]%
        {Heo2017_246}
\bibfield{author}{\bibinfo{person}{Kihong Heo}, \bibinfo{person}{Hakjoo Oh},
  {and} \bibinfo{person}{Kwangkeun Yi}.} \bibinfo{year}{2017}\natexlab{}.
\newblock \showarticletitle{Machine-Learning-Guided Selectively Unsound Static
  Analysis}. In \bibinfo{booktitle}{\emph{Proceedings of the 39th International
  Conference on Software Engineering}} (Buenos Aires, Argentina)
  \emph{(\bibinfo{series}{ICSE '17})}. \bibinfo{pages}{519–529}.
\newblock
\showISBNx{9781538638682}
\urldef\tempurl%
\url{https://doi.org/10.1109/ICSE.2017.54}
\showDOI{\tempurl}


\bibitem[\protect\citeauthoryear{Hoang, Kang, Lo, and Lawall}{Hoang
  et~al\mbox{.}}{2020}]%
        {Hoang2020_461}
\bibfield{author}{\bibinfo{person}{Thong Hoang}, \bibinfo{person}{Hong~Jin
  Kang}, \bibinfo{person}{David Lo}, {and} \bibinfo{person}{Julia Lawall}.}
  \bibinfo{year}{2020}\natexlab{}.
\newblock \showarticletitle{CC2Vec: Distributed Representations of Code
  Changes}. In \bibinfo{booktitle}{\emph{Proceedings of the ACM/IEEE 42nd
  International Conference on Software Engineering}} (Seoul, South Korea)
  \emph{(\bibinfo{series}{ICSE '20})}. \bibinfo{pages}{518–529}.
\newblock
\showISBNx{9781450371216}
\urldef\tempurl%
\url{https://doi.org/10.1145/3377811.3380361}
\showDOI{\tempurl}


\bibitem[\protect\citeauthoryear{Hort, Kechagia, Sarro, and Harman}{Hort
  et~al\mbox{.}}{2021}]%
        {hort21tse}
\bibfield{author}{\bibinfo{person}{Max Hort}, \bibinfo{person}{Maria Kechagia},
  \bibinfo{person}{Federica Sarro}, {and} \bibinfo{person}{Mark Harman}.}
  \bibinfo{year}{2021}\natexlab{}.
\newblock \showarticletitle{A Survey of Performance Optimization for Mobile
  Applications}.
\newblock \bibinfo{journal}{\emph{IEEE Transactions on Software Engineering
  (TSE)}} (\bibinfo{year}{2021}).
\newblock


\bibitem[\protect\citeauthoryear{Hou, Chang, Chen, Laih, and Chen}{Hou
  et~al\mbox{.}}{2010}]%
        {Hou2010_44}
\bibfield{author}{\bibinfo{person}{Yung-Tsung Hou}, \bibinfo{person}{Yimeng
  Chang}, \bibinfo{person}{Tsuhan Chen}, \bibinfo{person}{Chi-Sung Laih}, {and}
  \bibinfo{person}{Chia-Mei Chen}.} \bibinfo{year}{2010}\natexlab{}.
\newblock \showarticletitle{Malicious web content detection by machine
  learning}.
\newblock \bibinfo{journal}{\emph{Expert Systems with Applications}}
  \bibinfo{volume}{37}, \bibinfo{number}{1} (\bibinfo{year}{2010}),
  \bibinfo{pages}{55 -- 60}.
\newblock
\showISSN{0957-4174}
\urldef\tempurl%
\url{https://doi.org/10.1016/j.eswa.2009.05.023}
\showDOI{\tempurl}


\bibitem[\protect\citeauthoryear{Hu, Zhu, and Yang}{Hu et~al\mbox{.}}{2018b}]%
        {Hu2018_72}
\bibfield{author}{\bibinfo{person}{Gang Hu}, \bibinfo{person}{Linjie Zhu},
  {and} \bibinfo{person}{Junfeng Yang}.} \bibinfo{year}{2018}\natexlab{b}.
\newblock \showarticletitle{AppFlow: Using Machine Learning to Synthesize
  Robust, Reusable UI Tests}. In \bibinfo{booktitle}{\emph{Proceedings of the
  2018 26th ACM Joint Meeting on European Software Engineering Conference and
  Symposium on the Foundations of Software Engineering}} (Lake Buena Vista, FL,
  USA) \emph{(\bibinfo{series}{ESEC/FSE 2018})}. \bibinfo{pages}{269–282}.
\newblock
\showISBNx{9781450355735}
\urldef\tempurl%
\url{https://doi.org/10.1145/3236024.3236055}
\showDOI{\tempurl}


\bibitem[\protect\citeauthoryear{{Hu}, {Li}, {Xia}, {Lo}, and {Jin}}{{Hu}
  et~al\mbox{.}}{2018}]%
        {Hu2018_421}
\bibfield{author}{\bibinfo{person}{X. {Hu}}, \bibinfo{person}{G. {Li}},
  \bibinfo{person}{X. {Xia}}, \bibinfo{person}{D. {Lo}}, {and}
  \bibinfo{person}{Z. {Jin}}.} \bibinfo{year}{2018}\natexlab{}.
\newblock \showarticletitle{Deep Code Comment Generation}. In
  \bibinfo{booktitle}{\emph{2018 IEEE/ACM 26th International Conference on
  Program Comprehension (ICPC)}}. \bibinfo{pages}{200--20010}.
\newblock


\bibitem[\protect\citeauthoryear{Hu, Li, Xia, Lo, Lu, and Jin}{Hu
  et~al\mbox{.}}{2018a}]%
        {Xing18}
\bibfield{author}{\bibinfo{person}{Xing Hu}, \bibinfo{person}{Ge Li},
  \bibinfo{person}{Xin Xia}, \bibinfo{person}{David Lo}, \bibinfo{person}{Shuai
  Lu}, {and} \bibinfo{person}{Zhi Jin}.} \bibinfo{year}{2018}\natexlab{a}.
\newblock \showarticletitle{Summarizing Source Code with Transferred API
  Knowledge}. In \bibinfo{booktitle}{\emph{Proceedings of the Twenty-Seventh
  International Joint Conference on Artificial Intelligence, {IJCAI-18}}}.
  \bibinfo{publisher}{International Joint Conferences on Artificial
  Intelligence Organization}, \bibinfo{pages}{2269--2275}.
\newblock
\urldef\tempurl%
\url{https://doi.org/10.24963/ijcai.2018/314}
\showDOI{\tempurl}


\bibitem[\protect\citeauthoryear{Huang, Hu, Jia, Chen, Zheng, and Luo}{Huang
  et~al\mbox{.}}{2020a}]%
        {Huang2020_377}
\bibfield{author}{\bibinfo{person}{Yuan Huang}, \bibinfo{person}{Xinyu Hu},
  \bibinfo{person}{Nan Jia}, \bibinfo{person}{Xiangping Chen},
  \bibinfo{person}{Zibin Zheng}, {and} \bibinfo{person}{Xiapu Luo}.}
  \bibinfo{year}{2020}\natexlab{a}.
\newblock \showarticletitle{CommtPst: Deep learning source code for commenting
  positions prediction}.
\newblock \bibinfo{journal}{\emph{Journal of Systems and Software}}
  \bibinfo{volume}{170} (\bibinfo{year}{2020}), \bibinfo{pages}{110754}.
\newblock
\showISSN{0164-1212}
\urldef\tempurl%
\url{https://doi.org/10.1016/j.jss.2020.110754}
\showDOI{\tempurl}


\bibitem[\protect\citeauthoryear{Huang, Huang, Chen, Chen, Zheng, Luo, Jia, Hu,
  and Zhou}{Huang et~al\mbox{.}}{2020b}]%
        {1033_Huang2020}
\bibfield{author}{\bibinfo{person}{Yuan Huang}, \bibinfo{person}{Shaohao
  Huang}, \bibinfo{person}{Huanchao Chen}, \bibinfo{person}{Xiangping Chen},
  \bibinfo{person}{Zibin Zheng}, \bibinfo{person}{Xiapu Luo},
  \bibinfo{person}{Nan Jia}, \bibinfo{person}{Xinyu Hu}, {and}
  \bibinfo{person}{Xiaocong Zhou}.} \bibinfo{year}{2020}\natexlab{b}.
\newblock \showarticletitle{Towards automatically generating block comments for
  code snippets}.
\newblock \bibinfo{journal}{\emph{Information and Software Technology}}
  \bibinfo{volume}{127} (\bibinfo{year}{2020}), \bibinfo{pages}{106373}.
\newblock


\bibitem[\protect\citeauthoryear{Hussain, Huang, Zhou, and Wang}{Hussain
  et~al\mbox{.}}{2020}]%
        {Hussain2020_492}
\bibfield{author}{\bibinfo{person}{Yasir Hussain}, \bibinfo{person}{Zhiqiu
  Huang}, \bibinfo{person}{Yu Zhou}, {and} \bibinfo{person}{Senzhang Wang}.}
  \bibinfo{year}{2020}\natexlab{}.
\newblock \showarticletitle{CodeGRU: Context-aware deep learning with gated
  recurrent unit for source code modeling}.
\newblock \bibinfo{journal}{\emph{Information and Software Technology}}
  \bibinfo{volume}{125} (\bibinfo{year}{2020}), \bibinfo{pages}{106309}.
\newblock
\showISSN{0950-5849}
\urldef\tempurl%
\url{https://doi.org/10.1016/j.infsof.2020.106309}
\showDOI{\tempurl}


\bibitem[\protect\citeauthoryear{{Ivers}, {Ozkaya}, and {Nord}}{{Ivers}
  et~al\mbox{.}}{2019}]%
        {Ivers2019_395}
\bibfield{author}{\bibinfo{person}{J. {Ivers}}, \bibinfo{person}{I. {Ozkaya}},
  {and} \bibinfo{person}{R.~L. {Nord}}.} \bibinfo{year}{2019}\natexlab{}.
\newblock \showarticletitle{Can AI Close the Design-Code Abstraction Gap?}. In
  \bibinfo{booktitle}{\emph{2019 34th IEEE/ACM International Conference on
  Automated Software Engineering Workshop (ASEW)}}. \bibinfo{pages}{122--125}.
\newblock
\urldef\tempurl%
\url{https://doi.org/10.1109/ASEW.2019.00041}
\showDOI{\tempurl}


\bibitem[\protect\citeauthoryear{Iyer, Konstas, Cheung, and Zettlemoyer}{Iyer
  et~al\mbox{.}}{2016}]%
        {Iyer2016_440}
\bibfield{author}{\bibinfo{person}{Srinivasan Iyer}, \bibinfo{person}{Ioannis
  Konstas}, \bibinfo{person}{Alvin Cheung}, {and} \bibinfo{person}{Luke
  Zettlemoyer}.} \bibinfo{year}{2016}\natexlab{}.
\newblock \showarticletitle{Summarizing Source Code using a Neural Attention
  Model}. In \bibinfo{booktitle}{\emph{Proceedings of the 54th Annual Meeting
  of the Association for Computational Linguistics (Volume 1: Long Papers)}}.
  \bibinfo{pages}{2073--2083}.
\newblock
\urldef\tempurl%
\url{https://doi.org/10.18653/v1/P16-1195}
\showDOI{\tempurl}


\bibitem[\protect\citeauthoryear{Jain and Saha}{Jain and Saha}{2021}]%
        {824_Jain2021}
\bibfield{author}{\bibinfo{person}{Shivani Jain} {and} \bibinfo{person}{Anju
  Saha}.} \bibinfo{year}{2021}\natexlab{}.
\newblock \showarticletitle{Improving performance with hybrid feature selection
  and ensemble machine learning techniques for code smell detection}.
\newblock \bibinfo{journal}{\emph{Science of Computer Programming}}
  \bibinfo{volume}{212} (\bibinfo{year}{2021}), \bibinfo{pages}{102713}.
\newblock


\bibitem[\protect\citeauthoryear{{Ji}, {Pan}, {Chen}, and {Mao}}{{Ji}
  et~al\mbox{.}}{2018}]%
        {Ji2018_382}
\bibfield{author}{\bibinfo{person}{T. {Ji}}, \bibinfo{person}{J. {Pan}},
  \bibinfo{person}{L. {Chen}}, {and} \bibinfo{person}{X. {Mao}}.}
  \bibinfo{year}{2018}\natexlab{}.
\newblock \showarticletitle{Identifying Supplementary Bug-fix Commits}. In
  \bibinfo{booktitle}{\emph{2018 IEEE 42nd Annual Computer Software and
  Applications Conference (COMPSAC)}}, Vol.~\bibinfo{volume}{01}.
  \bibinfo{pages}{184--193}.
\newblock
\urldef\tempurl%
\url{https://doi.org/10.1109/COMPSAC.2018.00031}
\showDOI{\tempurl}


\bibitem[\protect\citeauthoryear{Jiang, Xiong, Zhang, Gao, and Chen}{Jiang
  et~al\mbox{.}}{2018}]%
        {Jiang2018_351}
\bibfield{author}{\bibinfo{person}{Jiajun Jiang}, \bibinfo{person}{Yingfei
  Xiong}, \bibinfo{person}{Hongyu Zhang}, \bibinfo{person}{Qing Gao}, {and}
  \bibinfo{person}{Xiangqun Chen}.} \bibinfo{year}{2018}\natexlab{}.
\newblock \showarticletitle{Shaping Program Repair Space with Existing Patches
  and Similar Code}. In \bibinfo{booktitle}{\emph{Proceedings of the 27th ACM
  SIGSOFT International Symposium on Software Testing and Analysis}}
  (Amsterdam, Netherlands) \emph{(\bibinfo{series}{ISSTA 2018})}.
  \bibinfo{pages}{298–309}.
\newblock
\showISBNx{9781450356992}
\urldef\tempurl%
\url{https://doi.org/10.1145/3213846.3213871}
\showDOI{\tempurl}


\bibitem[\protect\citeauthoryear{Jiang, Liu, and Jiang}{Jiang
  et~al\mbox{.}}{2019}]%
        {Jiang2019_384}
\bibfield{author}{\bibinfo{person}{Lin Jiang}, \bibinfo{person}{Hui Liu}, {and}
  \bibinfo{person}{He Jiang}.} \bibinfo{year}{2019}\natexlab{}.
\newblock \showarticletitle{Machine Learning Based Recommendation of Method
  Names: How Far Are We}. In \bibinfo{booktitle}{\emph{Proceedings of the 34th
  IEEE/ACM International Conference on Automated Software Engineering}} (San
  Diego, California) \emph{(\bibinfo{series}{ASE '19})}.
  \bibinfo{pages}{602–614}.
\newblock
\showISBNx{9781728125084}
\urldef\tempurl%
\url{https://doi.org/10.1109/ASE.2019.00062}
\showDOI{\tempurl}


\bibitem[\protect\citeauthoryear{Jiang, Lutellier, and Tan}{Jiang
  et~al\mbox{.}}{2021}]%
        {890_Jiang2021}
\bibfield{author}{\bibinfo{person}{Nan Jiang}, \bibinfo{person}{Thibaud
  Lutellier}, {and} \bibinfo{person}{Lin Tan}.}
  \bibinfo{year}{2021}\natexlab{}.
\newblock \showarticletitle{CURE: Code-aware neural machine translation for
  automatic program repair}. In \bibinfo{booktitle}{\emph{2021 IEEE/ACM 43rd
  International Conference on Software Engineering (ICSE)}}. IEEE,
  \bibinfo{pages}{1161--1173}.
\newblock


\bibitem[\protect\citeauthoryear{Jiang}{Jiang}{2019}]%
        {955_Jiang2019}
\bibfield{author}{\bibinfo{person}{Shuyao Jiang}.}
  \bibinfo{year}{2019}\natexlab{}.
\newblock \showarticletitle{Boosting neural commit message generation with code
  semantic analysis}. In \bibinfo{booktitle}{\emph{2019 34th IEEE/ACM
  International Conference on Automated Software Engineering (ASE)}}. IEEE,
  \bibinfo{pages}{1280--1282}.
\newblock


\bibitem[\protect\citeauthoryear{{Jiang}, {Armaly}, and {McMillan}}{{Jiang}
  et~al\mbox{.}}{2017}]%
        {Jiang2017_416}
\bibfield{author}{\bibinfo{person}{S. {Jiang}}, \bibinfo{person}{A. {Armaly}},
  {and} \bibinfo{person}{C. {McMillan}}.} \bibinfo{year}{2017}\natexlab{}.
\newblock \showarticletitle{Automatically generating commit messages from diffs
  using neural machine translation}. In \bibinfo{booktitle}{\emph{2017 32nd
  IEEE/ACM International Conference on Automated Software Engineering (ASE)}}.
  \bibinfo{pages}{135--146}.
\newblock
\urldef\tempurl%
\url{https://doi.org/10.1109/ASE.2017.8115626}
\showDOI{\tempurl}


\bibitem[\protect\citeauthoryear{Jiang and McMillan}{Jiang and
  McMillan}{2017}]%
        {952_Jiang2017}
\bibfield{author}{\bibinfo{person}{Siyuan Jiang} {and} \bibinfo{person}{Collin
  McMillan}.} \bibinfo{year}{2017}\natexlab{}.
\newblock \showarticletitle{Towards automatic generation of short summaries of
  commits}. In \bibinfo{booktitle}{\emph{2017 IEEE/ACM 25th International
  Conference on Program Comprehension (ICPC)}}. IEEE,
  \bibinfo{pages}{320--323}.
\newblock


\bibitem[\protect\citeauthoryear{Jie, Xiao-Hui, and Qiang}{Jie
  et~al\mbox{.}}{2016}]%
        {537_Jie2016}
\bibfield{author}{\bibinfo{person}{Gong Jie}, \bibinfo{person}{Kuang Xiao-Hui},
  {and} \bibinfo{person}{Liu Qiang}.} \bibinfo{year}{2016}\natexlab{}.
\newblock \showarticletitle{Survey on Software Vulnerability Analysis Method
  Based on Machine Learning}. In \bibinfo{booktitle}{\emph{2016 IEEE First
  International Conference on Data Science in Cyberspace (DSC)}}.
  \bibinfo{pages}{642--647}.
\newblock
\urldef\tempurl%
\url{https://doi.org/10.1109/DSC.2016.33}
\showDOI{\tempurl}


\bibitem[\protect\citeauthoryear{Jimenez, Rwemalika, Papadakis, Sarro,
  Le~Traon, and Harman}{Jimenez et~al\mbox{.}}{2019}]%
        {Jimenez2019}
\bibfield{author}{\bibinfo{person}{Matthieu Jimenez}, \bibinfo{person}{Renaud
  Rwemalika}, \bibinfo{person}{Mike Papadakis}, \bibinfo{person}{Federica
  Sarro}, \bibinfo{person}{Yves Le~Traon}, {and} \bibinfo{person}{Mark
  Harman}.} \bibinfo{year}{2019}\natexlab{}.
\newblock \showarticletitle{The Importance of Accounting for Real-World
  Labelling When Predicting Software Vulnerabilities}. In
  \bibinfo{booktitle}{\emph{Proceedings of the 2019 27th ACM Joint Meeting on
  European Software Engineering Conference and Symposium on the Foundations of
  Software Engineering}} (Tallinn, Estonia) \emph{(\bibinfo{series}{ESEC/FSE
  2019})}. \bibinfo{publisher}{Association for Computing Machinery},
  \bibinfo{address}{New York, NY, USA}, \bibinfo{pages}{695–705}.
\newblock
\showISBNx{9781450355728}
\urldef\tempurl%
\url{https://doi.org/10.1145/3338906.3338941}
\showDOI{\tempurl}


\bibitem[\protect\citeauthoryear{Jing, Ying, Zhang, Wu, and Liu}{Jing
  et~al\mbox{.}}{2014}]%
        {656_Jing2014}
\bibfield{author}{\bibinfo{person}{Xiao-Yuan Jing}, \bibinfo{person}{Shi Ying},
  \bibinfo{person}{Zhi-Wu Zhang}, \bibinfo{person}{Shan-Shan Wu}, {and}
  \bibinfo{person}{Jin Liu}.} \bibinfo{year}{2014}\natexlab{}.
\newblock \showarticletitle{Dictionary learning based software defect
  prediction}. In \bibinfo{booktitle}{\emph{Proceedings of the 36th
  international conference on software engineering}}.
  \bibinfo{pages}{414--423}.
\newblock


\bibitem[\protect\citeauthoryear{Just, Jalali, and Ernst}{Just
  et~al\mbox{.}}{2014}]%
        {Defects4J}
\bibfield{author}{\bibinfo{person}{Ren\'{e} Just}, \bibinfo{person}{Darioush
  Jalali}, {and} \bibinfo{person}{Michael~D. Ernst}.}
  \bibinfo{year}{2014}\natexlab{}.
\newblock \showarticletitle{Defects4J: A Database of Existing Faults to Enable
  Controlled Testing Studies for Java Programs}. In
  \bibinfo{booktitle}{\emph{Proceedings of the 2014 International Symposium on
  Software Testing and Analysis}} (San Jose, CA, USA)
  \emph{(\bibinfo{series}{ISSTA 2014})}. \bibinfo{publisher}{Association for
  Computing Machinery}, \bibinfo{address}{New York, NY, USA},
  \bibinfo{pages}{437–440}.
\newblock
\showISBNx{9781450326452}
\urldef\tempurl%
\url{https://doi.org/10.1145/2610384.2628055}
\showDOI{\tempurl}


\bibitem[\protect\citeauthoryear{Kanade, Maniatis, Balakrishnan, and
  Shi}{Kanade et~al\mbox{.}}{2020}]%
        {1087_Kanade2020}
\bibfield{author}{\bibinfo{person}{Aditya Kanade}, \bibinfo{person}{Petros
  Maniatis}, \bibinfo{person}{Gogul Balakrishnan}, {and}
  \bibinfo{person}{Kensen Shi}.} \bibinfo{year}{2020}\natexlab{}.
\newblock \showarticletitle{Learning and Evaluating Contextual Embedding of
  Source Code}. In \bibinfo{booktitle}{\emph{Proceedings of the 37th
  International Conference on Machine Learning}}
  \emph{(\bibinfo{series}{Proceedings of Machine Learning Research},
  Vol.~\bibinfo{volume}{119})}, \bibfield{editor}{\bibinfo{person}{Hal~Daumé
  III} {and} \bibinfo{person}{Aarti Singh}} (Eds.). \bibinfo{publisher}{PMLR},
  \bibinfo{pages}{5110--5121}.
\newblock
\urldef\tempurl%
\url{https://proceedings.mlr.press/v119/kanade20a.html}
\showURL{%
\tempurl}


\bibitem[\protect\citeauthoryear{Kang, Bissyandé, and Lo}{Kang
  et~al\mbox{.}}{2019}]%
        {1084_Kang2019}
\bibfield{author}{\bibinfo{person}{Hong~Jin Kang},
  \bibinfo{person}{Tegawendé~F. Bissyandé}, {and} \bibinfo{person}{David
  Lo}.} \bibinfo{year}{2019}\natexlab{}.
\newblock \showarticletitle{Assessing the Generalizability of Code2vec Token
  Embeddings}. In \bibinfo{booktitle}{\emph{2019 34th IEEE/ACM International
  Conference on Automated Software Engineering (ASE)}}. \bibinfo{pages}{1--12}.
\newblock
\urldef\tempurl%
\url{https://doi.org/10.1109/ASE.2019.00011}
\showDOI{\tempurl}


\bibitem[\protect\citeauthoryear{Karampatsis, Babii, Robbes, Sutton, and
  Janes}{Karampatsis et~al\mbox{.}}{2020}]%
        {Karampatsis2020_456}
\bibfield{author}{\bibinfo{person}{Rafael-Michael Karampatsis},
  \bibinfo{person}{Hlib Babii}, \bibinfo{person}{Romain Robbes},
  \bibinfo{person}{Charles Sutton}, {and} \bibinfo{person}{Andrea Janes}.}
  \bibinfo{year}{2020}\natexlab{}.
\newblock \showarticletitle{Big Code != Big Vocabulary: Open-Vocabulary Models
  for Source Code}. In \bibinfo{booktitle}{\emph{Proceedings of the ACM/IEEE
  42nd International Conference on Software Engineering}} (Seoul, South Korea)
  \emph{(\bibinfo{series}{ICSE '20})}. \bibinfo{pages}{1073–1085}.
\newblock
\showISBNx{9781450371216}
\urldef\tempurl%
\url{https://doi.org/10.1145/3377811.3380342}
\showDOI{\tempurl}


\bibitem[\protect\citeauthoryear{{Kaur}, {Jain}, and {Goel}}{{Kaur}
  et~al\mbox{.}}{2017}]%
        {Kaur2017_184}
\bibfield{author}{\bibinfo{person}{A. {Kaur}}, \bibinfo{person}{S. {Jain}},
  {and} \bibinfo{person}{S. {Goel}}.} \bibinfo{year}{2017}\natexlab{}.
\newblock \showarticletitle{A Support Vector Machine Based Approach for Code
  Smell Detection}. In \bibinfo{booktitle}{\emph{2017 International Conference
  on Machine Learning and Data Science (MLDS)}}. \bibinfo{pages}{9--14}.
\newblock
\urldef\tempurl%
\url{https://doi.org/10.1109/MLDS.2017.8}
\showDOI{\tempurl}


\bibitem[\protect\citeauthoryear{Kaur and Kaur}{Kaur and Kaur}{2015}]%
        {653_Kaur2015}
\bibfield{author}{\bibinfo{person}{Arvinder Kaur} {and}
  \bibinfo{person}{Kamaldeep Kaur}.} \bibinfo{year}{2015}\natexlab{}.
\newblock \showarticletitle{An empirical study of robustness and stability of
  machine learning classifiers in software defect prediction}.
\newblock In \bibinfo{booktitle}{\emph{Advances in intelligent informatics}}.
  \bibinfo{publisher}{Springer}, \bibinfo{pages}{383--397}.
\newblock


\bibitem[\protect\citeauthoryear{Kaur, Kaur, and Chopra}{Kaur
  et~al\mbox{.}}{2017}]%
        {Kaur2017_71}
\bibfield{author}{\bibinfo{person}{Arvinder Kaur}, \bibinfo{person}{Kamaldeep
  Kaur}, {and} \bibinfo{person}{Deepti Chopra}.}
  \bibinfo{year}{2017}\natexlab{}.
\newblock \showarticletitle{An empirical study of software entropy based bug
  prediction using machine learning}.
\newblock \bibinfo{journal}{\emph{International Journal of System Assurance
  Engineering and Management}} \bibinfo{volume}{8}, \bibinfo{number}{2}
  (\bibinfo{date}{November} \bibinfo{year}{2017}), \bibinfo{pages}{599--616}.
\newblock
\showISSN{0976-4348}
\urldef\tempurl%
\url{https://doi.org/10.1007/s13198-016-0479-2}
\showDOI{\tempurl}


\bibitem[\protect\citeauthoryear{Kaur and Kaur}{Kaur and Kaur}{2021}]%
        {814_Kaur2021}
\bibfield{author}{\bibinfo{person}{Inderpreet Kaur} {and}
  \bibinfo{person}{Arvinder Kaur}.} \bibinfo{year}{2021}\natexlab{}.
\newblock \showarticletitle{A novel four-way approach designed with ensemble
  feature selection for code smell detection}.
\newblock \bibinfo{journal}{\emph{IEEE Access}}  \bibinfo{volume}{9}
  (\bibinfo{year}{2021}), \bibinfo{pages}{8695--8707}.
\newblock


\bibitem[\protect\citeauthoryear{Khalid, Farooq, Iqbal, Alam, and
  Rasheed}{Khalid et~al\mbox{.}}{2019}]%
        {khalid_predicting_2019}
\bibfield{author}{\bibinfo{person}{Muhammad~Noman Khalid},
  \bibinfo{person}{Humera Farooq}, \bibinfo{person}{Muhammad Iqbal},
  \bibinfo{person}{Muhammad~Talha Alam}, {and} \bibinfo{person}{Kamran
  Rasheed}.} \bibinfo{year}{2019}\natexlab{}.
\newblock \showarticletitle{Predicting {Web} {Vulnerabilities} in {Web}
  {Applications} {Based} on {Machine} {Learning}}. In
  \bibinfo{booktitle}{\emph{Intelligent {Technologies} and {Applications}}}
  \emph{(\bibinfo{series}{Communications in {Computer} and {Information}
  {Science}})}, \bibfield{editor}{\bibinfo{person}{Imran~Sarwar Bajwa},
  \bibinfo{person}{Fairouz Kamareddine}, {and} \bibinfo{person}{Anna Costa}}
  (Eds.). \bibinfo{publisher}{Springer}, \bibinfo{address}{Singapore},
  \bibinfo{pages}{473--484}.
\newblock
\showISBNx{9789811360527}
\urldef\tempurl%
\url{https://doi.org/10.1007/978-981-13-6052-7_41}
\showDOI{\tempurl}


\bibitem[\protect\citeauthoryear{Khan, Iqbal, and Badshah}{Khan
  et~al\mbox{.}}{2020}]%
        {Khan2020_82}
\bibfield{author}{\bibinfo{person}{Bilal Khan}, \bibinfo{person}{Danish Iqbal},
  {and} \bibinfo{person}{Sher Badshah}.} \bibinfo{year}{2020}\natexlab{}.
\newblock \showarticletitle{Cross-Project Software Fault Prediction Using Data
  Leveraging Technique to Improve Software Quality}. In
  \bibinfo{booktitle}{\emph{Proceedings of the Evaluation and Assessment in
  Software Engineering}} (Trondheim, Norway) \emph{(\bibinfo{series}{EASE
  '20})}. \bibinfo{pages}{434–438}.
\newblock
\showISBNx{9781450377317}
\urldef\tempurl%
\url{https://doi.org/10.1145/3383219.3383281}
\showDOI{\tempurl}


\bibitem[\protect\citeauthoryear{Kim, Hubczenko, and Montague}{Kim
  et~al\mbox{.}}{2019}]%
        {Kim2019_59}
\bibfield{author}{\bibinfo{person}{Junae Kim}, \bibinfo{person}{David
  Hubczenko}, {and} \bibinfo{person}{Paul Montague}.}
  \bibinfo{year}{2019}\natexlab{}.
\newblock \showarticletitle{Towards Attention Based Vulnerability Discovery
  Using Source Code Representation}. In \bibinfo{booktitle}{\emph{Artificial
  Neural Networks and Machine Learning -- ICANN 2019: Text and Time Series}},
  \bibfield{editor}{\bibinfo{person}{Igor~V. Tetko},
  \bibinfo{person}{V{\v{e}}ra K{\r{u}}rkov{\'a}}, \bibinfo{person}{Pavel
  Karpov}, {and} \bibinfo{person}{Fabian Theis}} (Eds.).
  \bibinfo{pages}{731--746}.
\newblock
\showISBNx{978-3-030-30490-4}


\bibitem[\protect\citeauthoryear{{Kim}, {Kwon}, and {Yoo}}{{Kim}
  et~al\mbox{.}}{2018}]%
        {Kim2018_90}
\bibfield{author}{\bibinfo{person}{J. {Kim}}, \bibinfo{person}{M. {Kwon}},
  {and} \bibinfo{person}{S. {Yoo}}.} \bibinfo{year}{2018}\natexlab{}.
\newblock \showarticletitle{Generating Test Input with Deep Reinforcement
  Learning}. In \bibinfo{booktitle}{\emph{2018 IEEE/ACM 11th International
  Workshop on Search-Based Software Testing (SBST)}}. \bibinfo{pages}{51--58}.
\newblock


\bibitem[\protect\citeauthoryear{Kim, Hong, Oh, and Lee}{Kim
  et~al\mbox{.}}{2018}]%
        {kim_hong_oh_lee_2018}
\bibfield{author}{\bibinfo{person}{Sangwoo Kim}, \bibinfo{person}{Seokmyung
  Hong}, \bibinfo{person}{Jaesang Oh}, {and} \bibinfo{person}{Heejo Lee}.}
  \bibinfo{year}{2018}\natexlab{}.
\newblock \showarticletitle{Obfuscated VBA Macro Detection Using Machine
  Learning}. In \bibinfo{booktitle}{\emph{2018 48th Annual IEEE/IFIP
  International Conference on Dependable Systems and Networks (DSN)}}.
  \bibinfo{pages}{490--501}.
\newblock
\urldef\tempurl%
\url{https://doi.org/10.1109/DSN.2018.00057}
\showDOI{\tempurl}


\bibitem[\protect\citeauthoryear{{Kim}, {Jeong}, {Jeong}, and {Kim}}{{Kim}
  et~al\mbox{.}}{2009}]%
        {Kim2009_259}
\bibfield{author}{\bibinfo{person}{Y. {Kim}}, \bibinfo{person}{C. {Jeong}},
  \bibinfo{person}{A. {Jeong}}, {and} \bibinfo{person}{H.~S. {Kim}}.}
  \bibinfo{year}{2009}\natexlab{}.
\newblock \showarticletitle{Risky Module Estimation in Safety-Critical
  Software}. In \bibinfo{booktitle}{\emph{2009 Eighth IEEE/ACIS International
  Conference on Computer and Information Science}}. \bibinfo{pages}{967--970}.
\newblock
\urldef\tempurl%
\url{https://doi.org/10.1109/ICIS.2009.83}
\showDOI{\tempurl}


\bibitem[\protect\citeauthoryear{Knab, Pinzger, and Bernstein}{Knab
  et~al\mbox{.}}{2006}]%
        {Knab2006_118}
\bibfield{author}{\bibinfo{person}{Patrick Knab}, \bibinfo{person}{Martin
  Pinzger}, {and} \bibinfo{person}{Abraham Bernstein}.}
  \bibinfo{year}{2006}\natexlab{}.
\newblock \showarticletitle{Predicting Defect Densities in Source Code Files
  with Decision Tree Learners}. In \bibinfo{booktitle}{\emph{Proceedings of the
  2006 International Workshop on Mining Software Repositories}} (Shanghai,
  China) \emph{(\bibinfo{series}{MSR '06})}. \bibinfo{pages}{119–125}.
\newblock
\showISBNx{1595933972}
\urldef\tempurl%
\url{https://doi.org/10.1145/1137983.1138012}
\showDOI{\tempurl}


\bibitem[\protect\citeauthoryear{Kosker, Turhan, and Bener}{Kosker
  et~al\mbox{.}}{2009}]%
        {Kosker2009_161}
\bibfield{author}{\bibinfo{person}{Yasemin Kosker}, \bibinfo{person}{Burak
  Turhan}, {and} \bibinfo{person}{Ayse Bener}.}
  \bibinfo{year}{2009}\natexlab{}.
\newblock \showarticletitle{An expert system for determining candidate software
  classes for refactoring}.
\newblock \bibinfo{journal}{\emph{Expert Systems with Applications}}
  \bibinfo{volume}{36}, \bibinfo{number}{6} (\bibinfo{year}{2009}),
  \bibinfo{pages}{10000 -- 10003}.
\newblock
\showISSN{0957-4174}
\urldef\tempurl%
\url{https://doi.org/10.1016/j.eswa.2008.12.066}
\showDOI{\tempurl}


\bibitem[\protect\citeauthoryear{Kovalenko, Bogomolov, Bryksin, and
  Bacchelli}{Kovalenko et~al\mbox{.}}{2020}]%
        {Kovalenko2020_458}
\bibfield{author}{\bibinfo{person}{Vladimir Kovalenko}, \bibinfo{person}{Egor
  Bogomolov}, \bibinfo{person}{Timofey Bryksin}, {and} \bibinfo{person}{Alberto
  Bacchelli}.} \bibinfo{year}{2020}\natexlab{}.
\newblock \showarticletitle{Building Implicit Vector Representations of
  Individual Coding Style}. In \bibinfo{booktitle}{\emph{Proceedings of the
  IEEE/ACM 42nd International Conference on Software Engineering Workshops}}
  (Seoul, Republic of Korea) \emph{(\bibinfo{series}{ICSEW'20})}.
  \bibinfo{pages}{117–124}.
\newblock
\showISBNx{9781450379632}
\urldef\tempurl%
\url{https://doi.org/10.1145/3387940.3391494}
\showDOI{\tempurl}


\bibitem[\protect\citeauthoryear{Krasniqi and Cleland-Huang}{Krasniqi and
  Cleland-Huang}{2020}]%
        {770_Krasniqi2020}
\bibfield{author}{\bibinfo{person}{Rrezarta Krasniqi} {and}
  \bibinfo{person}{Jane Cleland-Huang}.} \bibinfo{year}{2020}\natexlab{}.
\newblock \showarticletitle{Enhancing Source Code Refactoring Detection with
  Explanations from Commit Messages}. In \bibinfo{booktitle}{\emph{2020 IEEE
  27th International Conference on Software Analysis, Evolution and
  Reengineering (SANER)}}. \bibinfo{pages}{512--516}.
\newblock
\urldef\tempurl%
\url{https://doi.org/10.1109/SANER48275.2020.770_Krasniqi2020}
\showDOI{\tempurl}


\bibitem[\protect\citeauthoryear{Krizhevsky, Sutskever, and Hinton}{Krizhevsky
  et~al\mbox{.}}{2012}]%
        {Krizhevsky2012}
\bibfield{author}{\bibinfo{person}{Alex Krizhevsky}, \bibinfo{person}{Ilya
  Sutskever}, {and} \bibinfo{person}{Geoffrey~E Hinton}.}
  \bibinfo{year}{2012}\natexlab{}.
\newblock \showarticletitle{ImageNet classification with deep convolutional
  neural networks}. In \bibinfo{booktitle}{\emph{Advances in neural information
  processing systems}}. \bibinfo{pages}{1097--1105}.
\newblock


\bibitem[\protect\citeauthoryear{Kronjee, Hommersom, and Vranken}{Kronjee
  et~al\mbox{.}}{2018}]%
        {Kronjee2018_26}
\bibfield{author}{\bibinfo{person}{Jorrit Kronjee}, \bibinfo{person}{Arjen
  Hommersom}, {and} \bibinfo{person}{Harald Vranken}.}
  \bibinfo{year}{2018}\natexlab{}.
\newblock \showarticletitle{Discovering Software Vulnerabilities Using
  Data-Flow Analysis and Machine Learning}. In
  \bibinfo{booktitle}{\emph{Proceedings of the 13th International Conference on
  Availability, Reliability and Security}} (Hamburg, Germany)
  \emph{(\bibinfo{series}{ARES 2018})}. Article \bibinfo{articleno}{6},
  \bibinfo{numpages}{10}~pages.
\newblock
\showISBNx{9781450364485}
\urldef\tempurl%
\url{https://doi.org/10.1145/3230833.3230856}
\showDOI{\tempurl}


\bibitem[\protect\citeauthoryear{Kumar, Rath, and Sureka}{Kumar
  et~al\mbox{.}}{2017}]%
        {742_Kumar2017}
\bibfield{author}{\bibinfo{person}{Lov Kumar}, \bibinfo{person}{Santanu~Kumar
  Rath}, {and} \bibinfo{person}{Ashish Sureka}.}
  \bibinfo{year}{2017}\natexlab{}.
\newblock \showarticletitle{Using source code metrics to predict change-prone
  web services: A case-study on ebay services}. In
  \bibinfo{booktitle}{\emph{2017 IEEE workshop on machine learning techniques
  for software quality evaluation (MaLTeSQuE)}}. IEEE, \bibinfo{pages}{1--7}.
\newblock


\bibitem[\protect\citeauthoryear{Kumar, Satapathy, and Murthy}{Kumar
  et~al\mbox{.}}{2019}]%
        {Kumar2019_169}
\bibfield{author}{\bibinfo{person}{Lov Kumar}, \bibinfo{person}{Shashank~Mouli
  Satapathy}, {and} \bibinfo{person}{Lalita~Bhanu Murthy}.}
  \bibinfo{year}{2019}\natexlab{}.
\newblock \showarticletitle{Method Level Refactoring Prediction on Five Open
  Source Java Projects Using Machine Learning Techniques}. In
  \bibinfo{booktitle}{\emph{Proceedings of the 12th Innovations on Software
  Engineering Conference (Formerly Known as India Software Engineering
  Conference)}} (Pune, India) \emph{(\bibinfo{series}{ISEC'19})}. Article
  \bibinfo{articleno}{7}, \bibinfo{numpages}{10}~pages.
\newblock
\showISBNx{9781450362153}
\urldef\tempurl%
\url{https://doi.org/10.1145/3299771.3299777}
\showDOI{\tempurl}


\bibitem[\protect\citeauthoryear{{Kumar} and {Sureka}}{{Kumar} and
  {Sureka}}{2017}]%
        {Kumar2017_162}
\bibfield{author}{\bibinfo{person}{L. {Kumar}} {and} \bibinfo{person}{A.
  {Sureka}}.} \bibinfo{year}{2017}\natexlab{}.
\newblock \showarticletitle{Application of LSSVM and SMOTE on Seven Open Source
  Projects for Predicting Refactoring at Class Level}. In
  \bibinfo{booktitle}{\emph{2017 24th Asia-Pacific Software Engineering
  Conference (APSEC)}}. \bibinfo{pages}{90--99}.
\newblock
\urldef\tempurl%
\url{https://doi.org/10.1109/APSEC.2017.15}
\showDOI{\tempurl}


\bibitem[\protect\citeauthoryear{{Kumar} and {Sureka}}{{Kumar} and
  {Sureka}}{2018}]%
        {Kumar2018_189}
\bibfield{author}{\bibinfo{person}{L. {Kumar}} {and} \bibinfo{person}{A.
  {Sureka}}.} \bibinfo{year}{2018}\natexlab{}.
\newblock \showarticletitle{An Empirical Analysis on Web Service Anti-pattern
  Detection Using a Machine Learning Framework}. In
  \bibinfo{booktitle}{\emph{2018 IEEE 42nd Annual Computer Software and
  Applications Conference (COMPSAC)}}, Vol.~\bibinfo{volume}{01}.
  \bibinfo{pages}{2--11}.
\newblock
\urldef\tempurl%
\url{https://doi.org/10.1109/COMPSAC.2018.00010}
\showDOI{\tempurl}


\bibitem[\protect\citeauthoryear{Kumar and Singh}{Kumar and Singh}{2012}]%
        {Kumar2012_77}
\bibfield{author}{\bibinfo{person}{Pradeep Kumar} {and} \bibinfo{person}{Yogesh
  Singh}.} \bibinfo{year}{2012}\natexlab{}.
\newblock \showarticletitle{Assessment of Software Testing Time Using Soft
  Computing Techniques}.
\newblock \bibinfo{journal}{\emph{SIGSOFT Softw. Eng. Notes}}
  \bibinfo{volume}{37}, \bibinfo{number}{1} (\bibinfo{date}{January}
  \bibinfo{year}{2012}), \bibinfo{pages}{1–6}.
\newblock
\showISSN{0163-5948}
\urldef\tempurl%
\url{https://doi.org/10.1145/2088883.2088895}
\showDOI{\tempurl}


\bibitem[\protect\citeauthoryear{Kurbatova, Veselov, Golubev, and
  Bryksin}{Kurbatova et~al\mbox{.}}{2020}]%
        {Kurbatova2020_172}
\bibfield{author}{\bibinfo{person}{Zarina Kurbatova}, \bibinfo{person}{Ivan
  Veselov}, \bibinfo{person}{Yaroslav Golubev}, {and} \bibinfo{person}{Timofey
  Bryksin}.} \bibinfo{year}{2020}\natexlab{}.
\newblock \showarticletitle{Recommendation of Move Method Refactoring Using
  Path-Based Representation of Code}. In \bibinfo{booktitle}{\emph{Proceedings
  of the IEEE/ACM 42nd International Conference on Software Engineering
  Workshops}} (Seoul, Republic of Korea) \emph{(\bibinfo{series}{ICSEW'20})}.
  \bibinfo{pages}{315–322}.
\newblock
\showISBNx{9781450379632}
\urldef\tempurl%
\url{https://doi.org/10.1145/3387940.3392191}
\showDOI{\tempurl}


\bibitem[\protect\citeauthoryear{{Lal} and {Pahwa}}{{Lal} and {Pahwa}}{2017}]%
        {Lal2017_446}
\bibfield{author}{\bibinfo{person}{H. {Lal}} {and} \bibinfo{person}{G.
  {Pahwa}}.} \bibinfo{year}{2017}\natexlab{}.
\newblock \showarticletitle{Code review analysis of software system using
  machine learning techniques}. In \bibinfo{booktitle}{\emph{2017 11th
  International Conference on Intelligent Systems and Control (ISCO)}}.
  \bibinfo{pages}{8--13}.
\newblock
\urldef\tempurl%
\url{https://doi.org/10.1109/ISCO.2017.7855962}
\showDOI{\tempurl}


\bibitem[\protect\citeauthoryear{Lal and Pahwa}{Lal and Pahwa}{2017}]%
        {1060_Lal2017}
\bibfield{author}{\bibinfo{person}{Harsh Lal} {and} \bibinfo{person}{Gaurav
  Pahwa}.} \bibinfo{year}{2017}\natexlab{}.
\newblock \showarticletitle{Root cause analysis of software bugs using machine
  learning techniques}. In \bibinfo{booktitle}{\emph{2017 7th International
  Conference on Cloud Computing, Data Science \& Engineering-Confluence}}.
  IEEE, \bibinfo{pages}{105--111}.
\newblock


\bibitem[\protect\citeauthoryear{Laradji, Alshayeb, and Ghouti}{Laradji
  et~al\mbox{.}}{2015}]%
        {663_Laradji2015}
\bibfield{author}{\bibinfo{person}{Issam~H Laradji}, \bibinfo{person}{Mohammad
  Alshayeb}, {and} \bibinfo{person}{Lahouari Ghouti}.}
  \bibinfo{year}{2015}\natexlab{}.
\newblock \showarticletitle{Software defect prediction using ensemble learning
  on selected features}.
\newblock \bibinfo{journal}{\emph{Information and Software Technology}}
  \bibinfo{volume}{58} (\bibinfo{year}{2015}), \bibinfo{pages}{388--402}.
\newblock


\bibitem[\protect\citeauthoryear{Law and Grépin}{Law and Grépin}{2010}]%
        {law_is_2010}
\bibfield{author}{\bibinfo{person}{Michael~R. Law} {and}
  \bibinfo{person}{Karen~A. Grépin}.} \bibinfo{year}{2010}\natexlab{}.
\newblock \showarticletitle{Is newer always better? {Re}-evaluating the
  benefits of newer pharmaceuticals}.
\newblock \bibinfo{journal}{\emph{Journal of Health Economics}}
  \bibinfo{volume}{29}, \bibinfo{number}{5} (\bibinfo{date}{Sept.}
  \bibinfo{year}{2010}), \bibinfo{pages}{743--750}.
\newblock
\showISSN{1879-1646}
\urldef\tempurl%
\url{https://doi.org/10.1016/j.jhealeco.2010.06.007}
\showDOI{\tempurl}


\bibitem[\protect\citeauthoryear{Le, Chen, and Babar}{Le et~al\mbox{.}}{2020}]%
        {Le2020_297}
\bibfield{author}{\bibinfo{person}{Triet H.~M. Le}, \bibinfo{person}{Hao Chen},
  {and} \bibinfo{person}{Muhammad~Ali Babar}.} \bibinfo{year}{2020}\natexlab{}.
\newblock \showarticletitle{Deep Learning for Source Code Modeling and
  Generation: Models, Applications, and Challenges}.
\newblock \bibinfo{journal}{\emph{ACM Comput. Surv.}} \bibinfo{volume}{53},
  \bibinfo{number}{3}, Article \bibinfo{articleno}{62} (\bibinfo{date}{June}
  \bibinfo{year}{2020}), \bibinfo{numpages}{38}~pages.
\newblock
\showISSN{0360-0300}
\urldef\tempurl%
\url{https://doi.org/10.1145/3383458}
\showDOI{\tempurl}


\bibitem[\protect\citeauthoryear{{Le}, {Le}, and {Lo}}{{Le}
  et~al\mbox{.}}{2015}]%
        {Le2015_352}
\bibfield{author}{\bibinfo{person}{X.~D. {Le}}, \bibinfo{person}{T.~B. {Le}},
  {and} \bibinfo{person}{D. {Lo}}.} \bibinfo{year}{2015}\natexlab{}.
\newblock \showarticletitle{Should fixing these failures be delegated to
  automated program repair?}. In \bibinfo{booktitle}{\emph{2015 IEEE 26th
  International Symposium on Software Reliability Engineering (ISSRE)}}.
  \bibinfo{pages}{427--437}.
\newblock
\urldef\tempurl%
\url{https://doi.org/10.1109/ISSRE.2015.7381836}
\showDOI{\tempurl}


\bibitem[\protect\citeauthoryear{Le~Goues, Holtschulte, Smith, Brun, Devanbu,
  Forrest, and Weimer}{Le~Goues et~al\mbox{.}}{2015}]%
        {LeGoues15}
\bibfield{author}{\bibinfo{person}{Claire Le~Goues}, \bibinfo{person}{Neal
  Holtschulte}, \bibinfo{person}{Edward~K. Smith}, \bibinfo{person}{Yuriy
  Brun}, \bibinfo{person}{Premkumar Devanbu}, \bibinfo{person}{Stephanie
  Forrest}, {and} \bibinfo{person}{Westley Weimer}.}
  \bibinfo{year}{2015}\natexlab{}.
\newblock \showarticletitle{The ManyBugs and IntroClass Benchmarks for
  Automated Repair of C Programs}.
\newblock \bibinfo{journal}{\emph{IEEE Transactions on Software Engineering}}
  \bibinfo{volume}{41}, \bibinfo{number}{12} (\bibinfo{year}{2015}),
  \bibinfo{pages}{1236--1256}.
\newblock
\urldef\tempurl%
\url{https://doi.org/10.1109/TSE.2015.2454513}
\showDOI{\tempurl}


\bibitem[\protect\citeauthoryear{LeClair, Bansal, and McMillan}{LeClair
  et~al\mbox{.}}{2021}]%
        {994_LeClair2021}
\bibfield{author}{\bibinfo{person}{Alexander LeClair}, \bibinfo{person}{Aakash
  Bansal}, {and} \bibinfo{person}{Collin McMillan}.}
  \bibinfo{year}{2021}\natexlab{}.
\newblock \showarticletitle{Ensemble Models for Neural Source Code
  Summarization of Subroutines}. In \bibinfo{booktitle}{\emph{2021 IEEE
  International Conference on Software Maintenance and Evolution (ICSME)}}.
  IEEE, \bibinfo{pages}{286--297}.
\newblock


\bibitem[\protect\citeauthoryear{LeClair, Haque, Wu, and McMillan}{LeClair
  et~al\mbox{.}}{2020}]%
        {LeClair2020_426}
\bibfield{author}{\bibinfo{person}{Alexander LeClair}, \bibinfo{person}{Sakib
  Haque}, \bibinfo{person}{Lingfei Wu}, {and} \bibinfo{person}{Collin
  McMillan}.} \bibinfo{year}{2020}\natexlab{}.
\newblock \showarticletitle{Improved Code Summarization via a Graph Neural
  Network}. In \bibinfo{booktitle}{\emph{Proceedings of the 28th International
  Conference on Program Comprehension}} (Seoul, Republic of Korea)
  \emph{(\bibinfo{series}{ICPC '20})}. \bibinfo{pages}{184–195}.
\newblock
\showISBNx{9781450379588}
\urldef\tempurl%
\url{https://doi.org/10.1145/3387904.3389268}
\showDOI{\tempurl}


\bibitem[\protect\citeauthoryear{LeClair, Jiang, and McMillan}{LeClair
  et~al\mbox{.}}{2019}]%
        {LeClair2019_407}
\bibfield{author}{\bibinfo{person}{Alexander LeClair}, \bibinfo{person}{Siyuan
  Jiang}, {and} \bibinfo{person}{Collin McMillan}.}
  \bibinfo{year}{2019}\natexlab{}.
\newblock \showarticletitle{A Neural Model for Generating Natural Language
  Summaries of Program Subroutines}. In \bibinfo{booktitle}{\emph{Proceedings
  of the 41st International Conference on Software Engineering}} (Montreal,
  Quebec, Canada) \emph{(\bibinfo{series}{ICSE '19})}.
  \bibinfo{pages}{795–806}.
\newblock
\urldef\tempurl%
\url{https://doi.org/10.1109/ICSE.2019.00087}
\showDOI{\tempurl}


\bibitem[\protect\citeauthoryear{LeClair and McMillan}{LeClair and
  McMillan}{2019}]%
        {Leclair2019}
\bibfield{author}{\bibinfo{person}{Alexander LeClair} {and}
  \bibinfo{person}{Collin McMillan}.} \bibinfo{year}{2019}\natexlab{}.
\newblock \bibinfo{title}{Recommendations for datasets for source code
  summarization}.
\newblock
\newblock


\bibitem[\protect\citeauthoryear{Lee, Lee, Lee, and Woo}{Lee
  et~al\mbox{.}}{2021}]%
        {Lee2021_495}
\bibfield{author}{\bibinfo{person}{Suin Lee}, \bibinfo{person}{Youngseok Lee},
  \bibinfo{person}{Chan-Gun Lee}, {and} \bibinfo{person}{Honguk Woo}.}
  \bibinfo{year}{2021}\natexlab{}.
\newblock \showarticletitle{Deep Learning-Based Logging Recommendation Using
  Merged Code Representation}. In \bibinfo{booktitle}{\emph{IT Convergence and
  Security}}, \bibfield{editor}{\bibinfo{person}{Hyuncheol Kim} {and}
  \bibinfo{person}{Kuinam~J. Kim}} (Eds.). \bibinfo{pages}{49--53}.
\newblock
\showISBNx{978-981-15-9354-3}


\bibitem[\protect\citeauthoryear{Lee, Yoon, and Cho}{Lee et~al\mbox{.}}{2017}]%
        {Lee2017}
\bibfield{author}{\bibinfo{person}{Song-Mi Lee}, \bibinfo{person}{Sang~Min
  Yoon}, {and} \bibinfo{person}{Heeryon Cho}.} \bibinfo{year}{2017}\natexlab{}.
\newblock \showarticletitle{Human activity recognition from accelerometer data
  using Convolutional Neural Network}. In \bibinfo{booktitle}{\emph{Big Data
  and Smart Computing (BigComp), 2017 IEEE International Conference on}}. IEEE,
  \bibinfo{pages}{131--134}.
\newblock


\bibitem[\protect\citeauthoryear{Levin and Yehudai}{Levin and Yehudai}{2017}]%
        {965_Levin2017}
\bibfield{author}{\bibinfo{person}{Stanislav Levin} {and}
  \bibinfo{person}{Amiram Yehudai}.} \bibinfo{year}{2017}\natexlab{}.
\newblock \showarticletitle{Boosting automatic commit classification into
  maintenance activities by utilizing source code changes}. In
  \bibinfo{booktitle}{\emph{Proceedings of the 13th International Conference on
  Predictive Models and Data Analytics in Software Engineering}}.
  \bibinfo{pages}{97--106}.
\newblock


\bibitem[\protect\citeauthoryear{Lewowski and Madeyski}{Lewowski and
  Madeyski}{2022}]%
        {828_Lewowski2022}
\bibfield{author}{\bibinfo{person}{Tomasz Lewowski} {and} \bibinfo{person}{Lech
  Madeyski}.} \bibinfo{year}{2022}\natexlab{}.
\newblock \showarticletitle{Code smells detection using artificial intelligence
  techniques: A business-driven systematic review}.
\newblock \bibinfo{journal}{\emph{Developments in Information I\& Knowledge
  Management for Business Applications}} (\bibinfo{year}{2022}),
  \bibinfo{pages}{285--319}.
\newblock


\bibitem[\protect\citeauthoryear{Li, Yan, Xia, Hu, Li, and Lo}{Li
  et~al\mbox{.}}{2020b}]%
        {Li2020_422}
\bibfield{author}{\bibinfo{person}{Boao Li}, \bibinfo{person}{Meng Yan},
  \bibinfo{person}{Xin Xia}, \bibinfo{person}{Xing Hu}, \bibinfo{person}{Ge
  Li}, {and} \bibinfo{person}{David Lo}.} \bibinfo{year}{2020}\natexlab{b}.
\newblock \showarticletitle{DeepCommenter: A Deep Code Comment Generation Tool
  with Hybrid Lexical and Syntactical Information}. In
  \bibinfo{booktitle}{\emph{Proceedings of the 28th ACM Joint Meeting on
  European Software Engineering Conference and Symposium on the Foundations of
  Software Engineering}} (Virtual Event, USA) \emph{(\bibinfo{series}{ESEC/FSE
  2020})}. \bibinfo{pages}{1571–1575}.
\newblock
\showISBNx{9781450370431}
\urldef\tempurl%
\url{https://doi.org/10.1145/3368089.3417926}
\showDOI{\tempurl}


\bibitem[\protect\citeauthoryear{Li, Li, Kim, Bissyand{\'e}, Lo, and
  Le~Traon}{Li et~al\mbox{.}}{2019a}]%
        {967_Li2019}
\bibfield{author}{\bibinfo{person}{Daoyuan Li}, \bibinfo{person}{Li Li},
  \bibinfo{person}{Dongsun Kim}, \bibinfo{person}{Tegawend{\'e}~F
  Bissyand{\'e}}, \bibinfo{person}{David Lo}, {and} \bibinfo{person}{Yves
  Le~Traon}.} \bibinfo{year}{2019}\natexlab{a}.
\newblock \showarticletitle{Watch out for this commit! a study of influential
  software changes}.
\newblock \bibinfo{journal}{\emph{Journal of Software: Evolution and Process}}
  \bibinfo{volume}{31}, \bibinfo{number}{12} (\bibinfo{year}{2019}),
  \bibinfo{pages}{e2181}.
\newblock


\bibitem[\protect\citeauthoryear{Li, He, Zhu, and Lyu}{Li
  et~al\mbox{.}}{2017}]%
        {652_Li2017}
\bibfield{author}{\bibinfo{person}{Jian Li}, \bibinfo{person}{Pinjia He},
  \bibinfo{person}{Jieming Zhu}, {and} \bibinfo{person}{Michael~R Lyu}.}
  \bibinfo{year}{2017}\natexlab{}.
\newblock \showarticletitle{Software defect prediction via convolutional neural
  network}. In \bibinfo{booktitle}{\emph{2017 IEEE International Conference on
  Software Quality, Reliability and Security (QRS)}}. IEEE,
  \bibinfo{pages}{318--328}.
\newblock


\bibitem[\protect\citeauthoryear{Li, Li, Li, Hu, Xia, and Jin}{Li
  et~al\mbox{.}}{2021a}]%
        {999_Li2021}
\bibfield{author}{\bibinfo{person}{Jia Li}, \bibinfo{person}{Yongmin Li},
  \bibinfo{person}{Ge Li}, \bibinfo{person}{Xing Hu}, \bibinfo{person}{Xin
  Xia}, {and} \bibinfo{person}{Zhi Jin}.} \bibinfo{year}{2021}\natexlab{a}.
\newblock \showarticletitle{EditSum: A Retrieve-and-Edit Framework for Source
  Code Summarization}. In \bibinfo{booktitle}{\emph{2021 36th IEEE/ACM
  International Conference on Automated Software Engineering (ASE)}}. IEEE,
  \bibinfo{pages}{155--166}.
\newblock


\bibitem[\protect\citeauthoryear{Li, Wang, Lyu, and King}{Li
  et~al\mbox{.}}{2018}]%
        {Li2018_491}
\bibfield{author}{\bibinfo{person}{Jian Li}, \bibinfo{person}{Yue Wang},
  \bibinfo{person}{Michael~R. Lyu}, {and} \bibinfo{person}{Irwin King}.}
  \bibinfo{year}{2018}\natexlab{}.
\newblock \showarticletitle{Code Completion with Neural Attention and Pointer
  Networks}. In \bibinfo{booktitle}{\emph{Proceedings of the 27th International
  Joint Conference on Artificial Intelligence}} (Stockholm, Sweden)
  \emph{(\bibinfo{series}{IJCAI'18})}. \bibinfo{pages}{4159–25}.
\newblock
\showISBNx{9780999241127}


\bibitem[\protect\citeauthoryear{Li, Zhang, Wu, and Zhou}{Li
  et~al\mbox{.}}{2011}]%
        {Li2011_124}
\bibfield{author}{\bibinfo{person}{M. Li}, \bibinfo{person}{H. Zhang},
  \bibinfo{person}{Rongxin Wu}, {and} \bibinfo{person}{Z. Zhou}.}
  \bibinfo{year}{2011}\natexlab{}.
\newblock \showarticletitle{Sample-based software defect prediction with active
  and semi-supervised learning}.
\newblock \bibinfo{journal}{\emph{Automated Software Engineering}}
  \bibinfo{volume}{19} (\bibinfo{year}{2011}), \bibinfo{pages}{201--230}.
\newblock


\bibitem[\protect\citeauthoryear{Li, Ma, and Jiao}{Li et~al\mbox{.}}{2015}]%
        {Li2015AHM}
\bibfield{author}{\bibinfo{person}{Yuancheng Li}, \bibinfo{person}{Rong Ma},
  {and} \bibinfo{person}{Runhai Jiao}.} \bibinfo{year}{2015}\natexlab{}.
\newblock \showarticletitle{A Hybrid Malicious Code Detection Method based on
  Deep Learning}.
\newblock \bibinfo{journal}{\emph{International journal of security and its
  applications}}  \bibinfo{volume}{9} (\bibinfo{year}{2015}),
  \bibinfo{pages}{205--216}.
\newblock


\bibitem[\protect\citeauthoryear{Li, Wang, and Nguyen}{Li
  et~al\mbox{.}}{2020a}]%
        {Li2020_303}
\bibfield{author}{\bibinfo{person}{Yi Li}, \bibinfo{person}{Shaohua Wang},
  {and} \bibinfo{person}{Tien~N. Nguyen}.} \bibinfo{year}{2020}\natexlab{a}.
\newblock \showarticletitle{DLFix: Context-Based Code Transformation Learning
  for Automated Program Repair}. In \bibinfo{booktitle}{\emph{Proceedings of
  the ACM/IEEE 42nd International Conference on Software Engineering}} (Seoul,
  South Korea) \emph{(\bibinfo{series}{ICSE '20})}. \bibinfo{pages}{602–614}.
\newblock
\showISBNx{9781450371216}
\urldef\tempurl%
\url{https://doi.org/10.1145/3377811.3380345}
\showDOI{\tempurl}


\bibitem[\protect\citeauthoryear{Li, Wang, and Nguyen}{Li
  et~al\mbox{.}}{2021b}]%
        {762_Li2021}
\bibfield{author}{\bibinfo{person}{Yi Li}, \bibinfo{person}{Shaohua Wang},
  {and} \bibinfo{person}{Tien~N Nguyen}.} \bibinfo{year}{2021}\natexlab{b}.
\newblock \showarticletitle{A Context-based Automated Approach for Method Name
  Consistency Checking and Suggestion}. In \bibinfo{booktitle}{\emph{2021
  IEEE/ACM 43rd International Conference on Software Engineering (ICSE)}}.
  IEEE, \bibinfo{pages}{574--586}.
\newblock


\bibitem[\protect\citeauthoryear{Li, Wang, Nguyen, and Van~Nguyen}{Li
  et~al\mbox{.}}{2019b}]%
        {Li2019_96}
\bibfield{author}{\bibinfo{person}{Yi Li}, \bibinfo{person}{Shaohua Wang},
  \bibinfo{person}{Tien~N. Nguyen}, {and} \bibinfo{person}{Son Van~Nguyen}.}
  \bibinfo{year}{2019}\natexlab{b}.
\newblock \showarticletitle{Improving Bug Detection via Context-Based Code
  Representation Learning and Attention-Based Neural Networks}.
\newblock \bibinfo{journal}{\emph{Proc. ACM Program. Lang.}}
  \bibinfo{volume}{3}, \bibinfo{number}{OOPSLA}, Article
  \bibinfo{articleno}{162} (\bibinfo{date}{October} \bibinfo{year}{2019}),
  \bibinfo{numpages}{30}~pages.
\newblock
\urldef\tempurl%
\url{https://doi.org/10.1145/3360588}
\showDOI{\tempurl}


\bibitem[\protect\citeauthoryear{{Li}, {Zou}, {Tang}, {Zhang}, {Sun}, and
  {Jin}}{{Li} et~al\mbox{.}}{2019}]%
        {Li2019_4}
\bibfield{author}{\bibinfo{person}{Z. {Li}}, \bibinfo{person}{D. {Zou}},
  \bibinfo{person}{J. {Tang}}, \bibinfo{person}{Z. {Zhang}},
  \bibinfo{person}{M. {Sun}}, {and} \bibinfo{person}{H. {Jin}}.}
  \bibinfo{year}{2019}\natexlab{}.
\newblock \showarticletitle{A Comparative Study of Deep Learning-Based
  Vulnerability Detection System}.
\newblock \bibinfo{journal}{\emph{IEEE Access}}  \bibinfo{volume}{7}
  (\bibinfo{year}{2019}), \bibinfo{pages}{103184--103197}.
\newblock
\urldef\tempurl%
\url{https://doi.org/10.1109/ACCESS.2019.2930578}
\showDOI{\tempurl}


\bibitem[\protect\citeauthoryear{Liang, Yu, Jiang, and Xie}{Liang
  et~al\mbox{.}}{2019}]%
        {1094_Liang2019}
\bibfield{author}{\bibinfo{person}{Hongliang Liang}, \bibinfo{person}{Yue Yu},
  \bibinfo{person}{Lin Jiang}, {and} \bibinfo{person}{Zhuosi Xie}.}
  \bibinfo{year}{2019}\natexlab{}.
\newblock \showarticletitle{Seml: A semantic LSTM model for software defect
  prediction}.
\newblock \bibinfo{journal}{\emph{IEEE Access}}  \bibinfo{volume}{7}
  (\bibinfo{year}{2019}), \bibinfo{pages}{83812--83824}.
\newblock


\bibitem[\protect\citeauthoryear{{Lim}}{{Lim}}{2018}]%
        {Lim2018_453}
\bibfield{author}{\bibinfo{person}{H. {Lim}}.} \bibinfo{year}{2018}\natexlab{}.
\newblock \showarticletitle{Applying Code Vectors for Presenting Software
  Features in Machine Learning}. In \bibinfo{booktitle}{\emph{2018 IEEE 42nd
  Annual Computer Software and Applications Conference (COMPSAC)}},
  Vol.~\bibinfo{volume}{01}. \bibinfo{pages}{803--804}.
\newblock
\urldef\tempurl%
\url{https://doi.org/10.1109/COMPSAC.2018.00128}
\showDOI{\tempurl}


\bibitem[\protect\citeauthoryear{{Lima}, {da Cruz}, and {Ribeiro}}{{Lima}
  et~al\mbox{.}}{2020}]%
        {Lima2020_74}
\bibfield{author}{\bibinfo{person}{R. {Lima}}, \bibinfo{person}{A.~M.~R. {da
  Cruz}}, {and} \bibinfo{person}{J. {Ribeiro}}.}
  \bibinfo{year}{2020}\natexlab{}.
\newblock \showarticletitle{Artificial Intelligence Applied to Software
  Testing: A Literature Review}. In \bibinfo{booktitle}{\emph{2020 15th Iberian
  Conference on Information Systems and Technologies (CISTI)}}.
  \bibinfo{pages}{1--6}.
\newblock
\urldef\tempurl%
\url{https://doi.org/10.23919/CISTI49556.2020.9141124}
\showDOI{\tempurl}


\bibitem[\protect\citeauthoryear{LIN, WANG, WEN, and MAO}{LIN
  et~al\mbox{.}}{2021}]%
        {933_LIN2021}
\bibfield{author}{\bibinfo{person}{BO LIN}, \bibinfo{person}{SHANGWEN WANG},
  \bibinfo{person}{MING WEN}, {and} \bibinfo{person}{XIAOGUANG MAO}.}
  \bibinfo{year}{2021}\natexlab{}.
\newblock \showarticletitle{Context-Aware Code Change Embedding for Better
  Patch Correctness Assessment}.
\newblock \bibinfo{journal}{\emph{J. ACM}} \bibinfo{volume}{1},
  \bibinfo{number}{1} (\bibinfo{year}{2021}).
\newblock


\bibitem[\protect\citeauthoryear{Lin, Ouyang, Zhuang, Chen, Li, and Wu}{Lin
  et~al\mbox{.}}{2021}]%
        {1006_Lin2021}
\bibfield{author}{\bibinfo{person}{Chen Lin}, \bibinfo{person}{Zhichao Ouyang},
  \bibinfo{person}{Junqing Zhuang}, \bibinfo{person}{Jianqiang Chen},
  \bibinfo{person}{Hui Li}, {and} \bibinfo{person}{Rongxin Wu}.}
  \bibinfo{year}{2021}\natexlab{}.
\newblock \showarticletitle{Improving code summarization with block-wise
  abstract syntax tree splitting}. In \bibinfo{booktitle}{\emph{2021 IEEE/ACM
  29th International Conference on Program Comprehension (ICPC)}}. IEEE,
  \bibinfo{pages}{184--195}.
\newblock


\bibitem[\protect\citeauthoryear{Lin, Xiao, Zhang, and Xiang}{Lin
  et~al\mbox{.}}{2020}]%
        {lin_deep_2020}
\bibfield{author}{\bibinfo{person}{Guanjun Lin}, \bibinfo{person}{Wei Xiao},
  \bibinfo{person}{Jun Zhang}, {and} \bibinfo{person}{Yang Xiang}.}
  \bibinfo{year}{2020}\natexlab{}.
\newblock \showarticletitle{Deep {Learning}-{Based} {Vulnerable} {Function}
  {Detection}: {A} {Benchmark}}. In \bibinfo{booktitle}{\emph{Information and
  {Communications} {Security}}} \emph{(\bibinfo{series}{Lecture {Notes} in
  {Computer} {Science}})}, \bibfield{editor}{\bibinfo{person}{Jianying Zhou},
  \bibinfo{person}{Xiapu Luo}, \bibinfo{person}{Qingni Shen}, {and}
  \bibinfo{person}{Zhen Xu}} (Eds.). \bibinfo{publisher}{Springer International
  Publishing}, \bibinfo{address}{Cham}, \bibinfo{pages}{219--232}.
\newblock
\showISBNx{978-3-030-41579-2}
\urldef\tempurl%
\url{https://doi.org/10.1007/978-3-030-41579-2_13}
\showDOI{\tempurl}


\bibitem[\protect\citeauthoryear{Lin, Zhang, Luo, Pan, Xiang, De~Vel, and
  Montague}{Lin et~al\mbox{.}}{2018a}]%
        {lin_zhang_luo_pan_xiang_de_montague_2018}
\bibfield{author}{\bibinfo{person}{Guanjun Lin}, \bibinfo{person}{Jun Zhang},
  \bibinfo{person}{Wei Luo}, \bibinfo{person}{Lei Pan}, \bibinfo{person}{Yang
  Xiang}, \bibinfo{person}{Olivier De~Vel}, {and} \bibinfo{person}{Paul
  Montague}.} \bibinfo{year}{2018}\natexlab{a}.
\newblock \showarticletitle{Cross-Project Transfer Representation Learning for
  Vulnerable Function Discovery}.
\newblock \bibinfo{journal}{\emph{IEEE Transactions on Industrial Informatics}}
  \bibinfo{volume}{14}, \bibinfo{number}{7} (\bibinfo{year}{2018}),
  \bibinfo{pages}{3289--3297}.
\newblock
\urldef\tempurl%
\url{https://doi.org/10.1109/TII.2018.2821768}
\showDOI{\tempurl}


\bibitem[\protect\citeauthoryear{Lin, Zhang, Luo, Pan, Xiang, De~Vel, and
  Montague}{Lin et~al\mbox{.}}{2018b}]%
        {trl_2018}
\bibfield{author}{\bibinfo{person}{Guanjun Lin}, \bibinfo{person}{Jun Zhang},
  \bibinfo{person}{Wei Luo}, \bibinfo{person}{Lei Pan}, \bibinfo{person}{Yang
  Xiang}, \bibinfo{person}{Olivier De~Vel}, {and} \bibinfo{person}{Paul
  Montague}.} \bibinfo{year}{2018}\natexlab{b}.
\newblock \showarticletitle{Cross-Project Transfer Representation Learning for
  Vulnerable Function Discovery}.
\newblock \bibinfo{journal}{\emph{IEEE Transactions on Industrial Informatics}}
  \bibinfo{volume}{14}, \bibinfo{number}{7} (\bibinfo{year}{2018}),
  \bibinfo{pages}{3289--3297}.
\newblock
\urldef\tempurl%
\url{https://doi.org/10.1109/TII.2018.2821768}
\showDOI{\tempurl}


\bibitem[\protect\citeauthoryear{Lin and Lu}{Lin and Lu}{2021}]%
        {527_Lin2021}
\bibfield{author}{\bibinfo{person}{Junhao Lin} {and} \bibinfo{person}{Lu Lu}.}
  \bibinfo{year}{2021}\natexlab{}.
\newblock \showarticletitle{Semantic feature learning via dual sequences for
  defect prediction}.
\newblock \bibinfo{journal}{\emph{IEEE Access}}  \bibinfo{volume}{9}
  (\bibinfo{year}{2021}), \bibinfo{pages}{13112--13124}.
\newblock


\bibitem[\protect\citeauthoryear{Ling, Lin, Zou, and Xie}{Ling
  et~al\mbox{.}}{2020}]%
        {Ling2020_515}
\bibfield{author}{\bibinfo{person}{Chunyang Ling}, \bibinfo{person}{Zeqi Lin},
  \bibinfo{person}{Yanzhen Zou}, {and} \bibinfo{person}{Bing Xie}.}
  \bibinfo{year}{2020}\natexlab{}.
\newblock \showarticletitle{Adaptive Deep Code Search}. In
  \bibinfo{booktitle}{\emph{Proceedings of the 28th International Conference on
  Program Comprehension}} \emph{(\bibinfo{series}{ICPC '20})}.
  \bibinfo{publisher}{Association for Computing Machinery},
  \bibinfo{pages}{48–--59}.
\newblock
\showISBNx{9781450379588}
\urldef\tempurl%
\url{https://doi.org/10.1145/3387904.3389278}
\showDOI{\tempurl}


\bibitem[\protect\citeauthoryear{{Linstead}, {Lopes}, and {Baldi}}{{Linstead}
  et~al\mbox{.}}{2008}]%
        {Linstead2008_371}
\bibfield{author}{\bibinfo{person}{E. {Linstead}}, \bibinfo{person}{C.
  {Lopes}}, {and} \bibinfo{person}{P. {Baldi}}.}
  \bibinfo{year}{2008}\natexlab{}.
\newblock \showarticletitle{An Application of Latent Dirichlet Allocation to
  Analyzing Software Evolution}. In \bibinfo{booktitle}{\emph{2008 Seventh
  International Conference on Machine Learning and Applications}}.
  \bibinfo{pages}{813--818}.
\newblock
\urldef\tempurl%
\url{https://doi.org/10.1109/ICMLA.2008.47}
\showDOI{\tempurl}


\bibitem[\protect\citeauthoryear{Liu, Wang, Zhang, Fan, Yin, and Deng}{Liu
  et~al\mbox{.}}{2019c}]%
        {Liu2019_408}
\bibfield{author}{\bibinfo{person}{Bohong Liu}, \bibinfo{person}{Tao Wang},
  \bibinfo{person}{Xunhui Zhang}, \bibinfo{person}{Qiang Fan},
  \bibinfo{person}{Gang Yin}, {and} \bibinfo{person}{Jinsheng Deng}.}
  \bibinfo{year}{2019}\natexlab{c}.
\newblock \showarticletitle{A Neural-Network Based Code Summarization Approach
  by Using Source Code and Its Call Dependencies}. In
  \bibinfo{booktitle}{\emph{Proceedings of the 11th Asia-Pacific Symposium on
  Internetware}} (Fukuoka, Japan) \emph{(\bibinfo{series}{Internetware '19})}.
  Article \bibinfo{articleno}{12}, \bibinfo{numpages}{10}~pages.
\newblock
\showISBNx{9781450377010}
\urldef\tempurl%
\url{https://doi.org/10.1145/3361242.3362774}
\showDOI{\tempurl}


\bibitem[\protect\citeauthoryear{Liu, Gao, Xia, Lo, Grundy, and Yang}{Liu
  et~al\mbox{.}}{2020b}]%
        {ChaoLiu2020}
\bibfield{author}{\bibinfo{person}{Chao Liu}, \bibinfo{person}{Cuiyun Gao},
  \bibinfo{person}{Xin Xia}, \bibinfo{person}{David Lo}, \bibinfo{person}{John
  Grundy}, {and} \bibinfo{person}{Xiaohu Yang}.}
  \bibinfo{year}{2020}\natexlab{b}.
\newblock \bibinfo{title}{On the Replicability and Reproducibility of Deep
  Learning in Software Engineering}.
\newblock
\newblock
\showeprint[arxiv]{2006.14244}~[cs.SE]


\bibitem[\protect\citeauthoryear{Liu, Yang, Tan, and Hafiz}{Liu
  et~al\mbox{.}}{2013}]%
        {975_Liu2013}
\bibfield{author}{\bibinfo{person}{Chen Liu}, \bibinfo{person}{Jinqiu Yang},
  \bibinfo{person}{Lin Tan}, {and} \bibinfo{person}{Munawar Hafiz}.}
  \bibinfo{year}{2013}\natexlab{}.
\newblock \showarticletitle{R2Fix: Automatically generating bug fixes from bug
  reports}. In \bibinfo{booktitle}{\emph{2013 IEEE Sixth international
  conference on software testing, verification and validation}}. IEEE,
  \bibinfo{pages}{282--291}.
\newblock


\bibitem[\protect\citeauthoryear{Liu, Li, Wei, Xia, Fu, and Jin}{Liu
  et~al\mbox{.}}{2020c}]%
        {Liu2020_485}
\bibfield{author}{\bibinfo{person}{Fang Liu}, \bibinfo{person}{Ge Li},
  \bibinfo{person}{Bolin Wei}, \bibinfo{person}{Xin Xia},
  \bibinfo{person}{Zhiyi Fu}, {and} \bibinfo{person}{Zhi Jin}.}
  \bibinfo{year}{2020}\natexlab{c}.
\newblock \showarticletitle{A Self-Attentional Neural Architecture for Code
  Completion with Multi-Task Learning}. In
  \bibinfo{booktitle}{\emph{Proceedings of the 28th International Conference on
  Program Comprehension}} (Seoul, Republic of Korea)
  \emph{(\bibinfo{series}{ICPC '20})}. \bibinfo{pages}{37–47}.
\newblock
\showISBNx{9781450379588}
\urldef\tempurl%
\url{https://doi.org/10.1145/3387904.3389261}
\showDOI{\tempurl}


\bibitem[\protect\citeauthoryear{{Liu}, {Li}, {Zhao}, and {Jin}}{{Liu}
  et~al\mbox{.}}{2020}]%
        {Liu2020_501}
\bibfield{author}{\bibinfo{person}{F. {Liu}}, \bibinfo{person}{G. {Li}},
  \bibinfo{person}{Y. {Zhao}}, {and} \bibinfo{person}{Z. {Jin}}.}
  \bibinfo{year}{2020}\natexlab{}.
\newblock \showarticletitle{Multi-task Learning based Pre-trained Language
  Model for Code Completion}. In \bibinfo{booktitle}{\emph{2020 35th IEEE/ACM
  International Conference on Automated Software Engineering (ASE)}}.
  \bibinfo{pages}{473--485}.
\newblock


\bibitem[\protect\citeauthoryear{Liu, Jin, Xu, Bu, Zou, and Zhang}{Liu
  et~al\mbox{.}}{2019a}]%
        {Liu2019_512}
\bibfield{author}{\bibinfo{person}{Hui Liu}, \bibinfo{person}{Jiahao Jin},
  \bibinfo{person}{Zhifeng Xu}, \bibinfo{person}{Yifan Bu},
  \bibinfo{person}{Yanzhen Zou}, {and} \bibinfo{person}{Lu Zhang}.}
  \bibinfo{year}{2019}\natexlab{a}.
\newblock \showarticletitle{Deep learning based code smell detection}.
\newblock \bibinfo{journal}{\emph{IEEE Transactions on Software Engineering}}
  (\bibinfo{year}{2019}).
\newblock


\bibitem[\protect\citeauthoryear{Liu, Wang, Koyuncu, Kim, Bissyand\'{e}, Kim,
  Wu, Klein, Mao, and Traon}{Liu et~al\mbox{.}}{2020}]%
        {Liu2020}
\bibfield{author}{\bibinfo{person}{Kui Liu}, \bibinfo{person}{Shangwen Wang},
  \bibinfo{person}{Anil Koyuncu}, \bibinfo{person}{Kisub Kim},
  \bibinfo{person}{Tegawend\'{e}~F. Bissyand\'{e}}, \bibinfo{person}{Dongsun
  Kim}, \bibinfo{person}{Peng Wu}, \bibinfo{person}{Jacques Klein},
  \bibinfo{person}{Xiaoguang Mao}, {and} \bibinfo{person}{Yves~Le Traon}.}
  \bibinfo{year}{2020}\natexlab{}.
\newblock \showarticletitle{On the Efficiency of Test Suite Based Program
  Repair: A Systematic Assessment of 16 Automated Repair Systems for {J}ava
  Programs}. In \bibinfo{booktitle}{\emph{Proceedings of the ACM/IEEE 42nd
  International Conference on Software Engineering}} (Seoul, South Korea)
  \emph{(\bibinfo{series}{ICSE '20})}. \bibinfo{pages}{615?--627}.
\newblock
\showISBNx{9781450371216}
\urldef\tempurl%
\url{https://doi.org/10.1145/3377811.3380338}
\showDOI{\tempurl}


\bibitem[\protect\citeauthoryear{Liu, Gao, Chen, Yiu, and Liu}{Liu
  et~al\mbox{.}}{2020a}]%
        {1010_Liu2020}
\bibfield{author}{\bibinfo{person}{Shangqing Liu}, \bibinfo{person}{Cuiyun
  Gao}, \bibinfo{person}{Sen Chen}, \bibinfo{person}{Nie~Lun Yiu}, {and}
  \bibinfo{person}{Yang Liu}.} \bibinfo{year}{2020}\natexlab{a}.
\newblock \showarticletitle{ATOM: Commit message generation based on abstract
  syntax tree and hybrid ranking}.
\newblock \bibinfo{journal}{\emph{IEEE Transactions on Software Engineering}}
  (\bibinfo{year}{2020}).
\newblock


\bibitem[\protect\citeauthoryear{Liu, Li, Prajapati, and Wu}{Liu
  et~al\mbox{.}}{2019b}]%
        {Liu2019_85}
\bibfield{author}{\bibinfo{person}{Xiao Liu}, \bibinfo{person}{Xiaoting Li},
  \bibinfo{person}{Rupesh Prajapati}, {and} \bibinfo{person}{Dinghao Wu}.}
  \bibinfo{year}{2019}\natexlab{b}.
\newblock \showarticletitle{DeepFuzz: Automatic Generation of Syntax Valid C
  Programs for Fuzz Testing}.
\newblock \bibinfo{journal}{\emph{Proceedings of the AAAI Conference on
  Artificial Intelligence}} \bibinfo{volume}{33}, \bibinfo{number}{01}
  (\bibinfo{date}{Jul.} \bibinfo{year}{2019}), \bibinfo{pages}{1044--1051}.
\newblock
\urldef\tempurl%
\url{https://doi.org/10.1609/aaai.v33i01.33011044}
\showDOI{\tempurl}


\bibitem[\protect\citeauthoryear{Liu, Xia, Hassan, Lo, Xing, and Wang}{Liu
  et~al\mbox{.}}{2018}]%
        {Liu2018_430}
\bibfield{author}{\bibinfo{person}{Zhongxin Liu}, \bibinfo{person}{Xin Xia},
  \bibinfo{person}{Ahmed~E. Hassan}, \bibinfo{person}{David Lo},
  \bibinfo{person}{Zhenchang Xing}, {and} \bibinfo{person}{Xinyu Wang}.}
  \bibinfo{year}{2018}\natexlab{}.
\newblock \showarticletitle{Neural-Machine-Translation-Based Commit Message
  Generation: How Far Are We?}. In \bibinfo{booktitle}{\emph{Proceedings of the
  33rd ACM/IEEE International Conference on Automated Software Engineering}}
  (Montpellier, France) \emph{(\bibinfo{series}{ASE 2018})}.
  \bibinfo{pages}{373–384}.
\newblock
\showISBNx{9781450359375}
\urldef\tempurl%
\url{https://doi.org/10.1145/3238147.3238190}
\showDOI{\tempurl}


\bibitem[\protect\citeauthoryear{Liu, Xia, Treude, Lo, and Li}{Liu
  et~al\mbox{.}}{2019d}]%
        {1014_Liu2019}
\bibfield{author}{\bibinfo{person}{Zhongxin Liu}, \bibinfo{person}{Xin Xia},
  \bibinfo{person}{Christoph Treude}, \bibinfo{person}{David Lo}, {and}
  \bibinfo{person}{Shanping Li}.} \bibinfo{year}{2019}\natexlab{d}.
\newblock \showarticletitle{Automatic generation of pull request descriptions}.
  In \bibinfo{booktitle}{\emph{2019 34th IEEE/ACM International Conference on
  Automated Software Engineering (ASE)}}. IEEE, \bibinfo{pages}{176--188}.
\newblock


\bibitem[\protect\citeauthoryear{Long and Rinard}{Long and Rinard}{2016}]%
        {Long2016_285}
\bibfield{author}{\bibinfo{person}{Fan Long} {and} \bibinfo{person}{Martin
  Rinard}.} \bibinfo{year}{2016}\natexlab{}.
\newblock \showarticletitle{Automatic Patch Generation by Learning Correct
  Code}. In \bibinfo{booktitle}{\emph{Proceedings of the 43rd Annual ACM
  SIGPLAN-SIGACT Symposium on Principles of Programming Languages}} (St.
  Petersburg, FL, USA) \emph{(\bibinfo{series}{POPL '16})}.
  \bibinfo{pages}{298–312}.
\newblock
\showISBNx{9781450335492}
\urldef\tempurl%
\url{https://doi.org/10.1145/2837614.2837617}
\showDOI{\tempurl}


\bibitem[\protect\citeauthoryear{Lopes, Bajracharya, Ossher, and Baldi}{Lopes
  et~al\mbox{.}}{2010}]%
        {Lopes2010}
\bibfield{author}{\bibinfo{person}{C. Lopes}, \bibinfo{person}{S. Bajracharya},
  \bibinfo{person}{J. Ossher}, {and} \bibinfo{person}{P. Baldi}.}
  \bibinfo{year}{2010}\natexlab{}.
\newblock \bibinfo{title}{{UCI} Source Code Data Sets}.
\newblock
\newblock
\urldef\tempurl%
\url{http://www.ics.uci.edu/$\sim$lopes/datasets/}
\showURL{%
\tempurl}


\bibitem[\protect\citeauthoryear{Lou, Ghanbari, Li, Zhang, Zhang, Hao, and
  Zhang}{Lou et~al\mbox{.}}{2020}]%
        {896_Lou2020}
\bibfield{author}{\bibinfo{person}{Yiling Lou}, \bibinfo{person}{Ali Ghanbari},
  \bibinfo{person}{Xia Li}, \bibinfo{person}{Lingming Zhang},
  \bibinfo{person}{Haotian Zhang}, \bibinfo{person}{Dan Hao}, {and}
  \bibinfo{person}{Lu Zhang}.} \bibinfo{year}{2020}\natexlab{}.
\newblock \showarticletitle{Can automated program repair refine fault
  localization? a unified debugging approach}. In
  \bibinfo{booktitle}{\emph{Proceedings of the 29th ACM SIGSOFT International
  Symposium on Software Testing and Analysis}}. \bibinfo{pages}{75--87}.
\newblock


\bibitem[\protect\citeauthoryear{Lu, Zhao, Li, and Jin}{Lu
  et~al\mbox{.}}{2017}]%
        {1031_Lu2017}
\bibfield{author}{\bibinfo{person}{Yangyang Lu}, \bibinfo{person}{Zelong Zhao},
  \bibinfo{person}{Ge Li}, {and} \bibinfo{person}{Zhi Jin}.}
  \bibinfo{year}{2017}\natexlab{}.
\newblock \showarticletitle{Learning to generate comments for api-based code
  snippets}.
\newblock In \bibinfo{booktitle}{\emph{Software Engineering and Methodology for
  Emerging Domains}}. \bibinfo{publisher}{Springer}, \bibinfo{pages}{3--14}.
\newblock


\bibitem[\protect\citeauthoryear{Luiz, de~Oliveira~Rodrigues, and
  Parreiras}{Luiz et~al\mbox{.}}{2019}]%
        {Luiz2019_242}
\bibfield{author}{\bibinfo{person}{Frederico~Caram Luiz},
  \bibinfo{person}{Bruno~Rafael de Oliveira~Rodrigues}, {and}
  \bibinfo{person}{Fernando~Silva Parreiras}.} \bibinfo{year}{2019}\natexlab{}.
\newblock \showarticletitle{Machine Learning Techniques for Code Smells
  Detection: An Empirical Experiment on a Highly Imbalanced Setup}. In
  \bibinfo{booktitle}{\emph{Proceedings of the XV Brazilian Symposium on
  Information Systems}} (Aracaju, Brazil) \emph{(\bibinfo{series}{SBSI'19})}.
  Article \bibinfo{articleno}{65}, \bibinfo{numpages}{8}~pages.
\newblock
\showISBNx{9781450372374}
\urldef\tempurl%
\url{https://doi.org/10.1145/3330204.3330275}
\showDOI{\tempurl}


\bibitem[\protect\citeauthoryear{Lujan, Pecorelli, Palomba, De~Lucia, and
  Lenarduzzi}{Lujan et~al\mbox{.}}{2020}]%
        {Lujan2020_181}
\bibfield{author}{\bibinfo{person}{Savanna Lujan}, \bibinfo{person}{Fabiano
  Pecorelli}, \bibinfo{person}{Fabio Palomba}, \bibinfo{person}{Andrea
  De~Lucia}, {and} \bibinfo{person}{Valentina Lenarduzzi}.}
  \bibinfo{year}{2020}\natexlab{}.
\newblock \showarticletitle{A Preliminary Study on the Adequacy of Static
  Analysis Warnings with Respect to Code Smell Prediction}. In
  \bibinfo{booktitle}{\emph{Proceedings of the 4th ACM SIGSOFT International
  Workshop on Machine-Learning Techniques for Software-Quality Evaluation}}
  (Virtual, USA) \emph{(\bibinfo{series}{MaLTeSQuE 2020})}.
  \bibinfo{pages}{1–6}.
\newblock
\showISBNx{9781450381246}
\urldef\tempurl%
\url{https://doi.org/10.1145/3416505.3423559}
\showDOI{\tempurl}


\bibitem[\protect\citeauthoryear{Luong, Brevdo, and Zhao}{Luong
  et~al\mbox{.}}{2017}]%
        {luong17}
\bibfield{author}{\bibinfo{person}{Minh{-}Thang Luong}, \bibinfo{person}{Eugene
  Brevdo}, {and} \bibinfo{person}{Rui Zhao}.} \bibinfo{year}{2017}\natexlab{}.
\newblock \showarticletitle{Neural Machine Translation (seq2seq) Tutorial}.
\newblock \bibinfo{journal}{\emph{https://github.com/tensorflow/nmt}}
  (\bibinfo{year}{2017}).
\newblock


\bibitem[\protect\citeauthoryear{Lutellier, Pham, Pang, Li, Wei, and
  Tan}{Lutellier et~al\mbox{.}}{2020}]%
        {Lutellier2020_287}
\bibfield{author}{\bibinfo{person}{Thibaud Lutellier},
  \bibinfo{person}{Hung~Viet Pham}, \bibinfo{person}{Lawrence Pang},
  \bibinfo{person}{Yitong Li}, \bibinfo{person}{Moshi Wei}, {and}
  \bibinfo{person}{Lin Tan}.} \bibinfo{year}{2020}\natexlab{}.
\newblock \showarticletitle{{CoCoNuT}: Combining Context-Aware Neural
  Translation Models Using Ensemble for Program Repair}. In
  \bibinfo{booktitle}{\emph{Proceedings of the 29th ACM SIGSOFT International
  Symposium on Software Testing and Analysis}} (Virtual Event, USA)
  \emph{(\bibinfo{series}{ISSTA 2020})}. \bibinfo{pages}{101–114}.
\newblock
\showISBNx{9781450380089}
\urldef\tempurl%
\url{https://doi.org/10.1145/3395363.3397369}
\showDOI{\tempurl}


\bibitem[\protect\citeauthoryear{Lyu}{Lyu}{1996}]%
        {MIS1996}
\bibfield{editor}{\bibinfo{person}{Michael~R. Lyu}} (Ed.).
  \bibinfo{year}{1996}\natexlab{}.
\newblock \bibinfo{booktitle}{\emph{Handbook of Software Reliability
  Engineering}}.
\newblock \bibinfo{publisher}{McGraw-Hill, Inc.}, \bibinfo{address}{USA}.
\newblock
\showISBNx{0070394008}


\bibitem[\protect\citeauthoryear{Ma, Fakhoury, Christensen, Arnaoudova, Zogaan,
  and Mirakhorli}{Ma et~al\mbox{.}}{2018}]%
        {Ma2018_373}
\bibfield{author}{\bibinfo{person}{Yuzhan Ma}, \bibinfo{person}{Sarah
  Fakhoury}, \bibinfo{person}{Michael Christensen}, \bibinfo{person}{Venera
  Arnaoudova}, \bibinfo{person}{Waleed Zogaan}, {and} \bibinfo{person}{Mehdi
  Mirakhorli}.} \bibinfo{year}{2018}\natexlab{}.
\newblock \showarticletitle{Automatic Classification of Software Artifacts in
  Open-Source Applications}. In \bibinfo{booktitle}{\emph{Proceedings of the
  15th International Conference on Mining Software Repositories}} (Gothenburg,
  Sweden) \emph{(\bibinfo{series}{MSR '18})}. \bibinfo{pages}{414–425}.
\newblock
\showISBNx{9781450357166}
\urldef\tempurl%
\url{https://doi.org/10.1145/3196398.3196446}
\showDOI{\tempurl}


\bibitem[\protect\citeauthoryear{Ma, Luo, Zeng, and Chen}{Ma
  et~al\mbox{.}}{2012}]%
        {Ma2012_147}
\bibfield{author}{\bibinfo{person}{Ying Ma}, \bibinfo{person}{Guangchun Luo},
  \bibinfo{person}{Xue Zeng}, {and} \bibinfo{person}{Aiguo Chen}.}
  \bibinfo{year}{2012}\natexlab{}.
\newblock \showarticletitle{Transfer learning for cross-company software defect
  prediction}.
\newblock \bibinfo{journal}{\emph{Information and Software Technology}}
  \bibinfo{volume}{54}, \bibinfo{number}{3} (\bibinfo{year}{2012}),
  \bibinfo{pages}{248 -- 256}.
\newblock
\showISSN{0950-5849}
\urldef\tempurl%
\url{https://doi.org/10.1016/j.infsof.2011.09.007}
\showDOI{\tempurl}


\bibitem[\protect\citeauthoryear{{Ma}, {Ge}, {Liu}, {Zhao}, and {Ma}}{{Ma}
  et~al\mbox{.}}{2019}]%
        {Ma2019_3}
\bibfield{author}{\bibinfo{person}{Z. {Ma}}, \bibinfo{person}{H. {Ge}},
  \bibinfo{person}{Y. {Liu}}, \bibinfo{person}{M. {Zhao}}, {and}
  \bibinfo{person}{J. {Ma}}.} \bibinfo{year}{2019}\natexlab{}.
\newblock \showarticletitle{A Combination Method for Android Malware Detection
  Based on Control Flow Graphs and Machine Learning Algorithms}.
\newblock \bibinfo{journal}{\emph{IEEE Access}}  \bibinfo{volume}{7}
  (\bibinfo{year}{2019}), \bibinfo{pages}{21235--21245}.
\newblock
\urldef\tempurl%
\url{https://doi.org/10.1109/ACCESS.2019.2896003}
\showDOI{\tempurl}


\bibitem[\protect\citeauthoryear{Madhavan and Whitehead}{Madhavan and
  Whitehead}{2007}]%
        {Madhavan2007_117}
\bibfield{author}{\bibinfo{person}{Janaki~T. Madhavan} {and}
  \bibinfo{person}{E.~James Whitehead}.} \bibinfo{year}{2007}\natexlab{}.
\newblock \showarticletitle{Predicting Buggy Changes inside an Integrated
  Development Environment}. In \bibinfo{booktitle}{\emph{Proceedings of the
  2007 OOPSLA Workshop on Eclipse Technology EXchange}} (Montreal, Quebec,
  Canada) \emph{(\bibinfo{series}{eclipse '07})}. \bibinfo{pages}{36–40}.
\newblock
\showISBNx{9781605580159}
\urldef\tempurl%
\url{https://doi.org/10.1145/1328279.1328287}
\showDOI{\tempurl}


\bibitem[\protect\citeauthoryear{Majd, Vahidi-Asl, Khalilian,
  Poorsarvi-Tehrani, and Haghighi}{Majd et~al\mbox{.}}{2020}]%
        {Majd2020_353}
\bibfield{author}{\bibinfo{person}{Amirabbas Majd}, \bibinfo{person}{Mojtaba
  Vahidi-Asl}, \bibinfo{person}{Alireza Khalilian}, \bibinfo{person}{Pooria
  Poorsarvi-Tehrani}, {and} \bibinfo{person}{Hassan Haghighi}.}
  \bibinfo{year}{2020}\natexlab{}.
\newblock \showarticletitle{{SLDeep}: Statement-level software defect
  prediction using deep-learning model on static code features}.
\newblock \bibinfo{journal}{\emph{Expert Systems with Applications}}
  \bibinfo{volume}{147} (\bibinfo{year}{2020}), \bibinfo{pages}{113156}.
\newblock
\showISSN{0957-4174}
\urldef\tempurl%
\url{https://doi.org/10.1016/j.eswa.2019.113156}
\showDOI{\tempurl}


\bibitem[\protect\citeauthoryear{Malhotra}{Malhotra}{2014}]%
        {Malhotra2014_81}
\bibfield{author}{\bibinfo{person}{Ruchika Malhotra}.}
  \bibinfo{year}{2014}\natexlab{}.
\newblock \showarticletitle{Comparative analysis of statistical and machine
  learning methods for predicting faulty modules}.
\newblock \bibinfo{journal}{\emph{Applied Soft Computing}}
  \bibinfo{volume}{21} (\bibinfo{year}{2014}), \bibinfo{pages}{286 -- 297}.
\newblock
\showISSN{1568-4946}
\urldef\tempurl%
\url{https://doi.org/10.1016/j.asoc.2014.03.032}
\showDOI{\tempurl}


\bibitem[\protect\citeauthoryear{{Malhotra}, {Bahl}, {Sehgal}, and
  {Priya}}{{Malhotra} et~al\mbox{.}}{2017}]%
        {Malhotra2017_88}
\bibfield{author}{\bibinfo{person}{R. {Malhotra}}, \bibinfo{person}{L. {Bahl}},
  \bibinfo{person}{S. {Sehgal}}, {and} \bibinfo{person}{P. {Priya}}.}
  \bibinfo{year}{2017}\natexlab{}.
\newblock \showarticletitle{Empirical comparison of machine learning algorithms
  for bug prediction in open source software}. In
  \bibinfo{booktitle}{\emph{2017 International Conference on Big Data Analytics
  and Computational Intelligence (ICBDAC)}}. \bibinfo{pages}{40--45}.
\newblock
\urldef\tempurl%
\url{https://doi.org/10.1109/ICBDACI.2017.8070806}
\showDOI{\tempurl}


\bibitem[\protect\citeauthoryear{Malhotra and Jain}{Malhotra and Jain}{2012}]%
        {925_Malhotra2012}
\bibfield{author}{\bibinfo{person}{Ruchika Malhotra} {and}
  \bibinfo{person}{Ankita Jain}.} \bibinfo{year}{2012}\natexlab{}.
\newblock \showarticletitle{Fault prediction using statistical and machine
  learning methods for improving software quality}.
\newblock \bibinfo{journal}{\emph{Journal of Information Processing Systems}}
  \bibinfo{volume}{8}, \bibinfo{number}{2} (\bibinfo{year}{2012}),
  \bibinfo{pages}{241--262}.
\newblock


\bibitem[\protect\citeauthoryear{Malhotra and Jangra}{Malhotra and
  Jangra}{2017}]%
        {Malhotra2017_121}
\bibfield{author}{\bibinfo{person}{R. Malhotra} {and} \bibinfo{person}{Rupender
  Jangra}.} \bibinfo{year}{2017}\natexlab{}.
\newblock \showarticletitle{Prediction \& Assessment of Change Prone Classes
  Using Statistical \& Machine Learning Techniques}.
\newblock \bibinfo{journal}{\emph{Journal of Information Processing Systems}}
  \bibinfo{volume}{13} (\bibinfo{date}{01} \bibinfo{year}{2017}),
  \bibinfo{pages}{778--804}.
\newblock
\urldef\tempurl%
\url{https://doi.org/10.3745/JIPS.04.0013}
\showDOI{\tempurl}


\bibitem[\protect\citeauthoryear{Malhotra and Kamal}{Malhotra and
  Kamal}{2019}]%
        {681_Malhotra2019}
\bibfield{author}{\bibinfo{person}{Ruchika Malhotra} {and}
  \bibinfo{person}{Shine Kamal}.} \bibinfo{year}{2019}\natexlab{}.
\newblock \showarticletitle{An empirical study to investigate oversampling
  methods for improving software defect prediction using imbalanced data}.
\newblock \bibinfo{journal}{\emph{Neurocomputing}}  \bibinfo{volume}{343}
  (\bibinfo{year}{2019}), \bibinfo{pages}{120--140}.
\newblock


\bibitem[\protect\citeauthoryear{Malhotra and Khanna}{Malhotra and
  Khanna}{2013}]%
        {918_Malhotra2013}
\bibfield{author}{\bibinfo{person}{Ruchika Malhotra} {and}
  \bibinfo{person}{Megha Khanna}.} \bibinfo{year}{2013}\natexlab{}.
\newblock \showarticletitle{Investigation of relationship between
  object-oriented metrics and change proneness}.
\newblock \bibinfo{journal}{\emph{International Journal of Machine Learning and
  Cybernetics}} \bibinfo{volume}{4}, \bibinfo{number}{4}
  (\bibinfo{year}{2013}), \bibinfo{pages}{273--286}.
\newblock


\bibitem[\protect\citeauthoryear{Malhotra and Singh}{Malhotra and
  Singh}{2011}]%
        {Malhotra2011_114}
\bibfield{author}{\bibinfo{person}{Ruchika Malhotra} {and}
  \bibinfo{person}{Yogesh Singh}.} \bibinfo{year}{2011}\natexlab{}.
\newblock \showarticletitle{On the applicability of machine learning techniques
  for object-oriented software fault prediction}.
\newblock \bibinfo{journal}{\emph{Software Engineering: An International
  Journal}}  \bibinfo{volume}{1} (\bibinfo{date}{01} \bibinfo{year}{2011}).
\newblock


\bibitem[\protect\citeauthoryear{Malhotra$^1$ and Chug}{Malhotra$^1$ and
  Chug}{2012}]%
        {924_Malhotra$^1$2012}
\bibfield{author}{\bibinfo{person}{Ruchika Malhotra$^1$} {and}
  \bibinfo{person}{Anuradha Chug}.} \bibinfo{year}{2012}\natexlab{}.
\newblock \showarticletitle{Software maintainability prediction using machine
  learning algorithms}.
\newblock \bibinfo{journal}{\emph{Software engineering: an international
  Journal (SeiJ)}} \bibinfo{volume}{2}, \bibinfo{number}{2}
  (\bibinfo{year}{2012}).
\newblock


\bibitem[\protect\citeauthoryear{{Malik}, {Patra}, and {Pradel}}{{Malik}
  et~al\mbox{.}}{2019}]%
        {Malik2019_389}
\bibfield{author}{\bibinfo{person}{R.~S. {Malik}}, \bibinfo{person}{J.
  {Patra}}, {and} \bibinfo{person}{M. {Pradel}}.}
  \bibinfo{year}{2019}\natexlab{}.
\newblock \showarticletitle{NL2Type: Inferring JavaScript Function Types from
  Natural Language Information}. In \bibinfo{booktitle}{\emph{2019 IEEE/ACM
  41st International Conference on Software Engineering (ICSE)}}.
  \bibinfo{pages}{304--315}.
\newblock
\urldef\tempurl%
\url{https://doi.org/10.1109/ICSE.2019.00045}
\showDOI{\tempurl}


\bibitem[\protect\citeauthoryear{Manjula and Florence}{Manjula and
  Florence}{2019}]%
        {684_Manjula2019}
\bibfield{author}{\bibinfo{person}{C Manjula} {and} \bibinfo{person}{Lilly
  Florence}.} \bibinfo{year}{2019}\natexlab{}.
\newblock \showarticletitle{Deep neural network based hybrid approach for
  software defect prediction using software metrics}.
\newblock \bibinfo{journal}{\emph{Cluster Computing}} \bibinfo{volume}{22},
  \bibinfo{number}{4} (\bibinfo{year}{2019}), \bibinfo{pages}{9847--9863}.
\newblock


\bibitem[\protect\citeauthoryear{Mariano, dos Santos, and Brandao}{Mariano
  et~al\mbox{.}}{2021}]%
        {970_Mariano2021}
\bibfield{author}{\bibinfo{person}{Richard~VR Mariano},
  \bibinfo{person}{Geanderson~E dos Santos}, {and}
  \bibinfo{person}{Wladmir~Cardoso Brandao}.} \bibinfo{year}{2021}\natexlab{}.
\newblock \showarticletitle{Improve Classification of Commits Maintenance
  Activities with Quantitative Changes in Source Code}.
\newblock  (\bibinfo{year}{2021}).
\newblock


\bibitem[\protect\citeauthoryear{Mariano, dos Santos, de~Almeida, and
  Brand{\\textasciitilde a}o}{Mariano et~al\mbox{.}}{2019}]%
        {971_Mariano2019}
\bibfield{author}{\bibinfo{person}{Richard~VR Mariano},
  \bibinfo{person}{Geanderson~E dos Santos}, \bibinfo{person}{Markos~V de
  Almeida}, {and} \bibinfo{person}{Wladmir~C Brand{\\textasciitilde a}o}.}
  \bibinfo{year}{2019}\natexlab{}.
\newblock \showarticletitle{Feature changes in source code for commit
  classification into maintenance activities}. In
  \bibinfo{booktitle}{\emph{2019 18th IEEE International Conference On Machine
  Learning And Applications (ICMLA)}}. IEEE, \bibinfo{pages}{515--518}.
\newblock


\bibitem[\protect\citeauthoryear{Mashhadi and Hemmati}{Mashhadi and
  Hemmati}{2021}]%
        {888_Mashhadi2021}
\bibfield{author}{\bibinfo{person}{Ehsan Mashhadi} {and} \bibinfo{person}{Hadi
  Hemmati}.} \bibinfo{year}{2021}\natexlab{}.
\newblock \showarticletitle{Applying codebert for automated program repair of
  java simple bugs}. In \bibinfo{booktitle}{\emph{2021 IEEE/ACM 18th
  International Conference on Mining Software Repositories (MSR)}}. IEEE,
  \bibinfo{pages}{505--509}.
\newblock


\bibitem[\protect\citeauthoryear{Mateless, Rejabek, Margalit, and
  Moskovitch}{Mateless et~al\mbox{.}}{2020}]%
        {mateless_decompiled_2020}
\bibfield{author}{\bibinfo{person}{Roni Mateless}, \bibinfo{person}{Daniel
  Rejabek}, \bibinfo{person}{Oded Margalit}, {and} \bibinfo{person}{Robert
  Moskovitch}.} \bibinfo{year}{2020}\natexlab{}.
\newblock \showarticletitle{Decompiled {APK} based malicious code
  classification}.
\newblock \bibinfo{journal}{\emph{Future Generation Computer Systems}}
  \bibinfo{volume}{110} (\bibinfo{year}{2020}), \bibinfo{pages}{135--147}.
\newblock
\showISSN{0167-739X}
\urldef\tempurl%
\url{https://doi.org/10.1016/j.future.2020.03.052}
\showDOI{\tempurl}


\bibitem[\protect\citeauthoryear{McCabe}{McCabe}{1976}]%
        {McCabe1976}
\bibfield{author}{\bibinfo{person}{Thomas~J McCabe}.}
  \bibinfo{year}{1976}\natexlab{}.
\newblock \showarticletitle{A complexity measure}.
\newblock \bibinfo{journal}{\emph{IEEE Transactions on software Engineering}}
  \bibinfo{number}{4} (\bibinfo{year}{1976}), \bibinfo{pages}{308--320}.
\newblock


\bibitem[\protect\citeauthoryear{Medeiros, Neves, and Correia}{Medeiros
  et~al\mbox{.}}{2016}]%
        {medeiros_neves_correia_2016}
\bibfield{author}{\bibinfo{person}{Ibéria Medeiros}, \bibinfo{person}{Nuno
  Neves}, {and} \bibinfo{person}{Miguel Correia}.}
  \bibinfo{year}{2016}\natexlab{}.
\newblock \showarticletitle{Detecting and Removing Web Application
  Vulnerabilities with Static Analysis and Data Mining}.
\newblock \bibinfo{journal}{\emph{IEEE Transactions on Reliability}}
  \bibinfo{volume}{65}, \bibinfo{number}{1} (\bibinfo{year}{2016}),
  \bibinfo{pages}{54--69}.
\newblock
\urldef\tempurl%
\url{https://doi.org/10.1109/TR.2015.2457411}
\showDOI{\tempurl}


\bibitem[\protect\citeauthoryear{Medeiros, Neves, and Correia}{Medeiros
  et~al\mbox{.}}{2013}]%
        {wap_2013}
\bibfield{author}{\bibinfo{person}{Iberia Medeiros}, \bibinfo{person}{Nuno~F.
  Neves}, {and} \bibinfo{person}{Miguel Correia}.}
  \bibinfo{year}{2013}\natexlab{}.
\newblock \showarticletitle{Securing energy metering software with automatic
  source code correction}. In \bibinfo{booktitle}{\emph{2013 11th {IEEE}
  International Conference on Industrial Informatics ({INDIN})}}.
\newblock
\urldef\tempurl%
\url{https://doi.org/10.1109/indin.2013.6622969}
\showDOI{\tempurl}


\bibitem[\protect\citeauthoryear{Medeiros, Neves, and Correia}{Medeiros
  et~al\mbox{.}}{2014}]%
        {Medeiros2014_17}
\bibfield{author}{\bibinfo{person}{Ib\'{e}ria Medeiros},
  \bibinfo{person}{Nuno~F. Neves}, {and} \bibinfo{person}{Miguel Correia}.}
  \bibinfo{year}{2014}\natexlab{}.
\newblock \showarticletitle{Automatic Detection and Correction of Web
  Application Vulnerabilities Using Data Mining to Predict False Positives}. In
  \bibinfo{booktitle}{\emph{Proceedings of the 23rd International Conference on
  World Wide Web}} (Seoul, Korea) \emph{(\bibinfo{series}{WWW '14})}.
  \bibinfo{pages}{63–74}.
\newblock
\showISBNx{9781450327442}
\urldef\tempurl%
\url{https://doi.org/10.1145/2566486.2568024}
\showDOI{\tempurl}


\bibitem[\protect\citeauthoryear{Meng, Jiang, and Zhong}{Meng
  et~al\mbox{.}}{2021}]%
        {964_Meng2021}
\bibfield{author}{\bibinfo{person}{Na Meng}, \bibinfo{person}{Zijian Jiang},
  {and} \bibinfo{person}{Hao Zhong}.} \bibinfo{year}{2021}\natexlab{}.
\newblock \showarticletitle{Classifying Code Commits with Convolutional Neural
  Networks}. In \bibinfo{booktitle}{\emph{2021 International Joint Conference
  on Neural Networks (IJCNN)}}. IEEE, \bibinfo{pages}{1--8}.
\newblock


\bibitem[\protect\citeauthoryear{Meqdadi, Alhindawi, Alsakran, Saifan, and
  Migdadi}{Meqdadi et~al\mbox{.}}{2019}]%
        {Meqdadi2019_387}
\bibfield{author}{\bibinfo{person}{Omar Meqdadi}, \bibinfo{person}{Nouh
  Alhindawi}, \bibinfo{person}{Jamal Alsakran}, \bibinfo{person}{Ahmad Saifan},
  {and} \bibinfo{person}{Hatim Migdadi}.} \bibinfo{year}{2019}\natexlab{}.
\newblock \showarticletitle{Mining software repositories for adaptive change
  commits using machine learning techniques}.
\newblock \bibinfo{journal}{\emph{Information and Software Technology}}
  \bibinfo{volume}{109} (\bibinfo{year}{2019}), \bibinfo{pages}{80 -- 91}.
\newblock
\showISSN{0950-5849}
\urldef\tempurl%
\url{https://doi.org/10.1016/j.infsof.2019.01.008}
\showDOI{\tempurl}


\bibitem[\protect\citeauthoryear{Mesbah, Rice, Johnston, Glorioso, and
  Aftandilian}{Mesbah et~al\mbox{.}}{2019}]%
        {Mesbah2019_300}
\bibfield{author}{\bibinfo{person}{Ali Mesbah}, \bibinfo{person}{Andrew Rice},
  \bibinfo{person}{Emily Johnston}, \bibinfo{person}{Nick Glorioso}, {and}
  \bibinfo{person}{Edward Aftandilian}.} \bibinfo{year}{2019}\natexlab{}.
\newblock \showarticletitle{{DeepDelta}: Learning to Repair Compilation
  Errors}. In \bibinfo{booktitle}{\emph{Proceedings of the 2019 27th ACM Joint
  Meeting on European Software Engineering Conference and Symposium on the
  Foundations of Software Engineering}} (Tallinn, Estonia)
  \emph{(\bibinfo{series}{ESEC/FSE 2019})}. \bibinfo{pages}{925–936}.
\newblock
\showISBNx{9781450355728}
\urldef\tempurl%
\url{https://doi.org/10.1145/3338906.3340455}
\showDOI{\tempurl}


\bibitem[\protect\citeauthoryear{Mhawish and Gupta}{Mhawish and Gupta}{2020}]%
        {Mhawish2020_252}
\bibfield{author}{\bibinfo{person}{Mohammad~Y. Mhawish} {and}
  \bibinfo{person}{Manjari Gupta}.} \bibinfo{year}{2020}\natexlab{}.
\newblock \showarticletitle{Predicting Code Smells and Analysis of Predictions:
  Using Machine Learning Techniques and Software Metrics}.
\newblock \bibinfo{journal}{\emph{J. Comput. Sci. Technol.}}
  \bibinfo{volume}{35} (\bibinfo{year}{2020}), \bibinfo{pages}{1428--1445}.
\newblock


\bibitem[\protect\citeauthoryear{Milosevic, Dehghantanha, and Choo}{Milosevic
  et~al\mbox{.}}{2017}]%
        {Milosevic2017_36}
\bibfield{author}{\bibinfo{person}{Nikola Milosevic}, \bibinfo{person}{Ali
  Dehghantanha}, {and} \bibinfo{person}{Kim-Kwang~Raymond Choo}.}
  \bibinfo{year}{2017}\natexlab{}.
\newblock \showarticletitle{Machine learning aided Android malware
  classification}.
\newblock \bibinfo{journal}{\emph{Computers \& Electrical Engineering}}
  \bibinfo{volume}{61} (\bibinfo{year}{2017}), \bibinfo{pages}{266 -- 274}.
\newblock
\showISSN{0045-7906}
\urldef\tempurl%
\url{https://doi.org/10.1016/j.compeleceng.2017.02.013}
\showDOI{\tempurl}


\bibitem[\protect\citeauthoryear{Moskovitch, Nissim, and Elovici}{Moskovitch
  et~al\mbox{.}}{2009}]%
        {Moskovitch2009_43}
\bibfield{author}{\bibinfo{person}{Robert Moskovitch}, \bibinfo{person}{Nir
  Nissim}, {and} \bibinfo{person}{Yuval Elovici}.}
  \bibinfo{year}{2009}\natexlab{}.
\newblock \showarticletitle{Malicious Code Detection Using Active Learning}. In
  \bibinfo{booktitle}{\emph{Privacy, Security, and Trust in KDD}},
  \bibfield{editor}{\bibinfo{person}{Francesco Bonchi}, \bibinfo{person}{Elena
  Ferrari}, \bibinfo{person}{Wei Jiang}, {and} \bibinfo{person}{Bradley Malin}}
  (Eds.). \bibinfo{pages}{74--91}.
\newblock
\showISBNx{978-3-642-01718-6}


\bibitem[\protect\citeauthoryear{Mostaeen, Roy, Roy, Schneider, and
  Svajlenko}{Mostaeen et~al\mbox{.}}{2020}]%
        {Mostaeen2020_178}
\bibfield{author}{\bibinfo{person}{Golam Mostaeen}, \bibinfo{person}{Banani
  Roy}, \bibinfo{person}{Chanchal~K. Roy}, \bibinfo{person}{Kevin Schneider},
  {and} \bibinfo{person}{Jeffrey Svajlenko}.} \bibinfo{year}{2020}\natexlab{}.
\newblock \showarticletitle{A machine learning based framework for code clone
  validation}.
\newblock \bibinfo{journal}{\emph{Journal of Systems and Software}}
  \bibinfo{volume}{169} (\bibinfo{year}{2020}), \bibinfo{pages}{110686}.
\newblock
\showISSN{0164-1212}
\urldef\tempurl%
\url{https://doi.org/10.1016/j.jss.2020.110686}
\showDOI{\tempurl}


\bibitem[\protect\citeauthoryear{Mostaeen, Svajlenko, Roy, Roy, and
  Schneider}{Mostaeen et~al\mbox{.}}{2018}]%
        {Mostaeen2018}
\bibfield{author}{\bibinfo{person}{Golam Mostaeen}, \bibinfo{person}{Jeffrey
  Svajlenko}, \bibinfo{person}{Banani Roy}, \bibinfo{person}{Chanchal Roy},
  {and} \bibinfo{person}{Kevin Schneider}.} \bibinfo{year}{2018}\natexlab{}.
\newblock \showarticletitle{[Research Paper] On the Use of Machine Learning
  Techniques Towards the Design of Cloud Based Automatic Code Clone Validation
  Tools}. \bibinfo{pages}{155--164}.
\newblock
\urldef\tempurl%
\url{https://doi.org/10.1109/SCAM.2018.00025}
\showDOI{\tempurl}


\bibitem[\protect\citeauthoryear{{Mostaeen}, {Svajlenko}, {Roy}, {Roy}, and
  {Schneider}}{{Mostaeen} et~al\mbox{.}}{2018}]%
        {Mostaeen2018_251}
\bibfield{author}{\bibinfo{person}{G. {Mostaeen}}, \bibinfo{person}{J.
  {Svajlenko}}, \bibinfo{person}{B. {Roy}}, \bibinfo{person}{C.~K. {Roy}},
  {and} \bibinfo{person}{K.~A. {Schneider}}.} \bibinfo{year}{2018}\natexlab{}.
\newblock \showarticletitle{[Research Paper] On the Use of Machine Learning
  Techniques Towards the Design of Cloud Based Automatic Code Clone Validation
  Tools}. In \bibinfo{booktitle}{\emph{2018 IEEE 18th International Working
  Conference on Source Code Analysis and Manipulation (SCAM)}}.
  \bibinfo{pages}{155--164}.
\newblock
\urldef\tempurl%
\url{https://doi.org/10.1109/SCAM.2018.00025}
\showDOI{\tempurl}


\bibitem[\protect\citeauthoryear{Mostaeen, Svajlenko, Roy, Roy, and
  Schneider}{Mostaeen et~al\mbox{.}}{2019}]%
        {Mostaeen2019_203}
\bibfield{author}{\bibinfo{person}{Golam Mostaeen}, \bibinfo{person}{Jeffrey
  Svajlenko}, \bibinfo{person}{Banani Roy}, \bibinfo{person}{Chanchal~K. Roy},
  {and} \bibinfo{person}{Kevin~A. Schneider}.} \bibinfo{year}{2019}\natexlab{}.
\newblock \showarticletitle{CloneCognition: Machine Learning Based Code Clone
  Validation Tool}. In \bibinfo{booktitle}{\emph{Proceedings of the 2019 27th
  ACM Joint Meeting on European Software Engineering Conference and Symposium
  on the Foundations of Software Engineering}} (Tallinn, Estonia)
  \emph{(\bibinfo{series}{ESEC/FSE 2019})}. \bibinfo{pages}{1105–1109}.
\newblock
\showISBNx{9781450355728}
\urldef\tempurl%
\url{https://doi.org/10.1145/3338906.3341182}
\showDOI{\tempurl}


\bibitem[\protect\citeauthoryear{Mou, Li, Zhang, Wang, and Jin}{Mou
  et~al\mbox{.}}{2016}]%
        {Mou2016_466}
\bibfield{author}{\bibinfo{person}{Lili Mou}, \bibinfo{person}{Ge Li},
  \bibinfo{person}{Lu Zhang}, \bibinfo{person}{Tao Wang}, {and}
  \bibinfo{person}{Zhi Jin}.} \bibinfo{year}{2016}\natexlab{}.
\newblock \showarticletitle{Convolutional Neural Networks over Tree Structures
  for Programming Language Processing}. In
  \bibinfo{booktitle}{\emph{Proceedings of the Thirtieth AAAI Conference on
  Artificial Intelligence}} (Phoenix, Arizona)
  \emph{(\bibinfo{series}{AAAI'16})}. \bibinfo{pages}{1287–1293}.
\newblock


\bibitem[\protect\citeauthoryear{Movshovitz-Attias and Cohen}{Movshovitz-Attias
  and Cohen}{2013}]%
        {1073_Movshovitz-Attias2013}
\bibfield{author}{\bibinfo{person}{Dana Movshovitz-Attias} {and}
  \bibinfo{person}{William Cohen}.} \bibinfo{year}{2013}\natexlab{}.
\newblock \showarticletitle{Natural Language Models for Predicting Programming
  Comments}.
\newblock \bibinfo{journal}{\emph{ACL 2013 - 51st Annual Meeting of the
  Association for Computational Linguistics, Proceedings of the Conference}}
  \bibinfo{volume}{2} (\bibinfo{date}{08} \bibinfo{year}{2013}),
  \bibinfo{pages}{35--40}.
\newblock


\bibitem[\protect\citeauthoryear{Nair, Meinke, and Eldh}{Nair
  et~al\mbox{.}}{2019}]%
        {Nair2019_101}
\bibfield{author}{\bibinfo{person}{Aravind Nair}, \bibinfo{person}{Karl
  Meinke}, {and} \bibinfo{person}{Sigrid Eldh}.}
  \bibinfo{year}{2019}\natexlab{}.
\newblock \showarticletitle{Leveraging Mutants for Automatic Prediction of
  Metamorphic Relations Using Machine Learning}. In
  \bibinfo{booktitle}{\emph{Proceedings of the 3rd ACM SIGSOFT International
  Workshop on Machine Learning Techniques for Software Quality Evaluation}}
  (Tallinn, Estonia) \emph{(\bibinfo{series}{MaLTeSQuE 2019})}.
  \bibinfo{pages}{1–6}.
\newblock
\showISBNx{9781450368551}
\urldef\tempurl%
\url{https://doi.org/10.1145/3340482.3342741}
\showDOI{\tempurl}


\bibitem[\protect\citeauthoryear{Narayanan, Chandramohan, Chen, and
  Liu}{Narayanan et~al\mbox{.}}{2018}]%
        {narayanan_chandramohan_chen_liu_2018}
\bibfield{author}{\bibinfo{person}{Annamalai Narayanan},
  \bibinfo{person}{Mahinthan Chandramohan}, \bibinfo{person}{Lihui Chen}, {and}
  \bibinfo{person}{Yang Liu}.} \bibinfo{year}{2018}\natexlab{}.
\newblock \showarticletitle{A Multi-View Context-Aware Approach to Android
  Malware Detection and Malicious Code Localization}.
\newblock \bibinfo{journal}{\emph{Empirical Softw. Engg.}}
  \bibinfo{volume}{23}, \bibinfo{number}{3} (\bibinfo{date}{jun}
  \bibinfo{year}{2018}), \bibinfo{pages}{1222–1274}.
\newblock
\showISSN{1382-3256}
\urldef\tempurl%
\url{https://doi.org/10.1007/s10664-017-9539-8}
\showDOI{\tempurl}


\bibitem[\protect\citeauthoryear{Nazar, Hu, and Jiang}{Nazar
  et~al\mbox{.}}{2016}]%
        {Nazar2016_439}
\bibfield{author}{\bibinfo{person}{N. Nazar}, \bibinfo{person}{Y. Hu}, {and}
  \bibinfo{person}{He Jiang}.} \bibinfo{year}{2016}\natexlab{}.
\newblock \showarticletitle{Summarizing Software Artifacts: A Literature
  Review}.
\newblock \bibinfo{journal}{\emph{Journal of Computer Science and Technology}}
  \bibinfo{volume}{31} (\bibinfo{year}{2016}), \bibinfo{pages}{883--909}.
\newblock


\bibitem[\protect\citeauthoryear{Nazar, Jiang, Gao, Zhang, Li, and Ren}{Nazar
  et~al\mbox{.}}{2015}]%
        {Nazar2015_437}
\bibfield{author}{\bibinfo{person}{N. Nazar}, \bibinfo{person}{He Jiang},
  \bibinfo{person}{Guojun Gao}, \bibinfo{person}{Tao Zhang},
  \bibinfo{person}{Xiaochen Li}, {and} \bibinfo{person}{Zhilei Ren}.}
  \bibinfo{year}{2015}\natexlab{}.
\newblock \showarticletitle{Source code fragment summarization with small-scale
  crowdsourcing based features}.
\newblock \bibinfo{journal}{\emph{Frontiers of Computer Science}}
  \bibinfo{volume}{10} (\bibinfo{year}{2015}), \bibinfo{pages}{504--517}.
\newblock


\bibitem[\protect\citeauthoryear{Ndichu, Kim, Ozawa, Misu, and
  Makishima}{Ndichu et~al\mbox{.}}{2019}]%
        {Ndichu2019_5}
\bibfield{author}{\bibinfo{person}{Samuel Ndichu}, \bibinfo{person}{Sangwook
  Kim}, \bibinfo{person}{Seiichi Ozawa}, \bibinfo{person}{Takeshi Misu}, {and}
  \bibinfo{person}{Kazuo Makishima}.} \bibinfo{year}{2019}\natexlab{}.
\newblock \showarticletitle{A machine learning approach to detection of
  JavaScript-based attacks using AST features and paragraph vectors}.
\newblock \bibinfo{journal}{\emph{Applied Soft Computing}}
  \bibinfo{volume}{84} (\bibinfo{year}{2019}), \bibinfo{pages}{105721}.
\newblock
\showISSN{1568-4946}
\urldef\tempurl%
\url{https://doi.org/10.1016/j.asoc.2019.105721}
\showDOI{\tempurl}


\bibitem[\protect\citeauthoryear{{Nguyen}, {Nguyen}, {Phan}, and
  {Nguyen}}{{Nguyen} et~al\mbox{.}}{2018}]%
        {Nguyen2018_448}
\bibfield{author}{\bibinfo{person}{A.~T. {Nguyen}}, \bibinfo{person}{T.~D.
  {Nguyen}}, \bibinfo{person}{H.~D. {Phan}}, {and} \bibinfo{person}{T.~N.
  {Nguyen}}.} \bibinfo{year}{2018}\natexlab{}.
\newblock \showarticletitle{A deep neural network language model with contexts
  for source code}. In \bibinfo{booktitle}{\emph{2018 IEEE 25th International
  Conference on Software Analysis, Evolution and Reengineering (SANER)}}.
  \bibinfo{pages}{323--334}.
\newblock
\urldef\tempurl%
\url{https://doi.org/10.1109/SANER.2018.8330220}
\showDOI{\tempurl}


\bibitem[\protect\citeauthoryear{Nguyen, Do, Huynh, Vo, and Ha}{Nguyen
  et~al\mbox{.}}{2019}]%
        {Nguyen2019_126}
\bibfield{author}{\bibinfo{person}{Duc-Man Nguyen}, \bibinfo{person}{Hoang-Nhat
  Do}, \bibinfo{person}{Quyet-Thang Huynh}, \bibinfo{person}{Dinh-Thien Vo},
  {and} \bibinfo{person}{Nhu-Hang Ha}.} \bibinfo{year}{2019}\natexlab{}.
\newblock \showarticletitle{Shinobi: A Novel Approach for Context-Driven
  Testing (CDT) Using Heuristics and Machine Learning for Web Applications}. In
  \bibinfo{booktitle}{\emph{Industrial Networks and Intelligent Systems}},
  \bibfield{editor}{\bibinfo{person}{Trung~Q Duong} {and}
  \bibinfo{person}{Nguyen-Son Vo}} (Eds.). \bibinfo{pages}{86--102}.
\newblock
\showISBNx{978-3-030-05873-9}


\bibitem[\protect\citeauthoryear{Nguyen, Nguyen, Nguyen, and Nguyen}{Nguyen
  et~al\mbox{.}}{2013}]%
        {1086_Nguyen2013}
\bibfield{author}{\bibinfo{person}{Tung~Thanh Nguyen},
  \bibinfo{person}{Anh~Tuan Nguyen}, \bibinfo{person}{Hoan~Anh Nguyen}, {and}
  \bibinfo{person}{Tien~N. Nguyen}.} \bibinfo{year}{2013}\natexlab{}.
\newblock \showarticletitle{A Statistical Semantic Language Model for Source
  Code}. In \bibinfo{booktitle}{\emph{Proceedings of the 2013 9th Joint Meeting
  on Foundations of Software Engineering}} (Saint Petersburg, Russia)
  \emph{(\bibinfo{series}{ESEC/FSE 2013})}. \bibinfo{publisher}{Association for
  Computing Machinery}, \bibinfo{address}{New York, NY, USA},
  \bibinfo{pages}{532–542}.
\newblock
\showISBNx{9781450322379}
\urldef\tempurl%
\url{https://doi.org/1086_Nguyen2013}
\showDOI{\tempurl}


\bibitem[\protect\citeauthoryear{Nie, Gao, Zhong, Lam, Liu, and Xu}{Nie
  et~al\mbox{.}}{2021}]%
        {1009_Nie2021}
\bibfield{author}{\bibinfo{person}{Lun~Yiu Nie}, \bibinfo{person}{Cuiyun Gao},
  \bibinfo{person}{Zhicong Zhong}, \bibinfo{person}{Wai Lam},
  \bibinfo{person}{Yang Liu}, {and} \bibinfo{person}{Zenglin Xu}.}
  \bibinfo{year}{2021}\natexlab{}.
\newblock \showarticletitle{CoreGen: Contextualized Code Representation
  Learning for Commit Message Generation}.
\newblock \bibinfo{journal}{\emph{Neurocomputing}}  \bibinfo{volume}{459}
  (\bibinfo{year}{2021}), \bibinfo{pages}{97--107}.
\newblock


\bibitem[\protect\citeauthoryear{{Nyamawe}, {Liu}, {Niu}, {Umer}, and
  {Niu}}{{Nyamawe} et~al\mbox{.}}{2019}]%
        {Nyamawe2019_163}
\bibfield{author}{\bibinfo{person}{A.~S. {Nyamawe}}, \bibinfo{person}{H.
  {Liu}}, \bibinfo{person}{N. {Niu}}, \bibinfo{person}{Q. {Umer}}, {and}
  \bibinfo{person}{Z. {Niu}}.} \bibinfo{year}{2019}\natexlab{}.
\newblock \showarticletitle{Automated Recommendation of Software Refactorings
  Based on Feature Requests}. In \bibinfo{booktitle}{\emph{2019 IEEE 27th
  International Requirements Engineering Conference (RE)}}.
  \bibinfo{pages}{187--198}.
\newblock
\urldef\tempurl%
\url{https://doi.org/10.1109/RE.2019.00029}
\showDOI{\tempurl}


\bibitem[\protect\citeauthoryear{Nyamawe, Liu, Niu, Umer, and Niu}{Nyamawe
  et~al\mbox{.}}{2020}]%
        {760_Nyamawe_2020}
\bibfield{author}{\bibinfo{person}{Ally~S. Nyamawe}, \bibinfo{person}{Hui Liu},
  \bibinfo{person}{Nan Niu}, \bibinfo{person}{Qasim Umer}, {and}
  \bibinfo{person}{Zhendong Niu}.} \bibinfo{year}{2020}\natexlab{}.
\newblock \showarticletitle{Feature Requests-Based Recommendation of Software
  Refactorings}.
\newblock \bibinfo{journal}{\emph{Empirical Softw. Engg.}}
  \bibinfo{volume}{25}, \bibinfo{number}{5} (\bibinfo{date}{sep}
  \bibinfo{year}{2020}), \bibinfo{pages}{4315–4347}.
\newblock
\showISSN{1382-3256}
\urldef\tempurl%
\url{https://doi.org/10.1007/s10664-020-09871-2}
\showDOI{\tempurl}


\bibitem[\protect\citeauthoryear{Ochodek, Hebig, Meding, Frost, and
  Staron}{Ochodek et~al\mbox{.}}{2019}]%
        {Ochodek2019_257}
\bibfield{author}{\bibinfo{person}{Miroslaw Ochodek}, \bibinfo{person}{Regina
  Hebig}, \bibinfo{person}{Wilhelm Meding}, \bibinfo{person}{Gert Frost}, {and}
  \bibinfo{person}{Miroslaw Staron}.} \bibinfo{year}{2019}\natexlab{}.
\newblock \showarticletitle{Recognizing lines of code violating
  company-specific coding guidelines using machine learning}.
\newblock \bibinfo{journal}{\emph{Empirical Software Engineering}}
  \bibinfo{volume}{25} (\bibinfo{year}{2019}), \bibinfo{pages}{220--265}.
\newblock


\bibitem[\protect\citeauthoryear{{Oda}, {Fudaba}, {Neubig}, {Hata}, {Sakti},
  {Toda}, and {Nakamura}}{{Oda} et~al\mbox{.}}{2015}]%
        {Oda2015_325}
\bibfield{author}{\bibinfo{person}{Y. {Oda}}, \bibinfo{person}{H. {Fudaba}},
  \bibinfo{person}{G. {Neubig}}, \bibinfo{person}{H. {Hata}},
  \bibinfo{person}{S. {Sakti}}, \bibinfo{person}{T. {Toda}}, {and}
  \bibinfo{person}{S. {Nakamura}}.} \bibinfo{year}{2015}\natexlab{}.
\newblock \showarticletitle{Learning to Generate Pseudo-Code from Source Code
  Using Statistical Machine Translation}. In \bibinfo{booktitle}{\emph{2015
  30th IEEE/ACM International Conference on Automated Software Engineering
  (ASE)}}. \bibinfo{pages}{574--584}.
\newblock
\urldef\tempurl%
\url{https://doi.org/10.1109/ASE.2015.36}
\showDOI{\tempurl}


\bibitem[\protect\citeauthoryear{Okutan and Y{\i}ld{\i}z}{Okutan and
  Y{\i}ld{\i}z}{2014}]%
        {661_Okutan2014}
\bibfield{author}{\bibinfo{person}{Ahmet Okutan} {and}
  \bibinfo{person}{Olcay~Taner Y{\i}ld{\i}z}.} \bibinfo{year}{2014}\natexlab{}.
\newblock \showarticletitle{Software defect prediction using Bayesian
  networks}.
\newblock \bibinfo{journal}{\emph{Empirical Software Engineering}}
  \bibinfo{volume}{19}, \bibinfo{number}{1} (\bibinfo{year}{2014}),
  \bibinfo{pages}{154--181}.
\newblock


\bibitem[\protect\citeauthoryear{Oliveira, Assun\c{c}\~{a}o, Souza, Oizumi,
  Garcia, and Fonseca}{Oliveira et~al\mbox{.}}{2020}]%
        {Oliveira2020_195}
\bibfield{author}{\bibinfo{person}{Daniel Oliveira}, \bibinfo{person}{Wesley
  K.~G. Assun\c{c}\~{a}o}, \bibinfo{person}{Leonardo Souza},
  \bibinfo{person}{Willian Oizumi}, \bibinfo{person}{Alessandro Garcia}, {and}
  \bibinfo{person}{Baldoino Fonseca}.} \bibinfo{year}{2020}\natexlab{}.
\newblock \showarticletitle{Applying Machine Learning to Customized Smell
  Detection: A Multi-Project Study} \emph{(\bibinfo{series}{SBES '20})}.
  \bibinfo{pages}{233–242}.
\newblock
\showISBNx{9781450387538}
\urldef\tempurl%
\url{https://doi.org/10.1145/3422392.3422427}
\showDOI{\tempurl}


\bibitem[\protect\citeauthoryear{Omri and Sinz}{Omri and Sinz}{2020}]%
        {Omri2020_83}
\bibfield{author}{\bibinfo{person}{Safa Omri} {and} \bibinfo{person}{Carsten
  Sinz}.} \bibinfo{year}{2020}\natexlab{}.
\newblock \showarticletitle{Deep Learning for Software Defect Prediction: A
  Survey}. In \bibinfo{booktitle}{\emph{Proceedings of the IEEE/ACM 42nd
  International Conference on Software Engineering Workshops}} (Seoul, Republic
  of Korea) \emph{(\bibinfo{series}{ICSEW'20})}. \bibinfo{pages}{209–214}.
\newblock
\showISBNx{9781450379632}
\urldef\tempurl%
\url{https://doi.org/10.1145/3387940.3391463}
\showDOI{\tempurl}


\bibitem[\protect\citeauthoryear{Padmanabhuni and Tan}{Padmanabhuni and
  Tan}{2015}]%
        {padmanabhuni_tan_2015}
\bibfield{author}{\bibinfo{person}{Bindu~Madhavi Padmanabhuni} {and}
  \bibinfo{person}{Hee Beng~Kuan Tan}.} \bibinfo{year}{2015}\natexlab{}.
\newblock \showarticletitle{Buffer Overflow Vulnerability Prediction from x86
  Executables Using Static Analysis and Machine Learning}. In
  \bibinfo{booktitle}{\emph{2015 IEEE 39th Annual Computer Software and
  Applications Conference}}, Vol.~\bibinfo{volume}{2}.
  \bibinfo{pages}{450--459}.
\newblock
\urldef\tempurl%
\url{https://doi.org/10.1109/COMPSAC.2015.78}
\showDOI{\tempurl}


\bibitem[\protect\citeauthoryear{Palomba, Zanoni, Fontana, De~Lucia, and
  Oliveto}{Palomba et~al\mbox{.}}{2016}]%
        {808_Palomba2016}
\bibfield{author}{\bibinfo{person}{Fabio Palomba}, \bibinfo{person}{Marco
  Zanoni}, \bibinfo{person}{Francesca~Arcelli Fontana}, \bibinfo{person}{Andrea
  De~Lucia}, {and} \bibinfo{person}{Rocco Oliveto}.}
  \bibinfo{year}{2016}\natexlab{}.
\newblock \showarticletitle{Smells like teen spirit: Improving bug prediction
  performance using the intensity of code smells}. In
  \bibinfo{booktitle}{\emph{2016 IEEE International Conference on Software
  Maintenance and Evolution (ICSME)}}. IEEE, \bibinfo{pages}{244--255}.
\newblock


\bibitem[\protect\citeauthoryear{Palomba, Zanoni, Fontana, De~Lucia, and
  Oliveto}{Palomba et~al\mbox{.}}{2017}]%
        {809_Palomba2017}
\bibfield{author}{\bibinfo{person}{Fabio Palomba}, \bibinfo{person}{Marco
  Zanoni}, \bibinfo{person}{Francesca~Arcelli Fontana}, \bibinfo{person}{Andrea
  De~Lucia}, {and} \bibinfo{person}{Rocco Oliveto}.}
  \bibinfo{year}{2017}\natexlab{}.
\newblock \showarticletitle{Toward a smell-aware bug prediction model}.
\newblock \bibinfo{journal}{\emph{IEEE Transactions on Software Engineering}}
  \bibinfo{volume}{45}, \bibinfo{number}{2} (\bibinfo{year}{2017}),
  \bibinfo{pages}{194--218}.
\newblock


\bibitem[\protect\citeauthoryear{Pan, Lu, Xu, and Gao}{Pan
  et~al\mbox{.}}{2019}]%
        {680_Pan2019}
\bibfield{author}{\bibinfo{person}{Cong Pan}, \bibinfo{person}{Minyan Lu},
  \bibinfo{person}{Biao Xu}, {and} \bibinfo{person}{Houleng Gao}.}
  \bibinfo{year}{2019}\natexlab{}.
\newblock \showarticletitle{An improved CNN model for within-project software
  defect prediction}.
\newblock \bibinfo{journal}{\emph{Applied Sciences}} \bibinfo{volume}{9},
  \bibinfo{number}{10} (\bibinfo{year}{2019}), \bibinfo{pages}{2138}.
\newblock


\bibitem[\protect\citeauthoryear{Pandey and Gupta}{Pandey and Gupta}{2018}]%
        {Pandey2018_137}
\bibfield{author}{\bibinfo{person}{A.~K. Pandey} {and} \bibinfo{person}{Manjari
  Gupta}.} \bibinfo{year}{2018}\natexlab{}.
\newblock \showarticletitle{Software fault classification using extreme
  learning machine: a cognitive approach}.
\newblock \bibinfo{journal}{\emph{Evolutionary Intelligence}}
  (\bibinfo{year}{2018}), \bibinfo{pages}{1--8}.
\newblock


\bibitem[\protect\citeauthoryear{Pandey, Mishra, and Tripathi}{Pandey
  et~al\mbox{.}}{2021}]%
        {702_Pandey2021}
\bibfield{author}{\bibinfo{person}{Sushant~Kumar Pandey},
  \bibinfo{person}{Ravi~Bhushan Mishra}, {and} \bibinfo{person}{Anil~Kumar
  Tripathi}.} \bibinfo{year}{2021}\natexlab{}.
\newblock \showarticletitle{Machine learning based methods for software fault
  prediction: A survey}.
\newblock \bibinfo{journal}{\emph{Expert Systems with Applications}}
  \bibinfo{volume}{172} (\bibinfo{year}{2021}), \bibinfo{pages}{114595}.
\newblock


\bibitem[\protect\citeauthoryear{{Pang}, {Xue}, and {Namin}}{{Pang}
  et~al\mbox{.}}{2016}]%
        {Pang2016_27}
\bibfield{author}{\bibinfo{person}{Y. {Pang}}, \bibinfo{person}{X. {Xue}},
  {and} \bibinfo{person}{A.~S. {Namin}}.} \bibinfo{year}{2016}\natexlab{}.
\newblock \showarticletitle{Early Identification of Vulnerable Software
  Components via Ensemble Learning}. In \bibinfo{booktitle}{\emph{2016 15th
  IEEE International Conference on Machine Learning and Applications (ICMLA)}}.
  \bibinfo{pages}{476--481}.
\newblock
\urldef\tempurl%
\url{https://doi.org/10.1109/ICMLA.2016.0084}
\showDOI{\tempurl}


\bibitem[\protect\citeauthoryear{Pang, Xue, and Wang}{Pang
  et~al\mbox{.}}{2017}]%
        {Pang_2017_vul}
\bibfield{author}{\bibinfo{person}{Yulei Pang}, \bibinfo{person}{Xiaozhen Xue},
  {and} \bibinfo{person}{Huaying Wang}.} \bibinfo{year}{2017}\natexlab{}.
\newblock \showarticletitle{Predicting Vulnerable Software Components through
  Deep Neural Network}. In \bibinfo{booktitle}{\emph{Proceedings of the 2017
  International Conference on Deep Learning Technologies}} (Chengdu, China)
  \emph{(\bibinfo{series}{ICDLT '17})}. \bibinfo{publisher}{Association for
  Computing Machinery}, \bibinfo{address}{New York, NY, USA},
  \bibinfo{pages}{6–10}.
\newblock
\showISBNx{9781450352321}
\urldef\tempurl%
\url{https://doi.org/10.1145/3094243.3094245}
\showDOI{\tempurl}


\bibitem[\protect\citeauthoryear{Panichella, Aponte, Di~Penta, Marcus, and
  Canfora}{Panichella et~al\mbox{.}}{2012}]%
        {Panichella2012}
\bibfield{author}{\bibinfo{person}{Sebastiano Panichella},
  \bibinfo{person}{Jairo Aponte}, \bibinfo{person}{Massimiliano Di~Penta},
  \bibinfo{person}{Andrian Marcus}, {and} \bibinfo{person}{Gerardo Canfora}.}
  \bibinfo{year}{2012}\natexlab{}.
\newblock \showarticletitle{Mining source code descriptions from developer
  communications}. In \bibinfo{booktitle}{\emph{2012 20th IEEE International
  Conference on Program Comprehension (ICPC)}}. \bibinfo{pages}{63--72}.
\newblock
\urldef\tempurl%
\url{https://doi.org/10.1109/ICPC.2012.6240510}
\showDOI{\tempurl}


\bibitem[\protect\citeauthoryear{Pascarella, Palomba, and Bacchelli}{Pascarella
  et~al\mbox{.}}{2018}]%
        {1065_Pascarella2018}
\bibfield{author}{\bibinfo{person}{Luca Pascarella}, \bibinfo{person}{Fabio
  Palomba}, {and} \bibinfo{person}{Alberto Bacchelli}.}
  \bibinfo{year}{2018}\natexlab{}.
\newblock \showarticletitle{Re-evaluating method-level bug prediction}. In
  \bibinfo{booktitle}{\emph{2018 IEEE 25th International Conference on Software
  Analysis, Evolution and Reengineering (SANER)}}. IEEE,
  \bibinfo{pages}{592--601}.
\newblock


\bibitem[\protect\citeauthoryear{Patel, Fogarty, Landay, and Harrison}{Patel
  et~al\mbox{.}}{2008}]%
        {Patel2008_397}
\bibfield{author}{\bibinfo{person}{Kayur Patel}, \bibinfo{person}{James
  Fogarty}, \bibinfo{person}{James~A. Landay}, {and} \bibinfo{person}{Beverly
  Harrison}.} \bibinfo{year}{2008}\natexlab{}.
\newblock \showarticletitle{Investigating Statistical Machine Learning as a
  Tool for Software Development}. In \bibinfo{booktitle}{\emph{Proceedings of
  the SIGCHI Conference on Human Factors in Computing Systems}} (Florence,
  Italy) \emph{(\bibinfo{series}{CHI '08})}. \bibinfo{pages}{667–676}.
\newblock
\showISBNx{9781605580111}
\urldef\tempurl%
\url{https://doi.org/10.1145/1357054.1357160}
\showDOI{\tempurl}


\bibitem[\protect\citeauthoryear{Pecorelli, Di~Nucci, De~Roover, and
  De~Lucia}{Pecorelli et~al\mbox{.}}{2019}]%
        {Pecorelli2019_250}
\bibfield{author}{\bibinfo{person}{Fabiano Pecorelli}, \bibinfo{person}{Dario
  Di~Nucci}, \bibinfo{person}{Coen De~Roover}, {and} \bibinfo{person}{Andrea
  De~Lucia}.} \bibinfo{year}{2019}\natexlab{}.
\newblock \showarticletitle{On the Role of Data Balancing for Machine
  Learning-Based Code Smell Detection}. In
  \bibinfo{booktitle}{\emph{Proceedings of the 3rd ACM SIGSOFT International
  Workshop on Machine Learning Techniques for Software Quality Evaluation}}
  (Tallinn, Estonia) \emph{(\bibinfo{series}{MaLTeSQuE 2019})}.
  \bibinfo{pages}{19–24}.
\newblock
\showISBNx{9781450368551}
\urldef\tempurl%
\url{https://doi.org/10.1145/3340482.3342744}
\showDOI{\tempurl}


\bibitem[\protect\citeauthoryear{{Pecorelli}, {Palomba}, {Di Nucci}, and {De
  Lucia}}{{Pecorelli} et~al\mbox{.}}{2019}]%
        {Pecorelli2019_212}
\bibfield{author}{\bibinfo{person}{F. {Pecorelli}}, \bibinfo{person}{F.
  {Palomba}}, \bibinfo{person}{D. {Di Nucci}}, {and} \bibinfo{person}{A. {De
  Lucia}}.} \bibinfo{year}{2019}\natexlab{}.
\newblock \showarticletitle{Comparing Heuristic and Machine Learning Approaches
  for Metric-Based Code Smell Detection}. In \bibinfo{booktitle}{\emph{2019
  IEEE/ACM 27th International Conference on Program Comprehension (ICPC)}}.
  \bibinfo{pages}{93--104}.
\newblock


\bibitem[\protect\citeauthoryear{{Pereira}, {Campos}, and {Vieira}}{{Pereira}
  et~al\mbox{.}}{2019}]%
        {Pereira2019_12}
\bibfield{author}{\bibinfo{person}{J.~D. {Pereira}}, \bibinfo{person}{J.~R.
  {Campos}}, {and} \bibinfo{person}{M. {Vieira}}.}
  \bibinfo{year}{2019}\natexlab{}.
\newblock \showarticletitle{An Exploratory Study on Machine Learning to Combine
  Security Vulnerability Alerts from Static Analysis Tools}. In
  \bibinfo{booktitle}{\emph{2019 9th Latin-American Symposium on Dependable
  Computing (LADC)}}. \bibinfo{pages}{1--10}.
\newblock
\urldef\tempurl%
\url{https://doi.org/10.1109/LADC48089.2019.8995685}
\showDOI{\tempurl}


\bibitem[\protect\citeauthoryear{Perl, Dechand, Smith, Arp, Yamaguchi, Rieck,
  Fahl, and Acar}{Perl et~al\mbox{.}}{2015}]%
        {Perl2015_61}
\bibfield{author}{\bibinfo{person}{Henning Perl}, \bibinfo{person}{Sergej
  Dechand}, \bibinfo{person}{Matthew Smith}, \bibinfo{person}{Daniel Arp},
  \bibinfo{person}{Fabian Yamaguchi}, \bibinfo{person}{Konrad Rieck},
  \bibinfo{person}{Sascha Fahl}, {and} \bibinfo{person}{Yasemin Acar}.}
  \bibinfo{year}{2015}\natexlab{}.
\newblock \showarticletitle{VCCFinder: Finding Potential Vulnerabilities in
  Open-Source Projects to Assist Code Audits}. In
  \bibinfo{booktitle}{\emph{Proceedings of the 22nd ACM SIGSAC Conference on
  Computer and Communications Security}} (Denver, Colorado, USA)
  \emph{(\bibinfo{series}{CCS '15})}. \bibinfo{pages}{426–437}.
\newblock
\showISBNx{9781450338325}
\urldef\tempurl%
\url{https://doi.org/10.1145/2810103.2813604}
\showDOI{\tempurl}


\bibitem[\protect\citeauthoryear{Phan and Jannesari}{Phan and
  Jannesari}{2020}]%
        {Phan2020_506}
\bibfield{author}{\bibinfo{person}{Hung Phan} {and} \bibinfo{person}{Ali
  Jannesari}.} \bibinfo{year}{2020}\natexlab{}.
\newblock \showarticletitle{Statistical Machine Translation Outperforms Neural
  Machine Translation in Software Engineering: Why and How}. In
  \bibinfo{booktitle}{\emph{Proceedings of the 1st ACM SIGSOFT International
  Workshop on Representation Learning for Software Engineering and Program
  Languages}} (Virtual, USA) \emph{(\bibinfo{series}{RL+SE\&amp;PL 2020})}.
  \bibinfo{pages}{3–12}.
\newblock
\showISBNx{9781450381253}
\urldef\tempurl%
\url{https://doi.org/10.1145/3416506.3423576}
\showDOI{\tempurl}


\bibitem[\protect\citeauthoryear{Pinconschi, Abreu, and Ad{\\textasciitilde
  a}o}{Pinconschi et~al\mbox{.}}{2021}]%
        {915_Pinconschi2021}
\bibfield{author}{\bibinfo{person}{Eduard Pinconschi}, \bibinfo{person}{Rui
  Abreu}, {and} \bibinfo{person}{Pedro Ad{\\textasciitilde a}o}.}
  \bibinfo{year}{2021}\natexlab{}.
\newblock \showarticletitle{A Comparative Study of Automatic Program Repair
  Techniques for Security Vulnerabilities}. In \bibinfo{booktitle}{\emph{2021
  IEEE 32nd International Symposium on Software Reliability Engineering
  (ISSRE)}}. IEEE, \bibinfo{pages}{196--207}.
\newblock


\bibitem[\protect\citeauthoryear{Piskachev, Do, and Bodden}{Piskachev
  et~al\mbox{.}}{2019}]%
        {Piskachev2019_21}
\bibfield{author}{\bibinfo{person}{Goran Piskachev}, \bibinfo{person}{Lisa
  Nguyen~Quang Do}, {and} \bibinfo{person}{Eric Bodden}.}
  \bibinfo{year}{2019}\natexlab{}.
\newblock \showarticletitle{Codebase-Adaptive Detection of Security-Relevant
  Methods}. In \bibinfo{booktitle}{\emph{Proceedings of the 28th ACM SIGSOFT
  International Symposium on Software Testing and Analysis}} (Beijing, China)
  \emph{(\bibinfo{series}{ISSTA 2019})}. \bibinfo{pages}{181–191}.
\newblock
\showISBNx{9781450362245}
\urldef\tempurl%
\url{https://doi.org/10.1145/3293882.3330556}
\showDOI{\tempurl}


\bibitem[\protect\citeauthoryear{Ponta, Plate, Sabetta, Bezzi, and
  Dangremont}{Ponta et~al\mbox{.}}{2019}]%
        {Ponta2019_400}
\bibfield{author}{\bibinfo{person}{Serena~E. Ponta}, \bibinfo{person}{Henrik
  Plate}, \bibinfo{person}{Antonino Sabetta}, \bibinfo{person}{Michele Bezzi},
  {and} \bibinfo{person}{C\'{e}dric Dangremont}.}
  \bibinfo{year}{2019}\natexlab{}.
\newblock \showarticletitle{A Manually-Curated Dataset of Fixes to
  Vulnerabilities of Open-Source Software}. In
  \bibinfo{booktitle}{\emph{Proceedings of the 16th International Conference on
  Mining Software Repositories}} (Montreal, Quebec, Canada)
  \emph{(\bibinfo{series}{MSR '19})}. \bibinfo{pages}{383–387}.
\newblock
\urldef\tempurl%
\url{https://doi.org/10.1109/MSR.2019.00064}
\showDOI{\tempurl}


\bibitem[\protect\citeauthoryear{Pour, Li, Ma, and Hemmati}{Pour
  et~al\mbox{.}}{2021}]%
        {780_Pour2021}
\bibfield{author}{\bibinfo{person}{Maryam~Vahdat Pour}, \bibinfo{person}{Zhuo
  Li}, \bibinfo{person}{Lei Ma}, {and} \bibinfo{person}{Hadi Hemmati}.}
  \bibinfo{year}{2021}\natexlab{}.
\newblock \showarticletitle{A search-based testing framework for deep neural
  networks of source code embedding}. In \bibinfo{booktitle}{\emph{2021 14th
  IEEE Conference on Software Testing, Verification and Validation (ICST)}}.
  IEEE, \bibinfo{pages}{36--46}.
\newblock


\bibitem[\protect\citeauthoryear{{Prabha} and {Shivakumar}}{{Prabha} and
  {Shivakumar}}{2020}]%
        {Prabha2020_133}
\bibfield{author}{\bibinfo{person}{C.~L. {Prabha}} {and} \bibinfo{person}{N.
  {Shivakumar}}.} \bibinfo{year}{2020}\natexlab{}.
\newblock \showarticletitle{Software Defect Prediction Using Machine Learning
  Techniques}. In \bibinfo{booktitle}{\emph{2020 4th International Conference
  on Trends in Electronics and Informatics (ICOEI)(48184)}}.
  \bibinfo{pages}{728--733}.
\newblock
\urldef\tempurl%
\url{https://doi.org/10.1109/ICOEI48184.2020.9142909}
\showDOI{\tempurl}


\bibitem[\protect\citeauthoryear{Pradel and Sen}{Pradel and Sen}{2018}]%
        {Pradel2018_84}
\bibfield{author}{\bibinfo{person}{Michael Pradel} {and}
  \bibinfo{person}{Koushik Sen}.} \bibinfo{year}{2018}\natexlab{}.
\newblock \showarticletitle{DeepBugs: A Learning Approach to Name-Based Bug
  Detection}.
\newblock \bibinfo{journal}{\emph{Proc. ACM Program. Lang.}}
  \bibinfo{volume}{2}, \bibinfo{number}{OOPSLA}, Article
  \bibinfo{articleno}{147} (\bibinfo{date}{October} \bibinfo{year}{2018}),
  \bibinfo{numpages}{25}~pages.
\newblock
\urldef\tempurl%
\url{https://doi.org/10.1145/3276517}
\showDOI{\tempurl}


\bibitem[\protect\citeauthoryear{Premalatha and Srikrishna}{Premalatha and
  Srikrishna}{2017}]%
        {866_Premalatha2017}
\bibfield{author}{\bibinfo{person}{Hosahalli~Mahalingappa Premalatha} {and}
  \bibinfo{person}{Chimanahalli~Venkateshavittalachar Srikrishna}.}
  \bibinfo{year}{2017}\natexlab{}.
\newblock \showarticletitle{Software Fault Prediction and Classification using
  Cost based Random Forest in Spiral Life Cycle Model}.
\newblock \bibinfo{journal}{\emph{system}}  \bibinfo{volume}{11}
  (\bibinfo{year}{2017}).
\newblock


\bibitem[\protect\citeauthoryear{Prince}{Prince}{2004}]%
        {Prince2004}
\bibfield{author}{\bibinfo{person}{Michael Prince}.}
  \bibinfo{year}{2004}\natexlab{}.
\newblock \showarticletitle{Does active learning work? A review of the
  research}.
\newblock \bibinfo{journal}{\emph{Journal of engineering education}}
  \bibinfo{volume}{93}, \bibinfo{number}{3} (\bibinfo{year}{2004}),
  \bibinfo{pages}{223--231}.
\newblock


\bibitem[\protect\citeauthoryear{{Pritam}, {Khari}, {Hoang Son}, {Kumar},
  {Jha}, {Priyadarshini}, {Abdel-Basset}, and {Viet Long}}{{Pritam}
  et~al\mbox{.}}{2019}]%
        {Pritam2019_196}
\bibfield{author}{\bibinfo{person}{N. {Pritam}}, \bibinfo{person}{M. {Khari}},
  \bibinfo{person}{L. {Hoang Son}}, \bibinfo{person}{R. {Kumar}},
  \bibinfo{person}{S. {Jha}}, \bibinfo{person}{I. {Priyadarshini}},
  \bibinfo{person}{M. {Abdel-Basset}}, {and} \bibinfo{person}{H. {Viet Long}}.}
  \bibinfo{year}{2019}\natexlab{}.
\newblock \showarticletitle{Assessment of Code Smell for Predicting Class
  Change Proneness Using Machine Learning}.
\newblock \bibinfo{journal}{\emph{IEEE Access}}  \bibinfo{volume}{7}
  (\bibinfo{year}{2019}), \bibinfo{pages}{37414--37425}.
\newblock
\urldef\tempurl%
\url{https://doi.org/10.1109/ACCESS.2019.2905133}
\showDOI{\tempurl}


\bibitem[\protect\citeauthoryear{Proksch, Lerch, and Mezini}{Proksch
  et~al\mbox{.}}{2015}]%
        {Proksch2015_498}
\bibfield{author}{\bibinfo{person}{Sebastian Proksch},
  \bibinfo{person}{Johannes Lerch}, {and} \bibinfo{person}{Mira Mezini}.}
  \bibinfo{year}{2015}\natexlab{}.
\newblock \showarticletitle{Intelligent Code Completion with Bayesian
  Networks}.
\newblock \bibinfo{journal}{\emph{ACM Trans. Softw. Eng. Methodol.}}
  \bibinfo{volume}{25}, \bibinfo{number}{1}, Article \bibinfo{articleno}{3}
  (\bibinfo{date}{December} \bibinfo{year}{2015}),
  \bibinfo{numpages}{31}~pages.
\newblock
\showISSN{1049-331X}
\urldef\tempurl%
\url{https://doi.org/10.1145/2744200}
\showDOI{\tempurl}


\bibitem[\protect\citeauthoryear{Psarras, Diamantopoulos, and
  Symeonidis}{Psarras et~al\mbox{.}}{2019}]%
        {998_Psarras2019}
\bibfield{author}{\bibinfo{person}{Christos Psarras},
  \bibinfo{person}{Themistoklis Diamantopoulos}, {and} \bibinfo{person}{Andreas
  Symeonidis}.} \bibinfo{year}{2019}\natexlab{}.
\newblock \showarticletitle{A mechanism for automatically summarizing software
  functionality from source code}. In \bibinfo{booktitle}{\emph{2019 IEEE 19th
  International Conference on Software Quality, Reliability and Security
  (QRS)}}. IEEE, \bibinfo{pages}{121--130}.
\newblock


\bibitem[\protect\citeauthoryear{Qiao, Li, Umer, and Guo}{Qiao
  et~al\mbox{.}}{2020}]%
        {682_Qiao2020}
\bibfield{author}{\bibinfo{person}{Lei Qiao}, \bibinfo{person}{Xuesong Li},
  \bibinfo{person}{Qasim Umer}, {and} \bibinfo{person}{Ping Guo}.}
  \bibinfo{year}{2020}\natexlab{}.
\newblock \showarticletitle{Deep learning based software defect prediction}.
\newblock \bibinfo{journal}{\emph{Neurocomputing}}  \bibinfo{volume}{385}
  (\bibinfo{year}{2020}), \bibinfo{pages}{100--110}.
\newblock


\bibitem[\protect\citeauthoryear{Rabin, Mukherjee, Gnawali, and Alipour}{Rabin
  et~al\mbox{.}}{2020}]%
        {Rabin2020_481}
\bibfield{author}{\bibinfo{person}{Md~Rafiqul~Islam Rabin},
  \bibinfo{person}{Arjun Mukherjee}, \bibinfo{person}{Omprakash Gnawali}, {and}
  \bibinfo{person}{Mohammad~Amin Alipour}.} \bibinfo{year}{2020}\natexlab{}.
\newblock \showarticletitle{Towards Demystifying Dimensions of Source Code
  Embeddings}. In \bibinfo{booktitle}{\emph{Proceedings of the 1st ACM SIGSOFT
  International Workshop on Representation Learning for Software Engineering
  and Program Languages}} (Virtual, USA) \emph{(\bibinfo{series}{RL+SE\&amp;PL
  2020})}. \bibinfo{pages}{29–38}.
\newblock
\showISBNx{9781450381253}
\urldef\tempurl%
\url{https://doi.org/10.1145/3416506.3423580}
\showDOI{\tempurl}


\bibitem[\protect\citeauthoryear{Rahman, Pradhan, Partho, and Williams}{Rahman
  et~al\mbox{.}}{2017}]%
        {Rahman2017_48}
\bibfield{author}{\bibinfo{person}{Akond Rahman}, \bibinfo{person}{Priysha
  Pradhan}, \bibinfo{person}{Asif Partho}, {and} \bibinfo{person}{Laurie
  Williams}.} \bibinfo{year}{2017}\natexlab{}.
\newblock \showarticletitle{Predicting Android Application Security and Privacy
  Risk with Static Code Metrics}. In \bibinfo{booktitle}{\emph{Proceedings of
  the 4th International Conference on Mobile Software Engineering and Systems}}
  (Buenos Aires, Argentina) \emph{(\bibinfo{series}{MOBILESoft '17})}.
  \bibinfo{pages}{149–153}.
\newblock
\showISBNx{9781538626696}
\urldef\tempurl%
\url{https://doi.org/10.1109/MOBILESoft.2017.14}
\showDOI{\tempurl}


\bibitem[\protect\citeauthoryear{Rahman, Watanobe, and Nakamura}{Rahman
  et~al\mbox{.}}{2020}]%
        {Rahman2020_484}
\bibfield{author}{\bibinfo{person}{M. Rahman}, \bibinfo{person}{Yutaka
  Watanobe}, {and} \bibinfo{person}{K. Nakamura}.}
  \bibinfo{year}{2020}\natexlab{}.
\newblock \showarticletitle{A Neural Network Based Intelligent Support Model
  for Program Code Completion}.
\newblock \bibinfo{journal}{\emph{Sci. Program.}}  \bibinfo{volume}{2020}
  (\bibinfo{year}{2020}), \bibinfo{pages}{7426461:1--7426461:18}.
\newblock
\urldef\tempurl%
\url{https://doi.org/10.1155/2020/7426461}
\showDOI{\tempurl}


\bibitem[\protect\citeauthoryear{Rahman, Roy, and Keivanloo}{Rahman
  et~al\mbox{.}}{2015}]%
        {scam2015masud}
\bibfield{author}{\bibinfo{person}{M.~M. Rahman}, \bibinfo{person}{C.~K. Roy},
  {and} \bibinfo{person}{I. Keivanloo}.} \bibinfo{year}{2015}\natexlab{}.
\newblock \showarticletitle{Recommending {I}nsightful {C}omments for {S}ource
  {C}ode using {C}rowdsourced {K}nowledge}. In \bibinfo{booktitle}{\emph{Proc.
  SCAM}}. \bibinfo{pages}{81--90}.
\newblock


\bibitem[\protect\citeauthoryear{Rathore and Kumar}{Rathore and Kumar}{2021}]%
        {725_Rathore2021}
\bibfield{author}{\bibinfo{person}{Santosh~S Rathore} {and}
  \bibinfo{person}{Sandeep Kumar}.} \bibinfo{year}{2021}\natexlab{}.
\newblock \showarticletitle{Software fault prediction based on the dynamic
  selection of learning technique: findings from the eclipse project study}.
\newblock \bibinfo{journal}{\emph{Applied Intelligence}} \bibinfo{volume}{51},
  \bibinfo{number}{12} (\bibinfo{year}{2021}), \bibinfo{pages}{8945--8960}.
\newblock


\bibitem[\protect\citeauthoryear{Raychev, Bielik, and Vechev}{Raychev
  et~al\mbox{.}}{2016}]%
        {Raychev2016_504}
\bibfield{author}{\bibinfo{person}{Veselin Raychev}, \bibinfo{person}{Pavol
  Bielik}, {and} \bibinfo{person}{Martin Vechev}.}
  \bibinfo{year}{2016}\natexlab{}.
\newblock \showarticletitle{Probabilistic Model for Code with Decision Trees}.
\newblock \bibinfo{journal}{\emph{SIGPLAN Not.}} \bibinfo{volume}{51},
  \bibinfo{number}{10} (\bibinfo{date}{October} \bibinfo{year}{2016}),
  \bibinfo{pages}{731–747}.
\newblock
\showISSN{0362-1340}
\urldef\tempurl%
\url{https://doi.org/10.1145/3022671.2984041}
\showDOI{\tempurl}


\bibitem[\protect\citeauthoryear{Reddivari and Raman}{Reddivari and
  Raman}{2019}]%
        {920_Reddivari2019}
\bibfield{author}{\bibinfo{person}{Sandeep Reddivari} {and}
  \bibinfo{person}{Jayalakshmi Raman}.} \bibinfo{year}{2019}\natexlab{}.
\newblock \showarticletitle{Software quality prediction: an investigation based
  on machine learning}. In \bibinfo{booktitle}{\emph{2019 IEEE 20th
  International Conference on Information Reuse and Integration for Data
  Science (IRI)}}. IEEE, \bibinfo{pages}{115--122}.
\newblock


\bibitem[\protect\citeauthoryear{Ren, Qin, Ma, and Luo}{Ren
  et~al\mbox{.}}{2014}]%
        {637_Ren2014}
\bibfield{author}{\bibinfo{person}{Jinsheng Ren}, \bibinfo{person}{Ke Qin},
  \bibinfo{person}{Ying Ma}, {and} \bibinfo{person}{Guangchun Luo}.}
  \bibinfo{year}{2014}\natexlab{}.
\newblock \showarticletitle{On software defect prediction using machine
  learning}.
\newblock \bibinfo{journal}{\emph{Journal of Applied Mathematics}}
  \bibinfo{volume}{2014} (\bibinfo{year}{2014}).
\newblock


\bibitem[\protect\citeauthoryear{Ren, Zheng, Liu, Wei, and Yan}{Ren
  et~al\mbox{.}}{2019}]%
        {ren_buffer_2019}
\bibfield{author}{\bibinfo{person}{Jiadong Ren}, \bibinfo{person}{Zhangqi
  Zheng}, \bibinfo{person}{Qian Liu}, \bibinfo{person}{Zhiyao Wei}, {and}
  \bibinfo{person}{Huaizhi Yan}.} \bibinfo{year}{2019}\natexlab{}.
\newblock \showarticletitle{A {Buffer} {Overflow} {Prediction} {Approach}
  {Based} on {Software} {Metrics} and {Machine} {Learning}}.
\newblock \bibinfo{journal}{\emph{Security and Communication Networks}}
  \bibinfo{volume}{2019} (\bibinfo{date}{March} \bibinfo{year}{2019}),
  \bibinfo{pages}{e8391425}.
\newblock
\showISSN{1939-0114}
\urldef\tempurl%
\url{https://doi.org/10.1155/2019/8391425}
\showDOI{\tempurl}
\newblock
\shownote{Publisher: Hindawi.}


\bibitem[\protect\citeauthoryear{Ren, Xiao, Chang, Huang, Li, Chen, and
  Wang}{Ren et~al\mbox{.}}{2020}]%
        {Ren2020survey}
\bibfield{author}{\bibinfo{person}{Pengzhen Ren}, \bibinfo{person}{Yun Xiao},
  \bibinfo{person}{Xiaojun Chang}, \bibinfo{person}{Po-Yao Huang},
  \bibinfo{person}{Zhihui Li}, \bibinfo{person}{Xiaojiang Chen}, {and}
  \bibinfo{person}{Xin Wang}.} \bibinfo{year}{2020}\natexlab{}.
\newblock \showarticletitle{A survey of deep active learning}.
\newblock \bibinfo{journal}{\emph{arXiv preprint arXiv:2009.00236}}
  (\bibinfo{year}{2020}).
\newblock


\bibitem[\protect\citeauthoryear{Renzullo, Weimer, and Forrest}{Renzullo
  et~al\mbox{.}}{2021}]%
        {939_Renzullo2021}
\bibfield{author}{\bibinfo{person}{Joseph Renzullo}, \bibinfo{person}{Westley
  Weimer}, {and} \bibinfo{person}{Stephanie Forrest}.}
  \bibinfo{year}{2021}\natexlab{}.
\newblock \showarticletitle{Multiplicative Weights Algorithms for Parallel
  Automated Software Repair}. In \bibinfo{booktitle}{\emph{2021 IEEE
  International Parallel and Distributed Processing Symposium (IPDPS)}}. IEEE,
  \bibinfo{pages}{984--993}.
\newblock


\bibitem[\protect\citeauthoryear{Ribeiro, Meirelles, Lago, and Kon}{Ribeiro
  et~al\mbox{.}}{2019}]%
        {Ribeiro2019_256}
\bibfield{author}{\bibinfo{person}{Athos Ribeiro}, \bibinfo{person}{Paulo
  Meirelles}, \bibinfo{person}{Nelson Lago}, {and} \bibinfo{person}{Fabio
  Kon}.} \bibinfo{year}{2019}\natexlab{}.
\newblock \showarticletitle{Ranking Warnings from Multiple Source Code Static
  Analyzers via Ensemble Learning}. In \bibinfo{booktitle}{\emph{Proceedings of
  the 15th International Symposium on Open Collaboration}} (Sk\"{o}vde, Sweden)
  \emph{(\bibinfo{series}{OpenSym '19})}. Article \bibinfo{articleno}{5},
  \bibinfo{numpages}{10}~pages.
\newblock
\showISBNx{9781450363198}
\urldef\tempurl%
\url{https://doi.org/10.1145/3306446.3340828}
\showDOI{\tempurl}


\bibitem[\protect\citeauthoryear{Rodriguez, Mateos, Listorti, Hammer, and
  Misra}{Rodriguez et~al\mbox{.}}{2019}]%
        {Rodriguez2019_159}
\bibfield{author}{\bibinfo{person}{Guillermo Rodriguez},
  \bibinfo{person}{Cristian Mateos}, \bibinfo{person}{Luciano Listorti},
  \bibinfo{person}{Brian Hammer}, {and} \bibinfo{person}{Sanjay Misra}.}
  \bibinfo{year}{2019}\natexlab{}.
\newblock \showarticletitle{A Novel Unsupervised Learning Approach for
  Assessing Web Services Refactoring}. In \bibinfo{booktitle}{\emph{Information
  and Software Technologies}}, \bibfield{editor}{\bibinfo{person}{Robertas
  Dama{\v{s}}evi{\v{c}}ius} {and} \bibinfo{person}{Giedr{\.{e}}
  Vasiljevien{\.{e}}}} (Eds.). \bibinfo{pages}{273--284}.
\newblock
\showISBNx{978-3-030-30275-7}


\bibitem[\protect\citeauthoryear{{Russell}, {Kim}, {Hamilton}, {Lazovich},
  {Harer}, {Ozdemir}, {Ellingwood}, and {McConley}}{{Russell}
  et~al\mbox{.}}{2018}]%
        {Russell2018_16}
\bibfield{author}{\bibinfo{person}{R. {Russell}}, \bibinfo{person}{L. {Kim}},
  \bibinfo{person}{L. {Hamilton}}, \bibinfo{person}{T. {Lazovich}},
  \bibinfo{person}{J. {Harer}}, \bibinfo{person}{O. {Ozdemir}},
  \bibinfo{person}{P. {Ellingwood}}, {and} \bibinfo{person}{M. {McConley}}.}
  \bibinfo{year}{2018}\natexlab{}.
\newblock \showarticletitle{Automated Vulnerability Detection in Source Code
  Using Deep Representation Learning}. In \bibinfo{booktitle}{\emph{2018 17th
  IEEE International Conference on Machine Learning and Applications (ICMLA)}}.
  \bibinfo{pages}{757--762}.
\newblock
\urldef\tempurl%
\url{https://doi.org/10.1109/ICMLA.2018.00120}
\showDOI{\tempurl}


\bibitem[\protect\citeauthoryear{Russell, Kim, Hamilton, Lazovich, Harer,
  Ozdemir, Ellingwood, and McConley}{Russell et~al\mbox{.}}{2018}]%
        {vdisc_2018}
\bibfield{author}{\bibinfo{person}{Rebecca Russell}, \bibinfo{person}{Louis
  Kim}, \bibinfo{person}{Lei Hamilton}, \bibinfo{person}{Tomo Lazovich},
  \bibinfo{person}{Jacob Harer}, \bibinfo{person}{Onur Ozdemir},
  \bibinfo{person}{Paul Ellingwood}, {and} \bibinfo{person}{Marc McConley}.}
  \bibinfo{year}{2018}\natexlab{}.
\newblock \showarticletitle{Automated Vulnerability Detection in Source Code
  Using Deep Representation Learning}. In \bibinfo{booktitle}{\emph{2018 17th
  IEEE International Conference on Machine Learning and Applications (ICMLA)}}.
  \bibinfo{pages}{757--762}.
\newblock
\urldef\tempurl%
\url{https://doi.org/10.1109/ICMLA.2018.00120}
\showDOI{\tempurl}


\bibitem[\protect\citeauthoryear{Sabetta and Bezzi}{Sabetta and Bezzi}{2018}]%
        {949_Sabetta2018}
\bibfield{author}{\bibinfo{person}{Antonino Sabetta} {and}
  \bibinfo{person}{Michele Bezzi}.} \bibinfo{year}{2018}\natexlab{}.
\newblock \showarticletitle{A practical approach to the automatic
  classification of security-relevant commits}. In
  \bibinfo{booktitle}{\emph{2018 IEEE International conference on software
  maintenance and evolution (ICSME)}}. IEEE, \bibinfo{pages}{579--582}.
\newblock


\bibitem[\protect\citeauthoryear{{Saccente}, {Dehlinger}, {Deng},
  {Chakraborty}, and {Xiong}}{{Saccente} et~al\mbox{.}}{2019}]%
        {Saccente2019_50}
\bibfield{author}{\bibinfo{person}{N. {Saccente}}, \bibinfo{person}{J.
  {Dehlinger}}, \bibinfo{person}{L. {Deng}}, \bibinfo{person}{S.
  {Chakraborty}}, {and} \bibinfo{person}{Y. {Xiong}}.}
  \bibinfo{year}{2019}\natexlab{}.
\newblock \showarticletitle{Project Achilles: A Prototype Tool for Static
  Method-Level Vulnerability Detection of Java Source Code Using a Recurrent
  Neural Network}. In \bibinfo{booktitle}{\emph{2019 34th IEEE/ACM
  International Conference on Automated Software Engineering Workshop (ASEW)}}.
  \bibinfo{pages}{114--121}.
\newblock
\urldef\tempurl%
\url{https://doi.org/10.1109/ASEW.2019.00040}
\showDOI{\tempurl}


\bibitem[\protect\citeauthoryear{Sachdev, Li, Luan, Kim, Sen, and
  Chandra}{Sachdev et~al\mbox{.}}{2018}]%
        {Sachdev2018_444}
\bibfield{author}{\bibinfo{person}{Saksham Sachdev}, \bibinfo{person}{Hongyu
  Li}, \bibinfo{person}{Sifei Luan}, \bibinfo{person}{Seohyun Kim},
  \bibinfo{person}{Koushik Sen}, {and} \bibinfo{person}{Satish Chandra}.}
  \bibinfo{year}{2018}\natexlab{}.
\newblock \showarticletitle{Retrieval on Source Code: A Neural Code Search}. In
  \bibinfo{booktitle}{\emph{Proceedings of the 2nd ACM SIGPLAN International
  Workshop on Machine Learning and Programming Languages}} (Philadelphia, PA,
  USA) \emph{(\bibinfo{series}{MAPL 2018})}. \bibinfo{pages}{31–41}.
\newblock
\showISBNx{9781450358347}
\urldef\tempurl%
\url{https://doi.org/10.1145/3211346.3211353}
\showDOI{\tempurl}


\bibitem[\protect\citeauthoryear{Sagar, AlOmar, Mkaouer, Ouni, and
  Newman}{Sagar et~al\mbox{.}}{2021}]%
        {733_Sagar2021}
\bibfield{author}{\bibinfo{person}{Priyadarshni~Suresh Sagar},
  \bibinfo{person}{Eman~Abdulah AlOmar}, \bibinfo{person}{Mohamed~Wiem
  Mkaouer}, \bibinfo{person}{Ali Ouni}, {and} \bibinfo{person}{Christian~D.
  Newman}.} \bibinfo{year}{2021}\natexlab{}.
\newblock \showarticletitle{Comparing Commit Messages and Source Code Metrics
  for the Prediction Refactoring Activities}.
\newblock \bibinfo{journal}{\emph{Algorithms}} \bibinfo{volume}{14},
  \bibinfo{number}{10} (\bibinfo{year}{2021}).
\newblock
\showISSN{1999-4893}
\urldef\tempurl%
\url{https://doi.org/10.3390/733_Sagar2021}
\showDOI{\tempurl}


\bibitem[\protect\citeauthoryear{{Saha}, {Lyu}, {Yoshida}, and {Prasad}}{{Saha}
  et~al\mbox{.}}{2017}]%
        {Saha2017_306}
\bibfield{author}{\bibinfo{person}{R.~K. {Saha}}, \bibinfo{person}{Y. {Lyu}},
  \bibinfo{person}{H. {Yoshida}}, {and} \bibinfo{person}{M.~R. {Prasad}}.}
  \bibinfo{year}{2017}\natexlab{}.
\newblock \showarticletitle{Elixir: Effective object-oriented program repair}.
  In \bibinfo{booktitle}{\emph{2017 32nd IEEE/ACM International Conference on
  Automated Software Engineering (ASE)}}. \bibinfo{pages}{648--659}.
\newblock
\urldef\tempurl%
\url{https://doi.org/10.1109/ASE.2017.8115675}
\showDOI{\tempurl}


\bibitem[\protect\citeauthoryear{{Saha}, k.~{Saha}, and r.~{Prasad}}{{Saha}
  et~al\mbox{.}}{2019}]%
        {Saha2019_315}
\bibfield{author}{\bibinfo{person}{S. {Saha}}, \bibinfo{person}{R. k. {Saha}},
  {and} \bibinfo{person}{M. r. {Prasad}}.} \bibinfo{year}{2019}\natexlab{}.
\newblock \showarticletitle{Harnessing Evolution for Multi-Hunk Program
  Repair}. In \bibinfo{booktitle}{\emph{2019 IEEE/ACM 41st International
  Conference on Software Engineering (ICSE)}}. \bibinfo{pages}{13--24}.
\newblock
\urldef\tempurl%
\url{https://doi.org/10.1109/ICSE.2019.00020}
\showDOI{\tempurl}


\bibitem[\protect\citeauthoryear{Saidani, Ouni, and Mkaouer}{Saidani
  et~al\mbox{.}}{2020}]%
        {787_Saidani2020}
\bibfield{author}{\bibinfo{person}{Islem Saidani}, \bibinfo{person}{Ali Ouni},
  {and} \bibinfo{person}{Mohamed~Wiem Mkaouer}.}
  \bibinfo{year}{2020}\natexlab{}.
\newblock \showarticletitle{Web Service API Anti-patterns Detection as a
  Multi-label Learning Problem}. In \bibinfo{booktitle}{\emph{International
  Conference on Web Services}}. Springer, \bibinfo{pages}{114--132}.
\newblock


\bibitem[\protect\citeauthoryear{Sainath, Kingsbury, Saon, Soltau, Mohamed,
  Dahl, and Ramabhadran}{Sainath et~al\mbox{.}}{2015}]%
        {Sainath2015}
\bibfield{author}{\bibinfo{person}{Tara~N Sainath}, \bibinfo{person}{Brian
  Kingsbury}, \bibinfo{person}{George Saon}, \bibinfo{person}{Hagen Soltau},
  \bibinfo{person}{Abdel-rahman Mohamed}, \bibinfo{person}{George Dahl}, {and}
  \bibinfo{person}{Bhuvana Ramabhadran}.} \bibinfo{year}{2015}\natexlab{}.
\newblock \showarticletitle{Deep convolutional neural networks for large-scale
  speech tasks}.
\newblock \bibinfo{journal}{\emph{Neural Networks}}  \bibinfo{volume}{64}
  (\bibinfo{year}{2015}), \bibinfo{pages}{39--48}.
\newblock


\bibitem[\protect\citeauthoryear{Sakkas, Endres, Cosman, Weimer, and
  Jhala}{Sakkas et~al\mbox{.}}{2020}]%
        {Sakkas2020_366}
\bibfield{author}{\bibinfo{person}{Georgios Sakkas}, \bibinfo{person}{Madeline
  Endres}, \bibinfo{person}{Benjamin Cosman}, \bibinfo{person}{Westley Weimer},
  {and} \bibinfo{person}{Ranjit Jhala}.} \bibinfo{year}{2020}\natexlab{}.
\newblock \showarticletitle{Type Error Feedback via Analytic Program Repair}.
  In \bibinfo{booktitle}{\emph{Proceedings of the 41st ACM SIGPLAN Conference
  on Programming Language Design and Implementation}} (London, UK)
  \emph{(\bibinfo{series}{PLDI 2020})}. \bibinfo{pages}{16–30}.
\newblock
\showISBNx{9781450376136}
\urldef\tempurl%
\url{https://doi.org/10.1145/3385412.3386005}
\showDOI{\tempurl}


\bibitem[\protect\citeauthoryear{Sankaran, Aralikatte, Mani, Khare, Panwar, and
  Gantayat}{Sankaran et~al\mbox{.}}{2017}]%
        {Sankaran2017}
\bibfield{author}{\bibinfo{person}{Anush Sankaran}, \bibinfo{person}{Rahul
  Aralikatte}, \bibinfo{person}{Senthil Mani}, \bibinfo{person}{Shreya Khare},
  \bibinfo{person}{Naveen Panwar}, {and} \bibinfo{person}{Neelamadhav
  Gantayat}.} \bibinfo{year}{2017}\natexlab{}.
\newblock \showarticletitle{{DARVIZ:} Deep Abstract Representation,
  Visualization, and Verification of Deep Learning Models}.
\newblock \bibinfo{journal}{\emph{CoRR}}  \bibinfo{volume}{abs/1708.04915}
  (\bibinfo{year}{2017}).
\newblock
\showeprint[arxiv]{1708.04915}
\urldef\tempurl%
\url{http://arxiv.org/abs/1708.04915}
\showURL{%
\tempurl}


\bibitem[\protect\citeauthoryear{{Santos}, {Campbell}, {Patel}, {Hindle}, and
  {Amaral}}{{Santos} et~al\mbox{.}}{2018}]%
        {Santos2018_358}
\bibfield{author}{\bibinfo{person}{E.~A. {Santos}}, \bibinfo{person}{J.~C.
  {Campbell}}, \bibinfo{person}{D. {Patel}}, \bibinfo{person}{A. {Hindle}},
  {and} \bibinfo{person}{J.~N. {Amaral}}.} \bibinfo{year}{2018}\natexlab{}.
\newblock \showarticletitle{Syntax and sensibility: Using language models to
  detect and correct syntax errors}. In \bibinfo{booktitle}{\emph{2018 IEEE
  25th International Conference on Software Analysis, Evolution and
  Reengineering (SANER)}}. \bibinfo{pages}{311--322}.
\newblock
\urldef\tempurl%
\url{https://doi.org/10.1109/SANER.2018.8330219}
\showDOI{\tempurl}


\bibitem[\protect\citeauthoryear{Santos, Devesa, Brezo, Nieves, and
  Bringas}{Santos et~al\mbox{.}}{2013}]%
        {Santos2013_46}
\bibfield{author}{\bibinfo{person}{Igor Santos}, \bibinfo{person}{Jaime
  Devesa}, \bibinfo{person}{F{\'e}lix Brezo}, \bibinfo{person}{Javier Nieves},
  {and} \bibinfo{person}{Pablo~Garcia Bringas}.}
  \bibinfo{year}{2013}\natexlab{}.
\newblock \showarticletitle{OPEM: A Static-Dynamic Approach for
  Machine-Learning-Based Malware Detection}. In
  \bibinfo{booktitle}{\emph{International Joint Conference
  CISIS'12-ICEUTE{\textasciiacute}12-SOCO{\textasciiacute}12 Special
  Sessions}}, \bibfield{editor}{\bibinfo{person}{{\'A}lvaro Herrero},
  \bibinfo{person}{V{\'a}clav Sn{\'a}{\v{s}}el}, \bibinfo{person}{Ajith
  Abraham}, \bibinfo{person}{Ivan Zelinka}, \bibinfo{person}{Bruno Baruque},
  \bibinfo{person}{H{\'e}ctor Quinti{\'a}n}, \bibinfo{person}{Jos{\'e}~Luis
  Calvo}, \bibinfo{person}{Javier Sedano}, {and} \bibinfo{person}{Emilio
  Corchado}} (Eds.). \bibinfo{pages}{271--280}.
\newblock
\showISBNx{978-3-642-33018-6}


\bibitem[\protect\citeauthoryear{Sarro, Di~Martino, Ferrucci, and
  Gravino}{Sarro et~al\mbox{.}}{2012}]%
        {Sarro2012}
\bibfield{author}{\bibinfo{person}{F. Sarro}, \bibinfo{person}{S. Di~Martino},
  \bibinfo{person}{F. Ferrucci}, {and} \bibinfo{person}{C. Gravino}.}
  \bibinfo{year}{2012}\natexlab{}.
\newblock \showarticletitle{A Further Analysis on the Use of Genetic Algorithm
  to Configure Support Vector Machines for Inter-Release Fault Prediction}. In
  \bibinfo{booktitle}{\emph{Proceedings of the 27th Annual ACM Symposium on
  Applied Computing}} (Trento, Italy) \emph{(\bibinfo{series}{SAC '12})}.
  \bibinfo{publisher}{Association for Computing Machinery},
  \bibinfo{address}{New York, NY, USA}, \bibinfo{pages}{1215–1220}.
\newblock
\showISBNx{9781450308571}
\urldef\tempurl%
\url{https://doi.org/10.1145/2245276.2231967}
\showDOI{\tempurl}


\bibitem[\protect\citeauthoryear{Sayyad~Shirabad and Menzies}{Sayyad~Shirabad
  and Menzies}{2005}]%
        {Promise_dataset_2005}
\bibfield{author}{\bibinfo{person}{J. Sayyad~Shirabad} {and}
  \bibinfo{person}{T.J. Menzies}.} \bibinfo{year}{2005}\natexlab{}.
\newblock \bibinfo{title}{{The {PROMISE} Repository of Software Engineering
  Databases.}}
\newblock \bibinfo{howpublished}{School of Information Technology and
  Engineering, University of Ottawa, Canada}.
\newblock
\urldef\tempurl%
\url{http://promise.site.uottawa.ca/SERepository}
\showURL{%
\tempurl}


\bibitem[\protect\citeauthoryear{Schumacher, Le, and Andrzejak}{Schumacher
  et~al\mbox{.}}{2020}]%
        {Schumacher2020_497}
\bibfield{author}{\bibinfo{person}{Max Eric~Henry Schumacher},
  \bibinfo{person}{Kim~Tuyen Le}, {and} \bibinfo{person}{Artur Andrzejak}.}
  \bibinfo{year}{2020}\natexlab{}.
\newblock \showarticletitle{Improving Code Recommendations by Combining Neural
  and Classical Machine Learning Approaches}. In
  \bibinfo{booktitle}{\emph{Proceedings of the IEEE/ACM 42nd International
  Conference on Software Engineering Workshops}} (Seoul, Republic of Korea)
  \emph{(\bibinfo{series}{ICSEW'20})}. \bibinfo{pages}{476–482}.
\newblock
\showISBNx{9781450379632}
\urldef\tempurl%
\url{https://doi.org/10.1145/3387940.3391489}
\showDOI{\tempurl}


\bibitem[\protect\citeauthoryear{Schuster, Song, Tromer, and
  Shmatikov}{Schuster et~al\mbox{.}}{2021}]%
        {Schuster2021_508}
\bibfield{author}{\bibinfo{person}{R. Schuster}, \bibinfo{person}{Congzheng
  Song}, \bibinfo{person}{Eran Tromer}, {and} \bibinfo{person}{Vitaly
  Shmatikov}.} \bibinfo{year}{2021}\natexlab{}.
\newblock \showarticletitle{You Autocomplete Me: Poisoning Vulnerabilities in
  Neural Code Completion}. In \bibinfo{booktitle}{\emph{30th {USENIX} Security
  Symposium ({USENIX} Security 21)}}.
\newblock


\bibitem[\protect\citeauthoryear{{Sethi} and {Gagandeep}}{{Sethi} and
  {Gagandeep}}{2016}]%
        {Sethi2016_93}
\bibfield{author}{\bibinfo{person}{T. {Sethi}} {and}
  \bibinfo{person}{{Gagandeep}}.} \bibinfo{year}{2016}\natexlab{}.
\newblock \showarticletitle{Improved approach for software defect prediction
  using artificial neural networks}. In \bibinfo{booktitle}{\emph{2016 5th
  International Conference on Reliability, Infocom Technologies and
  Optimization (Trends and Future Directions) (ICRITO)}}.
  \bibinfo{pages}{480--485}.
\newblock
\urldef\tempurl%
\url{https://doi.org/10.1109/ICRITO.2016.7785003}
\showDOI{\tempurl}


\bibitem[\protect\citeauthoryear{Settles}{Settles}{2009}]%
        {Settles2009}
\bibfield{author}{\bibinfo{person}{Burr Settles}.}
  \bibinfo{year}{2009}\natexlab{}.
\newblock \showarticletitle{Active learning literature survey}.
\newblock  (\bibinfo{year}{2009}).
\newblock


\bibitem[\protect\citeauthoryear{Shabtai, Moskovitch, Elovici, and
  Glezer}{Shabtai et~al\mbox{.}}{2009}]%
        {Shabtai2009_25}
\bibfield{author}{\bibinfo{person}{Asaf Shabtai}, \bibinfo{person}{Robert
  Moskovitch}, \bibinfo{person}{Yuval Elovici}, {and} \bibinfo{person}{Chanan
  Glezer}.} \bibinfo{year}{2009}\natexlab{}.
\newblock \showarticletitle{Detection of malicious code by applying machine
  learning classifiers on static features: A state-of-the-art survey}.
\newblock \bibinfo{journal}{\emph{Information Security Technical Report}}
  \bibinfo{volume}{14}, \bibinfo{number}{1} (\bibinfo{year}{2009}),
  \bibinfo{pages}{16 -- 29}.
\newblock
\showISSN{1363-4127}
\urldef\tempurl%
\url{https://doi.org/10.1016/j.istr.2009.03.003}
\showDOI{\tempurl}
\newblock
\shownote{Malware.}


\bibitem[\protect\citeauthoryear{{Shar}, {Briand}, and {Tan}}{{Shar}
  et~al\mbox{.}}{2015}]%
        {Shar2015_64}
\bibfield{author}{\bibinfo{person}{L.~K. {Shar}}, \bibinfo{person}{L.~C.
  {Briand}}, {and} \bibinfo{person}{H.~B.~K. {Tan}}.}
  \bibinfo{year}{2015}\natexlab{}.
\newblock \showarticletitle{Web Application Vulnerability Prediction Using
  Hybrid Program Analysis and Machine Learning}.
\newblock \bibinfo{journal}{\emph{IEEE Transactions on Dependable and Secure
  Computing}} \bibinfo{volume}{12}, \bibinfo{number}{6} (\bibinfo{year}{2015}),
  \bibinfo{pages}{688--707}.
\newblock
\urldef\tempurl%
\url{https://doi.org/10.1109/TDSC.2014.2373377}
\showDOI{\tempurl}


\bibitem[\protect\citeauthoryear{Sharma}{Sharma}{2018}]%
        {DesigniteJava}
\bibfield{author}{\bibinfo{person}{Tushar Sharma}.}
  \bibinfo{year}{2018}\natexlab{}.
\newblock \bibinfo{title}{{DesigniteJava}}.
\newblock
\newblock
\urldef\tempurl%
\url{https://doi.org/10.5281/zenodo.2566861}
\showDOI{\tempurl}
\newblock
\shownote{https://github.com/tushartushar/DesigniteJava.}


\bibitem[\protect\citeauthoryear{Sharma}{Sharma}{2019a}]%
        {CodeSplitCS}
\bibfield{author}{\bibinfo{person}{Tushar Sharma}.}
  \bibinfo{year}{2019}\natexlab{a}.
\newblock \bibinfo{title}{{CodeSplit for C\#}}.
\newblock
\newblock
\urldef\tempurl%
\url{https://doi.org/10.5281/zenodo.2566905}
\showDOI{\tempurl}


\bibitem[\protect\citeauthoryear{Sharma}{Sharma}{2019b}]%
        {CodeSplitJava}
\bibfield{author}{\bibinfo{person}{Tushar Sharma}.}
  \bibinfo{year}{2019}\natexlab{b}.
\newblock \bibinfo{title}{{CodeSplitJava}}.
\newblock
\newblock
\urldef\tempurl%
\url{https://doi.org/10.5281/zenodo.2566865}
\showDOI{\tempurl}
\newblock
\shownote{https://github.com/tushartushar/CodeSplitJava.}


\bibitem[\protect\citeauthoryear{Sharma, Efstathiou, Louridas, and
  Spinellis}{Sharma et~al\mbox{.}}{2021}]%
        {Sharma2021_510}
\bibfield{author}{\bibinfo{person}{Tushar Sharma}, \bibinfo{person}{Vasiliki
  Efstathiou}, \bibinfo{person}{Panos Louridas}, {and}
  \bibinfo{person}{Diomidis Spinellis}.} \bibinfo{year}{2021}\natexlab{}.
\newblock \showarticletitle{Code smell detection by deep direct-learning and
  transfer-learning}.
\newblock \bibinfo{journal}{\emph{Journal of Systems and Software}}
  \bibinfo{volume}{176} (\bibinfo{year}{2021}), \bibinfo{pages}{110936}.
\newblock
\showISSN{0164-1212}
\urldef\tempurl%
\url{https://doi.org/10.1016/j.jss.2021.110936}
\showDOI{\tempurl}


\bibitem[\protect\citeauthoryear{Sharma, Kechagia, Georgiou, Tiwari, Vats,
  Moazen, and Sarro}{Sharma et~al\mbox{.}}{2022}]%
        {Replication_ML4SCA}
\bibfield{author}{\bibinfo{person}{Tushar Sharma}, \bibinfo{person}{Maria
  Kechagia}, \bibinfo{person}{Stefanos Georgiou}, \bibinfo{person}{Rohit
  Tiwari}, \bibinfo{person}{Indira Vats}, \bibinfo{person}{Hadi Moazen}, {and}
  \bibinfo{person}{Federica Sarro}.} \bibinfo{year}{2022}\natexlab{}.
\newblock \bibinfo{booktitle}{\emph{{Replication package for Machine Learning
  for Source Code Analysis survey paper}}}.
\newblock
\urldef\tempurl%
\url{https://github.com/tushartushar/ML4SCA}
\showURL{%
\tempurl}


\bibitem[\protect\citeauthoryear{Sharma and Kessentini}{Sharma and
  Kessentini}{2021}]%
        {QScored}
\bibfield{author}{\bibinfo{person}{T. Sharma} {and} \bibinfo{person}{M.
  Kessentini}.} \bibinfo{year}{2021}\natexlab{}.
\newblock \showarticletitle{QScored: A Large Dataset of Code Smells and Quality
  Metrics}. In \bibinfo{booktitle}{\emph{2021 2021 IEEE/ACM 18th International
  Conference on Mining Software Repositories (MSR) (MSR)}}.
  \bibinfo{publisher}{IEEE Computer Society}, \bibinfo{address}{Los Alamitos,
  CA, USA}, \bibinfo{pages}{590--594}.
\newblock
\urldef\tempurl%
\url{https://doi.org/10.1109/MSR52588.2021.00080}
\showDOI{\tempurl}


\bibitem[\protect\citeauthoryear{Sharma, Mishra, and Tiwari}{Sharma
  et~al\mbox{.}}{2016}]%
        {Designite}
\bibfield{author}{\bibinfo{person}{Tushar Sharma}, \bibinfo{person}{Pratibha
  Mishra}, {and} \bibinfo{person}{Rohit Tiwari}.}
  \bibinfo{year}{2016}\natexlab{}.
\newblock \showarticletitle{{Designite --- A Software Design Quality Assessment
  Tool}}. In \bibinfo{booktitle}{\emph{Proceedings of the First International
  Workshop on Bringing Architecture Design Thinking into Developers' Daily
  Activities}} \emph{(\bibinfo{series}{BRIDGE '16})}.
\newblock
\urldef\tempurl%
\url{https://doi.org/10.1145/2896935.2896938}
\showDOI{\tempurl}


\bibitem[\protect\citeauthoryear{Sharma and Spinellis}{Sharma and
  Spinellis}{2018}]%
        {Sharma2018}
\bibfield{author}{\bibinfo{person}{Tushar Sharma} {and}
  \bibinfo{person}{Diomidis Spinellis}.} \bibinfo{year}{2018}\natexlab{}.
\newblock \showarticletitle{A survey on software smells}.
\newblock \bibinfo{journal}{\emph{Journal of Systems and Software}}
  \bibinfo{volume}{138} (\bibinfo{year}{2018}), \bibinfo{pages}{158--173}.
\newblock
\showISSN{0164-1212}
\urldef\tempurl%
\url{https://doi.org/10.1016/j.jss.2017.12.034}
\showDOI{\tempurl}


\bibitem[\protect\citeauthoryear{Shedko, Palachev, Kvochko, Semenov, and
  Sun}{Shedko et~al\mbox{.}}{2020}]%
        {Shedko2020_454}
\bibfield{author}{\bibinfo{person}{Andrey Shedko}, \bibinfo{person}{Ilya
  Palachev}, \bibinfo{person}{Andrey Kvochko}, \bibinfo{person}{Aleksandr
  Semenov}, {and} \bibinfo{person}{Kwangwon Sun}.}
  \bibinfo{year}{2020}\natexlab{}.
\newblock \showarticletitle{Applying Probabilistic Models to C++ Code on an
  Industrial Scale}. In \bibinfo{booktitle}{\emph{Proceedings of the IEEE/ACM
  42nd International Conference on Software Engineering Workshops}} (Seoul,
  Republic of Korea) \emph{(\bibinfo{series}{ICSEW'20})}.
  \bibinfo{pages}{595–602}.
\newblock
\showISBNx{9781450379632}
\urldef\tempurl%
\url{https://doi.org/10.1145/3387940.3391477}
\showDOI{\tempurl}


\bibitem[\protect\citeauthoryear{Shen and Chen}{Shen and Chen}{2020}]%
        {Shen2020_9}
\bibfield{author}{\bibinfo{person}{Zhidong Shen} {and} \bibinfo{person}{S.
  Chen}.} \bibinfo{year}{2020}\natexlab{}.
\newblock \showarticletitle{A Survey of Automatic Software Vulnerability
  Detection, Program Repair, and Defect Prediction Techniques}.
\newblock \bibinfo{journal}{\emph{Secur. Commun. Networks}}
  \bibinfo{volume}{2020} (\bibinfo{year}{2020}),
  \bibinfo{pages}{8858010:1--8858010:16}.
\newblock


\bibitem[\protect\citeauthoryear{{Sheneamer} and {Kalita}}{{Sheneamer} and
  {Kalita}}{2016}]%
        {Sheneamer2016_260}
\bibfield{author}{\bibinfo{person}{A. {Sheneamer}} {and} \bibinfo{person}{J.
  {Kalita}}.} \bibinfo{year}{2016}\natexlab{}.
\newblock \showarticletitle{Semantic Clone Detection Using Machine Learning}.
  In \bibinfo{booktitle}{\emph{2016 15th IEEE International Conference on
  Machine Learning and Applications (ICMLA)}}. \bibinfo{pages}{1024--1028}.
\newblock
\urldef\tempurl%
\url{https://doi.org/10.1109/ICMLA.2016.0185}
\showDOI{\tempurl}


\bibitem[\protect\citeauthoryear{Shi, Lu, Chang, and Wei}{Shi
  et~al\mbox{.}}{2020}]%
        {Shi2020_116}
\bibfield{author}{\bibinfo{person}{Ke Shi}, \bibinfo{person}{Yang Lu},
  \bibinfo{person}{Jingfei Chang}, {and} \bibinfo{person}{Zhen Wei}.}
  \bibinfo{year}{2020}\natexlab{}.
\newblock \showarticletitle{PathPair2Vec: An AST path pair-based code
  representation method for defect prediction}.
\newblock \bibinfo{journal}{\emph{Journal of Computer Languages}}
  \bibinfo{volume}{59} (\bibinfo{year}{2020}), \bibinfo{pages}{100979}.
\newblock
\showISSN{2590-1184}
\urldef\tempurl%
\url{https://doi.org/10.1016/j.cola.2020.100979}
\showDOI{\tempurl}


\bibitem[\protect\citeauthoryear{{Shido}, {Kobayashi}, {Yamamoto}, {Miyamoto},
  and {Matsumura}}{{Shido} et~al\mbox{.}}{2019}]%
        {Shido2019_415}
\bibfield{author}{\bibinfo{person}{Y. {Shido}}, \bibinfo{person}{Y.
  {Kobayashi}}, \bibinfo{person}{A. {Yamamoto}}, \bibinfo{person}{A.
  {Miyamoto}}, {and} \bibinfo{person}{T. {Matsumura}}.}
  \bibinfo{year}{2019}\natexlab{}.
\newblock \showarticletitle{Automatic Source Code Summarization with Extended
  Tree-LSTM}. In \bibinfo{booktitle}{\emph{2019 International Joint Conference
  on Neural Networks (IJCNN)}}. \bibinfo{pages}{1--8}.
\newblock
\urldef\tempurl%
\url{https://doi.org/10.1109/IJCNN.2019.8851751}
\showDOI{\tempurl}


\bibitem[\protect\citeauthoryear{{Shim}, {Patil}, {Yadav}, {Shinde}, and
  {Devale}}{{Shim} et~al\mbox{.}}{2020}]%
        {Shim2020_301}
\bibfield{author}{\bibinfo{person}{S. {Shim}}, \bibinfo{person}{P. {Patil}},
  \bibinfo{person}{R.~R. {Yadav}}, \bibinfo{person}{A. {Shinde}}, {and}
  \bibinfo{person}{V. {Devale}}.} \bibinfo{year}{2020}\natexlab{}.
\newblock \showarticletitle{{DeeperCoder}: Code Generation Using Machine
  Learning}. In \bibinfo{booktitle}{\emph{2020 10th Annual Computing and
  Communication Workshop and Conference (CCWC)}}. \bibinfo{pages}{0194--0199}.
\newblock
\urldef\tempurl%
\url{https://doi.org/10.1109/CCWC47524.2020.9031149}
\showDOI{\tempurl}


\bibitem[\protect\citeauthoryear{{Shimonaka}, {Sumi}, {Higo}, and
  {Kusumoto}}{{Shimonaka} et~al\mbox{.}}{2016}]%
        {Shimonaka2016_381}
\bibfield{author}{\bibinfo{person}{K. {Shimonaka}}, \bibinfo{person}{S.
  {Sumi}}, \bibinfo{person}{Y. {Higo}}, {and} \bibinfo{person}{S. {Kusumoto}}.}
  \bibinfo{year}{2016}\natexlab{}.
\newblock \showarticletitle{Identifying Auto-Generated Code by Using Machine
  Learning Techniques}. In \bibinfo{booktitle}{\emph{2016 7th International
  Workshop on Empirical Software Engineering in Practice (IWESEP)}}.
  \bibinfo{pages}{18--23}.
\newblock
\urldef\tempurl%
\url{https://doi.org/10.1109/IWESEP.2016.18}
\showDOI{\tempurl}


\bibitem[\protect\citeauthoryear{Shiqi, Shengwei, Long, Jiong, and Hua}{Shiqi
  et~al\mbox{.}}{2018}]%
        {shiqi_android_2018}
\bibfield{author}{\bibinfo{person}{L. Shiqi}, \bibinfo{person}{T. Shengwei},
  \bibinfo{person}{Y. Long}, \bibinfo{person}{Y. Jiong}, {and}
  \bibinfo{person}{S. Hua}.} \bibinfo{year}{2018}\natexlab{}.
\newblock \showarticletitle{Android malicious code {Classification} using
  {Deep} {Belief} {Network}}.
\newblock \bibinfo{journal}{\emph{KSII Transactions on Internet and Information
  Systems}}  \bibinfo{volume}{12} (\bibinfo{date}{Jan.} \bibinfo{year}{2018}),
  \bibinfo{pages}{454--475}.
\newblock
\urldef\tempurl%
\url{https://doi.org/10.3837/tiis.2018.01.022}
\showDOI{\tempurl}


\bibitem[\protect\citeauthoryear{Shuai, Xu, Liu, Yan, Xia, and Lei}{Shuai
  et~al\mbox{.}}{2020}]%
        {Shuai2020_442}
\bibfield{author}{\bibinfo{person}{Jianhang Shuai}, \bibinfo{person}{Ling Xu},
  \bibinfo{person}{Chao Liu}, \bibinfo{person}{Meng Yan}, \bibinfo{person}{Xin
  Xia}, {and} \bibinfo{person}{Yan Lei}.} \bibinfo{year}{2020}\natexlab{}.
\newblock \showarticletitle{Improving Code Search with Co-Attentive
  Representation Learning}. In \bibinfo{booktitle}{\emph{Proceedings of the
  28th International Conference on Program Comprehension}} (Seoul, Republic of
  Korea) \emph{(\bibinfo{series}{ICPC '20})}. \bibinfo{pages}{196–207}.
\newblock
\showISBNx{9781450379588}
\urldef\tempurl%
\url{https://doi.org/10.1145/3387904.3389269}
\showDOI{\tempurl}


\bibitem[\protect\citeauthoryear{Sidhu, Singh, and Sharma}{Sidhu
  et~al\mbox{.}}{2022}]%
        {734_Sidhu2022}
\bibfield{author}{\bibinfo{person}{Brahmaleen~Kaur Sidhu},
  \bibinfo{person}{Kawaljeet Singh}, {and} \bibinfo{person}{Neeraj Sharma}.}
  \bibinfo{year}{2022}\natexlab{}.
\newblock \showarticletitle{A machine learning approach to software model
  refactoring}.
\newblock \bibinfo{journal}{\emph{International Journal of Computers and
  Applications}} \bibinfo{volume}{44}, \bibinfo{number}{2}
  (\bibinfo{year}{2022}), \bibinfo{pages}{166--177}.
\newblock
\urldef\tempurl%
\url{https://doi.org/10.1080/1206212X.2020.1711616}
\showDOI{\tempurl}
\showeprint{https://doi.org/10.1080/1206212X.2020.1711616}


\bibitem[\protect\citeauthoryear{Singh, Bhatia, and Singhrova}{Singh
  et~al\mbox{.}}{2018}]%
        {710_Singh2018}
\bibfield{author}{\bibinfo{person}{Ajmer Singh}, \bibinfo{person}{Rajesh
  Bhatia}, {and} \bibinfo{person}{Anita Singhrova}.}
  \bibinfo{year}{2018}\natexlab{}.
\newblock \showarticletitle{Taxonomy of machine learning algorithms in software
  fault prediction using object oriented metrics}.
\newblock \bibinfo{journal}{\emph{Procedia computer science}}
  \bibinfo{volume}{132} (\bibinfo{year}{2018}), \bibinfo{pages}{993--1001}.
\newblock


\bibitem[\protect\citeauthoryear{{Singh} and {Chug}}{{Singh} and
  {Chug}}{2017}]%
        {Singh2017_130}
\bibfield{author}{\bibinfo{person}{P. {Singh}} {and} \bibinfo{person}{A.
  {Chug}}.} \bibinfo{year}{2017}\natexlab{}.
\newblock \showarticletitle{Software defect prediction analysis using machine
  learning algorithms}. In \bibinfo{booktitle}{\emph{2017 7th International
  Conference on Cloud Computing, Data Science Engineering - Confluence}}.
  \bibinfo{pages}{775--781}.
\newblock
\urldef\tempurl%
\url{https://doi.org/10.1109/CONFLUENCE.2017.7943255}
\showDOI{\tempurl}


\bibitem[\protect\citeauthoryear{{Singh} and {Malhotra}}{{Singh} and
  {Malhotra}}{2017}]%
        {Singh2017_76}
\bibfield{author}{\bibinfo{person}{P. {Singh}} {and} \bibinfo{person}{R.
  {Malhotra}}.} \bibinfo{year}{2017}\natexlab{}.
\newblock \showarticletitle{Assessment of machine learning algorithms for
  determining defective classes in an object-oriented software}. In
  \bibinfo{booktitle}{\emph{2017 6th International Conference on Reliability,
  Infocom Technologies and Optimization (Trends and Future Directions)
  (ICRITO)}}. \bibinfo{pages}{204--209}.
\newblock
\urldef\tempurl%
\url{https://doi.org/10.1109/ICRITO.2017.8342425}
\showDOI{\tempurl}


\bibitem[\protect\citeauthoryear{{Singh}, {Singh}, {Gill}, {Malhotra}, and
  {Garima}}{{Singh} et~al\mbox{.}}{2020}]%
        {Singh2020_146}
\bibfield{author}{\bibinfo{person}{R. {Singh}}, \bibinfo{person}{J. {Singh}},
  \bibinfo{person}{M.~S. {Gill}}, \bibinfo{person}{R. {Malhotra}}, {and}
  \bibinfo{person}{{Garima}}.} \bibinfo{year}{2020}\natexlab{}.
\newblock \showarticletitle{Transfer Learning Code Vectorizer based Machine
  Learning Models for Software Defect Prediction}. In
  \bibinfo{booktitle}{\emph{2020 International Conference on Computational
  Performance Evaluation (ComPE)}}. \bibinfo{pages}{497--502}.
\newblock
\urldef\tempurl%
\url{https://doi.org/10.1109/ComPE49325.2020.9200076}
\showDOI{\tempurl}


\bibitem[\protect\citeauthoryear{Soltanifar, Akbarinasaji, Caglayan, Bener,
  Filiz, and Kramer}{Soltanifar et~al\mbox{.}}{2016}]%
        {833_Soltanifar2016}
\bibfield{author}{\bibinfo{person}{Behjat Soltanifar}, \bibinfo{person}{Shirin
  Akbarinasaji}, \bibinfo{person}{Bora Caglayan}, \bibinfo{person}{Ayse~Basar
  Bener}, \bibinfo{person}{Asli Filiz}, {and} \bibinfo{person}{Bryan~M
  Kramer}.} \bibinfo{year}{2016}\natexlab{}.
\newblock \showarticletitle{Software analytics in practice: a defect prediction
  model using code smells}. In \bibinfo{booktitle}{\emph{Proceedings of the
  20th International Database Engineering \& Applications Symposium}}.
  \bibinfo{pages}{148--155}.
\newblock


\bibitem[\protect\citeauthoryear{Song, Guo, and Shepperd}{Song
  et~al\mbox{.}}{2019}]%
        {676_Song2019}
\bibfield{author}{\bibinfo{person}{Qinbao Song}, \bibinfo{person}{Yuchen Guo},
  {and} \bibinfo{person}{Martin Shepperd}.} \bibinfo{year}{2019}\natexlab{}.
\newblock \showarticletitle{A Comprehensive Investigation of the Role of
  Imbalanced Learning for Software Defect Prediction}.
\newblock \bibinfo{journal}{\emph{IEEE Transactions on Software Engineering}}
  \bibinfo{volume}{45}, \bibinfo{number}{12} (\bibinfo{year}{2019}),
  \bibinfo{pages}{1253--1269}.
\newblock
\urldef\tempurl%
\url{https://doi.org/10.1109/TSE.2018.2836442}
\showDOI{\tempurl}


\bibitem[\protect\citeauthoryear{{Soto} and {Le Goues}}{{Soto} and {Le
  Goues}}{2018}]%
        {Soto2018_292}
\bibfield{author}{\bibinfo{person}{M. {Soto}} {and} \bibinfo{person}{C. {Le
  Goues}}.} \bibinfo{year}{2018}\natexlab{}.
\newblock \showarticletitle{Common Statement Kind Changes to Inform Automatic
  Program Repair}. In \bibinfo{booktitle}{\emph{2018 IEEE/ACM 15th
  International Conference on Mining Software Repositories (MSR)}}.
  \bibinfo{pages}{102--105}.
\newblock


\bibitem[\protect\citeauthoryear{Sotto-Mayor and Kalech}{Sotto-Mayor and
  Kalech}{2021}]%
        {829_Sotto-Mayor2021}
\bibfield{author}{\bibinfo{person}{Bruno Sotto-Mayor} {and}
  \bibinfo{person}{Meir Kalech}.} \bibinfo{year}{2021}\natexlab{}.
\newblock \showarticletitle{Cross-project smell-based defect prediction}.
\newblock \bibinfo{journal}{\emph{Soft Computing}} \bibinfo{volume}{25},
  \bibinfo{number}{22} (\bibinfo{year}{2021}), \bibinfo{pages}{14171--14181}.
\newblock


\bibitem[\protect\citeauthoryear{Spreitzenbarth, Schreck, Echtler, Arp, and
  Hoffmann}{Spreitzenbarth et~al\mbox{.}}{2014}]%
        {Spreitzenbarth2014_248}
\bibfield{author}{\bibinfo{person}{Michael Spreitzenbarth},
  \bibinfo{person}{Thomas Schreck}, \bibinfo{person}{F. Echtler},
  \bibinfo{person}{D. Arp}, {and} \bibinfo{person}{Johannes Hoffmann}.}
  \bibinfo{year}{2014}\natexlab{}.
\newblock \showarticletitle{Mobile-Sandbox: combining static and dynamic
  analysis with machine-learning techniques}.
\newblock \bibinfo{journal}{\emph{International Journal of Information
  Security}}  \bibinfo{volume}{14} (\bibinfo{year}{2014}),
  \bibinfo{pages}{141--153}.
\newblock


\bibitem[\protect\citeauthoryear{Stapleton, Gambhir, LeClair, Eberhart, Weimer,
  Leach, and Huang}{Stapleton et~al\mbox{.}}{2020}]%
        {Stapleton2020_404}
\bibfield{author}{\bibinfo{person}{Sean Stapleton}, \bibinfo{person}{Yashmeet
  Gambhir}, \bibinfo{person}{Alexander LeClair}, \bibinfo{person}{Zachary
  Eberhart}, \bibinfo{person}{Westley Weimer}, \bibinfo{person}{Kevin Leach},
  {and} \bibinfo{person}{Yu Huang}.} \bibinfo{year}{2020}\natexlab{}.
\newblock \showarticletitle{A Human Study of Comprehension and Code
  Summarization}. In \bibinfo{booktitle}{\emph{Proceedings of the 28th
  International Conference on Program Comprehension}} (Seoul, Republic of
  Korea) \emph{(\bibinfo{series}{ICPC '20})}. \bibinfo{pages}{2–13}.
\newblock
\showISBNx{9781450379588}
\urldef\tempurl%
\url{https://doi.org/10.1145/3387904.3389258}
\showDOI{\tempurl}


\bibitem[\protect\citeauthoryear{Storey}{Storey}{2005}]%
        {Storey2005}
\bibfield{author}{\bibinfo{person}{M.-A. Storey}.}
  \bibinfo{year}{2005}\natexlab{}.
\newblock \showarticletitle{Theories, methods and tools in program
  comprehension: past, present and future}. In \bibinfo{booktitle}{\emph{13th
  International Workshop on Program Comprehension (IWPC'05)}}.
  \bibinfo{pages}{181--191}.
\newblock
\urldef\tempurl%
\url{https://doi.org/10.1109/WPC.2005.38}
\showDOI{\tempurl}


\bibitem[\protect\citeauthoryear{Sui, Cheng, Zhang, and Wang}{Sui
  et~al\mbox{.}}{2020}]%
        {Sui2020_469}
\bibfield{author}{\bibinfo{person}{Yulei Sui}, \bibinfo{person}{Xiao Cheng},
  \bibinfo{person}{Guanqin Zhang}, {and} \bibinfo{person}{Haoyu Wang}.}
  \bibinfo{year}{2020}\natexlab{}.
\newblock \showarticletitle{Flow2Vec: Value-Flow-Based Precise Code Embedding}.
\newblock \bibinfo{journal}{\emph{Proc. ACM Program. Lang.}}
  \bibinfo{volume}{4}, \bibinfo{number}{OOPSLA}, Article
  \bibinfo{articleno}{233} (\bibinfo{date}{November} \bibinfo{year}{2020}),
  \bibinfo{numpages}{27}~pages.
\newblock
\urldef\tempurl%
\url{https://doi.org/10.1145/3428301}
\showDOI{\tempurl}


\bibitem[\protect\citeauthoryear{Sui and Xue}{Sui and Xue}{2016}]%
        {sui2016svf}
\bibfield{author}{\bibinfo{person}{Yulei Sui} {and} \bibinfo{person}{Jingling
  Xue}.} \bibinfo{year}{2016}\natexlab{}.
\newblock \showarticletitle{SVF: interprocedural static value-flow analysis in
  LLVM}. In \bibinfo{booktitle}{\emph{Proceedings of the 25th international
  conference on compiler construction}}. ACM, \bibinfo{pages}{265--266}.
\newblock


\bibitem[\protect\citeauthoryear{Sultana}{Sultana}{2017}]%
        {sultana_kazi_2017}
\bibfield{author}{\bibinfo{person}{Kazi~Zakia Sultana}.}
  \bibinfo{year}{2017}\natexlab{}.
\newblock \showarticletitle{Towards a software vulnerability prediction model
  using traceable code patterns and software metrics}. In
  \bibinfo{booktitle}{\emph{2017 32nd IEEE/ACM International Conference on
  Automated Software Engineering (ASE)}}. \bibinfo{pages}{1022--1025}.
\newblock
\urldef\tempurl%
\url{https://doi.org/10.1109/ASE.2017.8115724}
\showDOI{\tempurl}


\bibitem[\protect\citeauthoryear{Sultana, Anu, and Chong}{Sultana
  et~al\mbox{.}}{2021}]%
        {sultana_using_2021}
\bibfield{author}{\bibinfo{person}{Kazi~Zakia Sultana},
  \bibinfo{person}{Vaibhav Anu}, {and} \bibinfo{person}{Tai-Yin Chong}.}
  \bibinfo{year}{2021}\natexlab{}.
\newblock \showarticletitle{Using software metrics for predicting vulnerable
  classes and methods in {Java} projects: {A} machine learning approach}.
\newblock \bibinfo{journal}{\emph{Journal of Software: Evolution and Process}}
  \bibinfo{volume}{33}, \bibinfo{number}{3} (\bibinfo{year}{2021}),
  \bibinfo{pages}{e2303}.
\newblock
\showISSN{2047-7481}
\urldef\tempurl%
\url{https://doi.org/10.1002/smr.2303}
\showDOI{\tempurl}


\bibitem[\protect\citeauthoryear{Sun, Song, and Zhu}{Sun et~al\mbox{.}}{2012}]%
        {690_Sun2012}
\bibfield{author}{\bibinfo{person}{Zhongbin Sun}, \bibinfo{person}{Qinbao
  Song}, {and} \bibinfo{person}{Xiaoyan Zhu}.} \bibinfo{year}{2012}\natexlab{}.
\newblock \showarticletitle{Using coding-based ensemble learning to improve
  software defect prediction}.
\newblock \bibinfo{journal}{\emph{IEEE Transactions on Systems, Man, and
  Cybernetics, Part C (Applications and Reviews)}} \bibinfo{volume}{42},
  \bibinfo{number}{6} (\bibinfo{year}{2012}), \bibinfo{pages}{1806--1817}.
\newblock


\bibitem[\protect\citeauthoryear{Suresh, Kumar, and Rath}{Suresh
  et~al\mbox{.}}{2014}]%
        {717_Suresh2014}
\bibfield{author}{\bibinfo{person}{Yeresime Suresh}, \bibinfo{person}{Lov
  Kumar}, {and} \bibinfo{person}{Santanu~Ku Rath}.}
  \bibinfo{year}{2014}\natexlab{}.
\newblock \showarticletitle{Statistical and machine learning methods for
  software fault prediction using CK metric suite: a comparative analysis}.
\newblock \bibinfo{journal}{\emph{International Scholarly Research Notices}}
  \bibinfo{volume}{2014} (\bibinfo{year}{2014}).
\newblock


\bibitem[\protect\citeauthoryear{Suryanarayana, Samarthyam, and
  Sharma}{Suryanarayana et~al\mbox{.}}{2014}]%
        {Suryanarayana2014}
\bibfield{author}{\bibinfo{person}{Girish Suryanarayana},
  \bibinfo{person}{Ganesh Samarthyam}, {and} \bibinfo{person}{Tushar Sharma}.}
  \bibinfo{year}{2014}\natexlab{}.
\newblock \bibinfo{booktitle}{\emph{{Refactoring for Software Design Smells:
  Managing Technical Debt}} (\bibinfo{edition}{1} ed.)}.
\newblock \bibinfo{publisher}{Morgan Kaufmann}.
\newblock
\showISBNx{0128013974}


\bibitem[\protect\citeauthoryear{Svajlenko, Islam, Keivanloo, Roy, and
  Mia}{Svajlenko et~al\mbox{.}}{2014}]%
        {Svajlenko2014}
\bibfield{author}{\bibinfo{person}{Jeffrey Svajlenko},
  \bibinfo{person}{Judith~F. Islam}, \bibinfo{person}{Iman Keivanloo},
  \bibinfo{person}{Chanchal~K. Roy}, {and} \bibinfo{person}{Mohammad~Mamun
  Mia}.} \bibinfo{year}{2014}\natexlab{}.
\newblock \showarticletitle{Towards a Big Data Curated Benchmark of
  Inter-project Code Clones}. In \bibinfo{booktitle}{\emph{2014 IEEE
  International Conference on Software Maintenance and Evolution}}.
  \bibinfo{pages}{476--480}.
\newblock
\urldef\tempurl%
\url{https://doi.org/10.1109/ICSME.2014.77}
\showDOI{\tempurl}


\bibitem[\protect\citeauthoryear{Svyatkovskiy, Deng, Fu, and
  Sundaresan}{Svyatkovskiy et~al\mbox{.}}{2020}]%
        {Svyatkovskiy2020_321}
\bibfield{author}{\bibinfo{person}{Alexey Svyatkovskiy},
  \bibinfo{person}{Shao~Kun Deng}, \bibinfo{person}{Shengyu Fu}, {and}
  \bibinfo{person}{Neel Sundaresan}.} \bibinfo{year}{2020}\natexlab{}.
\newblock \showarticletitle{IntelliCode Compose: Code Generation Using
  Transformer}. In \bibinfo{booktitle}{\emph{Proceedings of the 28th ACM Joint
  Meeting on European Software Engineering Conference and Symposium on the
  Foundations of Software Engineering}} (Virtual Event, USA)
  \emph{(\bibinfo{series}{ESEC/FSE 2020})}. \bibinfo{pages}{1433–1443}.
\newblock
\showISBNx{9781450370431}
\urldef\tempurl%
\url{https://doi.org/10.1145/3368089.3417058}
\showDOI{\tempurl}


\bibitem[\protect\citeauthoryear{Svyatkovskiy, Lee, Hadjitofi, Riechert,
  Franco, and Allamanis}{Svyatkovskiy et~al\mbox{.}}{2021}]%
        {1092_Svyatkovskiy2021}
\bibfield{author}{\bibinfo{person}{Alexey Svyatkovskiy},
  \bibinfo{person}{Sebastian Lee}, \bibinfo{person}{Anna Hadjitofi},
  \bibinfo{person}{Maik Riechert}, \bibinfo{person}{Juliana~Vicente Franco},
  {and} \bibinfo{person}{Miltiadis Allamanis}.}
  \bibinfo{year}{2021}\natexlab{}.
\newblock \showarticletitle{Fast and memory-efficient neural code completion}.
  In \bibinfo{booktitle}{\emph{2021 IEEE/ACM 18th International Conference on
  Mining Software Repositories (MSR)}}. IEEE, \bibinfo{pages}{329--340}.
\newblock


\bibitem[\protect\citeauthoryear{Svyatkovskiy, Zhao, Fu, and
  Sundaresan}{Svyatkovskiy et~al\mbox{.}}{2019}]%
        {Svyatkovskiy2019_505}
\bibfield{author}{\bibinfo{person}{Alexey Svyatkovskiy}, \bibinfo{person}{Ying
  Zhao}, \bibinfo{person}{Shengyu Fu}, {and} \bibinfo{person}{Neel
  Sundaresan}.} \bibinfo{year}{2019}\natexlab{}.
\newblock \showarticletitle{Pythia: AI-Assisted Code Completion System}. In
  \bibinfo{booktitle}{\emph{Proceedings of the 25th ACM SIGKDD International
  Conference on Knowledge Discovery \&amp; Data Mining}} (Anchorage, AK, USA)
  \emph{(\bibinfo{series}{KDD '19})}. \bibinfo{pages}{2727–2735}.
\newblock
\showISBNx{9781450362016}
\urldef\tempurl%
\url{https://doi.org/10.1145/3292500.3330699}
\showDOI{\tempurl}


\bibitem[\protect\citeauthoryear{Szegedy, Liu, Jia, Sermanet, Reed, Anguelov,
  Erhan, Vanhoucke, and Rabinovich}{Szegedy et~al\mbox{.}}{2015}]%
        {Szegedy2015}
\bibfield{author}{\bibinfo{person}{Christian Szegedy}, \bibinfo{person}{Wei
  Liu}, \bibinfo{person}{Yangqing Jia}, \bibinfo{person}{Pierre Sermanet},
  \bibinfo{person}{Scott Reed}, \bibinfo{person}{Dragomir Anguelov},
  \bibinfo{person}{Dumitru Erhan}, \bibinfo{person}{Vincent Vanhoucke}, {and}
  \bibinfo{person}{Andrew Rabinovich}.} \bibinfo{year}{2015}\natexlab{}.
\newblock \showarticletitle{Going deeper with convolutions}. In
  \bibinfo{booktitle}{\emph{Proceedings of the IEEE conference on computer
  vision and pattern recognition}}. \bibinfo{pages}{1--9}.
\newblock


\bibitem[\protect\citeauthoryear{Takahashi, Shiina, and Kobayashi}{Takahashi
  et~al\mbox{.}}{2019}]%
        {1034_Takahashi2019}
\bibfield{author}{\bibinfo{person}{Akiyoshi Takahashi},
  \bibinfo{person}{Hiromitsu Shiina}, {and} \bibinfo{person}{Nobuyuki
  Kobayashi}.} \bibinfo{year}{2019}\natexlab{}.
\newblock \showarticletitle{Automatic Generation of Program Comments based on
  Problem Statements for Computational Thinking}. In
  \bibinfo{booktitle}{\emph{2019 8th International Congress on Advanced Applied
  Informatics (IIAI-AAI)}}. IEEE, \bibinfo{pages}{629--634}.
\newblock


\bibitem[\protect\citeauthoryear{{Terada} and {Watanobe}}{{Terada} and
  {Watanobe}}{2019}]%
        {Terada2019_487}
\bibfield{author}{\bibinfo{person}{K. {Terada}} {and} \bibinfo{person}{Y.
  {Watanobe}}.} \bibinfo{year}{2019}\natexlab{}.
\newblock \showarticletitle{Code Completion for Programming Education based on
  Recurrent Neural Network}. In \bibinfo{booktitle}{\emph{2019 IEEE 11th
  International Workshop on Computational Intelligence and Applications
  (IWCIA)}}. \bibinfo{pages}{109--114}.
\newblock
\urldef\tempurl%
\url{https://doi.org/10.1109/IWCIA47330.2019.8955090}
\showDOI{\tempurl}


\bibitem[\protect\citeauthoryear{{Thaller}, {Linsbauer}, and {Egyed}}{{Thaller}
  et~al\mbox{.}}{2019}]%
        {Thaller2019_468}
\bibfield{author}{\bibinfo{person}{H. {Thaller}}, \bibinfo{person}{L.
  {Linsbauer}}, {and} \bibinfo{person}{A. {Egyed}}.}
  \bibinfo{year}{2019}\natexlab{}.
\newblock \showarticletitle{Feature Maps: A Comprehensible Software
  Representation for Design Pattern Detection}. In
  \bibinfo{booktitle}{\emph{2019 IEEE 26th International Conference on Software
  Analysis, Evolution and Reengineering (SANER)}}. \bibinfo{pages}{207--217}.
\newblock
\urldef\tempurl%
\url{https://doi.org/10.1109/SANER.2019.8667978}
\showDOI{\tempurl}


\bibitem[\protect\citeauthoryear{{Thongkum} and {Mekruksavanich}}{{Thongkum}
  and {Mekruksavanich}}{2020}]%
        {Thongkum2020_217}
\bibfield{author}{\bibinfo{person}{P. {Thongkum}} {and} \bibinfo{person}{S.
  {Mekruksavanich}}.} \bibinfo{year}{2020}\natexlab{}.
\newblock \showarticletitle{Design Flaws Prediction for Impact on Software
  Maintainability using Extreme Learning Machine}. In
  \bibinfo{booktitle}{\emph{2020 Joint International Conference on Digital
  Arts, Media and Technology with ECTI Northern Section Conference on
  Electrical, Electronics, Computer and Telecommunications Engineering (ECTI
  DAMT NCON)}}. \bibinfo{pages}{79--82}.
\newblock
\urldef\tempurl%
\url{https://doi.org/10.1109/ECTIDAMTNCON48261.2020.9090717}
\showDOI{\tempurl}


\bibitem[\protect\citeauthoryear{{Tian}, {Liu}, {Kaboré}, {Koyuncu}, {Li},
  {Klein}, and {Bissyandé}}{{Tian} et~al\mbox{.}}{2020}]%
        {Tian2020_309}
\bibfield{author}{\bibinfo{person}{H. {Tian}}, \bibinfo{person}{K. {Liu}},
  \bibinfo{person}{A.~K. {Kaboré}}, \bibinfo{person}{A. {Koyuncu}},
  \bibinfo{person}{L. {Li}}, \bibinfo{person}{J. {Klein}}, {and}
  \bibinfo{person}{T.~F. {Bissyandé}}.} \bibinfo{year}{2020}\natexlab{}.
\newblock \showarticletitle{Evaluating Representation Learning of Code Changes
  for Predicting Patch Correctness in Program Repair}. In
  \bibinfo{booktitle}{\emph{2020 35th IEEE/ACM International Conference on
  Automated Software Engineering (ASE)}}. \bibinfo{pages}{981--992}.
\newblock


\bibitem[\protect\citeauthoryear{Tollin, Fontana, Zanoni, and Roveda}{Tollin
  et~al\mbox{.}}{2017}]%
        {Tollin2017_374}
\bibfield{author}{\bibinfo{person}{Irene Tollin},
  \bibinfo{person}{Francesca~Arcelli Fontana}, \bibinfo{person}{Marco Zanoni},
  {and} \bibinfo{person}{Riccardo Roveda}.} \bibinfo{year}{2017}\natexlab{}.
\newblock \showarticletitle{Change Prediction through Coding Rules Violations}.
  In \bibinfo{booktitle}{\emph{Proceedings of the 21st International Conference
  on Evaluation and Assessment in Software Engineering}} (Karlskrona, Sweden)
  \emph{(\bibinfo{series}{EASE'17})}. \bibinfo{pages}{61–64}.
\newblock
\showISBNx{9781450348041}
\urldef\tempurl%
\url{https://doi.org/10.1145/3084226.3084282}
\showDOI{\tempurl}


\bibitem[\protect\citeauthoryear{Tsantalis, Ketkar, and Dig}{Tsantalis
  et~al\mbox{.}}{2020}]%
        {RefactoringMiner2.0}
\bibfield{author}{\bibinfo{person}{Nikolaos Tsantalis}, \bibinfo{person}{Ameya
  Ketkar}, {and} \bibinfo{person}{Danny Dig}.} \bibinfo{year}{2020}\natexlab{}.
\newblock \showarticletitle{RefactoringMiner 2.0}.
\newblock \bibinfo{journal}{\emph{IEEE Transactions on Software Engineering}}
  (\bibinfo{year}{2020}), \bibinfo{numpages}{21}~pages.
\newblock
\urldef\tempurl%
\url{https://doi.org/10.1109/TSE.2020.3007722}
\showDOI{\tempurl}


\bibitem[\protect\citeauthoryear{Tsintzira, Arvanitou, Ampatzoglou, and
  Chatzigeorgiou}{Tsintzira et~al\mbox{.}}{2020}]%
        {Tsintzira2020_194}
\bibfield{author}{\bibinfo{person}{Angeliki-Agathi Tsintzira},
  \bibinfo{person}{Elvira-Maria Arvanitou}, \bibinfo{person}{Apostolos
  Ampatzoglou}, {and} \bibinfo{person}{Alexander Chatzigeorgiou}.}
  \bibinfo{year}{2020}\natexlab{}.
\newblock \showarticletitle{Applying Machine Learning in Technical Debt
  Management: Future Opportunities and Challenges}. In
  \bibinfo{booktitle}{\emph{Quality of Information and Communications
  Technology}}, \bibfield{editor}{\bibinfo{person}{Martin Shepperd},
  \bibinfo{person}{Fernando Brito~e Abreu}, \bibinfo{person}{Alberto
  Rodrigues~da Silva}, {and} \bibinfo{person}{Ricardo P{\'e}rez-Castillo}}
  (Eds.). \bibinfo{pages}{53--67}.
\newblock
\showISBNx{978-3-030-58793-2}


\bibitem[\protect\citeauthoryear{{Tufano}, {Pantiuchina}, {Watson}, {Bavota},
  and {Poshyvanyk}}{{Tufano} et~al\mbox{.}}{2019}]%
        {Tufano2019_390}
\bibfield{author}{\bibinfo{person}{M. {Tufano}}, \bibinfo{person}{J.
  {Pantiuchina}}, \bibinfo{person}{C. {Watson}}, \bibinfo{person}{G. {Bavota}},
  {and} \bibinfo{person}{D. {Poshyvanyk}}.} \bibinfo{year}{2019}\natexlab{}.
\newblock \showarticletitle{On Learning Meaningful Code Changes Via Neural
  Machine Translation}. In \bibinfo{booktitle}{\emph{2019 IEEE/ACM 41st
  International Conference on Software Engineering (ICSE)}}.
  \bibinfo{pages}{25--36}.
\newblock
\urldef\tempurl%
\url{https://doi.org/10.1109/ICSE.2019.00021}
\showDOI{\tempurl}


\bibitem[\protect\citeauthoryear{Tufano, Watson, Bavota, Di~Penta, White, and
  Poshyvanyk}{Tufano et~al\mbox{.}}{2018}]%
        {Tufano2018_467}
\bibfield{author}{\bibinfo{person}{Michele Tufano}, \bibinfo{person}{Cody
  Watson}, \bibinfo{person}{Gabriele Bavota}, \bibinfo{person}{Massimiliano
  Di~Penta}, \bibinfo{person}{Martin White}, {and} \bibinfo{person}{Denys
  Poshyvanyk}.} \bibinfo{year}{2018}\natexlab{}.
\newblock \showarticletitle{Deep Learning Similarities from Different
  Representations of Source Code} \emph{(\bibinfo{series}{MSR '18})}.
  \bibinfo{pages}{542–553}.
\newblock
\showISBNx{9781450357166}
\urldef\tempurl%
\url{https://doi.org/10.1145/3196398.3196431}
\showDOI{\tempurl}


\bibitem[\protect\citeauthoryear{Tufano, Watson, Bavota, Di~Penta, White, and
  Poshyvanyk}{Tufano et~al\mbox{.}}{2019a}]%
        {1080_Tufano2019}
\bibfield{author}{\bibinfo{person}{Michele Tufano}, \bibinfo{person}{Cody
  Watson}, \bibinfo{person}{Gabriele Bavota}, \bibinfo{person}{Massimiliano
  Di~Penta}, \bibinfo{person}{Martin White}, {and} \bibinfo{person}{Denys
  Poshyvanyk}.} \bibinfo{year}{2019}\natexlab{a}.
\newblock \showarticletitle{Learning how to mutate source code from bug-fixes}.
  In \bibinfo{booktitle}{\emph{2019 IEEE International Conference on Software
  Maintenance and Evolution (ICSME)}}. IEEE, \bibinfo{pages}{301--312}.
\newblock


\bibitem[\protect\citeauthoryear{Tufano, Watson, Bavota, Penta, White, and
  Poshyvanyk}{Tufano et~al\mbox{.}}{2019b}]%
        {Tufano2019_279}
\bibfield{author}{\bibinfo{person}{Michele Tufano}, \bibinfo{person}{Cody
  Watson}, \bibinfo{person}{Gabriele Bavota}, \bibinfo{person}{Massimiliano~Di
  Penta}, \bibinfo{person}{Martin White}, {and} \bibinfo{person}{Denys
  Poshyvanyk}.} \bibinfo{year}{2019}\natexlab{b}.
\newblock \showarticletitle{An Empirical Study on Learning Bug-Fixing Patches
  in the Wild via Neural Machine Translation}.
\newblock \bibinfo{journal}{\emph{ACM Trans. Softw. Eng. Methodol.}}
  \bibinfo{volume}{28}, \bibinfo{number}{4}, Article \bibinfo{articleno}{19}
  (\bibinfo{date}{September} \bibinfo{year}{2019}),
  \bibinfo{numpages}{29}~pages.
\newblock
\showISSN{1049-331X}
\urldef\tempurl%
\url{https://doi.org/10.1145/3340544}
\showDOI{\tempurl}


\bibitem[\protect\citeauthoryear{Tummalapalli, Kumar, Murthy, and
  Krishna}{Tummalapalli et~al\mbox{.}}{2022}]%
        {784_Tummalapalli2022}
\bibfield{author}{\bibinfo{person}{Sahithi Tummalapalli}, \bibinfo{person}{Lov
  Kumar}, \bibinfo{person}{NL~Bhanu Murthy}, {and} \bibinfo{person}{Aneesh
  Krishna}.} \bibinfo{year}{2022}\natexlab{}.
\newblock \showarticletitle{Detection of Web Service Anti-Patterns Using
  Weighted Extreme Learning Machine}.
\newblock \bibinfo{journal}{\emph{Computer Standards \& Interfaces}}
  (\bibinfo{year}{2022}), \bibinfo{pages}{103621}.
\newblock


\bibitem[\protect\citeauthoryear{Tummalapalli, Kumar, and Murthy}{Tummalapalli
  et~al\mbox{.}}{2020a}]%
        {Tummalapalli2020_254}
\bibfield{author}{\bibinfo{person}{Sahithi Tummalapalli}, \bibinfo{person}{Lov
  Kumar}, {and} \bibinfo{person}{N.~L.~Bhanu Murthy}.}
  \bibinfo{year}{2020}\natexlab{a}.
\newblock \showarticletitle{Prediction of Web Service Anti-Patterns Using
  Aggregate Software Metrics and Machine Learning Techniques}. In
  \bibinfo{booktitle}{\emph{Proceedings of the 13th Innovations in Software
  Engineering Conference on Formerly Known as India Software Engineering
  Conference}} (Jabalpur, India) \emph{(\bibinfo{series}{ISEC 2020})}. Article
  \bibinfo{articleno}{8}, \bibinfo{numpages}{11}~pages.
\newblock
\showISBNx{9781450375948}
\urldef\tempurl%
\url{https://doi.org/10.1145/3385032.3385042}
\showDOI{\tempurl}


\bibitem[\protect\citeauthoryear{Tummalapalli, Kumar, Murthy~Neti, Kocher, and
  Padmanabhuni}{Tummalapalli et~al\mbox{.}}{2021a}]%
        {795_Tummalapalli2021}
\bibfield{author}{\bibinfo{person}{Sahithi Tummalapalli}, \bibinfo{person}{Lov
  Kumar}, \bibinfo{person}{Lalitha~Bhanu Murthy~Neti}, \bibinfo{person}{Vipul
  Kocher}, {and} \bibinfo{person}{Srinivas Padmanabhuni}.}
  \bibinfo{year}{2021}\natexlab{a}.
\newblock \showarticletitle{A Novel Approach for the Detection of Web Service
  Anti-Patterns Using Word Embedding Techniques}. In
  \bibinfo{booktitle}{\emph{International Conference on Computational Science
  and Its Applications}}. Springer, \bibinfo{pages}{217--230}.
\newblock


\bibitem[\protect\citeauthoryear{Tummalapalli, Kumar, and Neti}{Tummalapalli
  et~al\mbox{.}}{2019}]%
        {749_Tummalapalli2019}
\bibfield{author}{\bibinfo{person}{Sahithi Tummalapalli}, \bibinfo{person}{Lov
  Kumar}, {and} \bibinfo{person}{Lalita Bhanu~Murthy Neti}.}
  \bibinfo{year}{2019}\natexlab{}.
\newblock \showarticletitle{An empirical framework for web service anti-pattern
  prediction using machine learning techniques}. In
  \bibinfo{booktitle}{\emph{2019 9th Annual Information Technology,
  Electromechanical Engineering and Microelectronics Conference (IEMECON)}}.
  IEEE, \bibinfo{pages}{137--143}.
\newblock


\bibitem[\protect\citeauthoryear{Tummalapalli, Mittal, Kumar, Murthy~Neti, and
  Rath}{Tummalapalli et~al\mbox{.}}{2021b}]%
        {796_Tummalapalli2021}
\bibfield{author}{\bibinfo{person}{Sahithi Tummalapalli}, \bibinfo{person}{Juhi
  Mittal}, \bibinfo{person}{Lov Kumar}, \bibinfo{person}{Lalitha~Bhanu
  Murthy~Neti}, {and} \bibinfo{person}{Santanu~Kumar Rath}.}
  \bibinfo{year}{2021}\natexlab{b}.
\newblock \showarticletitle{An Empirical Analysis on the Prediction of Web
  Service Anti-patterns Using Source Code Metrics and Ensemble Techniques}. In
  \bibinfo{booktitle}{\emph{International Conference on Computational Science
  and Its Applications}}. Springer, \bibinfo{pages}{263--276}.
\newblock


\bibitem[\protect\citeauthoryear{Tummalapalli, Murthy, Krishna,
  et~al\mbox{.}}{Tummalapalli et~al\mbox{.}}{2020b}]%
        {793_Tummalapalli2020}
\bibfield{author}{\bibinfo{person}{Sahithi Tummalapalli}, \bibinfo{person}{NL
  Murthy}, \bibinfo{person}{Aneesh Krishna}, {et~al\mbox{.}}}
  \bibinfo{year}{2020}\natexlab{b}.
\newblock \showarticletitle{Detection of web service anti-patterns using neural
  networks with multiple layers}. In \bibinfo{booktitle}{\emph{International
  Conference on Neural Information Processing}}. Springer,
  \bibinfo{pages}{571--579}.
\newblock


\bibitem[\protect\citeauthoryear{Ucci, Aniello, and Baldoni}{Ucci
  et~al\mbox{.}}{2019}]%
        {Ucci2019_54}
\bibfield{author}{\bibinfo{person}{Daniele Ucci}, \bibinfo{person}{Leonardo
  Aniello}, {and} \bibinfo{person}{Roberto Baldoni}.}
  \bibinfo{year}{2019}\natexlab{}.
\newblock \showarticletitle{Survey of machine learning techniques for malware
  analysis}.
\newblock \bibinfo{journal}{\emph{Computers \& Security}}  \bibinfo{volume}{81}
  (\bibinfo{year}{2019}), \bibinfo{pages}{123 -- 147}.
\newblock
\showISSN{0167-4048}
\urldef\tempurl%
\url{https://doi.org/10.1016/j.cose.2018.11.001}
\showDOI{\tempurl}


\bibitem[\protect\citeauthoryear{Uchiyama, Kubo, Washizaki, and
  Fukazawa}{Uchiyama et~al\mbox{.}}{2014}]%
        {Uchiyama2014_379}
\bibfield{author}{\bibinfo{person}{S. Uchiyama}, \bibinfo{person}{A. Kubo},
  \bibinfo{person}{H. Washizaki}, {and} \bibinfo{person}{Y. Fukazawa}.}
  \bibinfo{year}{2014}\natexlab{}.
\newblock \showarticletitle{Detecting Design Patterns in Object-Oriented
  Program Source Code by Using Metrics and Machine Learning}.
\newblock \bibinfo{journal}{\emph{Journal of Software Engineering and
  Applications}}  \bibinfo{volume}{07} (\bibinfo{year}{2014}),
  \bibinfo{pages}{983--998}.
\newblock


\bibitem[\protect\citeauthoryear{Uch{\^o}a, Barbosa, Coutinho, Oizumi,
  Assun{\c{c}}ao, Vergilio, Pereira, Oliveira, and Garcia}{Uch{\^o}a
  et~al\mbox{.}}{2021}]%
        {836_Uchoa2021}
\bibfield{author}{\bibinfo{person}{Anderson Uch{\^o}a}, \bibinfo{person}{Caio
  Barbosa}, \bibinfo{person}{Daniel Coutinho}, \bibinfo{person}{Willian
  Oizumi}, \bibinfo{person}{Wesley~KG Assun{\c{c}}ao},
  \bibinfo{person}{Silvia~Regina Vergilio}, \bibinfo{person}{Juliana~Alves
  Pereira}, \bibinfo{person}{Anderson Oliveira}, {and}
  \bibinfo{person}{Alessandro Garcia}.} \bibinfo{year}{2021}\natexlab{}.
\newblock \showarticletitle{Predicting design impactful changes in modern code
  review: A large-scale empirical study}. In \bibinfo{booktitle}{\emph{2021
  IEEE/ACM 18th International Conference on Mining Software Repositories
  (MSR)}}. IEEE, \bibinfo{pages}{471--482}.
\newblock


\bibitem[\protect\citeauthoryear{Ugurel, Krovetz, and Giles}{Ugurel
  et~al\mbox{.}}{2002}]%
        {Ugurel2002_394}
\bibfield{author}{\bibinfo{person}{Secil Ugurel}, \bibinfo{person}{Robert
  Krovetz}, {and} \bibinfo{person}{C.~Lee Giles}.}
  \bibinfo{year}{2002}\natexlab{}.
\newblock \showarticletitle{What's the Code? Automatic Classification of Source
  Code Archives}. In \bibinfo{booktitle}{\emph{Proceedings of the Eighth ACM
  SIGKDD International Conference on Knowledge Discovery and Data Mining}}
  (Edmonton, Alberta, Canada) \emph{(\bibinfo{series}{KDD '02})}.
  \bibinfo{pages}{632–638}.
\newblock
\showISBNx{158113567X}
\urldef\tempurl%
\url{https://doi.org/10.1145/775047.775141}
\showDOI{\tempurl}


\bibitem[\protect\citeauthoryear{{Utting}, {Legeard}, {Dadeau}, {Tamagnan}, and
  {Bouquet}}{{Utting} et~al\mbox{.}}{2020}]%
        {Utting2020_92}
\bibfield{author}{\bibinfo{person}{M. {Utting}}, \bibinfo{person}{B.
  {Legeard}}, \bibinfo{person}{F. {Dadeau}}, \bibinfo{person}{F. {Tamagnan}},
  {and} \bibinfo{person}{F. {Bouquet}}.} \bibinfo{year}{2020}\natexlab{}.
\newblock \showarticletitle{Identifying and Generating Missing Tests using
  Machine Learning on Execution Traces}. In \bibinfo{booktitle}{\emph{2020 IEEE
  International Conference On Artificial Intelligence Testing (AITest)}}.
  \bibinfo{pages}{83--90}.
\newblock
\urldef\tempurl%
\url{https://doi.org/10.1109/AITEST49225.2020.00020}
\showDOI{\tempurl}


\bibitem[\protect\citeauthoryear{Van~Thuy, Anh, and Hoai}{Van~Thuy
  et~al\mbox{.}}{2018}]%
        {Hoang2018_280}
\bibfield{author}{\bibinfo{person}{Hoang Van~Thuy}, \bibinfo{person}{Phan~Viet
  Anh}, {and} \bibinfo{person}{Nguyen~Xuan Hoai}.}
  \bibinfo{year}{2018}\natexlab{}.
\newblock \showarticletitle{Automated Large Program Repair Based on Big Code}.
  In \bibinfo{booktitle}{\emph{Proceedings of the Ninth International Symposium
  on Information and Communication Technology}} (Danang City, Viet Nam)
  \emph{(\bibinfo{series}{SoICT 2018})}. \bibinfo{pages}{375?--381}.
\newblock
\showISBNx{9781450365390}
\urldef\tempurl%
\url{https://doi.org/10.1145/3287921.3287958}
\showDOI{\tempurl}


\bibitem[\protect\citeauthoryear{Vasic, Kanade, Maniatis, Bieber, and
  Singh}{Vasic et~al\mbox{.}}{2019}]%
        {Vasic2019_333}
\bibfield{author}{\bibinfo{person}{Marko Vasic}, \bibinfo{person}{Aditya
  Kanade}, \bibinfo{person}{Petros Maniatis}, \bibinfo{person}{David Bieber},
  {and} \bibinfo{person}{Rishabh Singh}.} \bibinfo{year}{2019}\natexlab{}.
\newblock \bibinfo{title}{Neural Program Repair by Jointly Learning to Localize
  and Repair}.
\newblock
\newblock


\bibitem[\protect\citeauthoryear{Vaswani, Shazeer, Parmar, Uszkoreit, Jones,
  Gomez, Kaiser, and Polosukhin}{Vaswani et~al\mbox{.}}{2017}]%
        {Vaswani2017}
\bibfield{author}{\bibinfo{person}{Ashish Vaswani}, \bibinfo{person}{Noam
  Shazeer}, \bibinfo{person}{Niki Parmar}, \bibinfo{person}{Jakob Uszkoreit},
  \bibinfo{person}{Llion Jones}, \bibinfo{person}{Aidan~N Gomez},
  \bibinfo{person}{\L~ukasz Kaiser}, {and} \bibinfo{person}{Illia Polosukhin}.}
  \bibinfo{year}{2017}\natexlab{}.
\newblock \showarticletitle{Attention is All you Need}. In
  \bibinfo{booktitle}{\emph{Advances in Neural Information Processing
  Systems}}, \bibfield{editor}{\bibinfo{person}{I.~Guyon},
  \bibinfo{person}{U.~V. Luxburg}, \bibinfo{person}{S.~Bengio},
  \bibinfo{person}{H.~Wallach}, \bibinfo{person}{R.~Fergus},
  \bibinfo{person}{S.~Vishwanathan}, {and} \bibinfo{person}{R.~Garnett}}
  (Eds.), Vol.~\bibinfo{volume}{30}. \bibinfo{publisher}{Curran Associates,
  Inc.}
\newblock
\urldef\tempurl%
\url{https://proceedings.neurips.cc/paper/2017/file/3f5ee243547dee91fbd053c1c4a845aa-Paper.pdf}
\showURL{%
\tempurl}


\bibitem[\protect\citeauthoryear{Vishnu and Jevitha}{Vishnu and
  Jevitha}{2014}]%
        {vishnu_jevitha_2014}
\bibfield{author}{\bibinfo{person}{B.~A. Vishnu} {and} \bibinfo{person}{K.~P.
  Jevitha}.} \bibinfo{year}{2014}\natexlab{}.
\newblock \showarticletitle{Prediction of Cross-Site Scripting Attack Using
  Machine Learning Algorithms}. In \bibinfo{booktitle}{\emph{Proceedings of the
  2014 International Conference on Interdisciplinary Advances in Applied
  Computing}} (Amritapuri, India) \emph{(\bibinfo{series}{ICONIAAC '14})}.
  \bibinfo{publisher}{Association for Computing Machinery},
  \bibinfo{address}{New York, NY, USA}, Article \bibinfo{articleno}{55},
  \bibinfo{numpages}{5}~pages.
\newblock
\showISBNx{9781450329088}
\urldef\tempurl%
\url{https://doi.org/10.1145/2660859.2660969}
\showDOI{\tempurl}


\bibitem[\protect\citeauthoryear{Viuginov and Filchenkov}{Viuginov and
  Filchenkov}{2019}]%
        {Viuginov2019_405}
\bibfield{author}{\bibinfo{person}{Nickolay Viuginov} {and}
  \bibinfo{person}{Andrey Filchenkov}.} \bibinfo{year}{2019}\natexlab{}.
\newblock \showarticletitle{A Machine Learning Based Automatic Folding of
  Dynamically Typed Languages}. In \bibinfo{booktitle}{\emph{Proceedings of the
  3rd ACM SIGSOFT International Workshop on Machine Learning Techniques for
  Software Quality Evaluation}} (Tallinn, Estonia)
  \emph{(\bibinfo{series}{MaLTeSQuE 2019})}. \bibinfo{pages}{31–36}.
\newblock
\showISBNx{9781450368551}
\urldef\tempurl%
\url{https://doi.org/10.1145/3340482.3342746}
\showDOI{\tempurl}


\bibitem[\protect\citeauthoryear{Wan, Shu, Sui, Xu, Zhao, Wu, and Yu}{Wan
  et~al\mbox{.}}{2019}]%
        {Wan2019_443}
\bibfield{author}{\bibinfo{person}{Yao Wan}, \bibinfo{person}{Jingdong Shu},
  \bibinfo{person}{Yulei Sui}, \bibinfo{person}{Guandong Xu},
  \bibinfo{person}{Zhou Zhao}, \bibinfo{person}{Jian Wu}, {and}
  \bibinfo{person}{Philip~S. Yu}.} \bibinfo{year}{2019}\natexlab{}.
\newblock \showarticletitle{Multi-Modal Attention Network Learning for Semantic
  Source Code Retrieval}. In \bibinfo{booktitle}{\emph{Proceedings of the 34th
  IEEE/ACM International Conference on Automated Software Engineering}} (San
  Diego, California) \emph{(\bibinfo{series}{ASE '19})}.
  \bibinfo{pages}{13–25}.
\newblock
\showISBNx{9781728125084}
\urldef\tempurl%
\url{https://doi.org/10.1109/ASE.2019.00012}
\showDOI{\tempurl}


\bibitem[\protect\citeauthoryear{Wan, Zhao, Yang, Xu, Ying, Wu, and Yu}{Wan
  et~al\mbox{.}}{2018}]%
        {Wan2018_427}
\bibfield{author}{\bibinfo{person}{Yao Wan}, \bibinfo{person}{Zhou Zhao},
  \bibinfo{person}{Min Yang}, \bibinfo{person}{Guandong Xu},
  \bibinfo{person}{Haochao Ying}, \bibinfo{person}{Jian Wu}, {and}
  \bibinfo{person}{Philip~S. Yu}.} \bibinfo{year}{2018}\natexlab{}.
\newblock \showarticletitle{Improving Automatic Source Code Summarization via
  Deep Reinforcement Learning}. In \bibinfo{booktitle}{\emph{Proceedings of the
  33rd ACM/IEEE International Conference on Automated Software Engineering}}
  (Montpellier, France) \emph{(\bibinfo{series}{ASE 2018})}.
  \bibinfo{pages}{397–407}.
\newblock
\showISBNx{9781450359375}
\urldef\tempurl%
\url{https://doi.org/10.1145/3238147.3238206}
\showDOI{\tempurl}


\bibitem[\protect\citeauthoryear{{Wan}, {Xia}, {Lo}, and {Murphy}}{{Wan}
  et~al\mbox{.}}{2019}]%
        {Wan2019_396}
\bibfield{author}{\bibinfo{person}{Z. {Wan}}, \bibinfo{person}{X. {Xia}},
  \bibinfo{person}{D. {Lo}}, {and} \bibinfo{person}{G.~C. {Murphy}}.}
  \bibinfo{year}{2019}\natexlab{}.
\newblock \showarticletitle{How does Machine Learning Change Software
  Development Practices?}
\newblock \bibinfo{journal}{\emph{IEEE Transactions on Software Engineering}}
  (\bibinfo{year}{2019}), \bibinfo{pages}{1--1}.
\newblock
\urldef\tempurl%
\url{https://doi.org/10.1109/TSE.2019.2937083}
\showDOI{\tempurl}


\bibitem[\protect\citeauthoryear{Wang, Dong, and Li}{Wang
  et~al\mbox{.}}{2020a}]%
        {Wang2020_450}
\bibfield{author}{\bibinfo{person}{Deze Wang}, \bibinfo{person}{Wei Dong},
  {and} \bibinfo{person}{Shanshan Li}.} \bibinfo{year}{2020}\natexlab{a}.
\newblock \showarticletitle{A Multi-Task Representation Learning Approach for
  Source Code}. In \bibinfo{booktitle}{\emph{Proceedings of the 1st ACM SIGSOFT
  International Workshop on Representation Learning for Software Engineering
  and Program Languages}} (Virtual, USA) \emph{(\bibinfo{series}{RL+SE\&amp;PL
  2020})}. \bibinfo{pages}{1–2}.
\newblock
\showISBNx{9781450381253}
\urldef\tempurl%
\url{https://doi.org/10.1145/3416506.3423575}
\showDOI{\tempurl}


\bibitem[\protect\citeauthoryear{Wang, Xia, Lo, He, Wang, and Grundy}{Wang
  et~al\mbox{.}}{2021}]%
        {1015_Wang2021}
\bibfield{author}{\bibinfo{person}{Haoye Wang}, \bibinfo{person}{Xin Xia},
  \bibinfo{person}{David Lo}, \bibinfo{person}{Qiang He},
  \bibinfo{person}{Xinyu Wang}, {and} \bibinfo{person}{John Grundy}.}
  \bibinfo{year}{2021}\natexlab{}.
\newblock \showarticletitle{Context-aware retrieval-based deep commit message
  Generation}.
\newblock \bibinfo{journal}{\emph{ACM Transactions on Software Engineering and
  Methodology (TOSEM)}} \bibinfo{volume}{30}, \bibinfo{number}{4}
  (\bibinfo{year}{2021}), \bibinfo{pages}{1--30}.
\newblock


\bibitem[\protect\citeauthoryear{{Wang}, {Zhang}, {Lu}, {Lyu}, and
  {Lyu}}{{Wang} et~al\mbox{.}}{2020a}]%
        {Wang2020_424}
\bibfield{author}{\bibinfo{person}{R. {Wang}}, \bibinfo{person}{H. {Zhang}},
  \bibinfo{person}{G. {Lu}}, \bibinfo{person}{L. {Lyu}}, {and}
  \bibinfo{person}{C. {Lyu}}.} \bibinfo{year}{2020}\natexlab{a}.
\newblock \showarticletitle{Fret: Functional Reinforced Transformer With BERT
  for Code Summarization}.
\newblock \bibinfo{journal}{\emph{IEEE Access}}  \bibinfo{volume}{8}
  (\bibinfo{year}{2020}), \bibinfo{pages}{135591--135604}.
\newblock
\urldef\tempurl%
\url{https://doi.org/10.1109/ACCESS.2020.3011744}
\showDOI{\tempurl}


\bibitem[\protect\citeauthoryear{Wang, Liu, Qiu, Ma, Liu, and Wu}{Wang
  et~al\mbox{.}}{2019}]%
        {Wang2019_494}
\bibfield{author}{\bibinfo{person}{Shuai Wang}, \bibinfo{person}{Jinyang Liu},
  \bibinfo{person}{Ye Qiu}, \bibinfo{person}{Zhiyi Ma}, \bibinfo{person}{Junfei
  Liu}, {and} \bibinfo{person}{Zhonghai Wu}.} \bibinfo{year}{2019}\natexlab{}.
\newblock \showarticletitle{Deep Learning Based Code Completion Models for
  Programming Codes}. In \bibinfo{booktitle}{\emph{Proceedings of the 2019 3rd
  International Symposium on Computer Science and Intelligent Control}}
  (Amsterdam, Netherlands) \emph{(\bibinfo{series}{ISCSIC 2019})}. Article
  \bibinfo{articleno}{16}, \bibinfo{numpages}{9}~pages.
\newblock
\showISBNx{9781450376617}
\urldef\tempurl%
\url{https://doi.org/10.1145/3386164.3389083}
\showDOI{\tempurl}


\bibitem[\protect\citeauthoryear{Wang, Liu, Nam, and Tan}{Wang
  et~al\mbox{.}}{2018}]%
        {526_Wang2018}
\bibfield{author}{\bibinfo{person}{Song Wang}, \bibinfo{person}{Taiyue Liu},
  \bibinfo{person}{Jaechang Nam}, {and} \bibinfo{person}{Lin Tan}.}
  \bibinfo{year}{2018}\natexlab{}.
\newblock \showarticletitle{Deep semantic feature learning for software defect
  prediction}.
\newblock \bibinfo{journal}{\emph{IEEE Transactions on Software Engineering}}
  \bibinfo{volume}{46}, \bibinfo{number}{12} (\bibinfo{year}{2018}),
  \bibinfo{pages}{1267--1293}.
\newblock


\bibitem[\protect\citeauthoryear{Wang, Liu, and Tan}{Wang
  et~al\mbox{.}}{2016a}]%
        {Wang2016_78}
\bibfield{author}{\bibinfo{person}{Song Wang}, \bibinfo{person}{Taiyue Liu},
  {and} \bibinfo{person}{Lin Tan}.} \bibinfo{year}{2016}\natexlab{a}.
\newblock \showarticletitle{Automatically Learning Semantic Features for Defect
  Prediction}. In \bibinfo{booktitle}{\emph{Proceedings of the 38th
  International Conference on Software Engineering}} (Austin, Texas)
  \emph{(\bibinfo{series}{ICSE '16})}. \bibinfo{pages}{297–308}.
\newblock
\showISBNx{9781450339001}
\urldef\tempurl%
\url{https://doi.org/10.1145/2884781.2884804}
\showDOI{\tempurl}


\bibitem[\protect\citeauthoryear{{Wang}, {Wen}, {Chen}, {Yi}, and {Mao}}{{Wang}
  et~al\mbox{.}}{2019}]%
        {Wang2019_316}
\bibfield{author}{\bibinfo{person}{S. {Wang}}, \bibinfo{person}{M. {Wen}},
  \bibinfo{person}{L. {Chen}}, \bibinfo{person}{X. {Yi}}, {and}
  \bibinfo{person}{X. {Mao}}.} \bibinfo{year}{2019}\natexlab{}.
\newblock \showarticletitle{How Different Is It Between Machine-Generated and
  Developer-Provided Patches? : An Empirical Study on the Correct Patches
  Generated by Automated Program Repair Techniques}. In
  \bibinfo{booktitle}{\emph{2019 ACM/IEEE International Symposium on Empirical
  Software Engineering and Measurement (ESEM)}}. \bibinfo{pages}{1--12}.
\newblock
\urldef\tempurl%
\url{https://doi.org/10.1109/ESEM.2019.8870172}
\showDOI{\tempurl}


\bibitem[\protect\citeauthoryear{{Wang} and {Yao}}{{Wang} and {Yao}}{2013}]%
        {Wang2013_150}
\bibfield{author}{\bibinfo{person}{S. {Wang}} {and} \bibinfo{person}{X.
  {Yao}}.} \bibinfo{year}{2013}\natexlab{}.
\newblock \showarticletitle{Using Class Imbalance Learning for Software Defect
  Prediction}.
\newblock \bibinfo{journal}{\emph{IEEE Transactions on Reliability}}
  \bibinfo{volume}{62}, \bibinfo{number}{2} (\bibinfo{year}{2013}),
  \bibinfo{pages}{434--443}.
\newblock
\urldef\tempurl%
\url{https://doi.org/10.1109/TR.2013.2259203}
\showDOI{\tempurl}


\bibitem[\protect\citeauthoryear{Wang, Zhang, Jing, and Zhang}{Wang
  et~al\mbox{.}}{2016b}]%
        {678_Wang2016}
\bibfield{author}{\bibinfo{person}{Tiejian Wang}, \bibinfo{person}{Zhiwu
  Zhang}, \bibinfo{person}{Xiaoyuan Jing}, {and} \bibinfo{person}{Liqiang
  Zhang}.} \bibinfo{year}{2016}\natexlab{b}.
\newblock \showarticletitle{Multiple kernel ensemble learning for software
  defect prediction}.
\newblock \bibinfo{journal}{\emph{Automated Software Engineering}}
  \bibinfo{volume}{23}, \bibinfo{number}{4} (\bibinfo{year}{2016}),
  \bibinfo{pages}{569--590}.
\newblock


\bibitem[\protect\citeauthoryear{Wang and Godfrey}{Wang and Godfrey}{2014}]%
        {757_Wang2014}
\bibfield{author}{\bibinfo{person}{Wei Wang} {and} \bibinfo{person}{Michael~W.
  Godfrey}.} \bibinfo{year}{2014}\natexlab{}.
\newblock \showarticletitle{Recommending Clones for Refactoring Using Design,
  Context, and History}. In \bibinfo{booktitle}{\emph{2014 IEEE International
  Conference on Software Maintenance and Evolution}}.
  \bibinfo{pages}{331--340}.
\newblock
\urldef\tempurl%
\url{https://doi.org/10.1109/ICSME.2014.55}
\showDOI{\tempurl}


\bibitem[\protect\citeauthoryear{Wang, Li, Shen, Xia, and Jin}{Wang
  et~al\mbox{.}}{2020b}]%
        {Wang2020_475}
\bibfield{author}{\bibinfo{person}{Wenhan Wang}, \bibinfo{person}{Ge Li},
  \bibinfo{person}{Sijie Shen}, \bibinfo{person}{Xin Xia}, {and}
  \bibinfo{person}{Zhi Jin}.} \bibinfo{year}{2020}\natexlab{b}.
\newblock \showarticletitle{Modular Tree Network for Source Code Representation
  Learning}.
\newblock \bibinfo{journal}{\emph{ACM Trans. Softw. Eng. Methodol.}}
  \bibinfo{volume}{29}, \bibinfo{number}{4}, Article \bibinfo{articleno}{31}
  (\bibinfo{date}{September} \bibinfo{year}{2020}),
  \bibinfo{numpages}{23}~pages.
\newblock
\showISSN{1049-331X}
\urldef\tempurl%
\url{https://doi.org/10.1145/3409331}
\showDOI{\tempurl}


\bibitem[\protect\citeauthoryear{{Wang}, {Zhang}, {Sui}, {Wan}, {Zhao}, {Wu},
  {Yu}, and {Xu}}{{Wang} et~al\mbox{.}}{2020b}]%
        {Wang2020_434}
\bibfield{author}{\bibinfo{person}{W. {Wang}}, \bibinfo{person}{Y. {Zhang}},
  \bibinfo{person}{Y. {Sui}}, \bibinfo{person}{Y. {Wan}}, \bibinfo{person}{Z.
  {Zhao}}, \bibinfo{person}{J. {Wu}}, \bibinfo{person}{P. {Yu}}, {and}
  \bibinfo{person}{G. {Xu}}.} \bibinfo{year}{2020}\natexlab{b}.
\newblock \showarticletitle{Reinforcement-Learning-Guided Source Code
  Summarization via Hierarchical Attention}.
\newblock \bibinfo{journal}{\emph{IEEE Transactions on Software Engineering}}
  (\bibinfo{year}{2020}), \bibinfo{pages}{1--1}.
\newblock
\urldef\tempurl%
\url{https://doi.org/10.1109/TSE.2020.2979701}
\showDOI{\tempurl}


\bibitem[\protect\citeauthoryear{Wang, Zhang, Sui, Wan, Zhao, Wu, Yu, and
  Xu}{Wang et~al\mbox{.}}{2020}]%
        {992_Wang2020}
\bibfield{author}{\bibinfo{person}{Wenhua Wang}, \bibinfo{person}{Yuqun Zhang},
  \bibinfo{person}{Yulei Sui}, \bibinfo{person}{Yao Wan}, \bibinfo{person}{Zhou
  Zhao}, \bibinfo{person}{Jian Wu}, \bibinfo{person}{Philip Yu}, {and}
  \bibinfo{person}{Guandong Xu}.} \bibinfo{year}{2020}\natexlab{}.
\newblock \showarticletitle{Reinforcement-learning-guided source code
  summarization via hierarchical attention}.
\newblock \bibinfo{journal}{\emph{IEEE Transactions on software Engineering}}
  (\bibinfo{year}{2020}).
\newblock


\bibitem[\protect\citeauthoryear{Wang, Wang, Sun, Batcheller, and Jajodia}{Wang
  et~al\mbox{.}}{2020d}]%
        {wang_want_sun_batcheller_jajodia_2020}
\bibfield{author}{\bibinfo{person}{Xinda Wang}, \bibinfo{person}{Shu Wang},
  \bibinfo{person}{Kun Sun}, \bibinfo{person}{Archer Batcheller}, {and}
  \bibinfo{person}{Sushil Jajodia}.} \bibinfo{year}{2020}\natexlab{d}.
\newblock \showarticletitle{A Machine Learning Approach to Classify Security
  Patches into Vulnerability Types}. In \bibinfo{booktitle}{\emph{2020 IEEE
  Conference on Communications and Network Security (CNS)}}.
  \bibinfo{pages}{1--9}.
\newblock
\urldef\tempurl%
\url{https://doi.org/10.1109/CNS48642.2020.9162237}
\showDOI{\tempurl}


\bibitem[\protect\citeauthoryear{Wang, Wang, Gao, and Wang}{Wang
  et~al\mbox{.}}{2020c}]%
        {Wang2020_473}
\bibfield{author}{\bibinfo{person}{Yu Wang}, \bibinfo{person}{Ke Wang},
  \bibinfo{person}{Fengjuan Gao}, {and} \bibinfo{person}{Linzhang Wang}.}
  \bibinfo{year}{2020}\natexlab{c}.
\newblock \showarticletitle{Learning Semantic Program Embeddings with Graph
  Interval Neural Network}.
\newblock \bibinfo{journal}{\emph{Proc. ACM Program. Lang.}}
  \bibinfo{volume}{4}, \bibinfo{number}{OOPSLA}, Article
  \bibinfo{articleno}{137} (\bibinfo{date}{November} \bibinfo{year}{2020}),
  \bibinfo{numpages}{27}~pages.
\newblock
\urldef\tempurl%
\url{https://doi.org/10.1145/3428205}
\showDOI{\tempurl}


\bibitem[\protect\citeauthoryear{Wei, Luo, Weng, Zhong, Zhang, and Yan}{Wei
  et~al\mbox{.}}{2017}]%
        {wei_luo_weng_zhong_zhang_yan_2017}
\bibfield{author}{\bibinfo{person}{Linfeng Wei}, \bibinfo{person}{Weiqi Luo},
  \bibinfo{person}{Jian Weng}, \bibinfo{person}{Yanjun Zhong},
  \bibinfo{person}{Xiaoqian Zhang}, {and} \bibinfo{person}{Zheng Yan}.}
  \bibinfo{year}{2017}\natexlab{}.
\newblock \showarticletitle{Machine Learning-Based Malicious Application
  Detection of Android}.
\newblock \bibinfo{journal}{\emph{IEEE Access}}  \bibinfo{volume}{5}
  (\bibinfo{year}{2017}), \bibinfo{pages}{25591--25601}.
\newblock
\urldef\tempurl%
\url{https://doi.org/10.1109/ACCESS.2017.2771470}
\showDOI{\tempurl}


\bibitem[\protect\citeauthoryear{White, Tufano, Martinez, Monperrus, and
  Poshyvanyk}{White et~al\mbox{.}}{2019}]%
        {889_White2019}
\bibfield{author}{\bibinfo{person}{Martin White}, \bibinfo{person}{Michele
  Tufano}, \bibinfo{person}{Matias Martinez}, \bibinfo{person}{Martin
  Monperrus}, {and} \bibinfo{person}{Denys Poshyvanyk}.}
  \bibinfo{year}{2019}\natexlab{}.
\newblock \showarticletitle{Sorting and transforming program repair ingredients
  via deep learning code similarities}. In \bibinfo{booktitle}{\emph{2019 IEEE
  26th International Conference on Software Analysis, Evolution and
  Reengineering (SANER)}}. IEEE, \bibinfo{pages}{479--490}.
\newblock


\bibitem[\protect\citeauthoryear{White, Tufano, Vendome, and Poshyvanyk}{White
  et~al\mbox{.}}{2016}]%
        {White2016_215}
\bibfield{author}{\bibinfo{person}{Martin White}, \bibinfo{person}{Michele
  Tufano}, \bibinfo{person}{Christopher Vendome}, {and} \bibinfo{person}{Denys
  Poshyvanyk}.} \bibinfo{year}{2016}\natexlab{}.
\newblock \showarticletitle{Deep Learning Code Fragments for Code Clone
  Detection}. In \bibinfo{booktitle}{\emph{Proceedings of the 31st IEEE/ACM
  International Conference on Automated Software Engineering}} (Singapore,
  Singapore) \emph{(\bibinfo{series}{ASE 2016})}. \bibinfo{pages}{87–98}.
\newblock
\showISBNx{9781450338455}
\urldef\tempurl%
\url{https://doi.org/10.1145/2970276.2970326}
\showDOI{\tempurl}


\bibitem[\protect\citeauthoryear{Wu, Li, Wu, and Zheng}{Wu
  et~al\mbox{.}}{2020}]%
        {Wu2020_519}
\bibfield{author}{\bibinfo{person}{Liwei Wu}, \bibinfo{person}{Fei Li},
  \bibinfo{person}{Youhua Wu}, {and} \bibinfo{person}{Tao Zheng}.}
  \bibinfo{year}{2020}\natexlab{}.
\newblock \showarticletitle{{GGF}: A Graph-Based Method for Programming
  Language Syntax Error Correction}. In \bibinfo{booktitle}{\emph{Proceedings
  of the 28th International Conference on Program Comprehension}}
  \emph{(\bibinfo{series}{ICPC '20})}. \bibinfo{publisher}{Association for
  Computing Machinery}, \bibinfo{pages}{139–--148}.
\newblock
\showISBNx{9781450379588}
\urldef\tempurl%
\url{https://doi.org/10.1145/3387904.3389252}
\showDOI{\tempurl}


\bibitem[\protect\citeauthoryear{Xiao, Miao, Shi, and Hong}{Xiao
  et~al\mbox{.}}{2020}]%
        {Xiao2020_102}
\bibfield{author}{\bibinfo{person}{L. Xiao}, \bibinfo{person}{HuaiKou Miao},
  \bibinfo{person}{Tingting Shi}, {and} \bibinfo{person}{Y. Hong}.}
  \bibinfo{year}{2020}\natexlab{}.
\newblock \showarticletitle{LSTM-based deep learning for spatial–temporal
  software testing}.
\newblock \bibinfo{journal}{\emph{Distributed and Parallel Databases}}
  (\bibinfo{year}{2020}), \bibinfo{pages}{1--26}.
\newblock


\bibitem[\protect\citeauthoryear{Xie, Ye, Sun, and Zhang}{Xie
  et~al\mbox{.}}{2021}]%
        {Xie2021_523}
\bibfield{author}{\bibinfo{person}{R. Xie}, \bibinfo{person}{W. Ye},
  \bibinfo{person}{J. Sun}, {and} \bibinfo{person}{S. Zhang}.}
  \bibinfo{year}{2021}\natexlab{}.
\newblock \showarticletitle{Exploiting Method Names to Improve Code
  Summarization: A Deliberation Multi-Task Learning Approach}. In
  \bibinfo{booktitle}{\emph{2021 2021 IEEE/ACM 29th International Conference on
  Program Comprehension (ICPC) (ICPC)}}. \bibinfo{pages}{138--148}.
\newblock
\urldef\tempurl%
\url{https://doi.org/10.1109/ICPC52881.2021.00022}
\showDOI{\tempurl}


\bibitem[\protect\citeauthoryear{Xiong, Wang, Fu, and Zang}{Xiong
  et~al\mbox{.}}{2018}]%
        {Xiong2018_327}
\bibfield{author}{\bibinfo{person}{Yingfei Xiong}, \bibinfo{person}{Bo Wang},
  \bibinfo{person}{Guirong Fu}, {and} \bibinfo{person}{Linfei Zang}.}
  \bibinfo{year}{2018}\natexlab{}.
\newblock \showarticletitle{Learning to Synthesize}. In
  \bibinfo{booktitle}{\emph{Proceedings of the 4th International Workshop on
  Genetic Improvement Workshop}} (Gothenburg, Sweden)
  \emph{(\bibinfo{series}{GI '18})}. \bibinfo{pages}{37–44}.
\newblock
\showISBNx{9781450357531}
\urldef\tempurl%
\url{https://doi.org/10.1145/3194810.3194816}
\showDOI{\tempurl}


\bibitem[\protect\citeauthoryear{Xu, Sivaraman, Khoo, and Xu}{Xu
  et~al\mbox{.}}{2017}]%
        {745_Xu2017}
\bibfield{author}{\bibinfo{person}{Sihan Xu}, \bibinfo{person}{Aishwarya
  Sivaraman}, \bibinfo{person}{Siau-Cheng Khoo}, {and} \bibinfo{person}{Jing
  Xu}.} \bibinfo{year}{2017}\natexlab{}.
\newblock \showarticletitle{GEMS: An Extract Method Refactoring Recommender}.
  In \bibinfo{booktitle}{\emph{2017 IEEE 28th International Symposium on
  Software Reliability Engineering (ISSRE)}}. \bibinfo{pages}{24--34}.
\newblock
\urldef\tempurl%
\url{https://doi.org/10.1109/ISSRE.2017.35}
\showDOI{\tempurl}


\bibitem[\protect\citeauthoryear{Xu, Zhang, Wang, Cao, Guo, and Xu}{Xu
  et~al\mbox{.}}{2019}]%
        {Xu2019_386}
\bibfield{author}{\bibinfo{person}{Sihan Xu}, \bibinfo{person}{Sen Zhang},
  \bibinfo{person}{Weijing Wang}, \bibinfo{person}{Xinya Cao},
  \bibinfo{person}{Chenkai Guo}, {and} \bibinfo{person}{Jing Xu}.}
  \bibinfo{year}{2019}\natexlab{}.
\newblock \showarticletitle{Method Name Suggestion with Hierarchical Attention
  Networks}. In \bibinfo{booktitle}{\emph{Proceedings of the 2019 ACM SIGPLAN
  Workshop on Partial Evaluation and Program Manipulation}} (Cascais, Portugal)
  \emph{(\bibinfo{series}{PEPM 2019})}. \bibinfo{pages}{10–21}.
\newblock
\showISBNx{9781450362269}
\urldef\tempurl%
\url{https://doi.org/10.1145/3294032.3294079}
\showDOI{\tempurl}


\bibitem[\protect\citeauthoryear{Yahav}{Yahav}{2018}]%
        {Yahav2018_470}
\bibfield{author}{\bibinfo{person}{Eran Yahav}.}
  \bibinfo{year}{2018}\natexlab{}.
\newblock \showarticletitle{From Programs to Interpretable Deep Models and
  Back}. In \bibinfo{booktitle}{\emph{Computer Aided Verification}},
  \bibfield{editor}{\bibinfo{person}{Hana Chockler} {and}
  \bibinfo{person}{Georg Weissenbacher}} (Eds.). \bibinfo{pages}{27--37}.
\newblock
\showISBNx{978-3-319-96145-3}


\bibitem[\protect\citeauthoryear{Yang, Li, Wu, Lu, and Han}{Yang
  et~al\mbox{.}}{2019b}]%
        {yang_li_wu_lu_han_2019}
\bibfield{author}{\bibinfo{person}{Hangfeng Yang}, \bibinfo{person}{Shudong
  Li}, \bibinfo{person}{Xiaobo Wu}, \bibinfo{person}{Hui Lu}, {and}
  \bibinfo{person}{Weihong Han}.} \bibinfo{year}{2019}\natexlab{b}.
\newblock \showarticletitle{A Novel Solutions for Malicious Code Detection and
  Family Clustering Based on Machine Learning}.
\newblock \bibinfo{journal}{\emph{IEEE Access}}  \bibinfo{volume}{7}
  (\bibinfo{year}{2019}), \bibinfo{pages}{148853--148860}.
\newblock
\urldef\tempurl%
\url{https://doi.org/10.1109/ACCESS.2019.2946482}
\showDOI{\tempurl}


\bibitem[\protect\citeauthoryear{Yang, Hotta, Higo, Igaki, and Kusumoto}{Yang
  et~al\mbox{.}}{2014}]%
        {Yang2014_202}
\bibfield{author}{\bibinfo{person}{Jiachen Yang}, \bibinfo{person}{K. Hotta},
  \bibinfo{person}{Yoshiki Higo}, \bibinfo{person}{H. Igaki}, {and}
  \bibinfo{person}{S. Kusumoto}.} \bibinfo{year}{2014}\natexlab{}.
\newblock \showarticletitle{Classification model for code clones based on
  machine learning}.
\newblock \bibinfo{journal}{\emph{Empirical Software Engineering}}
  \bibinfo{volume}{20} (\bibinfo{year}{2014}), \bibinfo{pages}{1095--1125}.
\newblock


\bibitem[\protect\citeauthoryear{Yang, Wu, Ji, Luo, and Wu}{Yang
  et~al\mbox{.}}{2018}]%
        {Yang2018_47}
\bibfield{author}{\bibinfo{person}{Mutian Yang}, \bibinfo{person}{Jingzheng
  Wu}, \bibinfo{person}{Shouling Ji}, \bibinfo{person}{Tianyue Luo}, {and}
  \bibinfo{person}{Yanjun Wu}.} \bibinfo{year}{2018}\natexlab{}.
\newblock \showarticletitle{Pre-Patch: Find Hidden Threats in Open Software
  Based on Machine Learning Method}. In \bibinfo{booktitle}{\emph{Services --
  SERVICES 2018}}, \bibfield{editor}{\bibinfo{person}{Alvin Yang},
  \bibinfo{person}{Siva Kantamneni}, \bibinfo{person}{Ying Li},
  \bibinfo{person}{Awel Dico}, \bibinfo{person}{Xiangang Chen},
  \bibinfo{person}{Rajesh Subramanyan}, {and} \bibinfo{person}{Liang-Jie
  Zhang}} (Eds.). \bibinfo{pages}{48--65}.
\newblock
\showISBNx{978-3-319-94472-2}


\bibitem[\protect\citeauthoryear{Yang, Chen, and Sun}{Yang
  et~al\mbox{.}}{2019a}]%
        {Yang2019_496}
\bibfield{author}{\bibinfo{person}{Yixiao Yang}, \bibinfo{person}{Xiang Chen},
  {and} \bibinfo{person}{Jiaguang Sun}.} \bibinfo{year}{2019}\natexlab{a}.
\newblock \showarticletitle{Improve Language Modeling for Code Completion
  Through Learning General Token Repetition of Source Code with Optimized
  Memory}.
\newblock \bibinfo{journal}{\emph{International Journal of Software Engineering
  and Knowledge Engineering}} \bibinfo{volume}{29}, \bibinfo{number}{11n12}
  (\bibinfo{year}{2019}), \bibinfo{pages}{1801--1818}.
\newblock
\urldef\tempurl%
\url{https://doi.org/10.1142/S0218194019400229}
\showDOI{\tempurl}
\showeprint{https://doi.org/10.1142/S0218194019400229}


\bibitem[\protect\citeauthoryear{Yang, Keung, Yu, Gu, Wei, Ma, and Zhang}{Yang
  et~al\mbox{.}}{2021}]%
        {Yang2021_522}
\bibfield{author}{\bibinfo{person}{Z. Yang}, \bibinfo{person}{J. Keung},
  \bibinfo{person}{X. Yu}, \bibinfo{person}{X. Gu}, \bibinfo{person}{Z. Wei},
  \bibinfo{person}{X. Ma}, {and} \bibinfo{person}{M. Zhang}.}
  \bibinfo{year}{2021}\natexlab{}.
\newblock \showarticletitle{A Multi-Modal Transformer-based Code Summarization
  Approach for Smart Contracts}. In \bibinfo{booktitle}{\emph{2021 2021
  IEEE/ACM 29th International Conference on Program Comprehension (ICPC)
  (ICPC)}}. \bibinfo{pages}{1--12}.
\newblock
\urldef\tempurl%
\url{https://doi.org/10.1109/ICPC52881.2021.00010}
\showDOI{\tempurl}


\bibitem[\protect\citeauthoryear{Yao, Peddamail, and Sun}{Yao
  et~al\mbox{.}}{2019}]%
        {Yao2019_418}
\bibfield{author}{\bibinfo{person}{Ziyu Yao},
  \bibinfo{person}{Jayavardhan~Reddy Peddamail}, {and} \bibinfo{person}{Huan
  Sun}.} \bibinfo{year}{2019}\natexlab{}.
\newblock \showarticletitle{CoaCor: Code Annotation for Code Retrieval with
  Reinforcement Learning}. In \bibinfo{booktitle}{\emph{The World Wide Web
  Conference}} (San Francisco, CA, USA) \emph{(\bibinfo{series}{WWW '19})}.
  \bibinfo{pages}{2203–2214}.
\newblock
\showISBNx{9781450366748}
\urldef\tempurl%
\url{https://doi.org/10.1145/3308558.3313632}
\showDOI{\tempurl}


\bibitem[\protect\citeauthoryear{Yao, Weld, Chen, and Sun}{Yao
  et~al\mbox{.}}{2018}]%
        {StaQC}
\bibfield{author}{\bibinfo{person}{Ziyu Yao}, \bibinfo{person}{Daniel~S. Weld},
  \bibinfo{person}{Wei-Peng Chen}, {and} \bibinfo{person}{Huan Sun}.}
  \bibinfo{year}{2018}\natexlab{}.
\newblock \showarticletitle{StaQC: A Systematically Mined Question-Code Dataset
  from Stack Overflow}. In \bibinfo{booktitle}{\emph{Proceedings of the 2018
  World Wide Web Conference}} (Lyon, France) \emph{(\bibinfo{series}{WWW
  '18})}. \bibinfo{publisher}{International World Wide Web Conferences Steering
  Committee}, \bibinfo{address}{Republic and Canton of Geneva, CHE},
  \bibinfo{pages}{1693–1703}.
\newblock
\showISBNx{9781450356398}
\urldef\tempurl%
\url{https://doi.org/10.1145/3178876.3186081}
\showDOI{\tempurl}


\bibitem[\protect\citeauthoryear{Ye, Xie, Zhang, Hu, Wang, and Zhang}{Ye
  et~al\mbox{.}}{2020}]%
        {Ye2020_428}
\bibfield{author}{\bibinfo{person}{Wei Ye}, \bibinfo{person}{Rui Xie},
  \bibinfo{person}{Jinglei Zhang}, \bibinfo{person}{Tianxiang Hu},
  \bibinfo{person}{Xiaoyin Wang}, {and} \bibinfo{person}{Shikun Zhang}.}
  \bibinfo{year}{2020}\natexlab{}.
\newblock \showarticletitle{Leveraging Code Generation to Improve Code
  Retrieval and Summarization via Dual Learning}. In
  \bibinfo{booktitle}{\emph{Proceedings of The Web Conference 2020}} (Taipei,
  Taiwan) \emph{(\bibinfo{series}{WWW '20})}. \bibinfo{pages}{2309–2319}.
\newblock
\showISBNx{9781450370233}
\urldef\tempurl%
\url{https://doi.org/10.1145/3366423.3380295}
\showDOI{\tempurl}


\bibitem[\protect\citeauthoryear{Yin, Deng, Chen, Vasilescu, and Neubig}{Yin
  et~al\mbox{.}}{2018}]%
        {YDC18}
\bibfield{author}{\bibinfo{person}{Pengcheng Yin}, \bibinfo{person}{Bowen
  Deng}, \bibinfo{person}{Edgar Chen}, \bibinfo{person}{Bogdan Vasilescu},
  {and} \bibinfo{person}{Graham Neubig}.} \bibinfo{year}{2018}\natexlab{}.
\newblock \showarticletitle{Learning to Mine Aligned Code and Natural Language
  Pairs from {Stack Overflow}}. In \bibinfo{booktitle}{\emph{Proceedings of the
  15th International Conference on Mining Software Repositories}} (Gothenburg,
  Sweden) \emph{(\bibinfo{series}{MSR '18})}. \bibinfo{publisher}{Association
  for Computing Machinery}, \bibinfo{address}{New York, NY, USA},
  \bibinfo{pages}{476--486}.
\newblock
\showISBNx{9781450357166}
\urldef\tempurl%
\url{https://doi.org/10.1145/3196398.3196408}
\showDOI{\tempurl}


\bibitem[\protect\citeauthoryear{Yohannese and Li}{Yohannese and Li}{2017}]%
        {705_Yohannese2017}
\bibfield{author}{\bibinfo{person}{Chubato~Wondaferaw Yohannese} {and}
  \bibinfo{person}{Tianrui Li}.} \bibinfo{year}{2017}\natexlab{}.
\newblock \showarticletitle{A combined-learning based framework for improved
  software fault prediction}.
\newblock \bibinfo{journal}{\emph{International Journal of Computational
  Intelligence Systems}} \bibinfo{volume}{10}, \bibinfo{number}{1}
  (\bibinfo{year}{2017}), \bibinfo{pages}{647}.
\newblock


\bibitem[\protect\citeauthoryear{Yosifova, Tasheva, and Trifonov}{Yosifova
  et~al\mbox{.}}{2021}]%
        {Yosifova_2021}
\bibfield{author}{\bibinfo{person}{Veneta Yosifova}, \bibinfo{person}{Antoniya
  Tasheva}, {and} \bibinfo{person}{Roumen Trifonov}.}
  \bibinfo{year}{2021}\natexlab{}.
\newblock \showarticletitle{Predicting Vulnerability Type in Common
  Vulnerabilities and Exposures (CVE) Database with Machine Learning
  Classifiers}. In \bibinfo{booktitle}{\emph{2021 12th National Conference with
  International Participation (ELECTRONICA)}}. \bibinfo{pages}{1--6}.
\newblock
\urldef\tempurl%
\url{https://doi.org/10.1109/ELECTRONICA52725.2021.9513723}
\showDOI{\tempurl}


\bibitem[\protect\citeauthoryear{Younis and Malaiya}{Younis and
  Malaiya}{2014}]%
        {younis_malaiya_2014}
\bibfield{author}{\bibinfo{person}{Awad~A. Younis} {and}
  \bibinfo{person}{Yashwant~K. Malaiya}.} \bibinfo{year}{2014}\natexlab{}.
\newblock \showarticletitle{Using Software Structure to Predict Vulnerability
  Exploitation Potential}. In \bibinfo{booktitle}{\emph{2014 IEEE Eighth
  International Conference on Software Security and Reliability-Companion}}.
  \bibinfo{pages}{13--18}.
\newblock
\urldef\tempurl%
\url{https://doi.org/10.1109/SERE-C.2014.17}
\showDOI{\tempurl}


\bibitem[\protect\citeauthoryear{{Yue}, {Gao}, {Meng}, {Xiong}, {Wang}, and
  {Morgenthaler}}{{Yue} et~al\mbox{.}}{2018}]%
        {Yue2018_164}
\bibfield{author}{\bibinfo{person}{R. {Yue}}, \bibinfo{person}{Z. {Gao}},
  \bibinfo{person}{N. {Meng}}, \bibinfo{person}{Y. {Xiong}},
  \bibinfo{person}{X. {Wang}}, {and} \bibinfo{person}{J.~D. {Morgenthaler}}.}
  \bibinfo{year}{2018}\natexlab{}.
\newblock \showarticletitle{Automatic Clone Recommendation for Refactoring
  Based on the Present and the Past}. In \bibinfo{booktitle}{\emph{2018 IEEE
  International Conference on Software Maintenance and Evolution (ICSME)}}.
  \bibinfo{pages}{115--126}.
\newblock
\urldef\tempurl%
\url{https://doi.org/10.1109/ICSME.2018.00021}
\showDOI{\tempurl}


\bibitem[\protect\citeauthoryear{Zanoni, Fontana, and Stella}{Zanoni
  et~al\mbox{.}}{2015}]%
        {1049_Zanoni2015}
\bibfield{author}{\bibinfo{person}{Marco Zanoni},
  \bibinfo{person}{Francesca~Arcelli Fontana}, {and} \bibinfo{person}{Fabio
  Stella}.} \bibinfo{year}{2015}\natexlab{}.
\newblock \showarticletitle{On applying machine learning techniques for design
  pattern detection}.
\newblock \bibinfo{journal}{\emph{Journal of Systems and Software}}
  \bibinfo{volume}{103} (\bibinfo{year}{2015}), \bibinfo{pages}{102--117}.
\newblock


\bibitem[\protect\citeauthoryear{Zhang and Tsai}{Zhang and Tsai}{2003}]%
        {Zhang2003_398}
\bibfield{author}{\bibinfo{person}{Du Zhang} {and} \bibinfo{person}{Jeffrey
  J.~P. Tsai}.} \bibinfo{year}{2003}\natexlab{}.
\newblock \showarticletitle{Machine Learning and Software Engineering}.
\newblock \bibinfo{journal}{\emph{Software Quality Journal}}
  \bibinfo{volume}{11}, \bibinfo{number}{2} (\bibinfo{date}{June}
  \bibinfo{year}{2003}), \bibinfo{pages}{87–119}.
\newblock
\showISSN{0963-9314}
\urldef\tempurl%
\url{https://doi.org/10.1023/A:1023760326768}
\showDOI{\tempurl}


\bibitem[\protect\citeauthoryear{Zhang and Khoo}{Zhang and Khoo}{2021}]%
        {753_Zhang2021}
\bibfield{author}{\bibinfo{person}{Fanlong Zhang} {and}
  \bibinfo{person}{Siau-cheng Khoo}.} \bibinfo{year}{2021}\natexlab{}.
\newblock \showarticletitle{An empirical study on clone consistency prediction
  based on machine learning}.
\newblock \bibinfo{journal}{\emph{Information and Software Technology}}
  \bibinfo{volume}{136} (\bibinfo{year}{2021}), \bibinfo{pages}{106573}.
\newblock


\bibitem[\protect\citeauthoryear{Zhang, Wang, Zhang, Sun, and Liu}{Zhang
  et~al\mbox{.}}{2020a}]%
        {Zhang2020_435}
\bibfield{author}{\bibinfo{person}{Jian Zhang}, \bibinfo{person}{Xu Wang},
  \bibinfo{person}{Hongyu Zhang}, \bibinfo{person}{Hailong Sun}, {and}
  \bibinfo{person}{Xudong Liu}.} \bibinfo{year}{2020}\natexlab{a}.
\newblock \showarticletitle{Retrieval-Based Neural Source Code Summarization}.
  In \bibinfo{booktitle}{\emph{Proceedings of the ACM/IEEE 42nd International
  Conference on Software Engineering}} (Seoul, South Korea)
  \emph{(\bibinfo{series}{ICSE '20})}. \bibinfo{pages}{1385–1397}.
\newblock
\showISBNx{9781450371216}
\urldef\tempurl%
\url{https://doi.org/10.1145/3377811.3380383}
\showDOI{\tempurl}


\bibitem[\protect\citeauthoryear{{Zhang}, {Wang}, {Zhang}, {Sun}, {Wang}, and
  {Liu}}{{Zhang} et~al\mbox{.}}{2019}]%
        {Zhang2019_451}
\bibfield{author}{\bibinfo{person}{J. {Zhang}}, \bibinfo{person}{X. {Wang}},
  \bibinfo{person}{H. {Zhang}}, \bibinfo{person}{H. {Sun}}, \bibinfo{person}{K.
  {Wang}}, {and} \bibinfo{person}{X. {Liu}}.} \bibinfo{year}{2019}\natexlab{}.
\newblock \showarticletitle{A Novel Neural Source Code Representation Based on
  Abstract Syntax Tree}. In \bibinfo{booktitle}{\emph{2019 IEEE/ACM 41st
  International Conference on Software Engineering (ICSE)}}.
  \bibinfo{pages}{783--794}.
\newblock
\urldef\tempurl%
\url{https://doi.org/10.1109/ICSE.2019.00086}
\showDOI{\tempurl}


\bibitem[\protect\citeauthoryear{Zhang, Xie, Ye, Zhang, and Zhang}{Zhang
  et~al\mbox{.}}{2020b}]%
        {Zhang2020_518}
\bibfield{author}{\bibinfo{person}{Jinglei Zhang}, \bibinfo{person}{Rui Xie},
  \bibinfo{person}{Wei Ye}, \bibinfo{person}{Yuhan Zhang}, {and}
  \bibinfo{person}{Shikun Zhang}.} \bibinfo{year}{2020}\natexlab{b}.
\newblock \showarticletitle{Exploiting Code Knowledge Graph for Bug
  Localization via Bi-Directional Attention}. In
  \bibinfo{booktitle}{\emph{Proceedings of the 28th International Conference on
  Program Comprehension}} \emph{(\bibinfo{series}{ICPC '20})}.
  \bibinfo{publisher}{Association for Computing Machinery},
  \bibinfo{pages}{219–--229}.
\newblock
\showISBNx{9781450379588}
\urldef\tempurl%
\url{https://doi.org/10.1145/3387904.3389281}
\showDOI{\tempurl}


\bibitem[\protect\citeauthoryear{Zhang and Harman}{Zhang and Harman}{2021}]%
        {Zhang2021}
\bibfield{author}{\bibinfo{person}{Jie~M. Zhang} {and} \bibinfo{person}{Mark
  Harman}.} \bibinfo{year}{2021}\natexlab{}.
\newblock \showarticletitle{"Ignorance and Prejudice" in Software Fairness}. In
  \bibinfo{booktitle}{\emph{2021 IEEE/ACM 43rd International Conference on
  Software Engineering (ICSE)}}. \bibinfo{pages}{1436--1447}.
\newblock
\urldef\tempurl%
\url{https://doi.org/10.1109/ICSE43902.2021.00129}
\showDOI{\tempurl}


\bibitem[\protect\citeauthoryear{{Zhang}, {Harman}, {Ma}, and {Liu}}{{Zhang}
  et~al\mbox{.}}{2020}]%
        {Zhang2020_157}
\bibfield{author}{\bibinfo{person}{J.~M. {Zhang}}, \bibinfo{person}{M.
  {Harman}}, \bibinfo{person}{L. {Ma}}, {and} \bibinfo{person}{Y. {Liu}}.}
  \bibinfo{year}{2020}\natexlab{}.
\newblock \showarticletitle{Machine Learning Testing: Survey, Landscapes and
  Horizons}.
\newblock \bibinfo{journal}{\emph{IEEE Transactions on Software Engineering}}
  (\bibinfo{year}{2020}), \bibinfo{pages}{1--1}.
\newblock
\urldef\tempurl%
\url{https://doi.org/10.1109/TSE.2019.2962027}
\showDOI{\tempurl}


\bibitem[\protect\citeauthoryear{{Zhang} and {Wu}}{{Zhang} and {Wu}}{2020}]%
        {Zhang2020_135}
\bibfield{author}{\bibinfo{person}{Q. {Zhang}} {and} \bibinfo{person}{B.
  {Wu}}.} \bibinfo{year}{2020}\natexlab{}.
\newblock \showarticletitle{Software Defect Prediction via Transformer}. In
  \bibinfo{booktitle}{\emph{2020 IEEE 4th Information Technology, Networking,
  Electronic and Automation Control Conference (ITNEC)}},
  Vol.~\bibinfo{volume}{1}. \bibinfo{pages}{874--879}.
\newblock
\urldef\tempurl%
\url{https://doi.org/10.1109/ITNEC48623.2020.9084745}
\showDOI{\tempurl}


\bibitem[\protect\citeauthoryear{Zhang and Dong}{Zhang and Dong}{2021}]%
        {835_Zhang2021}
\bibfield{author}{\bibinfo{person}{Yang Zhang} {and} \bibinfo{person}{Chunhao
  Dong}.} \bibinfo{year}{2021}\natexlab{}.
\newblock \showarticletitle{MARS: Detecting brain class/method code smell based
  on metric--attention mechanism and residual network}.
\newblock \bibinfo{journal}{\emph{Journal of Software: Evolution and Process}}
  (\bibinfo{year}{2021}), \bibinfo{pages}{e2403}.
\newblock


\bibitem[\protect\citeauthoryear{Zhang and Li}{Zhang and Li}{2020}]%
        {zhang_li_2020}
\bibfield{author}{\bibinfo{person}{Yu Zhang} {and} \bibinfo{person}{Binglong
  Li}.} \bibinfo{year}{2020}\natexlab{}.
\newblock \showarticletitle{Malicious Code Detection Based on Code Semantic
  Features}.
\newblock \bibinfo{journal}{\emph{IEEE Access}}  \bibinfo{volume}{8}
  (\bibinfo{year}{2020}), \bibinfo{pages}{176728--176737}.
\newblock
\urldef\tempurl%
\url{https://doi.org/10.1109/ACCESS.2020.3026052}
\showDOI{\tempurl}


\bibitem[\protect\citeauthoryear{Zhao and Huang}{Zhao and Huang}{2018}]%
        {Zhao2018_216}
\bibfield{author}{\bibinfo{person}{Gang Zhao} {and} \bibinfo{person}{Jeff
  Huang}.} \bibinfo{year}{2018}\natexlab{}.
\newblock \showarticletitle{DeepSim: Deep Learning Code Functional Similarity}.
  In \bibinfo{booktitle}{\emph{Proceedings of the 2018 26th ACM Joint Meeting
  on European Software Engineering Conference and Symposium on the Foundations
  of Software Engineering}} (Lake Buena Vista, FL, USA)
  \emph{(\bibinfo{series}{ESEC/FSE 2018})}. \bibinfo{pages}{141–151}.
\newblock
\showISBNx{9781450355735}
\urldef\tempurl%
\url{https://doi.org/10.1145/3236024.3236068}
\showDOI{\tempurl}


\bibitem[\protect\citeauthoryear{Zheng, Gao, Wu, Liu, Xun, Liu, and Chen}{Zheng
  et~al\mbox{.}}{2020}]%
        {Zheng2020_56}
\bibfield{author}{\bibinfo{person}{Wei Zheng}, \bibinfo{person}{Jialiang Gao},
  \bibinfo{person}{Xiaoxue Wu}, \bibinfo{person}{Fengyu Liu},
  \bibinfo{person}{Yuxing Xun}, \bibinfo{person}{Guoliang Liu}, {and}
  \bibinfo{person}{Xiang Chen}.} \bibinfo{year}{2020}\natexlab{}.
\newblock \showarticletitle{The impact factors on the performance of machine
  learning-based vulnerability detection: A comparative study}.
\newblock \bibinfo{journal}{\emph{Journal of Systems and Software}}
  \bibinfo{volume}{168} (\bibinfo{year}{2020}), \bibinfo{pages}{110659}.
\newblock
\showISSN{0164-1212}
\urldef\tempurl%
\url{https://doi.org/10.1016/j.jss.2020.110659}
\showDOI{\tempurl}


\bibitem[\protect\citeauthoryear{Zheng, Zhou, Li, and Wu}{Zheng
  et~al\mbox{.}}{2019}]%
        {1032_Zheng2019}
\bibfield{author}{\bibinfo{person}{Wenhao Zheng}, \bibinfo{person}{Hongyu
  Zhou}, \bibinfo{person}{Ming Li}, {and} \bibinfo{person}{Jianxin Wu}.}
  \bibinfo{year}{2019}\natexlab{}.
\newblock \showarticletitle{CodeAttention: translating source code to comments
  by exploiting the code constructs}.
\newblock \bibinfo{journal}{\emph{Frontiers of Computer Science}}
  \bibinfo{volume}{13}, \bibinfo{number}{3} (\bibinfo{year}{2019}),
  \bibinfo{pages}{565--578}.
\newblock


\bibitem[\protect\citeauthoryear{Zhong, Yang, and Sun}{Zhong
  et~al\mbox{.}}{2019}]%
        {Zhong2019_499}
\bibfield{author}{\bibinfo{person}{Chaoliang Zhong}, \bibinfo{person}{Ming
  Yang}, {and} \bibinfo{person}{Jun Sun}.} \bibinfo{year}{2019}\natexlab{}.
\newblock \showarticletitle{JavaScript Code Suggestion Based on Deep Learning}.
  In \bibinfo{booktitle}{\emph{Proceedings of the 2019 3rd International
  Conference on Innovation in Artificial Intelligence}} (Suzhou, China)
  \emph{(\bibinfo{series}{ICIAI 2019})}. \bibinfo{pages}{145–149}.
\newblock
\showISBNx{9781450361286}
\urldef\tempurl%
\url{https://doi.org/10.1145/3319921.3319922}
\showDOI{\tempurl}


\bibitem[\protect\citeauthoryear{Zhou and Jiang}{Zhou and Jiang}{2012}]%
        {genome_2012}
\bibfield{author}{\bibinfo{person}{Yajin Zhou} {and} \bibinfo{person}{Xuxian
  Jiang}.} \bibinfo{year}{2012}\natexlab{}.
\newblock \showarticletitle{Dissecting Android Malware: Characterization and
  Evolution}. In \bibinfo{booktitle}{\emph{Proceedings of the 2012 IEEE
  Symposium on Security and Privacy}} \emph{(\bibinfo{series}{SP '12})}.
  \bibinfo{pages}{95–109}.
\newblock
\showISBNx{9780769546810}
\urldef\tempurl%
\url{https://doi.org/10.1109/SP.2012.16}
\showDOI{\tempurl}


\bibitem[\protect\citeauthoryear{Zhou, Shen, Zhang, Yang, Han, and Chen}{Zhou
  et~al\mbox{.}}{2022}]%
        {982_Zhou2022}
\bibfield{author}{\bibinfo{person}{Yu Zhou}, \bibinfo{person}{Juanjuan Shen},
  \bibinfo{person}{Xiaoqing Zhang}, \bibinfo{person}{Wenhua Yang},
  \bibinfo{person}{Tingting Han}, {and} \bibinfo{person}{Taolue Chen}.}
  \bibinfo{year}{2022}\natexlab{}.
\newblock \showarticletitle{Automatic source code summarization with graph
  attention networks}.
\newblock \bibinfo{journal}{\emph{Journal of Systems and Software}}
  \bibinfo{volume}{188} (\bibinfo{year}{2022}), \bibinfo{pages}{111257}.
\newblock


\bibitem[\protect\citeauthoryear{Zhou, Yu, and Fan}{Zhou et~al\mbox{.}}{2021}]%
        {993_Zhou2021}
\bibfield{author}{\bibinfo{person}{Ziyi Zhou}, \bibinfo{person}{Huiqun Yu},
  {and} \bibinfo{person}{Guisheng Fan}.} \bibinfo{year}{2021}\natexlab{}.
\newblock \showarticletitle{Adversarial training and ensemble learning for
  automatic code summarization}.
\newblock \bibinfo{journal}{\emph{Neural Computing and Applications}}
  \bibinfo{volume}{33}, \bibinfo{number}{19} (\bibinfo{year}{2021}),
  \bibinfo{pages}{12571--12589}.
\newblock


\bibitem[\protect\citeauthoryear{Zhu, Sun, Xiao, Zhang, Yuan, Xiong, and
  Zhang}{Zhu et~al\mbox{.}}{2021}]%
        {934_Zhu2021}
\bibfield{author}{\bibinfo{person}{Qihao Zhu}, \bibinfo{person}{Zeyu Sun},
  \bibinfo{person}{Yuan-an Xiao}, \bibinfo{person}{Wenjie Zhang},
  \bibinfo{person}{Kang Yuan}, \bibinfo{person}{Yingfei Xiong}, {and}
  \bibinfo{person}{Lu Zhang}.} \bibinfo{year}{2021}\natexlab{}.
\newblock \showarticletitle{A syntax-guided edit decoder for neural program
  repair}. In \bibinfo{booktitle}{\emph{Proceedings of the 29th ACM Joint
  Meeting on European Software Engineering Conference and Symposium on the
  Foundations of Software Engineering}}. \bibinfo{pages}{341--353}.
\newblock


\bibitem[\protect\citeauthoryear{Zimmermann, Premraj, and Zeller}{Zimmermann
  et~al\mbox{.}}{2007}]%
        {Zimmermann2007}
\bibfield{author}{\bibinfo{person}{Thomas Zimmermann}, \bibinfo{person}{Rahul
  Premraj}, {and} \bibinfo{person}{Andreas Zeller}.}
  \bibinfo{year}{2007}\natexlab{}.
\newblock \showarticletitle{Predicting Defects for Eclipse}. In
  \bibinfo{booktitle}{\emph{Third International Workshop on Predictor Models in
  Software Engineering (PROMISE'07: ICSE Workshops 2007)}}.
  \bibinfo{pages}{9--9}.
\newblock
\urldef\tempurl%
\url{https://doi.org/10.1109/PROMISE.2007.10}
\showDOI{\tempurl}


\end{thebibliography}

\newpage
\appendix

\section{Abbreviations of ML techniques}

\topcaption{Machine learning techniques}
\rowcolors{2}{gray!25}{white}
  \begin{supertabular}{p{.15\textwidth}p{.8\textwidth}}
  	 
  	 \label{tab:appendix}
   \textbf{Acronym} & \textbf{Full form}  \\
A3C &  Asynchronous Advantage Actor-Critic 
\\
AB &AdaBoost
\\
AE & Autoencoder
\\
AIS & Artificial Immune Systems\\
ANFIS & Adaptive Neuro-Fuzzy Inference System\\
ANN & Artificial Neural Network\\
ARM & Association Rule Mining\\
B & Bagging\\
BDT & Boosted Decision Tree\\
BERT & Bidirectional Encoder Representations from Transformers\\
Bi-GRU & Bidirectional Gated Recurrent Unit\\
Bi-LSTM & Bi-Long Sort-Term Memory\\
Bi-RNN & Bidirectional Recurrent Neural Network\\
BiNN & Bilateral Neural Network\\
BMN & Best Matching Neighbours\\
BN & Bayes Net\\
BNB & Bernoulli Naive Bayes\\
BOW &	Bag of Words\\
BP-ANN & Back-propagation Artificial Neural Network\\
BR & Binary Relevance\\
CART & Classification and Regression Trees\\
CC & Classifier Chain\\
CCN & Cascade Correlation Network\\
CNN & Convolution Neural Network\\
COBWEB & COBWEB\\
Code2Vec & Code2Vec\\
CoForest-RF & Co-Forest Random Forest\\
CSC & Cost-Sensitive Classifier\\
DBN & Deep Belief Network\\
DDQN & Double Deep Q-Networks\\
DNN & Deep Neural Network\\
Doc2Vec & Doc2Vec\\
DR & Diverse Rank \\
DS & Decision Stump\\
DT & Decision Tree\\
EL &	Ensemble Learning\\
ELM & Extreme Learning Machine\\
EM & Expectation Minimization\\
EN-DE & Encoder-Decoder\\
FIS & Fuzzy Inference System\\
FL & Fuzzy Logic\\
FR-CNN & Faster R-Convolutional Neural Network\\
GAN & Generative Adversarial Network\\
GB & Gradient Boosting\\
GBDT & Gradient-Boosted Decision Tree\\
GBM & Gradient Boosting Machine\\
GBT & Gradient boosted trees\\
GCN & Graph convolutional networks\\
GD & Gradient Descent\\
GED & Gaussian Encoder-Decoder\\
GEP & Gene Expression Programming\\
GGNN & Gated Graph Neural Network \\
GINN & Graph Interval Neural Network\\
Glove & Global Vectors for Word Representation\\
GNB & Gaussian Naïve Bayes\\
GNN	&Graph Neural Network\\
GPT-C & Generative Pre-trained Transformer for Code \\
GRASSHOPER & Graph Random-walk with Absorbing StateS that HOPs among PEaks for Ranking \\
GRU & Gated Recurrent Unit\\
HAN & Hierarchical Attention Network\\
HC & Hierarchical Clustering\\
HMM & Hidden Markov Model\\
KM & KMeans\\
KNN & K Nearest Neighbours\\
KS & Kstar\\
LB & LogitBoost\\
LC & Label Combination\\
LCM & Log-bilinear Context Model\\
LDA & Linear Discriminant Analysis\\
LLR & Logistic Linear Regression\\
LMSR & Least Median Square Regression\\
LOG & Logistic regression\\
LR & Linear Regression\\
LSTM & Long Short Term Memory\\
MLP & Multi Level Perceptron\\
MMR & Maximal Marginal Relevance \\
MNB & Multinomial Naive Bayes\\
MNN & Memory Neural Network\\
MTN & Modular Tree-structured Recurrent Neural Network \\
MVE & Majority Voting Ensemble\\
NB & Naïve Bayes\\
NLM & Neural Language Model\\
NMT & Neural Machine Translation\\
NNC & Neural Network for Continuous goal\\
NND & Neural Network for Discrete goal\\
Node2Vec & Node2Vec\\
OCC & One Class Classifier\\
OR & OneRule\\
PN & Pointer Network\\
PNN & Probabilistic Neural Network\\
POLY & Polynomial regression\\
PR & Pace Regression\\
PSO & Particle Swarm Optimization\\
ReNN & Reverse NN\\
ResNet & Residual Neural Network\\
RF & Random Forrest\\
RGNN & Regression Neural Network\\
Ripper & Ripper\\
RL & Reinforcement Learning\\
RNN & Recuurent Neural Network\\
RT & RandomTree\\
SA & Simulated Annealing\\
Seq2Seq & Sequence-to-Sequence\\
SMO & Sequential Minimal Optimization \\
SMT & Statistical Machine Translation\\
SOM & Self Organizing Map\\
SVE & Soft Voting Ensemble\\
SVLR & Support Vector Logistic Regression\\
SVM & Support Vector Machine\\
SVR & Support Vector Regression\\
TF & Transformer\\
TNB & Transfer Naïve Bayes\\
V  & Voting\\
VSL & Version Space Learning\\
Word2Vec & Word2Vec\\
XG &	XGBoost\\
\hline
 \end{supertabular}
%\end{table}
\end{document}